\documentclass[twoside,12pt]{article}
\usepackage{epsfig}

\usepackage{amsmath,amssymb,graphicx}
\usepackage{floatflt}
\usepackage [small] {caption}
\usepackage{amsfonts}
\usepackage{color}
\usepackage{epstopdf}

\usepackage[T2A]{fontenc}
\usepackage[cp1251]{inputenc}
\usepackage[english]{babel}

\numberwithin{equation}{section}
\newcommand{\ra}{\rightarrow}
\newcommand{\bq}{\begin{eqnarray}}
\newcommand{\eq}{\end{eqnarray}}
\newcommand{\ov}{\overline}

\newcommand{\be}{\begin{equation}}
\newcommand{\ee}{\end{equation}}
\newcommand{\nn}{\nonumber}
\newcommand{\bo}{\rm b_o}
\renewcommand{\mathbf}{\vec}

\begin{document}
\vspace*{4cm}
\begin{center}{\bf{\Large Hard exclusive two photon processes  in QCD}}\end{center}
\vspace*{1cm}
\begin{center}{\bf{Victor L. Chernyak $^{1,2}$,\,\, Simon I. Eidelman $^{1,2}$}}\end{center}

\begin{center}(v.l.chernyak@inp.nsk.su;\, s.i.eidelman@inp.nsk.su) \end{center}
\vspace*{1mm}
\begin{center} $^1$ Budker Institute of Nuclear Physics SB RAS \end{center}
\begin{center} $^2$ Novosibirsk State University \end{center}
\begin{center} 630090 Novosibirsk, Russian Federation \end{center}
\vspace*{3mm}

\begin{center} (submitted to "Progress in Particle and Nuclear Physics" on 31 May 2014)\end{center}
\vspace*{1cm}
 
\begin{center}{\bf Abstract}\end{center}

This is a short review of some hard two-photon processes: a)\, $\gamma\gamma\to {\ov P}_1 P_2,\,\, {\ov P}_1 P_2= \{\pi^+\pi^-,\,K^+ K^-,\\ K_S K_S,\, \pi^o\pi^o,\, \pi^o\eta\}$,\,\, b)\, $\gamma\gamma\to V_1 V_2, \,\,V_1 V_2=\{\rho^o\rho^o,\, \phi\phi,\, \omega\phi,\, \omega\omega \}$,\,\, c)\, $\gamma\gamma\to {\rm baryon-antibaryon}$,\\ d)\, $\gamma^*\gamma\to P^o,\,\, P^o=\{\pi^o,\, \eta,\, \eta^\prime,\, \eta_c \}$.

The available experimental data are presented. A number of theoretical approaches to calculation of these processes is described, both those based mainly on QCD and more phenomenological (the handbag model, the diquark model, etc). Some theoretical questions tightly connected with this subject are discussed, in particular: the applications of various types of QCD sum rules, the endpoint behavior of the leading twist meson wave functions, etc.

\newpage
\tableofcontents

\newpage
\section{Introduction}

\hspace*{3mm} The general approach to calculations of hard exclusive processes in QCD was developed in \cite{c-1,c-2} and \cite{ER-80-1,ER-80-2} (the operator expansions and summation of Feynman diagrams in the covariant perturbation theory), and in \cite{LB-80} (summation of Feynman diagrams in the non-covariant
light-front perturbation theory in the special axial gauge and in the basis of free on mass-shell quarks and gluons). The overall review is \cite{cz-rev}.

\hspace*{3mm} As was first obtained in \cite{c-2} on the example of the pion form factor $F_{\pi}(Q^2)$, the contributions from short and large distances factorize in $F_{\pi}(Q^2)$ at large $Q^2$ due to color neutrality of pions, and the logarithmic evolution of the leading power term in $Q^2 F_{\pi}(Q^2)$ is determined by renormalization factors of leading twist local operators of the type $O^{(k)}_{\rm m}=
\partial^k(\,{\ov d}\, (\overleftrightarrow {D})^m\,\gamma_{\mu}\gamma_5\, u)\,$.
\footnote{\,
As was emphasized in the first paper in \cite{c-2}, separate diagrams in the covariant gauge give double logarithmic contributions $\sim (\alpha_s\ln^2 Q)^n$ at the n-th order, but all such terms down to $\sim (\alpha_s\ln Q)^n \ln Q$ cancel finally in the sum of the n-th order diagrams due to the pion neutrality in color, so that the leading n-th order contribution is $\sim (\alpha_s\ln Q)^n $.
}
The anomalous dimensions $\gamma_n$ of multiplicatively renormalized leading twist operators of this type are the same as in the deep inelastic scattering (see the last paper in \cite{c-2} and section 3 in \cite{cz-rev} for all details):
\be
Q^2F_{\pi}(Q^2)\ra \sum_{n_1}\sum_{n_2}\langle\pi^{+}(p_2)|O^\dagger_{{\rm n}_2}(x=0)_{Q}|0\rangle C_{{\rm n}_1 {\rm n}_2}\Bigl (\alpha_s(Q^2) \Bigr )\langle 0|O_{{\rm n}_1}(x=0)_{Q}|\pi^{+}(p_1)\rangle\,,
\ee
\be
O_{\rm n}={\ov d}\,\gamma_{\mu}\gamma_5\, C_{\rm n}^{3/2}(\xi)\, u\,,\quad \xi=\overleftrightarrow {D}/\partial\,,\quad C_{{\rm n}_1 {\rm n}_2}\sim \alpha_s(Q^2)\Bigl [1+O\Bigl (\alpha_s(Q^2)\Bigr ) \Bigr ]\,,
\nn
\ee
\be
O_{\rm n}(x=0)_{Q}=Z_{\rm n}(Q^2,\mu^2_o)\,O_{\rm n}(x=0)_{\mu_o}\,,\quad Z_{\rm n}(Q^2,\mu^2_o)=\Biggl (\frac{\alpha_s(Q^2)}{\alpha_s(\mu_o^2)}\Biggr )^{\gamma_{n}/\bo}\,,\quad \bo=11-\frac{2}{3}n_f\,,\nn
\ee
\be
\gamma_n=C_F\Bigl [\, 1-\frac{2}{(n+1)(n+2)}+4\sum_{j=2}^{n+1}\frac{1}{j}\, \Bigr ],\quad C_F=\frac{N^2_c-1}{2N_c}=\frac{4}{3}\,,\nn
\ee
where $C_{\rm n}^{3/2}(\xi)$ are the Gegenbauer polynomials. Because $\gamma_{n+1}>\gamma_n\geq 0$, the leading contribution to (1.1) at $Q^2\ra\infty$ for the pion target originates from all local operators  $\partial^k(\,{\ov d}\,\gamma_{\mu}\gamma_5\, u)$ with $\rm m=0$ in the basis $O^{(k)}_{\rm m}=\partial^k(\,{\ov d}\, (\overleftrightarrow {D})^m\,\gamma_{\mu}\gamma_5\, u)$ or from $O_{{\rm n}=0}$ only in the basis $O_{\rm n}={\ov d}\,\gamma_{\mu}\gamma_5\, C_{\rm n}^{3/2}(\xi)\, u$.

Therefore, {\it the strict} QCD prediction for $Q^2 F_{\pi}(Q^2)$ in the formal limit $Q^2\to \infty$ looks as (see the first paper in \cite{c-2}) :
\be
F_{\pi}(Q^2)\to \frac{32\pi^2|f_{\pi}|^2}{{\rm b}_o Q^2\ln Q^2}I_{\rm corr},\,\,\,
f_{\pi}\simeq 130.4\,MeV,\,\,\, I_{\rm corr}=\Biggl (1+O\Bigl (\frac{\alpha_s(Q)}{\alpha_s(\mu_o)}\Bigr )^{\frac{50}{9\,{\rm b}_o}}+\cdots\Biggr ).
\ee

Similarly to the deep-inelastic scattering where the operator expansion of two electromagnetic currents can be written in a compact form introducing the parton distribution functions of the target, $q(x,\mu)$, the operator expansion (1.1) can also be rewritten in a compact form introducing the leading twist pion wave function (=distribution amplitude) $\phi_{\pi}(x,\mu)$, $\mu$ is the normalization scale. It is defined as the matrix element of the bilocal operator
\be
\langle 0|\,\Bigl [{\ov d}(z){\cal P}(z,-z)\gamma_{\nu}\gamma_5\,u(-z)\Bigr ]_{\mu}\,|\pi^{+}(p)\rangle=i f_{\pi}p_{\nu}\int_{0}^{1}dx\, e^{i(2x-1)pz}\phi_{\pi}(x,\mu)\,,
\ee
where ${\cal P}(z,-z)$ is the straight line gauge link, while $0\leq x\leq 1$ is the momentum fraction carried by quark in the pion.  In terms $\phi_{\pi}(x,\mu\sim Q)$ the operator expansion (1.1) looks as
\be
Q^2F_{\pi}(Q^2)\to \frac{8\pi\alpha_s(Q^2)|f_{\pi}|^2}{9}\int_0^1 dx\int_0^1 dy\, \phi_{\pi}(x,\mu\sim Q)
\Biggl [T_{\rm hard}=\frac{1}{x y}\Bigl (1+O(\alpha_s(Q^2))\Bigr )\Biggr ]\phi_{\pi}(y,\mu\sim Q),
\ee
while the double sum in (1.1) originates from the expansions of the initial and final pion wave functions
\be
\phi_\pi(x,\mu\sim Q)=6x(1-x)\sum_n a_n(\mu_o)Z_n(Q^2,\mu^2_o) C_n^{3/2}(2x-1)
\ee
(this is the expansion of the bilocal operator in (1.3) over the multiplicatively renormalized operators $O_{\rm n}$ in (1.1)). Because $Z_{n>0}(Q^2\ra \infty,\mu^2_o)\ra 0$, only the term with $n=0$ survives
in (1.5) in the formal limit $Q^2\ra\infty$, so that the leading twist pion wave function $\phi_\pi(x,\mu\to \infty)$ evolves to its universal asymptotic form
\be
\phi_\pi(x,\mu\to \infty)\ra \phi^{\rm asy}(x)\Biggl [1+O\Bigl (\alpha^{\frac{50}{9{\rm b}_o}}_s(Q^2)\Bigl ) \Biggr ]\,,\quad \phi^{\rm asy}(x)=6x(1-x)
\ee
{\it independently of its form $\phi_\pi(x,\mu_o\sim 1\,GeV)$ at low scale}. As it is seen from (1.2),(1.5),(1.6), the logarithmic evolution of $Q^2F_{\pi}(Q^2)$ at large $Q^2$ is very slow.

Both approaches, the covariant operator expansions \cite{c-1,c-2},\,\cite{ER-80-1,ER-80-2} and the non-covariant light-front formalism with the on mass shell quark-gluon basic states \cite{LB-80},
were applied then for calculation of asymptotic behavior of many other exclusive processes, see the review \cite{cz-rev}.\\

The simplest exclusive process involving only one hadron is "$\gamma^{*}\gamma^{*}\to P$", where $P$ is the pseudoscalar meson, e.g. $\pi^0,\,\eta,\,\eta^\prime,\, \eta_c$\,. Its amplitude involves only one form factor $F_{\gamma P}(q_1^2,q_2^2)$. The asymptotic behavior of $F_{\gamma P}(Q_1^2=-q_1^2>0,q_2^2=0)$ at $Q_1^2\gg 1\,GeV^2$ has been measured in a number of experiments using the process "$e^+ e^-\to e^+ e^- +P$", while the behavior of $F_{\gamma P}(q_1^2>0,q_2^2=0)$ at $q_1^2=s\gg 1\,GeV^2$ was measured in the process "$e^+ e^-\to \gamma+P$". From the theory side, this form factor was studied in a large number of papers using various approaches and different models for the pseudoscalar meson wave functions, $\phi_{P}(x,\mu)$. The behavior of $F_{\gamma P}$ at large $|q^2|$ is considered at present as one of the most reliable and clear ways to obtain information on properties of the pseudoscalar meson leading twist wave function $\phi_{P}(x,\mu)$.  We present the experimental data on $F_{\gamma P}$ and discuss in some detail various theoretical approaches and results (and some other points tightly connected with them) in Section 6.

The available experimental data on the high energy and large scattering angle cross sections "$\gamma\gamma\to {\ov P} P$",\,\,$P=\{\pi^\pm,\,K^\pm,\,\pi^o,\, K^o,\, \eta\}$ are presented in Section 2 together with some leading term QCD calculations. Most of the data for these processes originate from the Belle Collaboration. The energy dependences and angular distributions of these processes are described and compared, when possible, with some leading term QCD predictions. The values of these cross sections are sufficiently sensitive, in principle, to the forms of the leading twist meson wave functions. The main problem here at present is that power corrections to the leading term QCD predictions are not under real control, while the cross sections are measured at energies which do not look high enough to ensure that power corrections are really sufficiently small.

We compare in Section 3 the leading term QCD predictions for these cross sections with those of the phenomenological handbag model. The underlying idea of this last approach is just that the present day energies are not sufficiently high for the QCD leading terms to be really dominant. It is assumed that the formally power suppressed nonleading terms are really sufficiently large numerically to dominate amplitudes of all these processes at present energies. As for a concrete dynamical mechanism responsible for the presence of these numerically large nonleading power terms, it is assumed that this is the well known Feynman mechanism of the endpoint region contributions (see Section 3 for details). The main deficiency of this approach at present is that it did not allow up to now to make definite predictions, so that the theoretical description uses a number of free parameters which are simply fitted to data.

There appeared recently the Belle Collaboration data on the cross sections $\gamma\gamma\to V^o_1 V^o_2$\,, where $V^o=\omega,\,\phi$. The comparison of leading term QCD predictions with these data is presented in Section 4.

And finally, we present in Section 5 the comparison of the leading term QCD predictions and fits from some other phenomenological models with the data on the large angle cross sections "$\gamma\gamma\to {\ov B} B$"\,, where $B$ is the baryon: $B=p,\,\Lambda,\,\Sigma$\,.

\section{\hspace*{2.5cm} Data on \boldmath{$\gamma\gamma\to {\ov P}P$} cross  sections \\ and comparison with leading term QCD predictions}

\subsection{Some experimental details for charged mesons}

\hspace*{3mm} In this section we describe, as typical examples, some details of the experimental measurements of two charged pseudoscalar particles production in two-photon collisions performed at Belle~\cite{Nakaz}. In all studies of the process $\gamma\gamma \to M_1 M_2$ the analysis was performed in the ``zero-tag'' mode, where neither the recoil electron nor the positron is detected.  The virtuality of the incident photons is restricted to be small by imposing strict transverse-momentum balance with respect to the beam axis for the final-state hadronic system.

We first describe typical criteria applied in the Belle experiment to select production of two charged particles in $\gamma\gamma$ collisions using an example of the study of  $\gamma\gamma \to \pi^+\pi^-$
and  $\gamma\gamma \to K^+K^-$. The signal events are collected predominantly by a trigger that requires two
charged tracks, with an opening angle in the $r\varphi$ plane (perpendicular to the $z$ axis) of at least $135^{\circ}$.

Signal candidates are selected according to the following criteria.  There must be exactly two oppositely-charged reconstructed tracks satisfying the following conditions:  $-0.47 \le \cos\;\theta_{\rm lab} \le 0.82$ for the polar angle $\theta_{\rm lab}$ of each track; $p_t > 0.8$ GeV/$c$ for the
momentum component in the $r\varphi$ plane of each track; $dr \le 1$ cm and $|dz| < 2$ cm for the
origin of each track relative to the nominal $e^+e^-$ collision point; and $|dz_1 - dz_2| \le 1$ cm for the two tracks' origin difference along the $z$ axis, where the origin is defined by the closest approach of the track to the nominal collision point in the $r\varphi$ plane. The event is vetoed if it contains any other reconstructed charged track with transverse momentum above 0.1 GeV/$c$.

Cosmic rays are suppressed by demanding that the opening angle $\alpha$ between the two tracks satisfies $\cos\alpha \geq -0.997$.  The signal is enriched relative to other backgrounds by requiring that the scalar sum of the momentum of the two tracks be below 6 GeV/$c$, the total energy deposited in the electromagnetic calorimeter (ECL) be below 6 GeV, the magnitude of the net transverse momentum of the above-selected two charged tracks in the $e^+e^-$ c.m.\ frame be below 0.2 GeV/$c$, the invariant mass of these two tracks be below 4.5 GeV/$c^2$, and that the squared missing mass of the event be above 2 GeV$^2$/$c^4$. Here, the two tracks are assumed to be massless particles. The latter two requirements eliminate radiative Bhabha and initial state radiation events.  The remaining events consist of two-photon production of $e^+e^-$, $\mu^+\mu^-$, $\pi^+\pi^-$, $K^+K^-$, and $p\bar{p}$ final states as well as unvetoed $e^+e^- \to \tau^+\tau^-$ events according to a Monte Carlo (MC) study.

The information about the interaction of charged tracks with a material of various detector subsystems is used to identify tracks and ascribe each event to some specific process. After removing events that appear to arise from two-photon production of $\mu^+\mu^-$, $e^+e^-$, and $p\bar{p}$ according to the above criteria, the remaining sample consists of two-photon production of $K^+K^-$, $\pi^+\pi^-$, and residual $\mu^+\mu^-$, as well as  $e^+e^- \to \tau^+\tau^-$ production where each $\tau$ lepton decays to a single pion or muon. Then pion and kaon events  are additionally separated from each other.

The $\pi^+\pi^-$ sample is somewhat contaminated by non-exclusive two-photon background $\gamma\gamma \to \pi^+\pi^- X$ as well as the $e^+e^- \to \tau^+\tau^-$ process, in roughly equal proportion. These backgrounds appear at high values of the magnitude of the net transverse momentum $p_{t,{\rm bal}}=
|\vec{\textbf{p}}_{t}^{+} + \vec{\textbf{p}}_{t}^{-}|$  in the $e^+e^-$ c.m.\ frame, and are often accompanied by photons from the prompt decay of a neutral pion in the final state.  Therefore, events are rejected in the $\pi^+\pi^-$ sample that contains a photon with energy above 400 MeV ($E_\gamma$-veto).  The
distributions of $p_{t,{\rm bal}}$ for the $\pi^+\pi^-$ candidates before and after application of this veto are shown as the histograms in Fig.1a.

The yields of the $\pi^+\pi^-$ and $K^+K^-$ events are expressed as functions of three variables: $W$ derived from the invariant mass of the two mesons, $\cos{\theta}^*$, cosine of the $\gamma\gamma$ c.m.\ scattering angle and $p_{t,{\rm bal}}$. Eighty-five 20 MeV wide bins in $W$ times six bins in the cosine of the $\gg$ c.m.\ scattering angle $\theta^*$ times twenty bins in net transverse momentum are used in the ranges
2.4~GeV~$<~W~<$~4.1~GeV,  $|\cos{\theta}^*|<0.6$, and $p_{t,{\rm bal}} < 0.2\,{\rm GeV}/c$.

The spectrum of the residual $\gamma\gamma \to \mu^+\mu^-$ background within the $\pi^+\pi^-$ sample is obtained from a MC simulation, based on a full $\mathcal{O}(\alpha^4)$ QED calculation~\cite{aafhb}, with a data sample corresponding to an integrated luminosity of $174.2\ \rm fb^{-1}$ that is processed by the full detector simulation program and then subjected to trigger simulation and the above event selection criteria.
After calibration of the muon identification efficiency to match that in the data using identified $\gamma\gamma\to \mu^+\mu^-$ events, the residual $\mu^+\mu^-$ background is scaled by the integrated luminosity ratio and then subtracted.

The excess in the $E_{\gamma}$-vetoed histogram of Fig.1a above the smooth curve from the signal MC, described in more detail below, is attributed to non-exclusive $\gamma\gamma \to \pi^+\pi^- X$ events that are not rejected by the $E_\gamma$-veto; most of the $e^+e^- \to \tau^+\tau^-$ events are rejected by this  veto. A similar excess appears in Fig.1b for the $\gamma\gamma\to K^+K^-$ process. Assuming that this remaining background is proportional to net transverse momentum, the slope is determined using the difference between data and MC in the range $0 < p_{t,{\rm bal}} < 0.17$ GeV/$c$ for each 200 MeV wide $W$ bin, then the so-determined slopes are smoothened by fitting them to a cubic polynomial in $W$. Then it is  verified that there is no dependence on the scattering angle $\theta^*$. Using the smoothed slope, the estimated non-exclusive background is subtracted from each bin. Finally, the signal region is restricted to net transverse momentum below $0.05 (0.10) GeV/c$ for $\gamma\gamma \to \pi^+\pi^-$
($\gamma\gamma \to K^+K^-$).

\begin{figure}[th]
\includegraphics[width=0.87\textwidth]{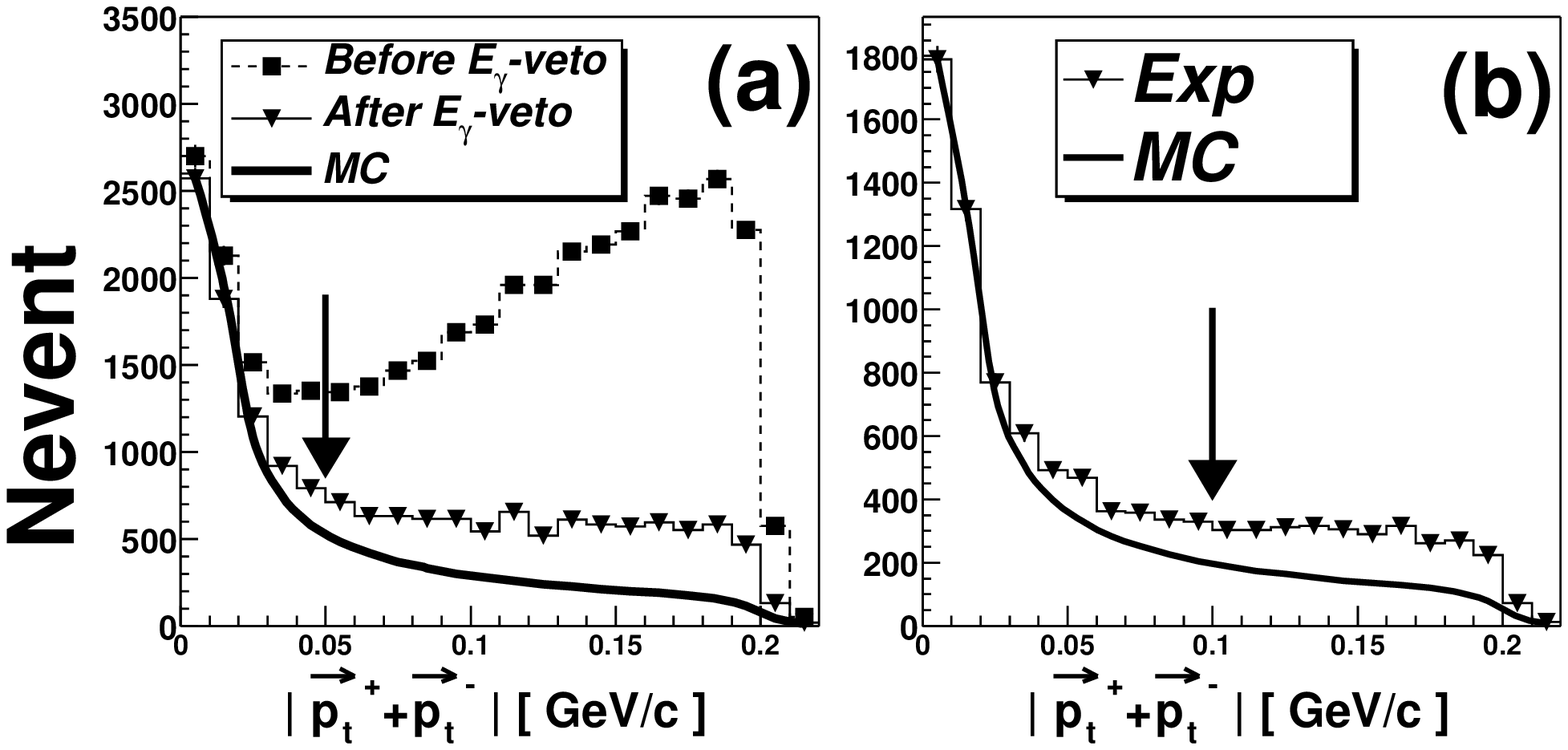}
\end{figure}
\hspace*{3mm} Fig.1.\,\,\,$|\vec{\textbf{p}}_{t}^{+}+\vec{\textbf{p}}_{t}^{-}|$ distribution for $\pi^+\pi^-$(a) and $K^+K^-$(b) candidates. The dashed and solid histograms in $\pi^+\pi^-$ indicate the distribution of events before and after $E_\gamma$-veto (which is not applied to the $K^+K^-$ candidates), respectively. The arrows indicate the upper boundaries of $|\vec{\textbf{p}}_{t}^{+} + \vec{\textbf{p}}_{t}^{-}|$ for the signal. The residual muon background has been subtracted from the $\pi^+\pi^-$ distribution. The curves show the signal MC distribution which is normalized to the signal candidates at the leftmost bin.
\label{fg:ptb}

\subsection{Leading term QCD vs data for $\gamma\gamma\to\pi^+\pi^-$ and $\gamma\gamma\to K^+K^-$}

\hspace*{3mm} The calculation of the large angle scattering amplitudes $M(\gamma\gamma\to M_2 M_1)$,  see Fig.2, was considered first in \cite{BL-81} (for symmetric meson wave functions only, $\phi_M(x)=
\phi_M(1-x)$) and later in \cite{Maurice-Ch} (for general wave functions and with account for $SU(3)$ symmetry breaking). The meson wave functions used in numerical calculations were obtained in \cite{cz-82,czz-82} using the method of QCD sum rules \cite{SVZ}, see the reviews \cite{Sh-92,Sh-98,Colan-00} for the QCD sum rules. The NLO perturbative corrections to  $M(\gamma\gamma\to M_2 M_1)$ were calculated in \cite{DN}.

The leading term QCD expressions for cross sections look as \cite{BL-81,Maurice-Ch} (the example is for $\gamma\gamma\to K^+K^-$)\,:
\be
\frac{d\sigma(\gamma\gamma\ra K^+K^-)}{d\cos\theta}=\frac{1}{32\pi W^2}\,\frac{1}{4}\sum_{
\lambda_1, \lambda_2=\pm 1}\,\,\Bigl | M_{\lambda_1\lambda_2}\Bigr |^2\,, \nonumber
\ee
\be
M^{(lead)}_{\lambda_1\lambda_2}(W,\theta)=\frac{64\pi^2}{9W^2}\,\alpha \,{\ov \alpha}_s \,f_P^2\int_0^1 dx_s\, \phi_{K}(x_s,\ov\mu)\int_0^1 dy_s\, \phi_{K}(y_s,\ov\mu)\,T_{\lambda_1\lambda_2}
(x_s,\,y_s,\,\theta)\,,\nonumber
\ee
\be
T_{++}=T_{--}=(e_u-e_s)^2\,\frac{1}{\sin^2\theta}\,\frac{A}{D}\,,\quad T_{+-}=T_{-+}\,, \nonumber
\ee
\be
T_{+-}=\frac{1}{D}\Biggl [ \frac{(e_u-e_s)^2}{\sin^2\theta}(1-A)+e_u e_s \frac
{A C}{A^2-B^2\cos^2\theta}+\frac{(e_u^2-e_s^2)}{2}(x_u-y_s)\Biggr ]\,,
\ee
\be
A=(x_sy_u+x_uy_s),\,\, B=(x_sy_u-x_uy_s),\,\, C=(x_sx_u+y_sy_u),\,\, D=x_u x_sy_u y_s\,.\nonumber
\ee
Here: $W$ is the total energy in the $\gamma\gamma$-c.m.s., $\lambda_1$ and $\lambda_2$ are the photon helicities, $x_s$ is the meson momentum fraction carried by the s-quark in the $K^{+}$-meson,
$x_s+x_u=1\,,\quad e_u=2/3, \quad e_s=e_d=-1/3$ are the quark charges, $\ov\alpha_s$ is the effective value of the coupling, $\phi_K(x,\ov\mu)$ is the leading twist K- meson wave function (= distribution amplitude), $\ov\mu$ is its appropriate normalization scale, $f_K$ is the decay constant\,:\,$f_K\simeq 157\,{\rm MeV},\, f_{\pi}\simeq 130.4\,{\rm MeV}.$
\vspace*{3mm}

\begin{minipage}{0.65\textwidth}
\hspace*{-1cm}
\includegraphics[width=0.8\textwidth]{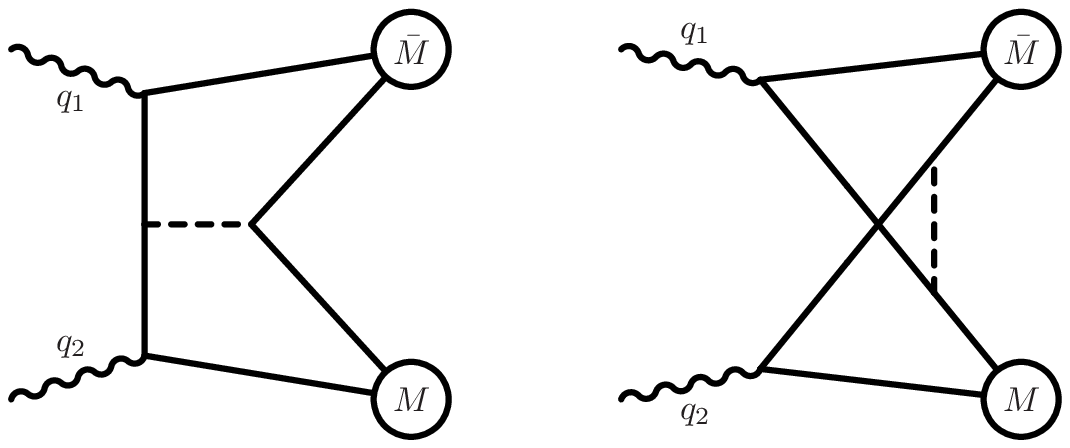}
\end{minipage}
\begin{minipage}{0.25\textwidth}
\hspace*{-2cm}
{Fig.2. \,\, Two typical lowest order Feynman diagrams for the leading term hard QCD contributions to the large angle amplitude $\gamma\gamma\ra {\ov M}_2 M_1$ (the dashed line is the hard
gluon exchange).}
\end{minipage}
\vspace*{3mm}

\hspace*{3mm} The leading contribution to  $d\sigma(\gamma\gamma\to\pi^+\pi^-)$ can be written as\,:
\be
\frac{s^3}{16\pi\alpha^2}\,\frac{d\sigma(\gamma\gamma\ra\pi^+\pi^-)}{d |\cos\theta |}
\equiv \frac{|\Phi^{(eff)}_{\pi}(s,\theta)|^2}{\sin ^4 \theta}
=\frac{|s F_{\pi}^{(lead)}(s)|^2}{\sin ^4 \theta}|1- \upsilon(\theta)|^2\,,
\ee
where
$F_{\pi}^{(lead)}(s)$ is the leading term of the pion form factor \,:
\be
|s F^{(lead)}_{\pi}(s)|= \frac{8\pi\,{\ov\alpha}_s f_{\pi}^2}{9}\,\Bigl
|\int_0^1 \frac{dx}{x}\,\phi_{\pi}(x,\,{\ov \mu})\Bigr |^2\,,\quad s=W^2
\ee
and $\upsilon(\theta)$ is due to the term $\sim AC$ in (2.1). As characteristic examples, we will compare below the predictions of two frequently considered models for $\phi_{\pi}(x)$: \, a) $\phi^{asy}(x)=6x(1-x)$ as a representative of the "normal" wave function, and b)\, $\phi_{\pi}^{CZ}(x,\mu_o)=30x(1-x)(2x-1)^2,\, \mu_o\sim 1{\rm GeV}$ - the CZ-model proposed in \cite{cz-82} and based on using the QCD sum rules \cite{SVZ} for calculation of the wave function moments, this one as a representative of the wide wave function, see Fig.3 (other examples of wide wave functions will be also considered below).

While the numerical value of $|s F^{(lead)}_{\pi}(s)|$ is highly sensitive to the form of $\phi_{\pi}(x,\ov\mu)$, the function $\upsilon(\theta)$ is only weakly dependent of $\theta$ at $|\cos\theta|<0.6$ and, as emphasized in \cite{BL-81}, is weakly sensitive to the form of $\phi_{\pi}(x,\ov\mu)$:\, ${\ov\upsilon}(\theta)\simeq 0.12$ for the above two very different pion wave functions, $\phi^{CZ}_{\pi}(x)$ and $\phi^{asy}_{\pi}(x)$.~
\vspace*{5mm}

~\begin{minipage}{0.45\textwidth}\hspace*{-5mm}
\includegraphics[width=0.75\textwidth,clip=true]{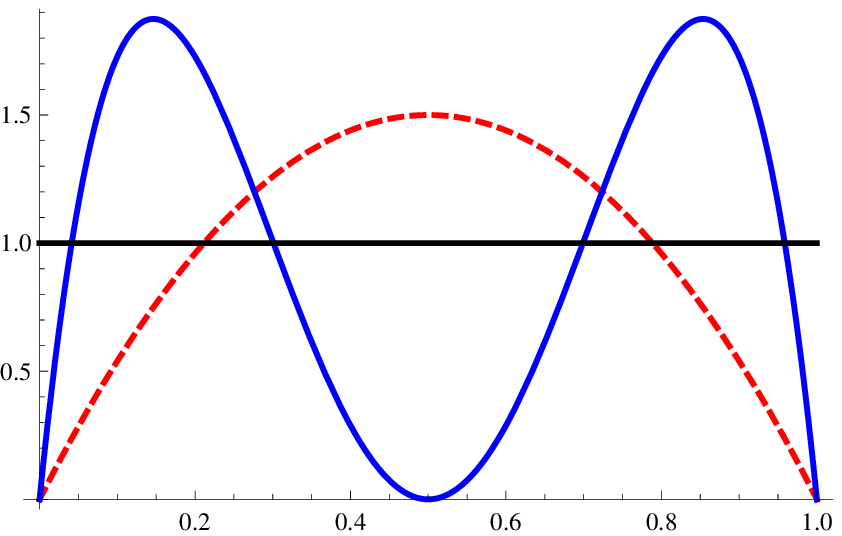}
\put (-193,40) {\rotatebox{90} {$\phi_{\pi}(x,\mu=1\,GeV)$}} \put (5,3) {$x$}
\end{minipage}
\begin{minipage}{0.45\textwidth}\vspace{1mm}

Fig.3.\, Some different models for the leading twist pion wave function, $\,\,\int_0^1 dx\, \phi_{\pi}(x, \mu)=1$\,.

Dashed line - the asymptotic wave function $\phi_{\pi}^{\rm asy}(x)=6x(1-x)$\,; solid line - the CZ wave function at low normalization point:\, $\phi^{\rm CZ}_{\pi}(x,\mu\simeq 1\,GeV)=
30x(1-x)(2x-1)^2$\,; straight line - the flat wave function $\phi_{\pi}(x,\mu\simeq 1\,GeV)=1$.\\
\end{minipage}
\vspace*{-1mm}

\begin{minipage}{0.47\textwidth}\hspace*{-4mm}
\includegraphics[width=0.75\textwidth]{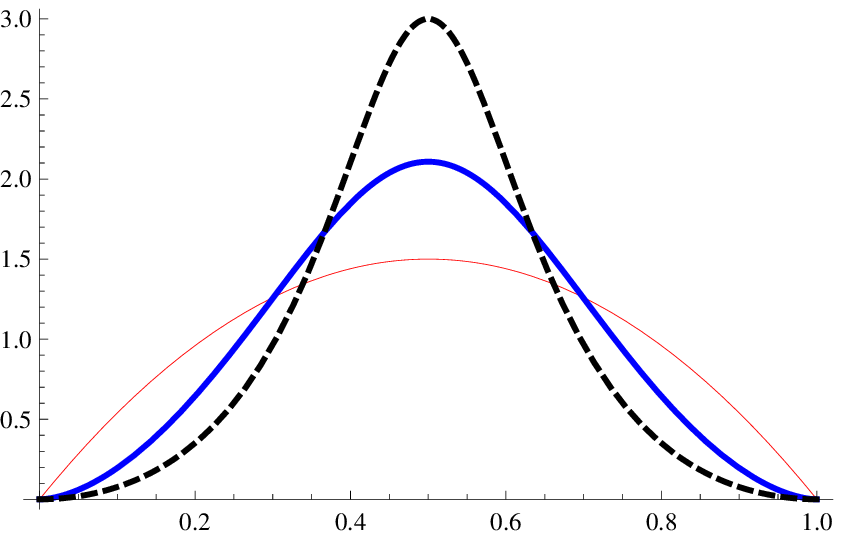}
\put (-203,50) {\rotatebox{90} {$\phi(x,{\it v}^2)$}} \put (5,3) {$x$}
\end{minipage}
\hspace*{-3mm}
\begin{minipage}{0.5\textwidth}
\vspace*{2mm}

Fig.4.\quad The characteristic shape of the quarkonium model wave function $\phi(x, v^2)$. Thin solid line:\, $v^2=1$ - massless quarks (asymptotic). Thick solid line:\, $v^2=0.3$ for the ground state charmonium $J/\psi,\,\eta_c\,$.  Dashed line:\, $v^2=0.1$ for the ground state bottomonium $\Upsilon(1S),\, \eta_b\,$. The width of the distribution is parametrically $\sim v^2$. Therefore, {\it the heavier is quark the narrower is wave function},\\
\hspace*{-3mm} $\int_0^1 dx\, \phi(x, v^2)=1,\quad \phi(x,v^2\ra 0)\ra \delta (x-0.5)$.
\end{minipage}
\vspace*{2mm}

\hspace*{3mm} As it follows from (2.1)-(2.3), the leading term QCD predictions for charged mesons $\pi^+\pi^-$ and $K^+K^-$ look as
\be
\frac{d\sigma}{d\cos\theta}\sim \frac{1}{W^6}\,\,\frac{1}{\sin^4\theta}\,.
\ee

The recent data from Belle \cite{Nakaz}, see Figs.5a-d, agree with the $\sim 1/\sin^4\theta $ dependence at $W\geq 3\,GeV$, while the angular distribution is somewhat steeper at lower energies. The energy dependence at $2.4\,{\rm GeV} < W < 4.1\, {\rm GeV}$ was fitted in \cite{Nakaz} as: $\sigma_o(\pi^+\pi^-, |\cos\theta|<0.6)=\int_{-0.6}^{+0.6}dz (d\sigma/dz) \sim W^{-n}\,,\,\, \,n=(7.9\pm 0.4_{\rm stat}\pm 1.5_{\rm syst})$ for $\pi^+\pi^-$, and $n=(7.3\pm 0.3_{\rm stat}\pm 1.5_{\rm syst})$ for $K^+K^-$. However, the overall value $n\simeq 6 $ is also acceptable, see the Belle fit in Fig.5\,.

As for the absolute normalization, the $\pi^+\pi^-$ data are fitted \cite{Nakaz} with, see (2.2)\,:
\be
|\Phi_{\pi}^{(eff)} (s,\theta)|=(0.503\pm 0.007_{\rm stat}\pm 0.035_{\rm syst})\,{\rm GeV}^2\,.
\ee

Clearly, in addition to the leading terms $M^{(lead)}_{\lambda_1\lambda_2}(W,\theta)$, this experimental value includes also all non-leading loop corrections \cite{DN} and all power corrections to the $\gamma\gamma\ra \pi^+\pi^- $ amplitudes, $M=M^{(lead)}+\delta M$. These are different, strictly speaking, from corrections $\delta F_{\pi}$ to the genuine pion form factor,\,\, $F_{\pi}=\, F_{\pi}^{(lead)}+\delta F_{\pi}$, see (2.3). So, the direct connection between the leading terms of $d\sigma(\pi^+\pi^-)$ and $|F_{\pi}|^2$ in~ (2.2) and (2.3) does not hold on account of loop and power corrections.

\hspace*{3mm} The experimental value :\, $|\Phi_{\pi}^{(eff)}(s,\theta)|=|1- \upsilon(\theta)||s F_{\pi}^{(lead)}(s)|\simeq 0.88 |s F_{\pi}^{(lead)}(s)|\simeq 0.5\,{\rm GeV}^2$, see (2.5), can be compared with the theoretical expression (2.3) for various model pion wave functions. The problem here is to determine the reasonable effective values of the coupling $\ov\alpha_s$ and the wave function normalization scale $\ov\mu$ in (2.3) for a given amplitude, such that non-leading logarithmic loop corrections will be sufficiently small.
\newpage
\begin{minipage}[c]{.5\textwidth}
\includegraphics[trim=0mm 0mm 0mm 0mm, width=0.75\textwidth,clip=true]{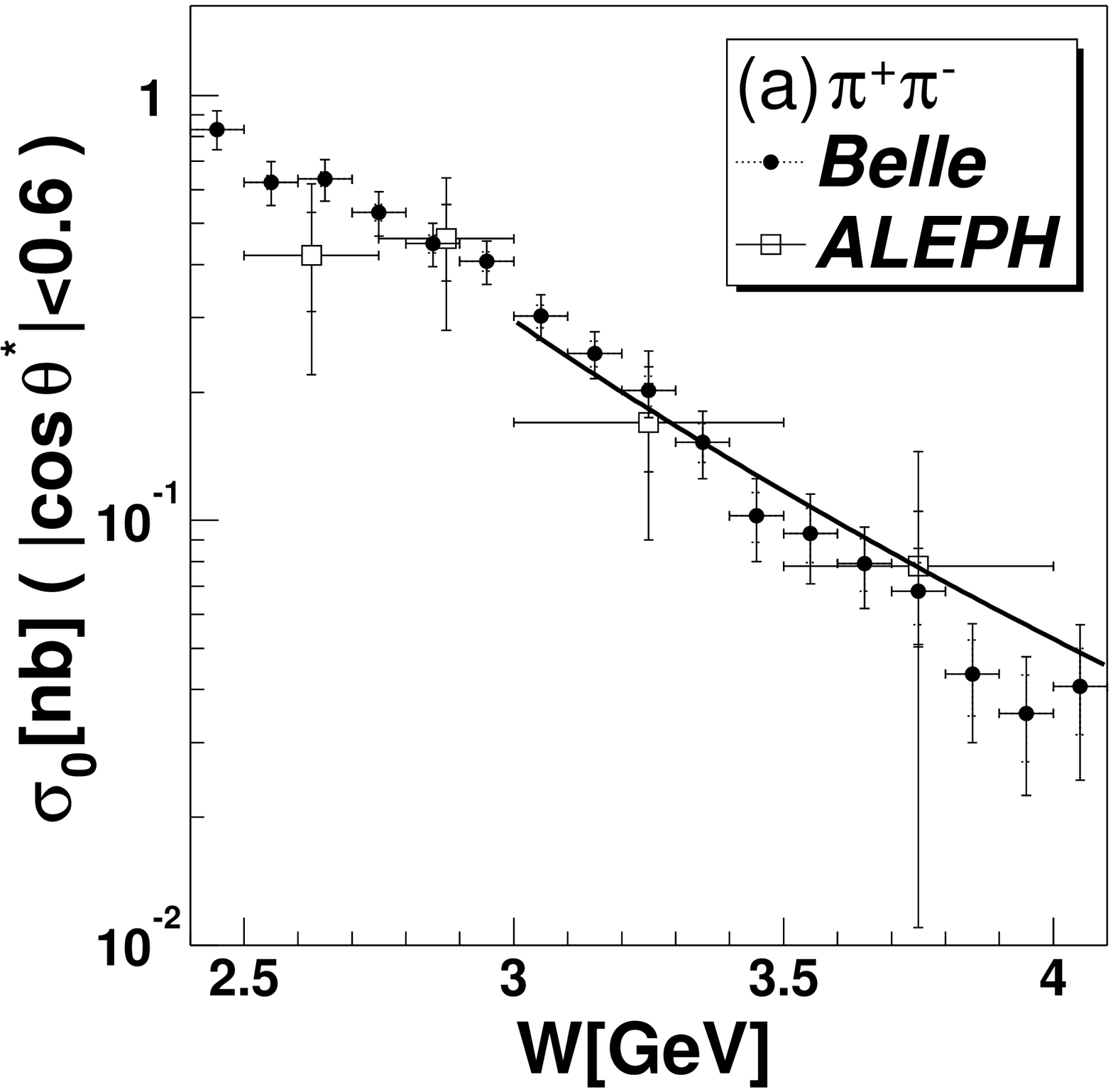}
\end{minipage}~
\begin{minipage}[c]{.5\textwidth}
\includegraphics[trim=0mm 0mm 0mm 0mm, width=0.75\textwidth,clip=true]{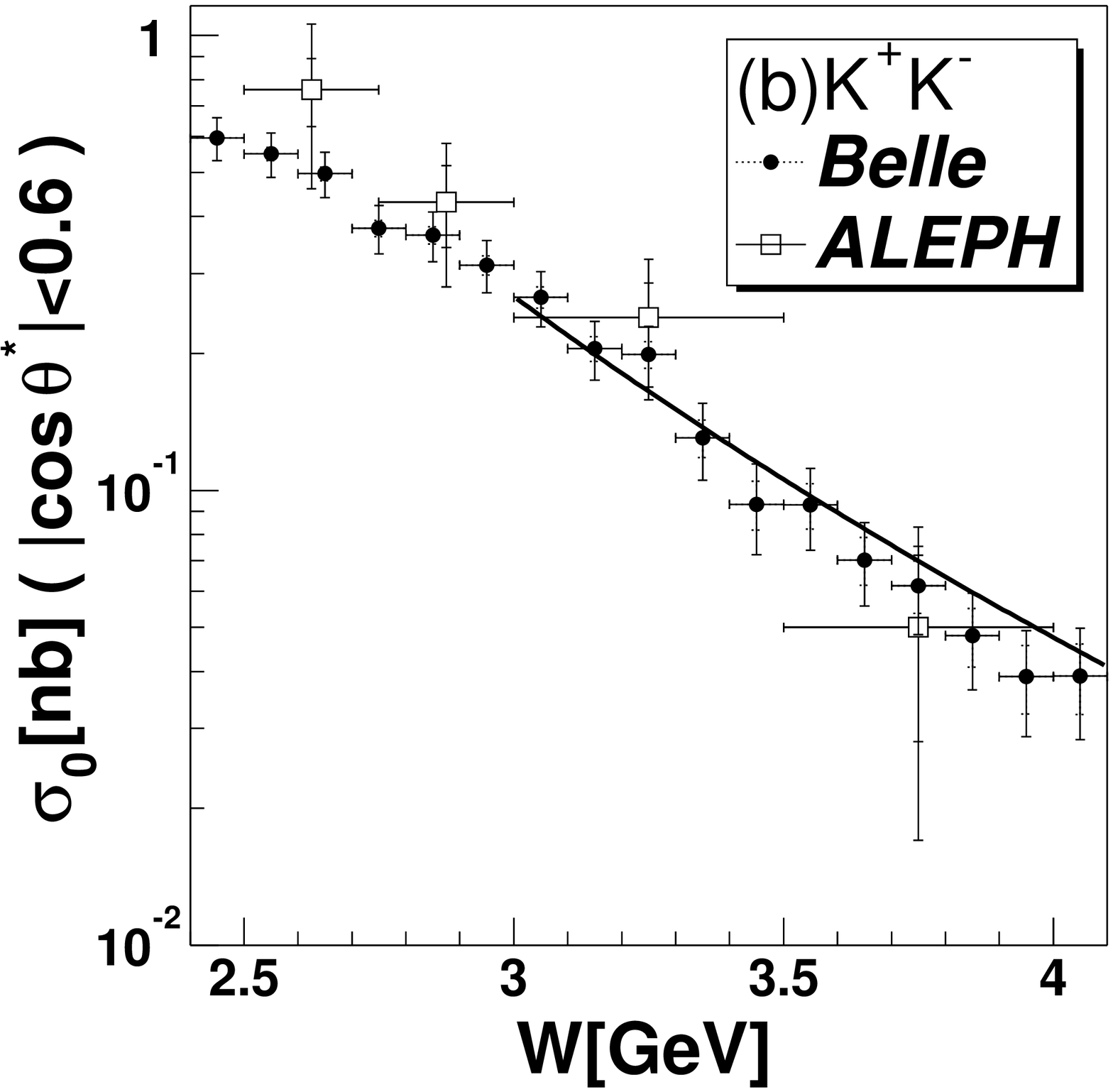}
\end{minipage}

\begin{minipage}[c]{0.9\textwidth}
\vspace{2mm}

\hspace*{4cm}
\includegraphics[width=0.55\textwidth]{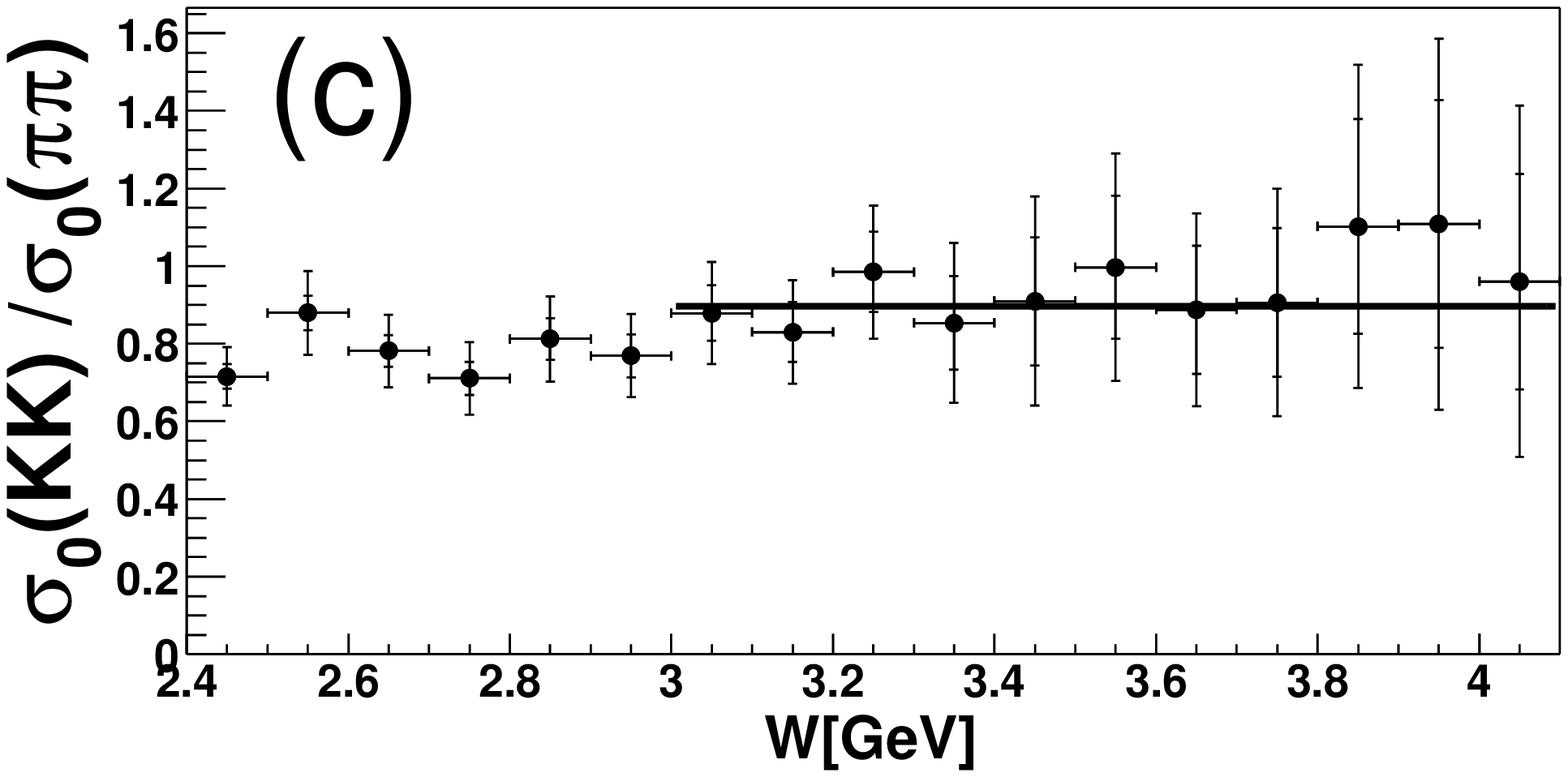}
\end{minipage}
\vspace*{1mm}

Fig.5\,\cite{Nakaz}.\, a,\,b)\, Cross sections $\sigma_{o}(\gamma\gamma\to \pi^+\pi^-)$ and $\sigma_{o}(\gamma\gamma\to K^+ K^-)$ integrated over the angular region $|\cos\theta|<0.6$ (see Table 1 in \cite{Nakaz} for numerical values), together with the  $1/W^6$ dependence~ line.\\
c)\, the cross section ratio\, $R_{\rm exp}=\sigma_{o}(K^+K^-)/\sigma_{o}(\pi^+
\pi^-)\simeq 0.9$ (the errors indicated by short ticks are statistical only).  Compare $R_{\rm exp}\simeq 0.9$ with the naive prediction for $\phi_{\pi}(x)=\phi_K (x)\,: \,R=(f_K/f_{\pi})^4\simeq 2.1$\,.\\

\begin{minipage}[c]{.9\textwidth}
\includegraphics[trim=0mm 0mm 0mm 0mm, width=1.0\textwidth,clip=true]{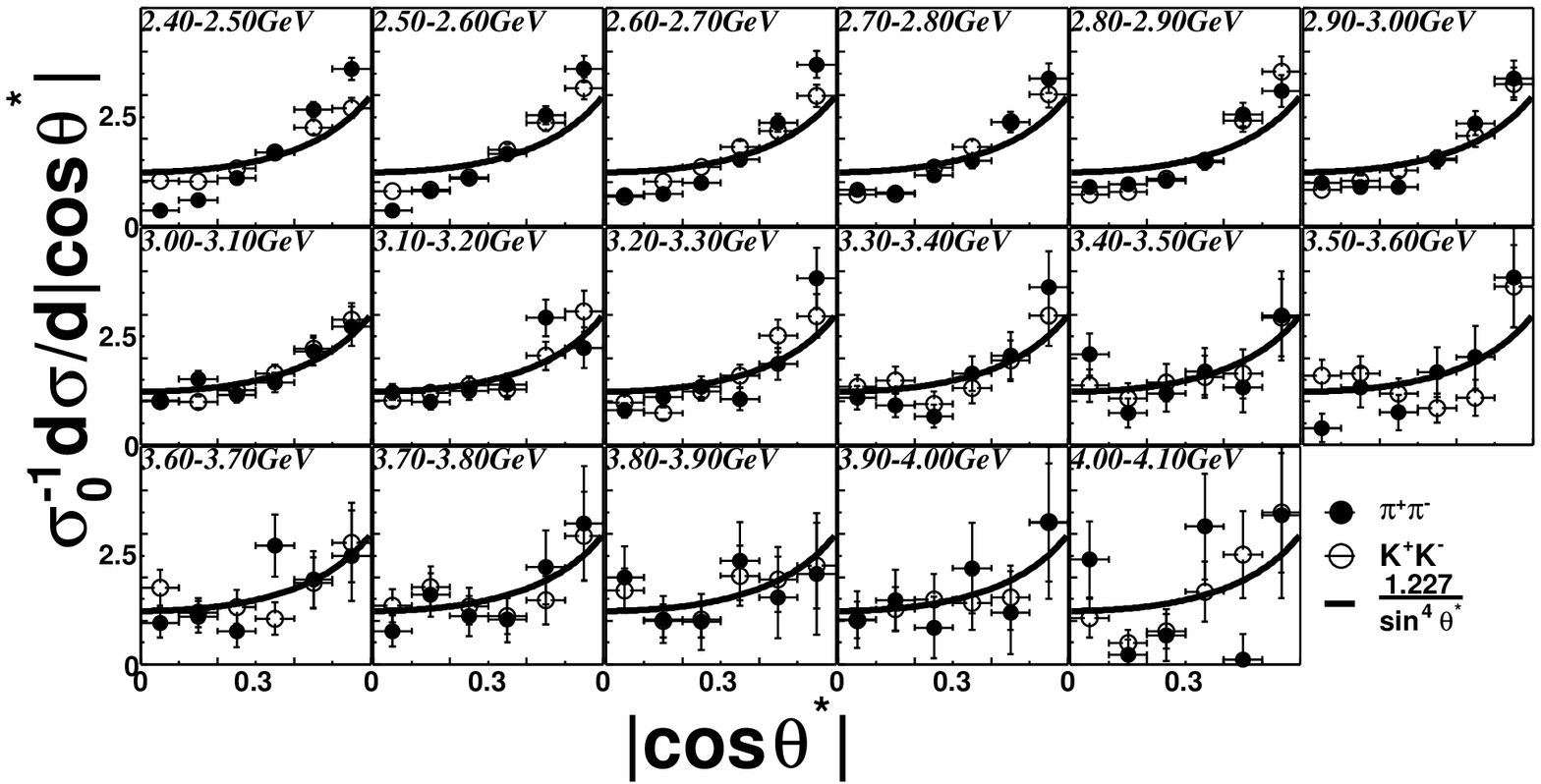}
\end{minipage}

\vspace*{1mm}
\hspace*{1cm}Fig.5d\,\cite{Nakaz}. Angular distributions of $\pi^+\pi^-$ and $K^+ K^-$ (the errors are statistical only).\\

In practice, however, only the first non-leading loop correction was calculated for a few amplitudes and the numerical results depend then significantly on the renormalization scheme used. It seems clear qualitatively that the reasonable characteristic scale $\ov{\mu}$ in the corresponding Born diagrams for the form factor $F_{\pi}^{(lead)}(Q^2)$ (as well as in Fig.2 diagrams) is determined by the typical gluon virtuality, $\ov{k^2}=\ov{xy}\, Q^2$. Therefore, it is something like $\ov\mu^2\simeq Q^2/4$ for $\phi^{asy}(x)$. But because $\phi^{asy}(x)$ does not run with $\mu$ in the leading logarithmic approximation (LO) and runs very slow in NLO, the value of $\ov\mu$ for $\phi^{asy}(x)$ is of no real importance. At the same time, the mean value of the gluon virtuality, $(\ov\mu)^2$, is much smaller for the wide pion wave function like $\phi^{CZ}_{\pi}(x)$. Really, $(\ov\mu)^2\simeq \ov{k^2}\lesssim 1\,GeV^2$ at $Q^2\sim 10-16\,GeV^2$ for $\phi^{CZ}_{\pi}(x)$, see \cite{cz-82,cz-rev}. Here and in what follows therefore, for rough estimates in the region $Q^2\sim 10-16\,GeV^2$, we will use the values $\ov\mu\sim 1\,GeV$ for the wave function scale and $\ov\alpha_s\simeq 0.3$ for the effective value of $\alpha_s$. Then from (2.3) for the wide pion wave function $\phi^{CZ}_{\pi}(x,\ov\mu\simeq 1\,GeV)$\,:\,$0.88 |s\, F_{\pi}^{(lead, CZ)}
(s)|\simeq 0.32\,{\rm GeV}^2$, while $0.88|s\,F_{\pi}^{(lead, asy)}(s)|\simeq 0.11\,{\rm GeV}^2$ for the narrower wave function $\phi^{asy}_{\pi}(x)$.

Therefore, for the pion wave function $\phi_{\pi}(x, {\ov\mu}\sim 1\,GeV)$ close to $\phi^{\rm asy}(x)$ the leading term (i.e. without power corrections) calculation predicts the cross section which is $\simeq 20$ times smaller than the data. It seems that at $s=W^2=10-16\,{\rm GeV}^2$ the power corrections can not cure so large difference. Moreover, if power corrections were dominant numerically at these energies then the cross sections $\sigma_o(\pi^+\pi^-)$ and $\sigma_o(K^+K^-)$ will decay more like $\sim 1/W^{10}$, rather than $\sim 1/W^6$, in contradiction with data.

We conclude that the Belle data \cite{Nakaz} (as well as many data for other hard exclusive processes) favor the leading twist pion wave function $\phi_{\pi}(x, {\ov \mu}\sim 1\,GeV)$ to be much wider than $\phi^{asy}(x)$.\\

The SU(3)-symmetry breaking, $d\sigma(K^+K^-)\neq d\sigma(\pi^+\pi^-),$ originates not only from different meson couplings, $f_K \,>\, f_{\pi}$, but also from symmetry breaking effects in meson wave functions, $\phi_{K}(x)\neq \phi_{\pi}(x)$ \cite{czz-82}, see Fig.6. (Let us recall: {\it the\, heavier\, is\, quark\, the\, narrower\, is\, wave\, function}). These two effects tend to cancel each other when using for the K-meson the wave function close to $\phi_{K}(x_s)$ obtained in \cite{czz-82} from the QCD sum rules. So, instead of the naive original prediction $\simeq (f_K/f_{\pi})^4\simeq 2.1$ with $\phi_{K}(x)=\phi_{\pi}(x)$ in \cite{BL-81}, the prediction in \cite{Maurice-Ch} for this ratio is close to unity, and this is in better agreement with the data from Belle \cite{Nakaz}, see Fig.5c :
\vspace*{-1mm}
\be
\frac{\sigma_o (\gamma\gamma\ra K^+K^-)}{\sigma_o (\gamma\gamma\ra \pi^+\pi^-)}=\left \{
\begin{array}[c]{ll}\displaystyle (f_K/f_{\pi})^4\simeq 2.1 \scriptstyle &
\begin{array}[c]{l}\hspace*{-4.0 cm}\rm{ Brodsky,\, Lepage} \,\, {\rm\cite{BL-81}} \end{array}
\\ & \vspace*{-3mm}\\ \hline \vspace*{-3mm} & \\ \simeq 1.06 & \begin{array}[c]{l} \hspace*{-4.2 cm} \rm{Benayoun,\,Chernyak} \,\,{\rm\cite{Maurice-Ch}}\end{array} \\ & \vspace*{-3mm}\\ \hline & \vspace*{-3mm}\\ (0.89\pm 0.04\pm 0.15)\hspace*{3.0 cm}  {\rm Belle} \,\,\, {\rm\cite{Nakaz}}\vspace*{-3mm}
\end{array} \right. \nonumber
\ee

\begin{minipage}[c]{.5\textwidth}\hspace*{-7mm}
\includegraphics[trim=0mm 0mm 0mm 0mm, width=0.7\textwidth,clip=true]{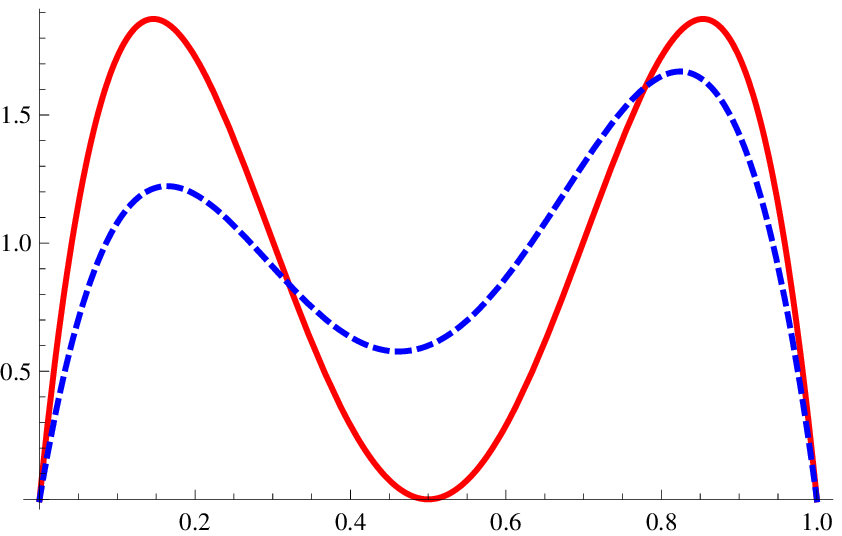}
\put (-200,50) {\rotatebox{90} {$\phi_{\pi,K}(x)$}} \put (5,3) {$x$}
\end{minipage}
\begin{minipage}[c]{.5\textwidth}
\vspace*{2mm}

\hspace*{-3cm} Fig.6.\,\, The kaon wave function is somewhat narrower than \\
\hspace*{-3cm} the pion one and asymmetric : the s-quark in the $K^-$ meson \\
\hspace*{-3cm} carries a larger part of the momentum fraction than $\ov u$-quark.\\
\hspace*{-3cm} Solid line\,: \, the model pion wave function \\
\hspace*{-2cm}$\phi^{CZ}_{\pi}(x,\mu\simeq 1\,GeV)=30\,x_d x_u(x_d-x_u)^2$.\\
\hspace*{-3cm} Dashed line\,:\, the model kaon wave function $\phi_K(x,\mu\simeq 1\,GeV)=$\\
\hspace*{-2.3cm}\vspace*{1mm} $30\,x_sx_u\Bigl [0.6(x_s-x_u)^2+0.08+0.08(x_s-x_u)\Bigr ]$.
\vspace*{0.1mm}
\end{minipage}
\vspace*{1mm}

\subsection{Some experimental details for neutral mesons}

\hspace*{3mm} We use an example of the process $\gamma \gamma \to \pi^0 \pi^0$ studied by Belle \cite{Ue-pio} to present some general conditions used in the analysis of the neutral particle production. The selection conditions for $\gamma \gamma \to \pi^0 \pi^0$ signal candidates are the following. All the variables in criteria (1)-(6) are measured in the laboratory frame: (1) there is no good track that satisfies $dr < 5$~cm, $dz < 5$~cm and $p_t > 0.1$~GeV/$c$, where $dr$ and $dz$ are the radial and axial distances, respectively, of closest approach (as seen in the $r\varphi$ plane) to the nominal collision point, and  the $p_t$ is the transverse momentum measured in the laboratory frame with respect to the $z$ axis; (3) there are two or more photons whose energies are greater than 100~MeV; (4) there are exactly two $\pi^0$'s, each $\pi^0$ having a transverse momentum greater than 0.15~GeV/$c$ with each of the decay-product photons having an energy greater than 70~MeV; (5) the two photons' momenta are recalculated using a $\pi^0$-mass-constrained fit, and required to have a minimum $\chi^2$ value for the fit (there was a negligible fraction of events with ambiguous photon combinations); (6) the total energy deposit in the ECL is smaller than 5.7~GeV.

The transverse momentum in the $e^+e^-$ c.m. frame ($|\Sigma \mbox{\boldmath$p$}_t^*|$) of the two-pion system is then calculated. For further analysis, (7) events with $|\Sigma \mbox{\boldmath$p$}_t^*| < 50$~MeV/$c$ are used as the signal candidates.

In order to reduce uncertainty from the efficiency of the hardware ECL triggers, the authors set offline selection criteria that emulate the hardware trigger conditions as follows: (8) the ECL energy sum within the triggerable region is greater than 1.25~GeV, {\it or}  all four photons composing the two $\pi^0$ are contained in the triggerable acceptance region. Here,  the triggerable acceptance region is defined as
the polar-angle ($\theta$) range in the laboratory system $17.0^\circ < \theta < 128.7^\circ$.

The $p_t$-balance distribution, i.e., the distribution in $|\Sigma \mbox{\boldmath$p$}_t^*|$, is used to separate the signal and background components. The signal Monte Carlo (MC) shows that the signal component peaks around 10-20~MeV/$c$ in this distribution. In the experimental data, however, in addition to the
signal component, some contributions  from $p_t$-unbalanced components are found in the low-$W$ region.
Such $p_t$-unbalanced backgrounds might originate from processes such as $\pi^0\pi^0\pi^0$, etc. However, the background found in the experimental data is very large only in the low-$W$ region where the $\pi^0\pi^0
\pi^0$ contribution is expected to be much smaller than $\pi^0\pi^0$ in two-photon collisions.
(Note that a $C=-1$ system cannot decay to $\pi^0\pi^0$.) The authors believe that the backgrounds are dominated by beam-background photons (or neutral pions from secondary interactions) or spurious hits in the detector.

Figures 7a and 7b show the $p_t$-balance distributions in the low $W$ region. With the fit described below, the signal components are separated from the background. In the intermediate or higher energy regions, the $p_t$ unbalanced backgrounds are either less than 1\%, buried under the $f_2(1270)$ peak
(Fig. 7c, or consistent with zero within statistical errors. For the highest energy region 3.6~GeV$\leq W \leq 4.0$~GeV, a 3\% background is subtracted from the yield in each bin  to account for the background from the $p_t$-unbalanced components and assign a systematic error of the same size, although it is not statistically significant even there, see Fig. 7d.

A fit to the $p_t$-balance distribution is performed in the region $|\sum \mbox{\boldmath$p$}_t^*| \leq$ 0.2~GeV$/c$ to separate the signal and background components for the $W$ region below 1.2~GeV.  The fit function is a sum of the signal and background components. The signal component is an empirical function
reproducing the shape of the signal MC, $y=Ax/(x^{2.1}+B+Cx)$, where $x \equiv |\sum \mbox{\boldmath$p$}_
t^*|$, $A$, $B$ and $C$ are the fitting parameters, and $y$ is the distribution. This function has a peak at $x=(\frac{B}{1.1})^{\frac{1}{2.1}}$ and vanishes at $x=0$ and as $x \to \infty$. The shape of the background is taken as a linear function $y=ax$ for $x<0.05$~GeV/$c$, which is smoothly connected to a quadratic function above $x>0.05$~GeV/$c$.

The background yields obtained from the fits are fitted to a smooth two-dimensional function of ($W$, $|\cos \theta^*|$), in order not to introduce statistical fluctuations. The backgrounds are then subtracted from the experimental yield distribution. The magnitude of the vector sum of the transverse momenta of the final particles in the $e^+e^-$ center-of-mass frame, $|\sum {\vec P}_t^{\ast}|$, which approximates the transverse momentum of the two-photon-collision system, is used as a discriminating variable to separate signal from background. The signal tends to accumulate at small $|\sum {\vec P}_t^{\ast}|$ values while the non-$\gamma \gamma$ background is distributed over a wider range. The number of $VV$ events in each $VV$ invariant mass bin is obtained by fitting the $|\sum {\vec P}_t^{\ast}|$ distribution between zero and 0.9 GeV/$c$. The signal shape is from MC simulation of the signal mode and the background shape is parameterized as a second-order Chebyshev polynomial. In order to control the background shape, the coefficients of the background polynomials are restricted in nearby invariant mass bins to vary smoothly.

\newpage
\begin{figure}[ht]
\centering
\includegraphics[width=10cm]{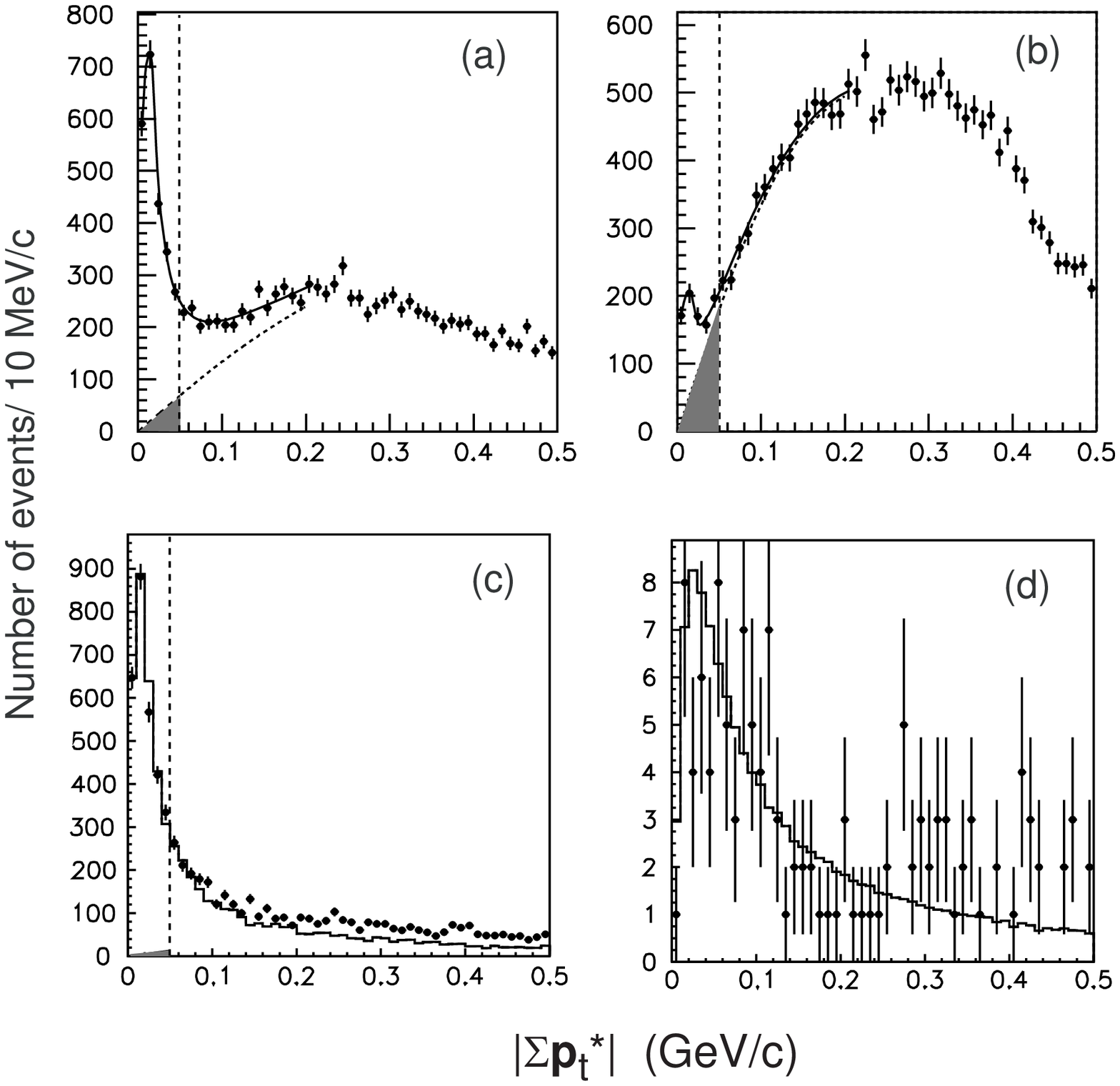}
\end{figure}~

~\hspace*{3mm} Fig.7.\,\,\,Distribution of imbalance in $|\sum \mbox{\boldmath$p$}^*_t|$ for candidate events. \,(a) In the bin centered at $W=0.90~$GeV and $|\cos \theta^*|=0.05$ (the bin width is 0.04~GeV and 0.1 in the $W$ and $|\cos \theta^*|$ directions, respectively, in (a)-(c)), the experimental distribution (dots with error bars) is fitted with the sum of signal and background components (curves). The gray region shows the estimated background contamination in the signal region. \,(b) The same as (a) for the bin centered at $W=0.66~$GeV. \,(c)  In the $W=1.18$~GeV, $|\cos \theta^*|=0.65$ bin the experimental
distribution is compared with the signal MC (histogram). \,(d) The same as (c) for 3.6~GeV $\leq W \leq$ 4.0~GeV and  $|\cos \theta^*| \leq 0.4$.\\

\subsection{Leading term QCD vs data for $\gamma\gamma\to K_S K_S $}

\hspace*{3mm} The leading terms in cross sections for neutral particles originating from the standard diagrams like those in Fig.1 are much smaller than for charged ones, see (2.1). For instance, it was obtained in \cite{Maurice-Ch} that the ratio $d\sigma^{(lead)}(\pi^o\pi^o)/d\sigma^{(lead)}(\pi^+\pi^-)$ varies from
$\simeq 0.07$ at $\cos \theta =0$ to $\simeq 0.04$ at $\cos\theta =0.6$. (However, there are additional contributions to $\sigma^{(lead)}(\pi^o\pi^o)$, see Fig.11 below. Such additional contributions are absent for $\sigma (\ov{ K^o} K^o)$ and for charged particles, $\sigma(\pi^+\pi^-),\, \sigma(K^+K^-)\,$).

At the same time, it was obtained in \cite{Maurice-Ch} for the ratios
\be
\frac{d\sigma (\ov{ K^o} K^o)^{(lead)}}{d\sigma (\pi^o\pi^o)^{(lead)}}\simeq 1.3\cdot(4/25)\simeq 0.2\,,\quad \frac{\sigma_{o}^{(lead)}(K_S K_S)}{\sigma_{o}^{(lead)}(K^+K^-)}\simeq 0.005\,.
\ee

It is seen that the leading contribution to $\sigma_o(K_SK_S)$ is very small. This implies that it is not yet dominant at present energies $ W^2 < 16\,{\rm GeV}^2$. That is, the amplitude $M(\gamma\gamma\ra K_S K_S)
= a(s,\theta)+b(s,\theta)$ is dominated at these energies by the non-leading term $b(s,\theta)\sim \varrho(
\theta)(s_o/s)^2$, while the formally leading term $a(s,\theta)\sim C_o f_{BC}(\theta)(s_o/s)$ has so small coefficient $|C_o|\ll 1$, that $|b(s,\theta)| > |a(s,\theta)|$ at, say, $s=W^2 < 12\,{\rm GeV}^2$.

Therefore, it has no much meaning to compare the leading term prediction of \cite{Maurice-Ch}, i.e.
\be
\frac{d\sigma(K_S K_S)}{d \cos\theta}\sim |f_{BC}(\theta)|^2/W^{6}\quad  {\rm at}\quad s=W^2\ra \infty
\ee
for the energy and angular dependences of $d\sigma(K_S K_S)$ with the data from Belle \cite{Chen, Ue-13} at $W^2<15\,GeV^2$. Really, the only QCD prediction for, say, $6\,{\rm GeV}^2 < W^2 < 12\,{\rm GeV}^2$ is the expected energy dependence
\be
\frac{d\sigma(K_S K_S)}{d\cos\theta} \sim \frac{|b(s,\theta)|^2}{s}\sim
\frac{|\varrho(\theta)|^2}{W^{10}}\,,
\ee
while the angular dependence $|\varrho(\theta)|^2$ and the absolute normalization are unknown. This energy dependence agrees with the results from  Belle \cite{Chen,Ue-13}, see Table 1 and Figs.8\,.

\begin{center}
\begin{table*}[t]
{Table 1. Belle results for the slope parameter $"n"$ from the power fit $\sigma \sim W^{-n}$ for $\gamma \gamma \to K_S K_S$ in different fit ranges. The first and second errors are statistical and systematic, respectively.}\\
\small
\vspace*{3mm}

{\hspace*{3cm}\begin{tabular}{cccc}
\hline
\hline
$W$ range (GeV) & $|\cos \theta^*|$ range & $n$ & Ref \\
\hline
$2.4 - 4.0$ (excluding $3.3 - 3.6$) & $<0.6$ & $10.5 \pm 0.6 \pm 0.5$ &
\cite{Chen}\\
\hline
$2.6 - 4.0$ (excluding $3.3 - 3.6$) & $<0.8$ & $11.0 \pm 0.4 \pm 0.4$ & \cite{Ue-13}\\
$2.6 - 3.3$ & $<0.8$ & $10.0 \pm 0.5 \pm 0.4$ & \cite{Ue-13}\\
$2.6 - 3.3$ & $<0.6$ & $11.8 \pm 0.6 \pm 0.4$ & \cite{Ue-13}\\
\hline
\hline
\end{tabular}}
\end{table*}
\end{center}

\begin{minipage}[c]{.5\textwidth}\vspace*{1mm}\hspace*{-0.8cm}
\includegraphics
[trim=0mm 0mm 0mm 0mm, width=0.95\textwidth,clip=true]{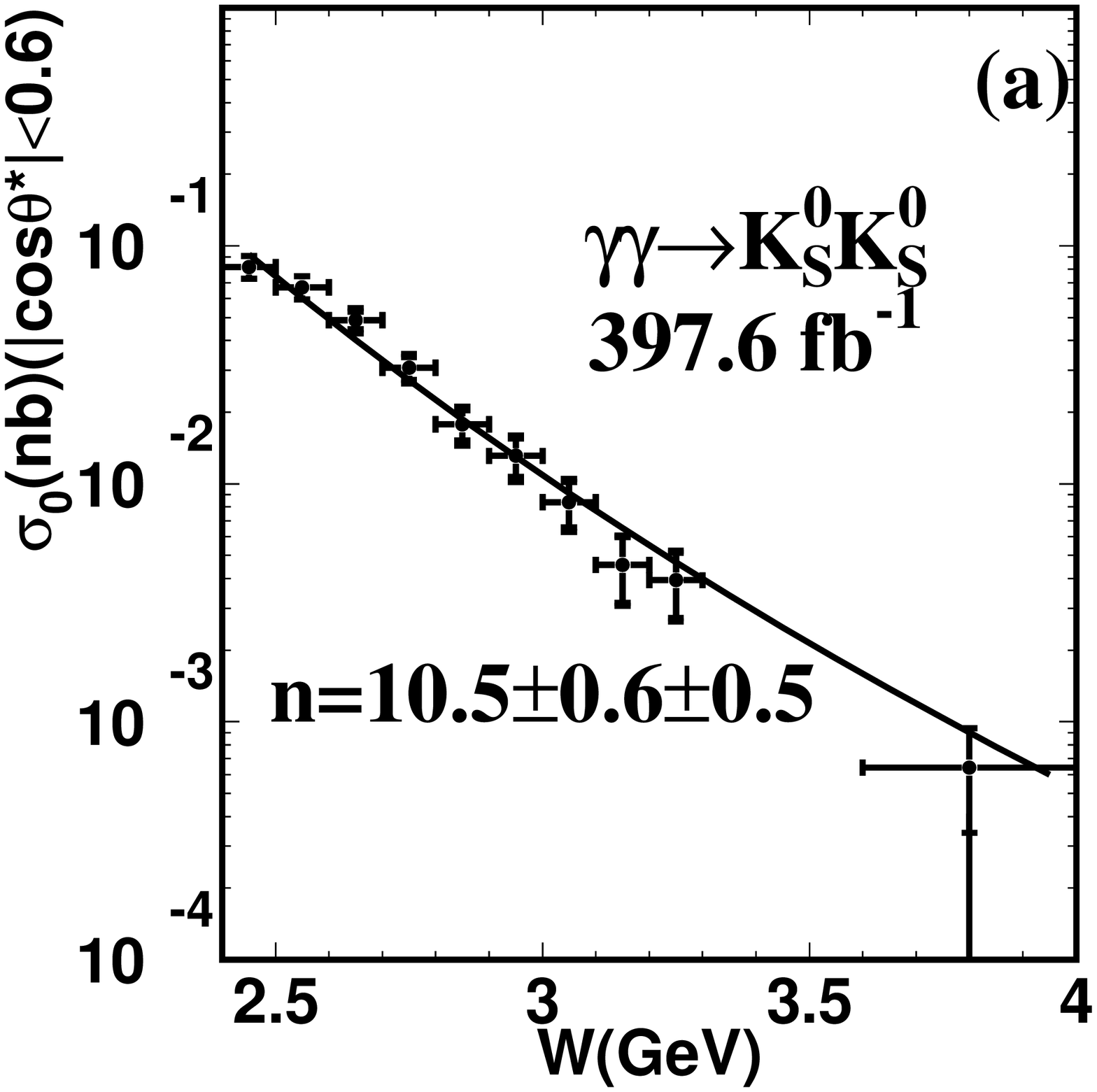}
\end{minipage}~
\begin{minipage}[c]{.5\textwidth}\vspace*{1mm}\hspace*{-0.8cm}
\includegraphics
[trim=0mm 0mm 0mm 0mm, width=0.95\textwidth,clip=true]{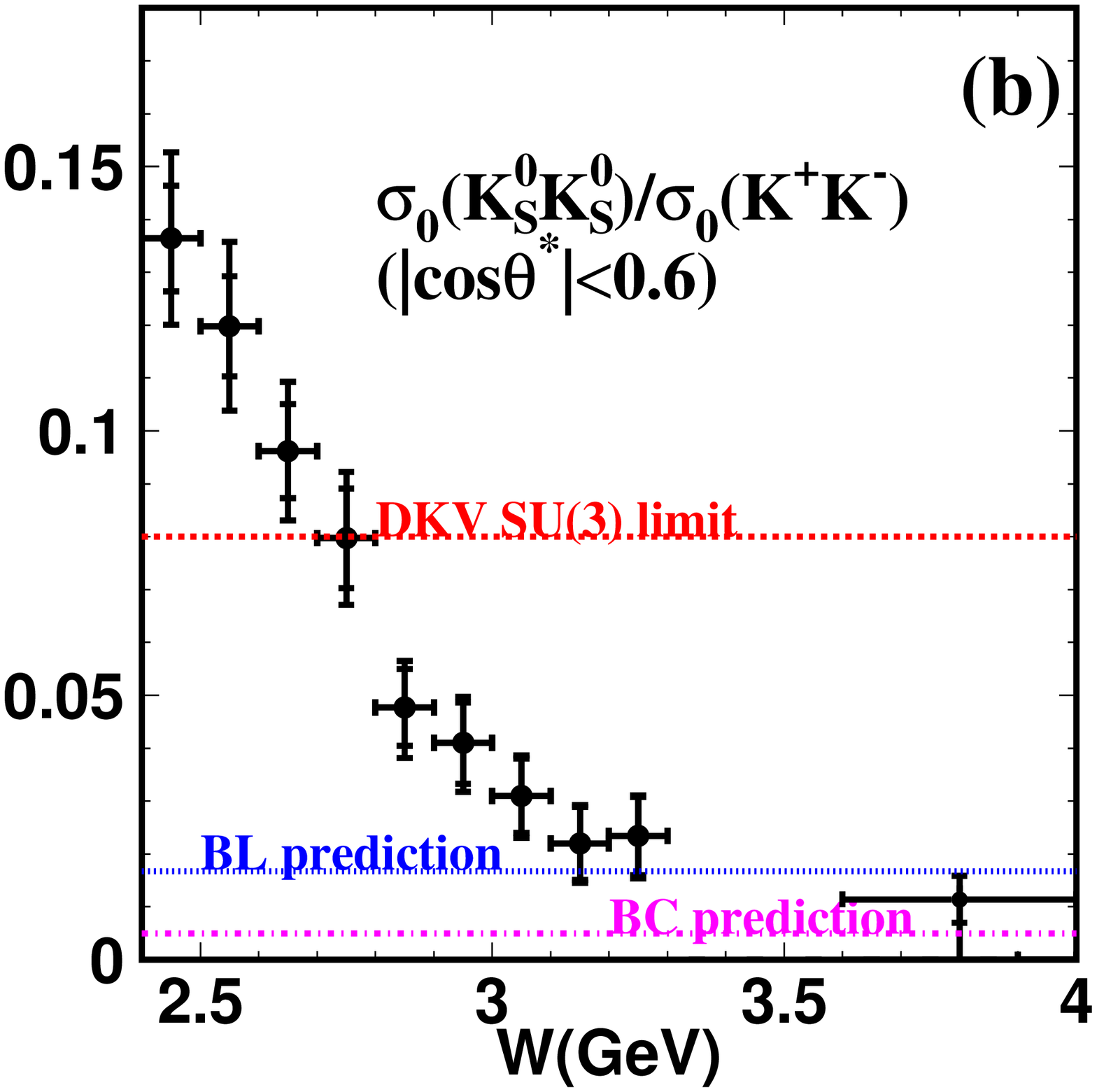}
\end{minipage}
\vspace*{1cm}

\hspace*{4mm} Fig. 8a.\, The total cross section $\sigma_o(\gamma\gamma\ra K_SK_S)$ in the c.m. angular region $|\cos\theta|<0.6$\,\,\cite{Chen}.\\  Here $n=(10.5\pm 0.6\pm 0.5)$ is the $W$-dependence: $\sigma_{o}(W)\sim 1/W^{n}$\,.

\hspace*{3mm}
Fig. 8b.\, The ratio\, $\sigma_0(K_SK_S)/\sigma_0(K^+K^-)$ versus $W$\,\,\, \cite{Chen}.
The dotted line DKV (Diehl-Kroll-Vogt) is the valence handbag model prediction in the $~SU(3)$ symmetry limit \cite{DKV} (see sect.3 below); the dashed and dash-dotted lines are the leading term QCD predictions (for sufficiently large energy $W$) from Brodsky-Lepage \cite{BL-81} ( with $\phi_K(x)=\phi^{\rm asy}(x)$) and from Benayoun-Chernyak \cite{Maurice-Ch} ( with $\phi_K(x)$ from \cite{czz-82}).
\vspace*{1mm}

\newpage
\begin{minipage}[c]{.5\textwidth}\hspace*{-0.8cm}\includegraphics
[trim=0mm 0mm 0mm 0mm, width=0.8\textwidth,clip=true]{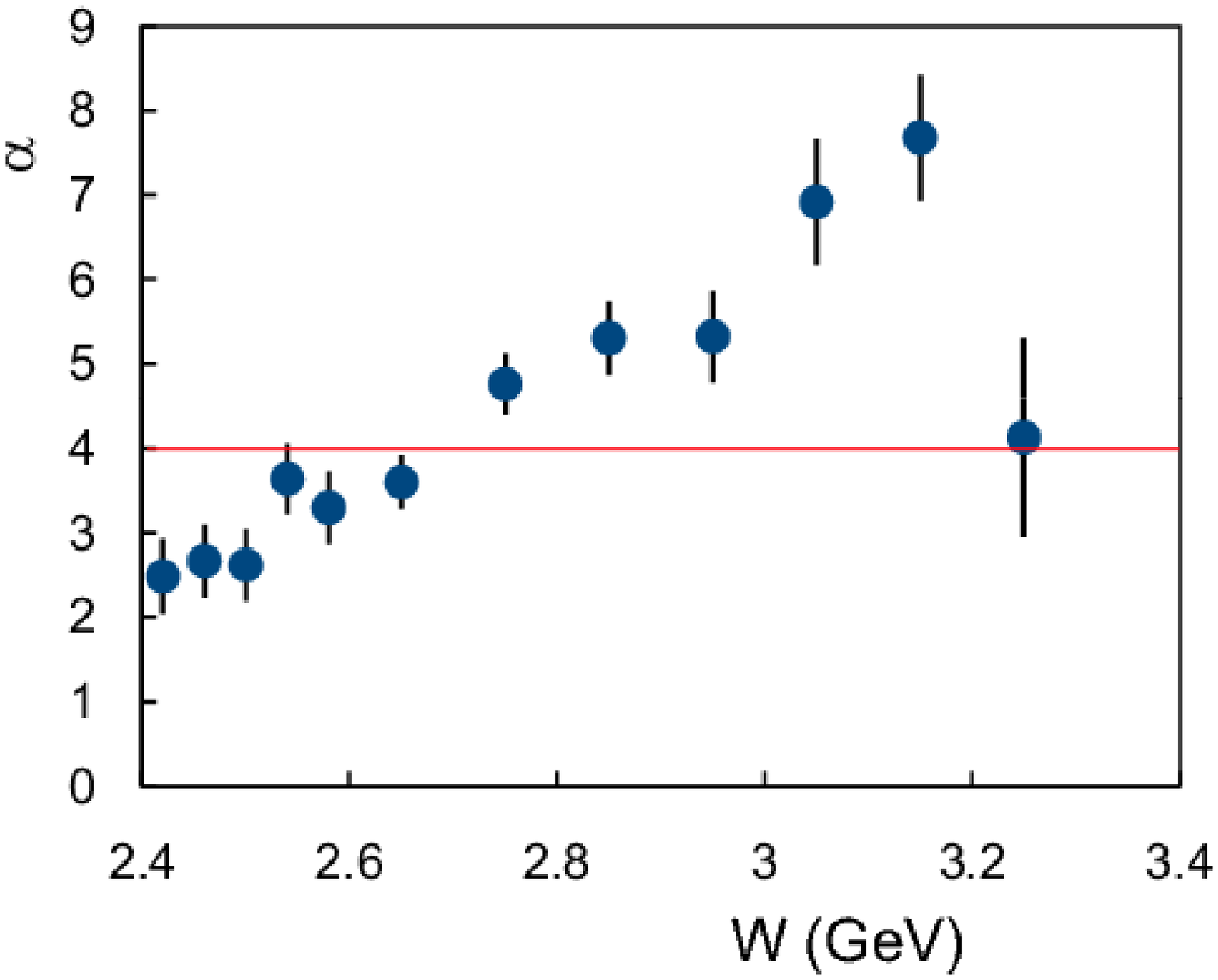}
\end{minipage}~
\begin{minipage}[c]{.5\textwidth}\hspace*{-2.5cm}\includegraphics
[trim=0mm 0mm 0mm 0mm, width=1.15\textwidth,clip=true]{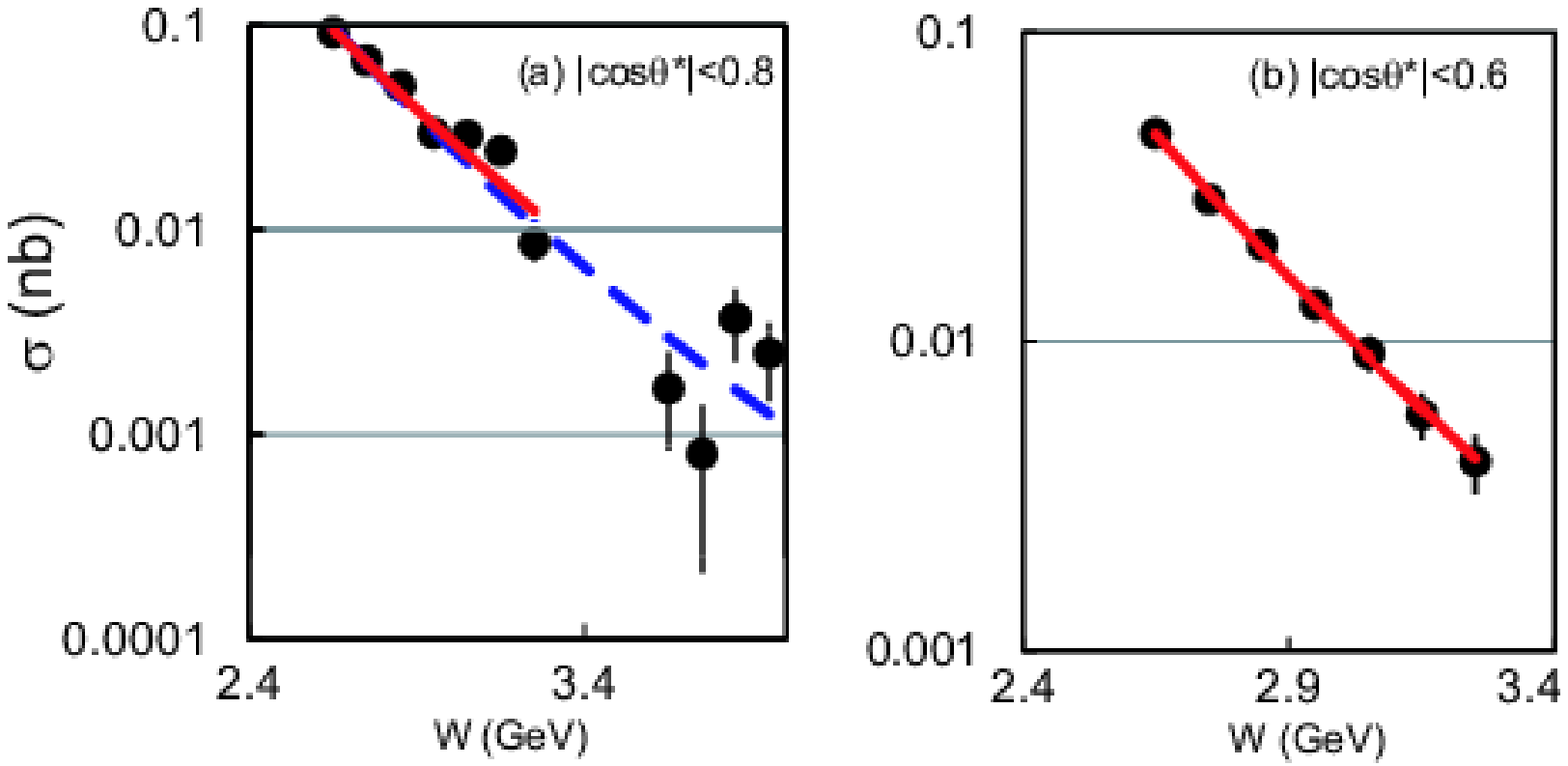}
\end{minipage}
\vspace*{1mm}

Fig.8c (left) \cite{Ue-13}.\, $W$ dependence of the parameter $\alpha$, which characterizes the angular dependence $d\sigma(\gamma\gamma\to K_SK_S)/d\cos\theta\sim (1/\sin\theta)^{\alpha}$ of the differential cross section. The horizontal line at $\alpha=4$ corresponds to the claim of the handbag model \cite{DKV}.

Figs.8c (right) \cite{Ue-13}.\, Results for the cross section $\sigma(\gamma\gamma\to K_SK_S)$ integrated over $|\cos \theta|$ regions: (a$^\prime$) - below 0.8 and (b$^\prime$) - below 0.6. The $W$ dependence is fitted to $W^{-n}$ in the different $W$ regions: 2.6 -- 4.0~GeV excluding 3.3 -- 3.6~GeV (dashed line) and 2.6 -- 3.3~GeV (solid line), see Table 1.\\

\hspace*{1.5cm} The angular distributions of $K_S K_S$ measured in \cite{Ue-13} are shown in Fig.8d.\\

\begin{minipage}{0.54\textwidth}
\hspace*{-1cm}\includegraphics[width=0.7\textwidth]{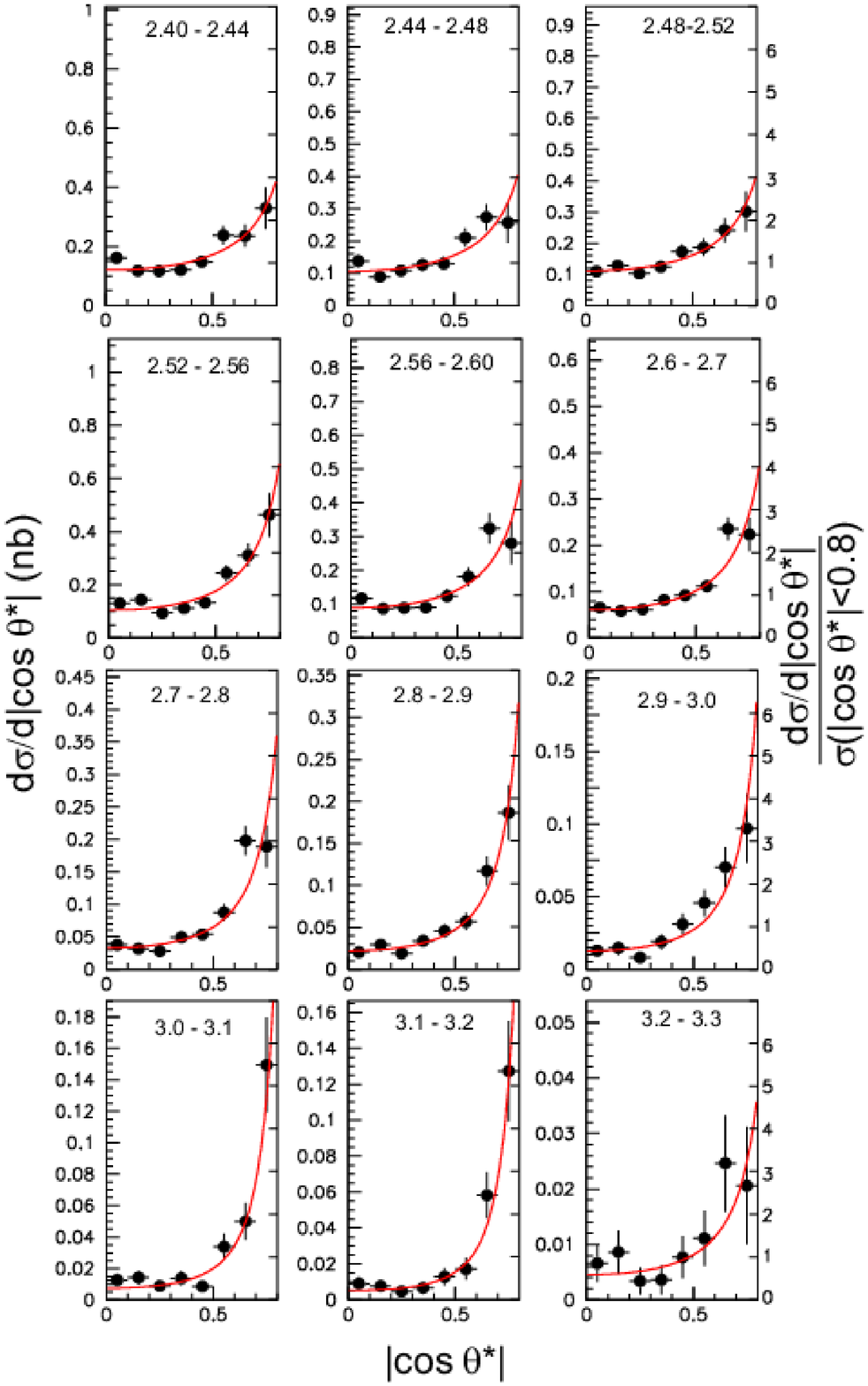}
\end{minipage}
\begin{minipage}{0.45\textwidth}

\hspace*{-2cm} Fig.8d \cite{Ue-13}.\, Data for the $\cos \theta$ dependence of the \\
\hspace*{-2.3cm} differential cross section $d\sigma(\gamma\gamma\to K_SK_S)/d\cos\theta$ \\
\hspace*{-2.3cm} and the results of the fits performed with the \\
\hspace*{-2.3cm} function proportional to  $1/\sin^\alpha \theta$ (solid curve). The \\
\hspace*{-2.3cm} numbers in each panel show the $W$ region in GeV. \\
\hspace*{-2.3cm} The left (right) vertical scale of each subfigure \\
\hspace*{-2.3cm} corresponds to the absolute scale (normalized in \\
\hspace*{-2.3cm} such a way that the average is 1.25, as described \\
\hspace*{-2.3cm} in the text \cite{Ue-13}) of the differential cross section.
\end{minipage}

\subsection{Leading term QCD vs data for $\gamma\gamma\to \pi^o\pi^o$ and $\gamma\gamma\to \eta\pi^o$}

\hspace*{3mm} The cross sections of other neutral particle productions were also measured by Belle Collaboration, in particular $\gamma\gamma\to \pi^o\pi^o$ and $\gamma\gamma\to \eta\pi^o$ \cite{Ue-pio, Ue-eta}, see Figs.9,\,10\,.\\

\begin{minipage}[c]{.5\textwidth}
\includegraphics[trim=0mm 0mm 0mm 0mm, width=0.65\textwidth,clip=true]{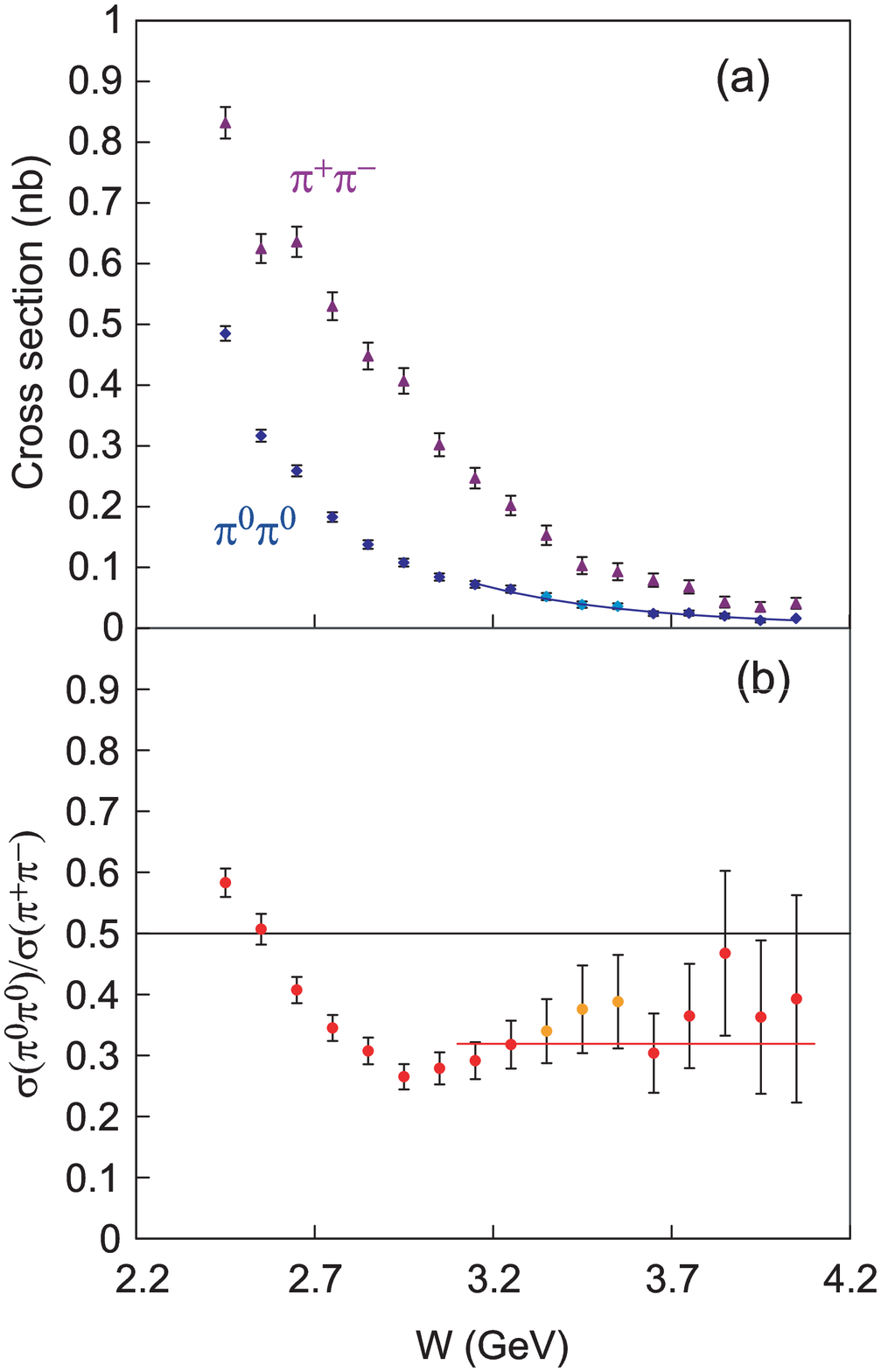}
\end{minipage}~
\begin{minipage}[c]{.5\textwidth}
\hspace*{-20mm} Fig. 9a.\quad Cross sections $\sigma_{o}(\gamma\gamma\to \pi^0\pi^0)$ \cite{Ue-pio}\\ and $\sigma_{o}(\gamma\gamma\to \pi^+\pi^-)$ for $|\cos \theta|<0.6$\,\cite{Nakaz};

\hspace*{-20mm} Fig. 9b\,\cite{Ue-pio}.\quad Their ratio.  The lines are fits to the \\ results in the
energy region indicated.\\

\hspace*{-15mm} The QCD predictions for this range of energies:
\vspace*{-2mm}
\be
\vspace*{-2mm}{\hspace*{-2cm}}\sigma(\pi^+\pi^-)\sim 1/W^6\,,\nn
\ee
{\hspace*{-3cm}} while the expected behavior of $\sigma(\pi^o\pi^o)$ if the higher twist\\
{\hspace*{-3cm}} terms are still dominant at $3<W<4\,GeV$ (and up to\\
{\hspace*{-3cm}} the odderon contribution, see below) is $~\sigma(\pi^o\pi^o)\sim 1/W^{10}$.\\

{\hspace*{-2cm}} The handbag model prediction \cite{DKV} (see below) is\,:\\
$${\hspace*{-4cm}} R=\sigma(\pi^o\pi^o)/\sigma(\pi^+\pi^-)=0.5\,$$.\nn
\end{minipage}

\begin{minipage}[c]{.45\textwidth}\hspace*{-2mm}
\includegraphics[trim=0mm 0mm 0mm 0mm, width=1.3\textwidth,clip=true]{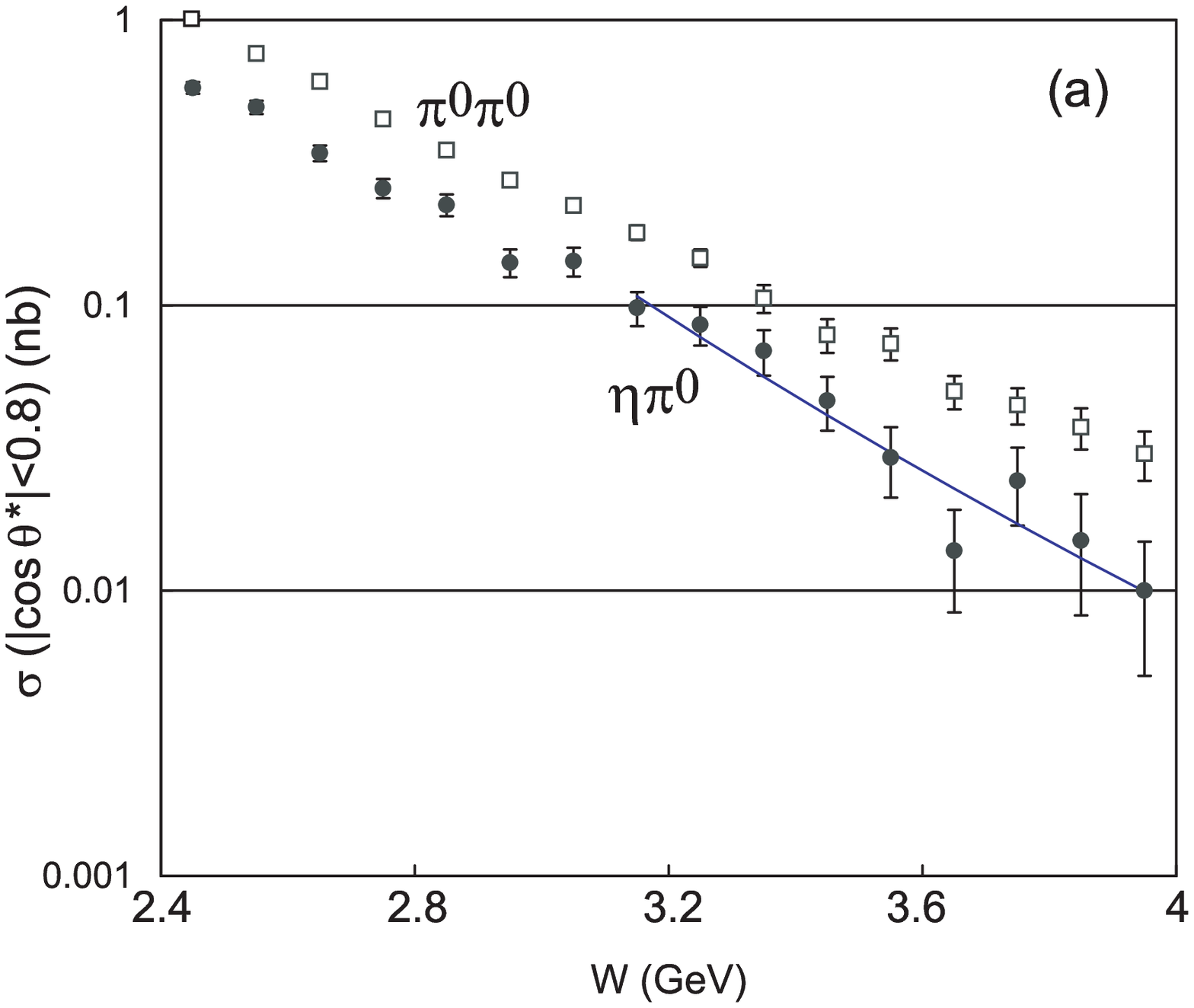}
\end{minipage}~
\begin{minipage}[c]{.55\textwidth}\hspace*{1mm}
\includegraphics[trim=0mm 0mm 0mm 0mm, width=0.85\textwidth,clip=true]{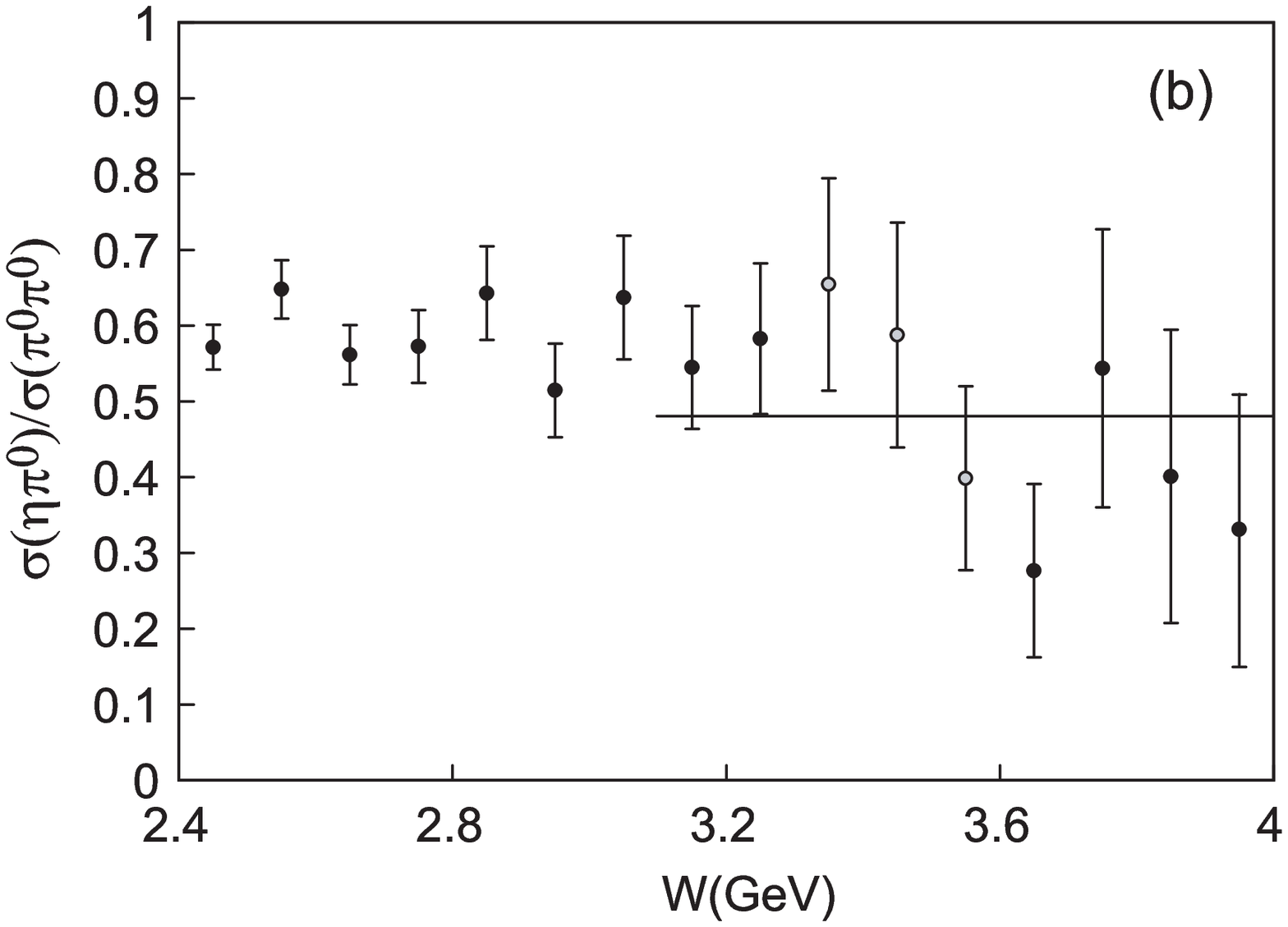}
\end{minipage}
\vspace{1mm}

Fig.10a. \quad  $W$ - dependence of  cross sections $\sigma(\gamma\gamma\ra \pi^{o}\pi^{o})$ and $\sigma(\gamma\gamma\ra \eta\pi^o )\,,\, \,\, |\cos \theta|<0.8$.\\
{\hspace*{4cm}}The power law fit \,: $\sigma(\eta\pi^o)\sim (1/W)^n,\,\,~ n=(10.5\pm 1.2\pm 0.5)$\,\,\cite{Ue-eta}\\
\vspace{0.1cm}
{\hspace*{5mm}}Fig.10b. \quad The ratio of cross sections $\sigma(\eta\pi^o)/\sigma(\pi^o\pi^o)$\,\,\cite{Ue-pio,Ue-eta}\\

\begin{minipage}[c]{.65\textwidth}\hspace*{-0cm}{\includegraphics
[trim=0mm 0mm 0mm 0mm, width=0.9\textwidth,clip=true]{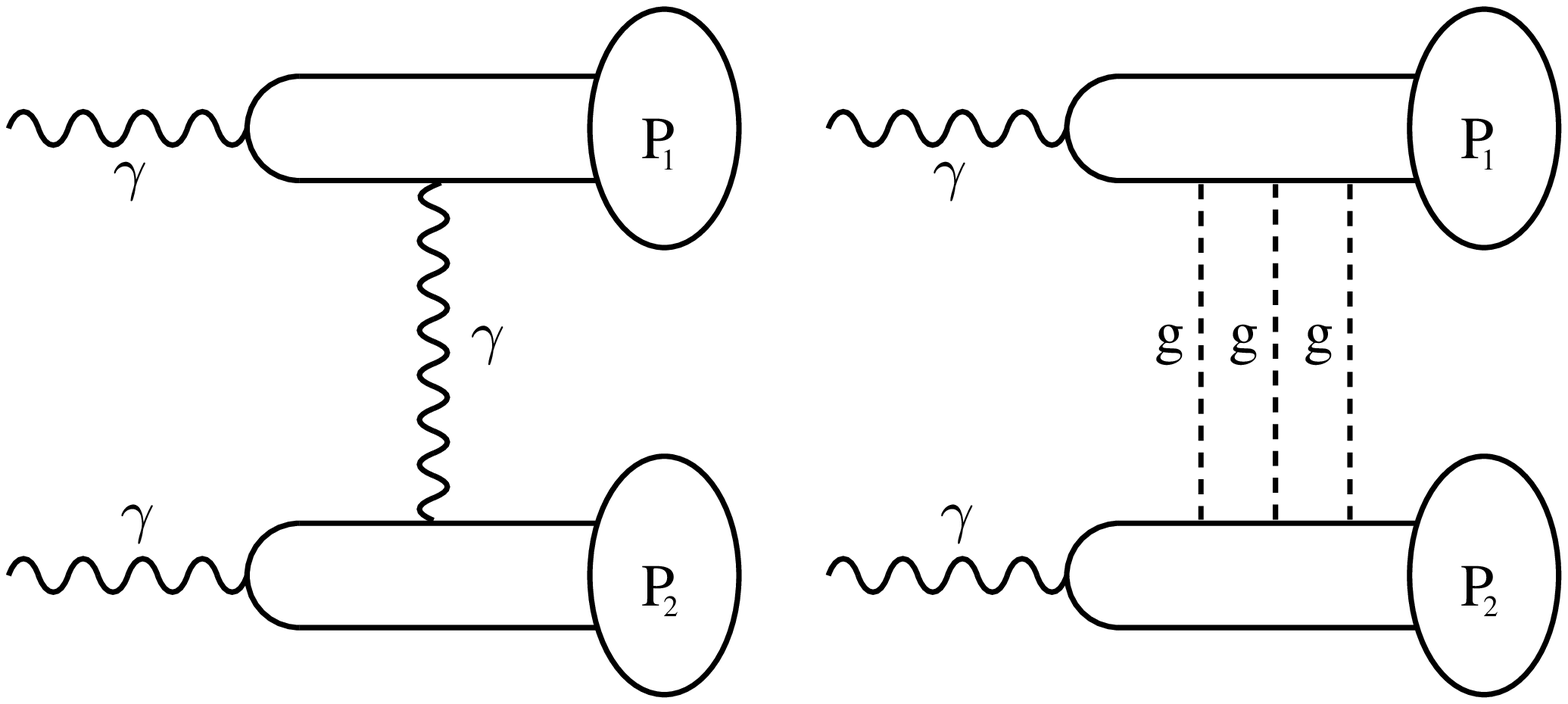}}
\end{minipage}
\begin{minipage}[c]{.35\textwidth}\hspace*{-1cm}
{Fig.11.\quad  $P_1,\, P_2=\pi^o,\,\eta,\,\eta^\prime,\,\eta_c$ }\\

\hspace*{-2cm}{\hspace*{1cm} The additional hard contributions \\\hspace*{-0.6cm}for neutral
pseudoscalar mesons}
\end{minipage}

\hspace*{3mm} The measured energy dependence of the $\pi^0\pi^0$ cross section is similar to $K_SK_S$ and
$\eta\pi^o$ cross sections at $6<W^2<9\, GeV^2$, see Figs.9b,10a, but behaves "abnormally"\, in the energy interval $9<W^2<16\,GeV^2$. In attempt to understand this "abnormal" behavior of the $\pi^0\pi^0$ cross section we can recall that, unlike the $\sigma(\ov{K^o}K^o)$ cross section, there are additional contributions to the $\sigma(\pi^o\pi^o)$ and $\sigma(\eta\pi^o)$ cross sections as shown in Fig.11 (the odderon contribution becomes the leading one at sufficiently large energies and small fixed angles).

The contribution of the diagram with the photon exchange to the amplitudes $M_{\gamma}(\gamma\gamma
\to\pi^o\pi^o)$ is readily calculated and the helicity amplitudes look as
\be
M_{\gamma}^{\pm\pm}=(4\pi\alpha)^2(e^2_u-e^2_d)^2\,\frac{2f^2_{\pi}}{s}\,\Bigl [\,\int_0^1\frac{\phi_{\pi}(x)}{x}\,\Bigr ]^2\,\Phi_{++}(\theta),\quad
\Phi_{++}(\theta)=\frac{2(1+z^2)}{(1-z^2)^2}\,,
\ee
\be
M_{\gamma}^{\pm\mp}=(4\pi\alpha)^2(e^2_u-e^2_d)^2\,\frac{2f^2_{\pi}}{s}\,\Bigl [\,\int_0^1\frac{\phi_{\pi}(x)}{x}\,
\Bigr ]^2\,\Phi_{+-}(\theta),\quad \Phi_{+-}(\theta)=\frac{1+3z^2}{(1-z^2)^2}\,,
\quad z=\cos\theta\,.\nonumber
\ee
As a result, using the pion wave function $\phi^{\rm CZ}_{\pi}(x)=30x(1-x)(2x-1)^2$ the ratio is (really,
this ratio is only weakly dependent on the pion wave function form)
\be
\frac{\sigma_{\gamma}(\pi^o\pi^o,|\cos\theta|<0.8)}{\sigma(\pi^+\pi^-,|\cos\theta|<0.8)}\simeq 1\cdot\Bigl (\frac{\alpha}{{\ov\alpha}_s}\Bigr )^2\sim 0.5\cdot 10^{-3}\,,
\ee
so that this contribution is very small and does not help.\\

The odderon contribution in Fig.11 has been calculated in \cite{Ginz-92} and looks at $s\gg |t|\gg \mu^2_o$ as
\be
M_{++}=M_{--}\simeq -2.5 M_{+-}=-2.5 M_{-+}\simeq 4\pi\alpha\Bigl (\, 4\pi\ov{\alpha}_s\,
\Bigr )^3\,\frac{5\, s f^2_{\pi}}{108\pi^2}\,\Bigl (\frac{1}{t^2}+\frac{1}{u^2}\Bigr )\, I_{\pi\pi}\nonumber
\ee
\be
\hspace*{3cm} I_{\pi\pi}=\int_{-1}^{1}
d\xi_1\,\frac{\phi_{\pi}(\xi_1)}{(1-\xi^2_1)}\int_{-1}^{1}d\xi_2\,\frac{\phi_{\pi}(\xi_2)}{(1-\xi^2_2)}\,
T_{\pi\pi}(\xi_1,\xi_2)\,,\quad \int^{1}_{-1}d\xi\, \phi_{\pi}(\xi)=1\,,
\ee
\be
T_{\pi\pi}(\xi_1,\xi_2)=\Biggl [\,\ln \Big |\frac{\xi_1+\xi_2}{1-\xi_1}\Big |\ln \Big |\frac{\xi_1+\xi_2}{1+
\xi_2}\Big |+(\xi_1\ra -\xi_1)\,\Biggr ],\quad\xi_1=x_1-x_2\,,\quad \xi_2=y_1-y_2\,.\nonumber
\ee
The numerical value of $I_{\pi\pi}$ in (2.11) is $I^{(CZ)}_{\pi\pi}\simeq 26.8$ \cite{Ginz-92} for $\phi_{\pi}(\xi)
=\phi^{\rm CZ}_{\pi}(\xi)$. Therefore, with $\phi_{\pi}(\xi)=\phi^{\rm CZ}_{\pi}(\xi)$\,:
\be
\frac{d\sigma^{(3\,\rm gl)}}{dt}(\gamma\gamma\to\pi^o \pi^o)\simeq 0.7\,nb\,GeV^6 \Bigl (\,\frac{{\ov\alpha}^{\,2}_s}{0.1}\,\Bigr )^3 \Bigl (\frac{1}{{t^2}}+\frac{1}{u^2} \Bigr )^2\,.
\ee
It is seen from (2.12) that $\sigma^{(3\,\rm gl)}(\pi^o\pi^o,\,|\cos\theta|<0.8)$ is dominated by the regions $|t|\simeq |t_{\rm min}|$ and $|u|\simeq |u_{\rm min}|$. We obtain with ${\ov\alpha}_s\simeq 0.3$\,:
\be
\sigma^{(3\,\rm gl)}(\pi^o\pi^o,\,|\cos\theta|<0.8)\simeq \left\{\begin{array}{l l l}
\,\, 23\cdot 10^{-2}\,nb &  [\,\rm experiment:\, 30\cdot 10^{-2}\,nb\,]\,\, {\rm at}\,\, W=3\,GeV
\\
\,\, 9\cdot 10^{-2}\,nb & [\,\rm experiment:\,\,\, 9\cdot 10^{-2}\,nb\,] \,\, {\rm at}\,\,\,\, W=3.5\,GeV
\\
\,\, 4\cdot 10^{-2}\,nb & [\,\rm experiment:\,\, 3.5\cdot 10^{-2}\,nb\,] \,\, {\rm at}\,\,\,\, W=4\,GeV
\\
\end{array}\right.
\ee

Hence, according to these estimates with $\phi_{\pi}(\xi)=\phi^{\rm CZ}_{\pi}(\xi)$, the odderon
contribution is sufficiently large and may well be responsible for a change of the behavior of  $\sigma(\pi^o\pi^o)$ at $W>3$\,GeV. At the same time the numerical value of $I_{\pi\pi}$ in (2.11) for $\phi_{\pi}(\xi)=\phi^{\rm asy}(\xi)$ is $I^{(asy)}_{\pi\pi}\simeq 7.4$, and so the value of $\sigma^{(3\,\rm gl)}(\pi^o\pi^o,\,|\cos\theta|<0.8)$ with $\phi_{\pi}(\xi)=\phi^{\rm asy}(\xi)$ will be $\simeq 13$ times smaller.\\

In the $SU(3)$ symmetry limit $\sigma^{(3\,\rm gl)}(\eta_8\pi^o)/\sigma^{(3\,\rm gl)}(\pi^o\pi^o)=2/3$. To estimate the effects of $SU(3)$ symmetry breaking we use the same model wave function of $\eta$ as those used in \cite{ch-09}, i.e. $|\eta\rangle=\cos\, \phi|n\rangle-\sin\phi\,|s\rangle,\, |n\rangle=|({\ov u}u+{\ov d}d)/\sqrt 2\rangle, \,|s\rangle=|{\ov s}s\rangle$, and with taking into account the $SU(3)$ symmetry breaking effects distinguishing $|n\rangle=|({\ov u}u+{\ov d}d)/\sqrt 2\,\rangle$ and $|{\ov s}s\rangle$ wave functions (let us recall: the heavier is quark the narrower is wave function). Then, instead of $\phi^{\rm CZ}_{\pi}(\xi_1)$ in (2.11) one has to substitute
\be
\phi^{\rm CZ}_{\pi}(\xi_1)\ra \phi_{\eta}(\xi_1)=\Bigl [\,\frac{\cos\phi}{3}\,\phi^{\rm CZ}_\pi(\xi_1)+\frac{f_s}{f_\pi}\frac{\sqrt 2\,\sin\phi}{3}\,\phi^{\rm asy}(\xi_1)\,\Bigr ]\,.
\ee

Then, with $\phi\simeq 38^o$ and $f_s/f_\pi\simeq 1.3$, see \cite{FKS-1,FKS-2,ch-09},
instead of $I^{(CZ)}_{\pi\pi}=26.8$ with $\phi_\pi(\xi_1)=\phi^{\rm CZ}_\pi(\xi_1)$, the corresponding integral in (2.11) will be $I_{\pi\eta}\simeq 11$ and $\sigma^{(3\,\rm gl)}(\pi^o\eta)/\sigma^{(3\,\rm gl)}(\pi^o\pi^o)\simeq 1/3$. It seems, this additional suppression may be a reason why the odderon contribution is still not seen clearly in $\sigma(\pi^o\eta)$ at $3<W<4\,GeV$ and $|\cos\theta|<0.8$, see Fig.10. The prediction is that it will be seen at somewhat higher energies.\\

The energy dependences of various cross sections measured and fitted by the Belle Collaboration are collected in Table 2.\\
\vspace*{1mm}

\begin{minipage}[c]{.99\textwidth}
\begin{center}
Table 2. \,\, The value of \, "n" \, in $\sigma_{\rm tot}\sim 1/W^{n}$ in
various reactions \\ fitted in the $W$ and $|\cos\theta|$ ranges indicated
\end{center}
\hspace*{2.5mm}
\begin{tabular}{l|c|c|c|c|c|c} \hline \hline
Process & n - experiment & $W$ range (GeV) & $|\cos\theta|$ & n - QCD & n - handbag & Ref\\
\hline\hline
$\pi^+\pi^-$ & $7.9 \pm 0.4 \pm 1.5$ & $3.0 - 4.1$ & <0.6 & $\simeq 6$ & $\simeq 10$ & \cite{Nakaz} \\
\hline
$K^+K^-$  & $7.3 \pm 0.3 \pm 1.5$ & $3.0 - 4.1$ & <0.6 & $\simeq 6$ & $\simeq 10$ & \cite{Nakaz} \\
\hline
$K^0_S K^0_S$  & $10.5 \pm 0.6 \pm 0.5$ & 2.4 -- 4.0 & <0.6 & $\simeq 10$ & $\simeq 10$ &\cite{Chen} \\
\hline
$K^0_S K^0_S$  & $11.8 \pm 0.6 \pm 0.4$ & 2.6 -- 3.3 & <0.6 & $\simeq 10$ & $\simeq 10$ &\cite{Ue-13} \\
\hline
$K^0_S K^0_S$  & $10.0 \pm 0.5 \pm 0.4$ & 2.6 -- 3.3 & <0.8 & $\simeq 10$ & $\simeq 10$ &\cite{Ue-13} \\
\hline
$K^0_S K^0_S$  & $11.0 \pm 0.4 \pm 0.4$ & 2.6 -- 4.0 & <0.8 & $\simeq 10$ & $\simeq 10$ &\cite{Ue-13} \\
\hline
$\eta \pi^0$ & $ 10.5 \pm 1.2 \pm 0.5 $ & 3.1 -- 4.1 & <0.8 & $\simeq 10$ & $\simeq 10$ &
\cite{Ue-eta} \\
\hline
$\pi^0\pi^0$ & $\simeq 10$ & 2.5 -- 3.0 & <0.8 & $\simeq 10$  & $\simeq 10$ &
\cite{Ue-pio} \\
\hline
$\pi^0\pi^0$ & $8.0 \pm 0.5 \pm 0.4$\,\, & 3.1 -- 4.1 & <0.8 & $\simeq 6$  & $\simeq 10$ &
\cite{Ue-pio}\\
\hline
$\pi^0\pi^0$ & $\simeq 10$\,\, & 2.5 -- 3.0 & <0.6 & $\simeq 10$  & $\simeq 10$ &
\cite{Ue-pio}\\
\hline
$\pi^0\pi^0$ & $6.9 \pm 0.6 \pm 0.7$\,\, & 3.1 -- 4.1 & <0.6 & $\simeq 6$  & $\simeq 10$ &
\cite{Ue-pio}\\
\hline
$\eta\eta$ & $7.8\pm 0.6\pm 0.4$ & 2.4 -- 3.3 & <0.8 & $\simeq 10$ & $\simeq 10$ &
\cite{Ue-ee}\\
\hline\hline
\end{tabular}
\end{minipage}

\section{Leading term QCD vs handbag model}

\hspace*{3mm} The handbag model \cite{DKV} is a definite application of the general idea which {\it assumes} that present day energies are insufficient for the leading terms QCD to be the main ones. Instead,  it is supposed that the soft nonperturbative contributions {\it dominate} the amplitudes. More definitely, the handbag model is a generalization of the old Feynman mechanism which assumes that at intermediate scales the amplitudes are dominated by the end-point contributions in which one common quark in both hadrons carries nearly whole hadron momenta, while all other quarks are wee partons. Therefore, the leading contributions originate in the Feynman mechanism from the hand-bag like diagrams, see Fig.12a. The handbag model \cite{DKV} tries to realize applications of this Feynman mechanism to a description of the large angle cross sections $d\sigma(\gamma \gamma\ra {\ov M}M)$ at present day energies $W<4\,GeV$.

As it is formulated in \cite{DKV}, the handbag model assumes that the hard QCD contributions described above in sect.2.2 are really dominant at very high energies only, while the main contributions at present energies originate from the Fig.12a diagram.\\

\begin{minipage}{1.\textwidth}
\hspace*{3cm}\includegraphics[width=0.6\textwidth]{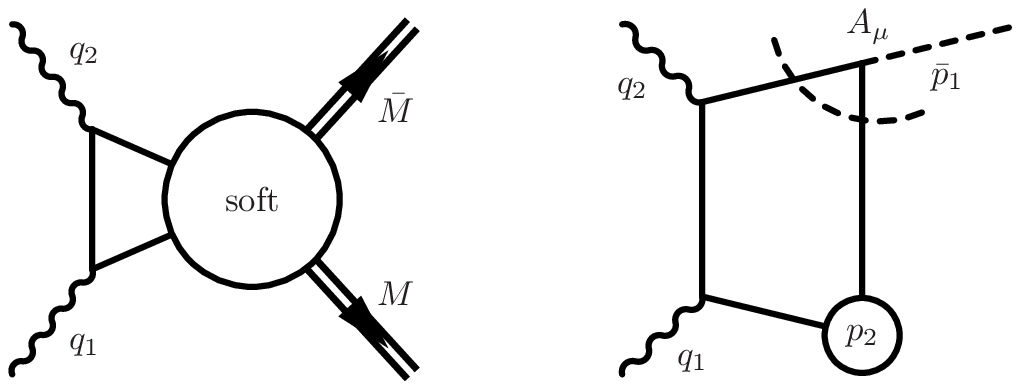}
\end{minipage}
\vspace{0.1mm}

Fig.12a.\quad The overall picture of the handbag model contribution \cite{DKV}\\
\vspace{1mm}
\hspace*{3mm} Fig.12b.\quad  The standard lowest order Feynman diagram for the light cone QCD sum rule \\
to calculate the soft valence handbag amplitude ${M^{\rm val}_{\rm handbag}(\gamma\gamma\to \pi^+\pi^-)}$\,
\cite{cher-06} ($A_\mu={\ov u}\gamma_\mu\gamma_5 d$)\\
\vspace*{1mm}

Here, {\it the two photons interact with the same quark only, and these two active ${\ov q}$ and $q$ quarks carry nearly the whole meson momenta, while the additional passive\, ${\ov q^\prime}$ and $q^\prime $ quarks are wee partons which are picked out from the vacuum by soft nonperturbative interactions} \cite{DKV}. Therefore, these soft form factors $R_{MM}(s)$ will be power suppressed in QCD at sufficiently large $s\,: R_{MM}(s) \sim 1/s^2\,,\, $ in comparison with the leading meson form factors, $F_{M}(s)\sim 1/s\,$. But, nevertheless, it is assumed that they are {\it numerically dominant} at present energies for both charged and neutral mesons.\\

The energy dependence and the absolute normalization of the handbag amplitude $M_{\rm handbag}(\gamma\gamma\\\ra {\ov M}_2 M_1)$ is not predicted in \cite{DKV} but fitted to the data. As for the angular dependence, it was also not really predicted in \cite{DKV} in a model independent way. The reason is that a number of special approximate relations were used in \cite{DKV} at intermediate steps to calculate the angular dependence of the handbag amplitude. All these relations were valid, at best, for the leading term only. But it turned out finally that their would be leading term gives zero contribution to the amplitude, and the whole answer is due to next power corrections, $\sim \Lambda_{QCD}^2/s$, which were not under control in \cite{DKV}. The "result"\, $M_{\rm handbag}(\gamma\gamma\to M_2 M_1)\sim 1/\sin^{2}\theta$ for the handbag amplitude in \cite{DKV} is completely due to the one especially (and arbitrary) chosen definite power suppressed term in the amplitude while ignoring all other power corrections of the same order of smallness.

The authors were fully aware of this arbitrariness \cite{DKV}: {\it "We  must then at this stage consider our result $M_{\rm handbag}\sim 1/\sin^{2}\theta$ as a model or a partial calculation of the soft handbag contribution"}.

Hence, finally, the approach in \cite{DKV} predicts neither energy and angular dependences nor the normalization of cross sections in a model independent way.

Therefore, what only remains are the specific predictions of the handbag model for the ratios of cross sections in the $SU(3)$ symmetry limit\,: {\it there is only one common valence handbag amplitude} $M^{\rm val}_{\rm handbag}$ (the soft non-valence handbag amplitudes are small, see below)\,:
\be
M(\pi^+\pi^-)=M(\pi^o\pi^o)=M(K^+K^-)=\frac{5}{2}\, M(\ov{K^o} K^o)=\frac{5}{2}\, M(K_S K_S)=M^{\rm val}_{\rm handbag}\,,
\ee
\be
\frac{5}{\sqrt 3}\, M(\pi^o\eta_8)=\frac{5}{\sqrt 6}\, M(\pi^o\eta_o)=\frac{5}{3}\, M(\eta_8\eta_8)=\frac{5 }{2}\, M(\ov{K^o} K^o)=M^{\rm val}_{\rm handbag}\,.\nonumber
\ee

Due to these relations, the predictions of the handbag model for the ratios of cross sections in comparison with the data look as ( all ratios in (3.2), except for ${\sigma (K^+K^-)}/{\sigma (\pi^+\pi^-)}=1$, are specific predictions of the valence handbag model)\,:
\be
\frac{\sigma (K^+K^-)}{\sigma (\pi^+\pi^-)}=1\,(0.89 \pm 0.04 \pm 0.15)_{\rm exp}\,,\quad \frac{\sigma (\pi^o\pi^o)}{\sigma (\pi^+\pi^-)}=\frac{1}{2} \,(0.32\pm 0.03\pm 0.05)_{\rm exp}\,,
\ee
\be
\hspace*{-0.5cm}\frac{\sigma (K_S K_S)}{\sigma (K^+K^-)}=0.08\,\,(0.13\,\ra\,0.01,\, \rm{see\, Fig.6})_{\rm exp}\,, \quad\frac{\sigma (K_S K_S)}{\sigma(\pi^o\eta_8)}=\frac{2}{3}\,\,
(\,\sim 0.1\,)_{\rm exp}\,\,, \nonumber
\ee
\be
\frac{\sigma (\eta_8\eta_8)}{\sigma (\pi^o\pi^o)}= 0.36\, (0.37 \pm 0.04)_{\rm exp}\,,\quad \frac{\sigma (\pi^o\eta_8)}{\sigma (\pi^o\pi^o)}=0.24\,(0.48 \pm 0.06)_{\rm exp}\,.\nonumber
\ee
\vspace*{0.5mm}

Recalling that the angular dependence of the handbag amplitude $M_{\rm handbag}\sim 1/\sin^2 \theta$ used in \cite{DKV} was a model form, it looks not so surprising that the explicit calculation of the valence handbag amplitude $M^{\rm val}_{\rm handbag}(W,\theta)$ in \cite{cher-06} (see also \cite{ch-09}) via the light cone QCD sum rules \cite{Balitsky,Br-Fil-89,cz-90} gave a different angular dependence, $M^{\rm val}_{\rm handbag}\sim {\rm const}$. These soft valence handbag contributions to the cross sections calculated explicitly from the light cone QCD sum rules in \cite{cher-06}, see Fig. 12b, {\it are definite functions of the energy and scattering angle}, and look as
\be
\frac{d\sigma_{\rm handbag}(\gamma\gamma\to {\ov M}_2  M_1)}{d\cos\theta} \sim \frac{{\rm const}}{W^{10}}
=\frac{{\rm const}}{s^{5}}
\ee
for all mesons, both charged and neutral.

Unfortunately, this angular behavior, $d\sigma_{\rm handbag}(\gamma\gamma\to {\ov M}_2  M_1)/d\cos\theta\sim \rm const$, disagrees with all data which behave similar to $\sim 1/\sin^{4}\theta$, and the energy behavior  $\sigma_{\rm handbag}(\gamma\gamma\to {\ov M}_2  M_1)\sim 1/W^{10}$ disagrees with the data for charged mesons $\pi^+\pi^-$ and $K^+ K^-$, compatible with $\sim 1/W^6$.

The above energy behavior $\sim 1/W^{10}$  is as expected (up to Sudakov effects from loop corrections) in QCD for {\it soft} valence power corrections to the leading terms due to the {\it Feynman end-point mechanism}.
\footnote{\,
But one should remember that there is also a number of {\it hard} valence power corrections in QCD with the same energy dependence $\sim 1/W^{10}$ as in (3.3), but with possibly different angular dependences.
}
\vspace*{-2mm}\hspace*{1.5cm}
\begin{minipage}[c]{0.6\textwidth}
\hspace*{-2cm}\includegraphics[trim=0mm 0mm 0mm 0mm, width=0.99\textwidth,clip=true]{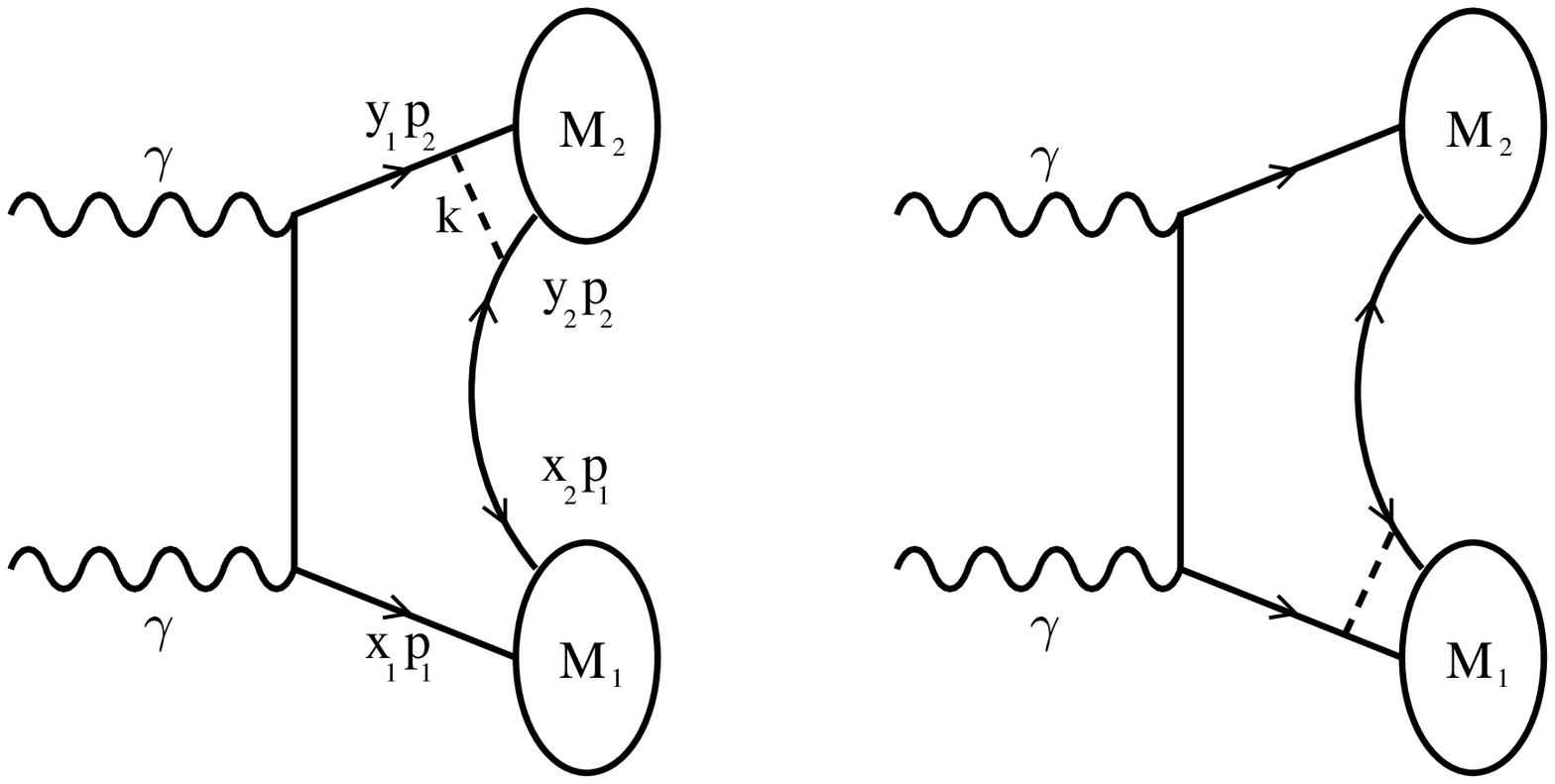}
\end{minipage}
\vspace*{0.5mm}
\begin{minipage}[c]{0.4\textwidth}
\hspace*{-3cm} Fig.13.\quad The diagrams to calculate the energy \\\hspace*{-3cm} and angular dependences of soft handbag\\ \hspace*{-3cm} amplitudes (the valence Feynman mechanism)
\end{minipage}

The end-point region contributions of only two diagrams shown in Fig.13 (the valence Feynman mechanism) are relevant for the standard valence handbag model as it is formulated in\,\,\,\cite{DKV}\,:
\be
x_1,\, y_1\ra 1,\,\,\,\quad k^2=x_2 y_2 s\sim \Lambda^2_{QCD}\,\,
\ra\,\, x_2\sim y_2\sim \delta=(\Lambda^2_{QCD}/s)^{1/2}=\Lambda_{QCD}/W\ll 1\,.
\ee
The direct calculation of these diagrams in Fig.13 gives for the hard kernel \cite{ch-12}
\be
T^{(lead)}=const\, \frac{f^2_{\pi}}{s}\,\, \frac{(e_1 e_2)(x_1+y_1)+2 x_1 y_1}{x_1 x_2 y_1 y_2} \quad
\ra \quad const\, \frac{f^2_{\pi}}{s}\,\, \frac{(e_1 e_2)+1}{x_2 y_2}\,,
\ee
($e_1$ and $e_2$ are the photon polarization vectors), and for the soft end-point region contributions to the whole valence handbag amplitude
\be
M_{\rm handbag}^{\rm val,\,\pm\mp}\sim \frac{f^2_{\pi}}{s}\Biggl [\int_0^{\delta}d x_2\Bigl (\frac{\phi_{\pi}(x)\sim x_1 x_2}{x_2}\sim const\Bigr )\Biggr ]^2\sim \frac{f^2_{\pi}}{s}\,\delta^2\sim \frac{f^2_{\pi}\Lambda^2_{QCD}}{s^2=W^4}\gg M_{\rm handbag}^{\rm val,\,\pm\pm}\,.
\ee

Therefore, the angular and energy dependences  of all cross sections resulting from these soft valence endpoint contributions (the Feynman mechanism) look as
\be
d\sigma_{handbag}/d\cos\theta\sim |M|^2/W^2\sim {\rm const}/W^{10}\,.
\ee

The expressions (3.6)-(3.7) agree with the predictions of the light cone QCD sum rules (3.3) \cite{cher-06} not only in the energy and angular dependences but also in dependences on photon helicities.\\

If we are interested only in the energy dependence of such soft end-point region contributions, there is
a simpler way to obtain it which does not require the direct calculation of Feynman diagrams. This can be done as follows \cite{ch-12}.

1)\,\, There is the hard part of any such diagram and it is the amplitude of the annihilation of two photons into a pair of active near mass-shell quarks with each one carrying nearly the whole meson momenta, $A_{\rm hard}(\gamma\gamma\to {\ov q} q)$. From the dimensional reasons it is: $A_{\rm hard}(\gamma\gamma\to {\ov q} q)\sim 1$.

2)\,\, All other parts of the Feynman diagrams are soft and, parametrically, depend on the scale $\Lambda_{QCD}$ only. So, the energy dependence of the soft valence end-point region contributions (i.e. the valence Feynman mechanism) looks here as follows ($\,\phi_2(x_1,x_2)\sim x_1 x_2,\,\, x_1+x_2=1,\, x_1, y_1\ra 1,\, 0\leq x_2, y_2\leq \delta\,$)
\be
R^{(\rm v,\,2)}_{MM}(W)\sim M_{2,\,\rm endpoint}(\gamma\gamma\to {\ov M} M)\sim \int_0^{\delta}dx_2\,\phi_2(x)\int_0^{\delta}dy_2\,\phi_2(y)\Bigl [A_{\rm hard}(\gamma\gamma\to {\ov q} q)\sim 1 \Bigr ]\sim \nonumber
\ee
\be
\sim \Bigl [\,\int_0^{\delta}dx_2\, x_2\int_0^{\delta}dy_2\, y_2\,\Bigr ]^{\rm n_{\rm wee}=1}\sim \Bigl (\,\delta^4\,\Bigr )^{\rm n_{\rm wee}=1}\sim\Lambda_{QCD}^4/W^4\,.
\ee

This method of obtaining the energy dependence of soft end-point contributions will be used below to calculate the energy dependence of {\it soft non-valence} handbag form factors originating from the 4-particle components of meson wave functions (see also Section 5).\\

The updated predictions of the handbag model for the $\gamma\gamma\to {\ov M_2}M_1$ cross sections were given in the next paper of the same authors \cite{Kroll-09}. In comparison with the original paper \cite{DKV}, the main new element in \cite{Kroll-09} is that (in the $SU(3)$ symmetry limit used in both \cite{Kroll-09} and \cite{DKV}) the sizable soft non-valence form factor $R^{\,\rm nv}_{\ov M M}(s)$ is introduced now, in addition to the soft valence one $R^{\,\rm v}_{\ov M M}(s)$ (the soft non-valence contributions were neglected previously in \cite{DKV}). Both functions, $R^{\,\rm v}_{{\ov M}M}(s)$ and $R^{\,\rm nv}_{{\ov M}M}(s)$, are parameterized then in arbitrary forms with a number of free parameters which are fitted in \cite{Kroll-09} to the data.
\footnote{\,
The form factors $R^{\,\rm u}_{2\pi}(\rm s)$ and $R^{\,\rm s}_{2\pi}(s)$ used in \cite{Kroll-09} are connected with those from \cite{ch-09} as: $R^{\,\rm u}_{2\pi}(s)=R^{\,\rm v}_{2\pi}(s)+R^{\,\rm nv}_{2\pi}(s)$,\,\, $R^{\,\rm s}_{2\pi}(s)=R^{\,\rm nv}_{2\pi}(s)$\,.\\
}
\\

As for the soft valence contributions to the cross sections and the soft valence form factors $R^{\,\rm v}_{{\ov M}M}(s)$, these were estimated numerically in \cite{cher-06} via the standard light cone QCD sum rules and were found smaller (and with the expected suppressed power behavior $R^{\,\rm v}_{{\ov M}M}(s)\sim 1/s^2$) than the values fitted to data in \cite{DKV} and \cite{Kroll-09}.\\

As for the non-valence contributions, the two types of such contributions are presented in Fig.14\, \cite{ch-09}.\\

\begin{minipage}[c]{1.1\textwidth}\hspace*{0.5cm}\includegraphics
[trim=0mm 0mm 0mm 0mm, width=0.6\textwidth, clip=true]{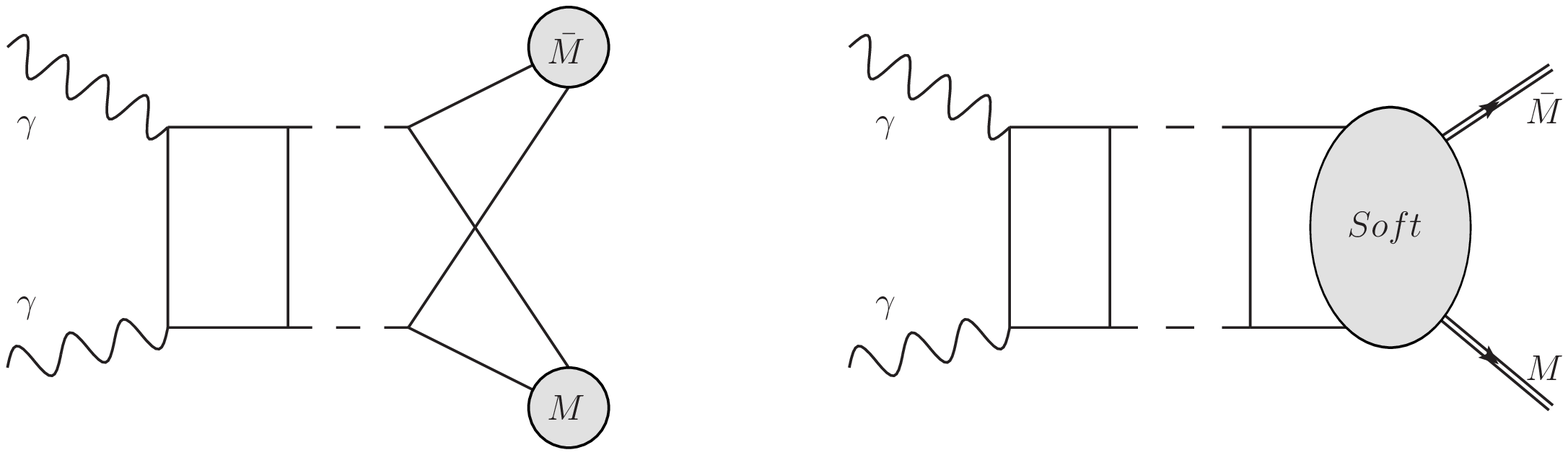}
\end{minipage}~
\vspace*{1mm}

\hspace*{2.5mm} Fig.14a\,. \quad The leading power {\it hard} non-valence one-loop correction. \\
\vspace*{1mm}
\hspace*{7mm} Fig.14b\,. \quad The leading contribution to the {\it soft} non-valence handbag form factor $R^{\,\rm nv}_{{\ov M}M}(s)$.\\
\vspace*{1mm}
\hspace*{15mm} The solid and dashed lines represent quarks and gluons.\\
\vspace*{3mm}

It is worth noting that both non-valence contributions in Fig.14 are $SU(3)$-flavor singlets in the $SU(3)$-symmetry limit. So, they contribute equally to the amplitudes $\pi^+\pi^-,\,
\pi^0\pi^0,\,K^+K^-,\, \ov{K^0}K^0$ and $\eta_8\eta_8$, and do not contribute to $\eta_8\pi^0$.\\

The diagrams in Fig.14a constitute a small subset of all one-loop corrections to the leading power contributions from the Born diagrams like those shown in Fig.2\,. If these leading power one-loop non-valence corrections to the Born contributions were really significant, this will contradict then the data on $K_SK_S$,\, see Fig.8\,.

In particular, this hard non-valence one-loop correction was calculated, among all others, in \cite{DN}. Its contribution to the cross section $\sigma(\gamma\gamma\to K^+K^-)$ (integrated over $|\cos\theta|<0.6$, and with $\phi_K(x)=\phi^{\rm asy}(x)$ ) is \cite{Dupl}\,:
\be
\frac{\delta\sigma^{\rm nv}}{\sigma} \simeq -\frac{{\ov\alpha}_s}{3\pi}\simeq -3\%\,,
\ee
i.e., its contribution to the amplitude is: $\delta{\ov A}^{\,\rm nv}/\,{\ov A}(K^+K^-)\simeq -1.5\%$.
The leading term amplitude $|{\ov A}(K_S K_S)|\simeq 0.15\, |{\ov A}(K^+K^-)|$,\, see Fig.8. Hence, the rough estimate of this non-valence one loop correction to the ${\ov A}(K_S K_S)$ amplitude is\,: $|\delta{\ov A}^{\,\rm nv}/\,{\ov A}(K_S K_S)|\simeq 10\%$.\\

As for the soft non-valence handbag form factor $R^{\,\rm nv}_{{\ov M}M}(s)$, it seems sufficient to say that the leading contribution to it originates first from the Fig.14b two-loop correction (without large logarithms) \cite{ch-09}, so that an estimate looks as\,:
\be
\frac{R^{(\rm nv,\,2)}_{M M}(s)}{R^{\,\rm v}_{M M}(s)}\sim\Biggl (\frac{{\ov\alpha}_s}{\pi}\Biggr )^2\sim 10^{-2}\,.
\ee

Besides, there are also soft non-valence contributions from the 4-quark components of the meson wave functions. For instance, the typical contribution of the pion 4-quark components, $|\pi^+\rangle_4\sim |
({\ov s}s+{\ov u}u+{\ov d}d)\,{\ov d}u\rangle$, is shown in Fig.15. The energy dependence of such contributions can be obtained in the same way as for the diagram in Fig.13 \cite{ch-12}:

\begin{minipage}[c]{.5\textwidth}
\includegraphics[trim=0mm 0mm 0mm 0mm, width=0.75\textwidth,clip=true]{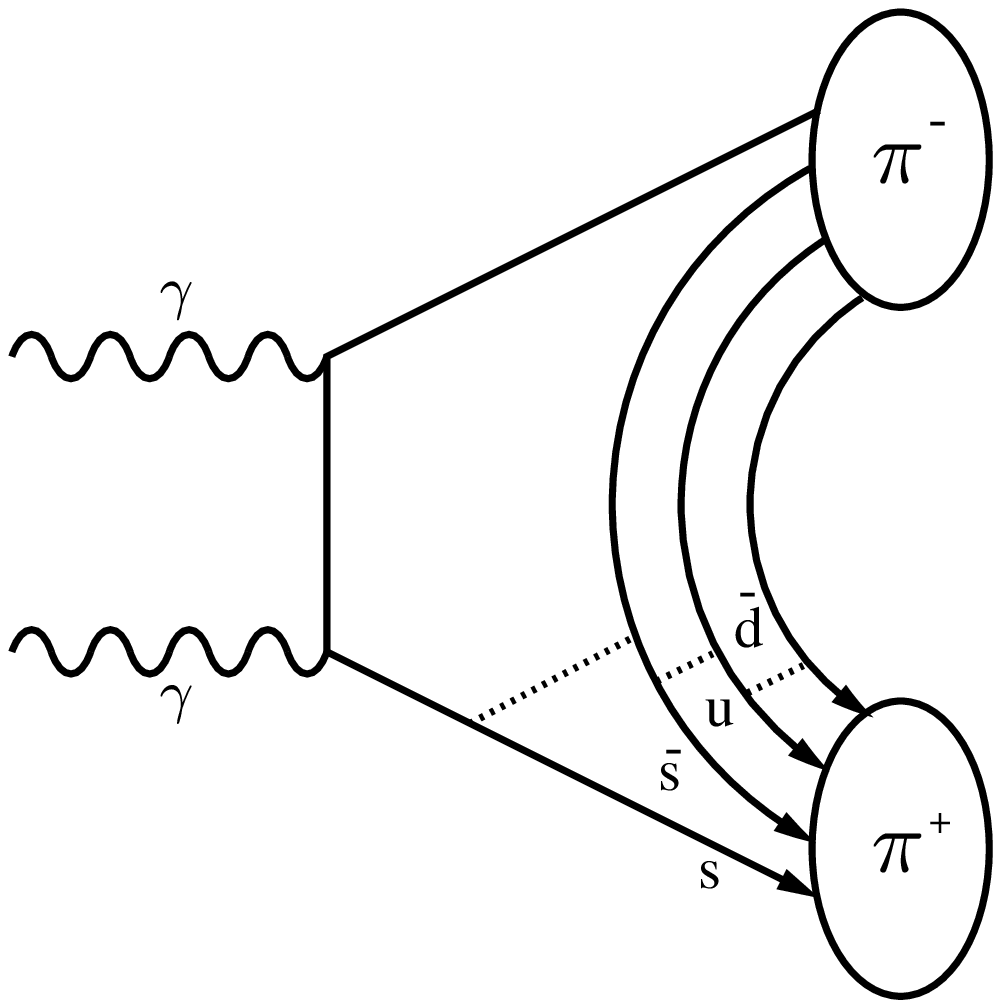}
\end{minipage}
\begin{minipage}[c]{0.4\textwidth}\vspace*{-5mm}

Fig.15 \cite{ch-12}.\quad The additional contribution $R^{(\rm nv,\,4)}_{2\pi}(W)$ to the soft non-valence handbag form factor. Here, one strange quark in $\pi^+$ and $\pi^-$ carries nearly the whole pion momentum, while three other quarks are wee partons ($\,\phi_4(x)\sim x_1 x_2 x_3 x_4,\,\sum x_i=1,\, \, x_1\ra 1,\,\,\, 0\leq x_{2,3,4}\leq\delta\sim \Lambda_{QCD}/W$)
\end{minipage}
\vspace*{1mm}
\be
R^{(\rm nv,\,4)}_{2\pi}(W)\sim M_{4,\,\rm endpoint}(\gamma\gamma\to\pi^+ \pi^-)\sim\int_0^{\delta}d x_2 d x_3 d x_4\,\phi_4(x)\int_0^{\delta}d y_2 d y_3 d y_4\,\phi_4(y)\Bigl [A_{\rm hard}(\gamma\gamma\to{\ov s}s)\sim 1 \Bigr ] \nonumber
\ee
\be
\sim \Bigl [\,\int_0^{\delta}dx\, x\int_0^{\delta}dy\, y\,\Bigr ]^{\rm n_{\rm wee}=3}\sim \Bigl (\,\delta^{4}\,\Bigr )^{n_{\rm wee}=3}\sim\Lambda_{QCD}^{12}/W^{12}\,, \nonumber
\ee
\be
\hspace*{-2cm}
\frac{R^{(\,\rm nv,\,4)}_{2\pi}(W)}{R^{\,\rm v}_{2\pi}(W)}\sim \Bigl (\frac{1\,GeV^2}{W^2}\Bigr )^{4}\sim 1.5\cdot 10^{-4}\,\, \,\,{\rm at}\,\,s=W^2=9\,GeV^2\,.
\ee

Clearly, so small soft non-valence contributions, $R^{(\rm nv,\,2)}_{M M}/R^{\,\rm v}_{M M}\sim 10^{-2}$
and $R^{(\rm nv,\,4)}_{M M}/R^{\,\rm v}_{M M}\sim 10^{-4}$ at $W^2\simeq 9\,GeV^2$, can be safely neglected and will not help.\\
\vspace*{1mm}

In comparison, $|R^{\,\rm nv}_{2\pi}(W^2=9\, GeV^2)|/|R^{\,\rm v}_{2\pi}(W^2=9\, GeV^2)|\simeq 0.3$
was used in \cite{Kroll-09} to fit the data.\\

\section{The cross sections \boldmath{$\gamma\gamma\ra V_1^o V_2^o$}}

The contribution of diagrams like Fig.2 to the amplitude $\gamma\gamma\to\rho^o\rho^o$ is also strongly suppressed numerically in comparison with $\gamma\gamma\to\rho^+\rho^-$. But there is the additional hard leading twist one-loop contributions to $\gamma\gamma\to\rho^o_L \rho^o_L$ ($L$ means the helicity zero state), see Fig.16, and it looks at $s\gg |t|,|u|$ as \cite{cz-83, GPS-87}\,:
\be
M_{++}=M_{--}\simeq -2M_{+-}=-2M_{-+}\simeq i\,s f^2_{\rho}\Bigl (\frac{1}{t^2}+\frac{1}{u^2}\Bigr )\,
(4\pi\alpha)\Bigl (4\pi{\ov\alpha}_s\Bigr )^2\frac{4}{9\pi}\,I_{\rho\rho}\,,
\ee
\be
I_{\rho\rho}=\int_{-1}^{1} d\xi_1\,\frac{\phi_{\rho}(\xi_1)}{(1-\xi^2_1)}\int_{-1}^{1}d\xi_2
\,\frac{\phi_{\rho}(\xi_2)}{(1-\xi^2_2)}\, T_{\rho\rho}(\xi_1,\xi_2)\,,\quad \quad \int^{1}_{-1}d\xi\, \phi_{\rho}(\xi)=1\,,\nonumber
\ee
\be
T_{\rho\rho}(\xi_1,\xi_2)=\frac{1}{2}\xi_1\xi_2 \ln \Bigl |\frac{\xi_1+\xi_2}{\xi_1-\xi_2}\Bigr |\,\,,
\quad \xi_1=x_1-x_2=2x_1-1,\quad \xi_2=y_1-y_2=2y_1-1\,\,\,.\nonumber
\ee
We obtain from (4.1) with $\ov\alpha_s\simeq 0.3$\,:
\footnote{\,
the number in (4.2) is for the $\rho$-meson wave function (4.5)
}
\be
\frac{d\sigma}{dt}(\gamma\gamma\to\rho^o_L \rho^o_L)\simeq 0.56\,nb\,GeV^6
\Bigl (\frac{1}{t^2}+\frac{1}{u^2}\Bigr )^2\,.
\ee
It is seen from (4.2) that $\sigma(\rho\rho,|\cos\theta|\leq 0.8)$ is dominated for this Pomeron contribution by the regions $|t|\simeq |t_{\rm min}|$ and $|u|\simeq |u_{\rm min}|$, similarly to the Odderon case (2.12). From (4.2)\,:
\be
\sigma(\rho^o\rho^o,|\cos\theta|\leq 0.8)\simeq\left\{\begin{array}{l l l}
\,\, 34\cdot10^{-2}\,nb &   \,\, {\rm at}\quad W=3\,GeV \\
\,\, 13\cdot10^{-2}\,nb &   \,\, {\rm at}\quad W=3.5\,GeV \\
\,\,\,\,\, 5.6\cdot10^{-2}\,nb &   \,\, {\rm at}\quad W=4\,GeV \\
\end{array}\right.
\ee

The relative contributions to the cross sections $\sigma_o(\rho^o\omega)$ and $\sigma_o(\omega\omega)$
from Fig.16 diagram (the wave functions of the $\rho$ and $\omega$ mesons are the same) look as
\be
\sigma_o(\rho^o\rho^o): \sigma_o(\rho^o\omega): \sigma_o(\omega\omega)\simeq 1:\frac{1}{5}:\frac{1}{80}\,.
\ee
\vspace*{1mm}

To estimate the numerical contributions to the cross sections $\gamma\gamma\to V_1^o V_2^o$ from diagrams in Fig.16 we use the model leading twist $V_{L}=V_{\lambda=0}$ wave functions taken in the form \cite{czz-82}, see Fig.17,
\be
\phi_{\rho}(\xi,\mu\simeq 1\,GeV)\simeq \phi_{\omega}(\xi,\mu_o\simeq 1\,GeV)\simeq \phi^{\rm asy}(\xi)
\Bigl (1+0.2\, C^{3/2}_2(\xi)\Bigr )=\phi^{\rm asy}(\xi)\Bigl (0.70+1.5\,\xi^2\Bigr )\,,
\ee
\be
\phi_{\phi}(\xi,\mu_o\simeq 1\,GeV)\sim \phi^{\rm asy}(\xi)
\Bigl (1+0.1\, C^{3/2}_2(\xi)\Bigr )=\phi^{\rm asy}(\xi)\Bigl (0.85+0.75\,\xi^2\Bigr )\,, \nonumber
\ee
\be
\phi^{\rm asy}(\xi)=\frac{3}{4}(1-\xi^2),\quad
f_{\rho}\simeq f_{\omega}\simeq 210\,MeV\,,\quad f_{\phi}\simeq 230\,MeV\,.\nonumber
\ee

\begin{minipage}[c]{0.6\textwidth}\vspace*{-4mm}
\includegraphics
[trim=0mm 0mm 0mm 0mm, width=0.85\textwidth, clip=true]{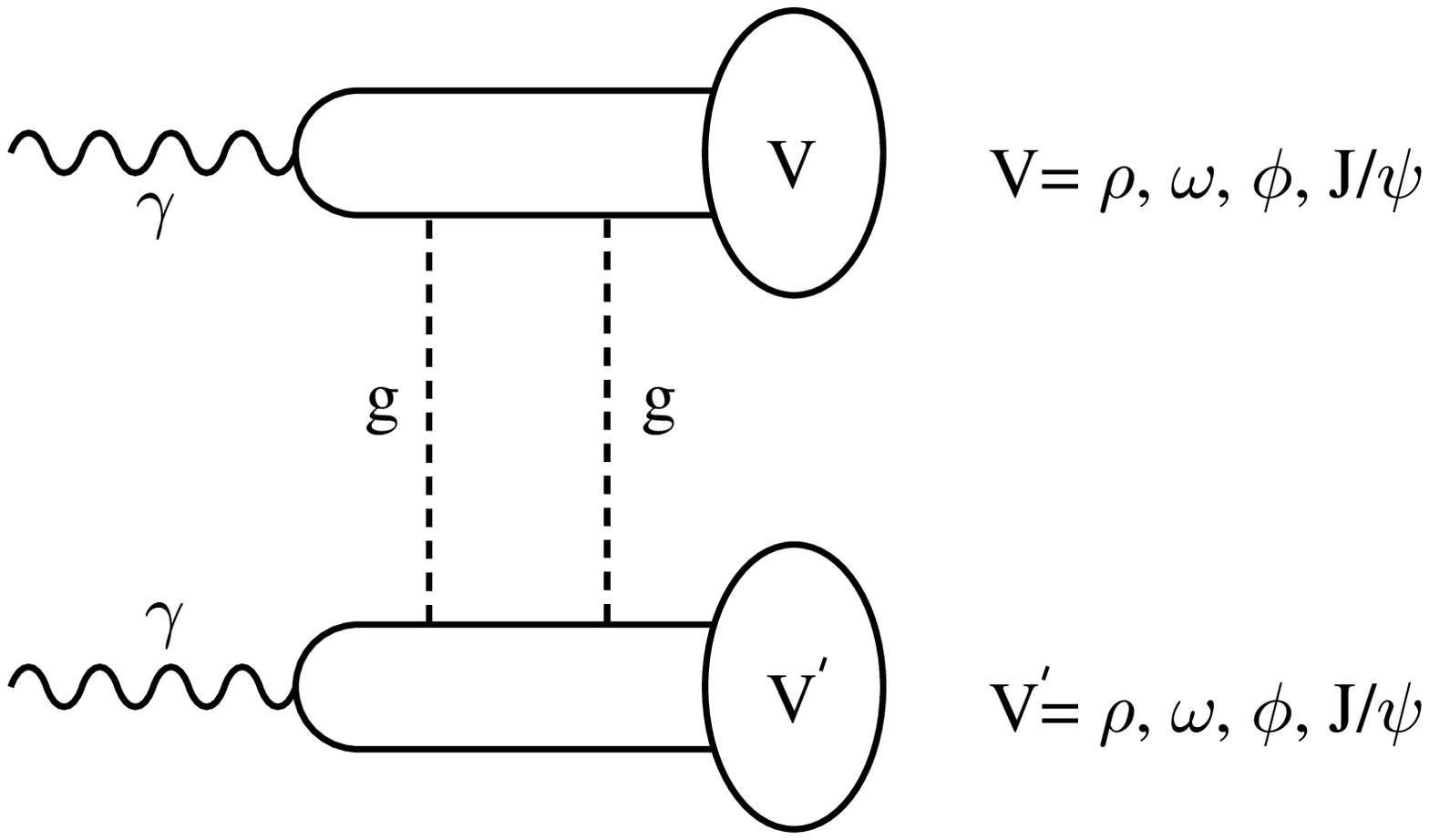}
\end{minipage}
\begin{minipage}[c]{0.4\textwidth \hspace*{-8mm}}
Fig.16.\\ The additional hard one-loop contribution to the amplitude $\gamma\gamma\ra V_1^o V_2^o$. It becomes dominant at high energies and small fixed angles for longitudinally polarized vector mesons \cite{cz-83}.
\end{minipage}

\begin{minipage}[c]{0.45\textwidth}\vspace*{3mm}
\includegraphics [trim=0mm 0mm 0mm 0mm, width=0.85\textwidth,clip=true]{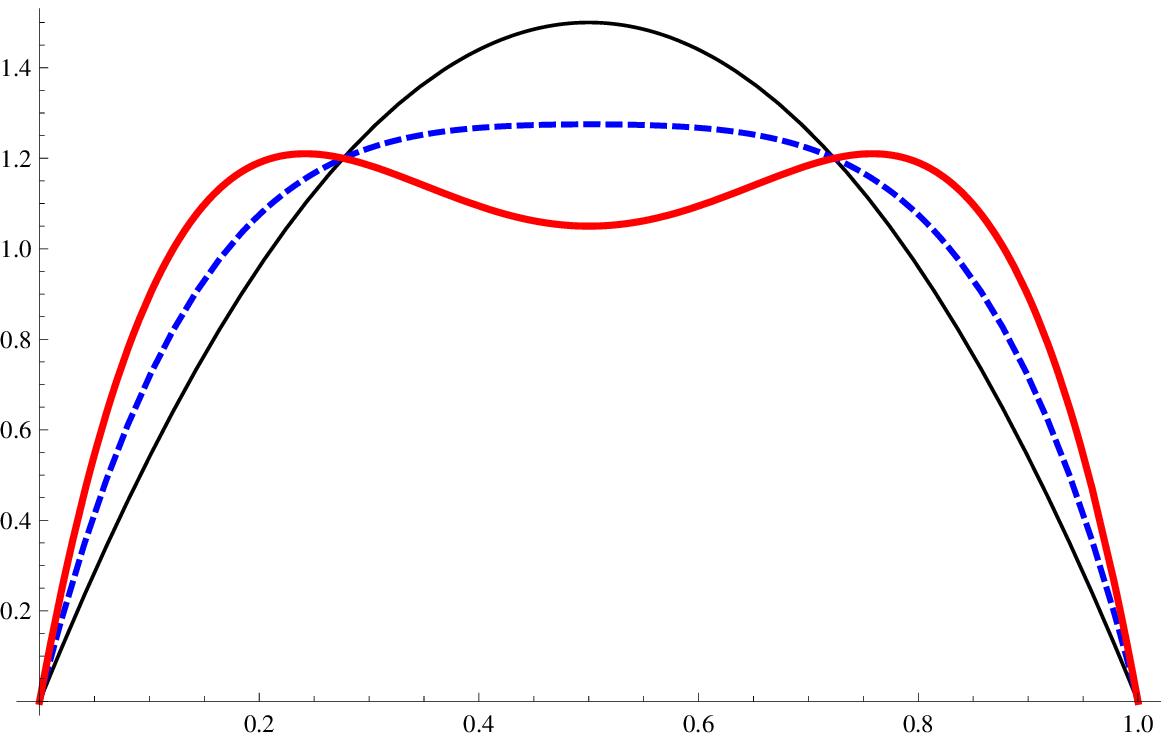}
\put (-225,50) {\rotatebox{90} {$\phi_{V}(x)$}} \put (5,3) {$x$}
\end{minipage}
\begin{minipage}[c]{0.5\textwidth \hspace*{0.5cm}}
\hspace*{5mm} Fig.17.\,\, The model wave functions. \\ Thick solid line\,:\,\,  the $\rho_L$-meson wave function.\\
Dashed line\,:\,\,  the $\phi_L$-meson wave function.\\
Thin solid line\,:\,\,  the asymptotic wave function.
\end{minipage}
\vspace*{2mm}

The cross sections $\sigma(\omega\phi),\, \sigma(\phi\phi)$ and $\sigma(\omega\omega)$ have been
measured recently by the Belle Collaboration \cite{VV-12}, see Figs.18-20.\\
\vspace*{1mm}

\begin{minipage}[c]{.6\textwidth}\vspace*{1mm}
\includegraphics [trim=0mm 0mm 0mm 0mm, width=0.9\textwidth,angle=-90,clip=true]{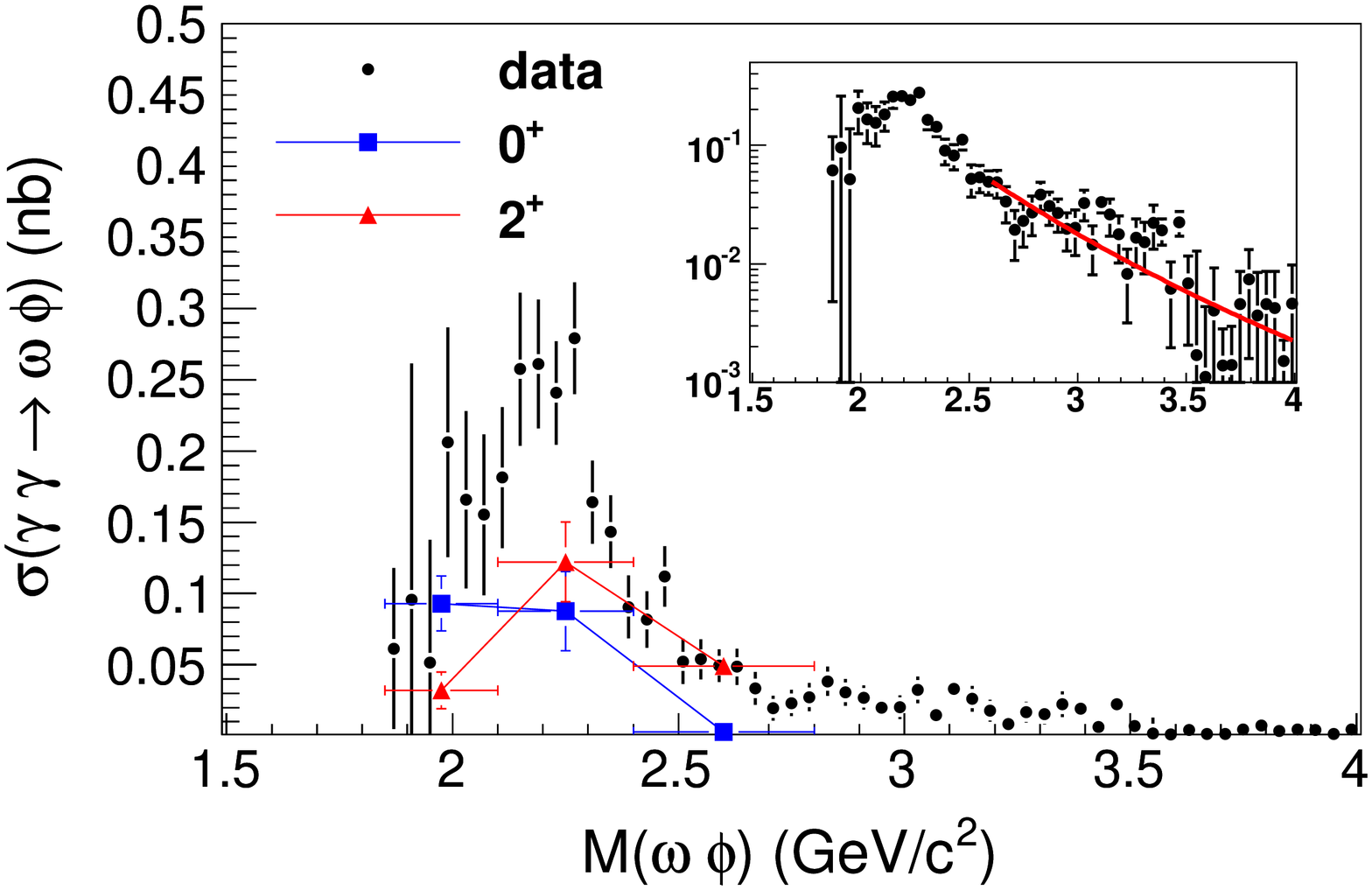}
\end{minipage}

\vspace*{6mm}\hspace*{1cm} Fig.18. \quad   The total cross section $\sigma(\omega\phi)$ in
the c.m. angular region $|\cos\theta|< 0.8$ \cite{VV-12}.\\
$M(\omega\phi)=W$-dependence\,:\, $\sigma_{\omega\phi}(W)\sim W^{-n},\,\, n=(7.2\pm 0.6_{stat}),\,\, \sigma_{\omega\phi}(W=4\,GeV)\simeq 2\cdot 10^{-3}\,nb$\,.\\
\hspace*{0.8cm} Theory:\,\, $\sigma_{\omega\phi}(W=4\,GeV,\,|\cos\theta|\leq 0.8)\simeq 2\cdot 10^{-3}\,nb$\,.

\newpage

\begin{minipage}[c]{.6\textwidth}
\includegraphics[trim=0mm 0mm 0mm 0mm, width=0.8\textwidth,angle=-90,clip=true]{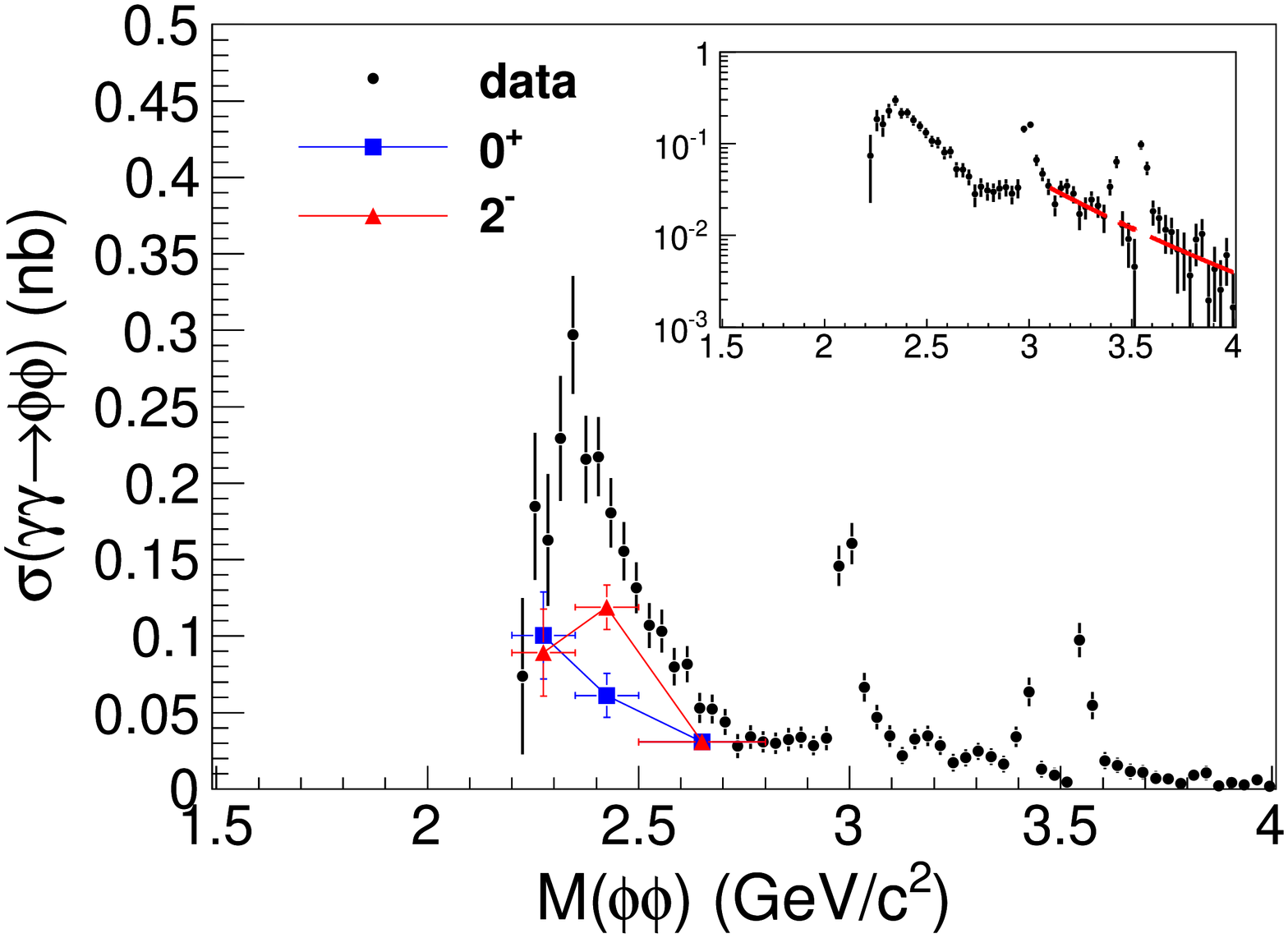}
\vspace*{1mm}
\end{minipage}~

\hspace*{1cm} Fig.19.\quad  The total cross section $\sigma(\phi\phi)$ in the c.m. angular region $|\cos\theta|< 0.8$ \cite{VV-12}.\\
W-dependence:\, $\sigma_{\phi\phi}(W)\sim W^{-n},\,\, n=(9.1\pm 0.6_{stat}),\,\,\sigma_{\phi\phi}(W=4\,GeV)\simeq 3\cdot 10^{-3}\,nb$\,.\\
\hspace*{0.6cm} ~Theory:\, $\sigma_{\phi\phi}(W=4\,GeV,\,|\cos\theta|\leq 0.8)\simeq 1.5\cdot 10^{-3}\,nb$\,.
\vspace*{5mm}

\begin{minipage}[c]{.6\textwidth}\includegraphics
[trim=0mm 0mm 0mm 0mm, width=0.8\textwidth,angle=-90,clip=true]{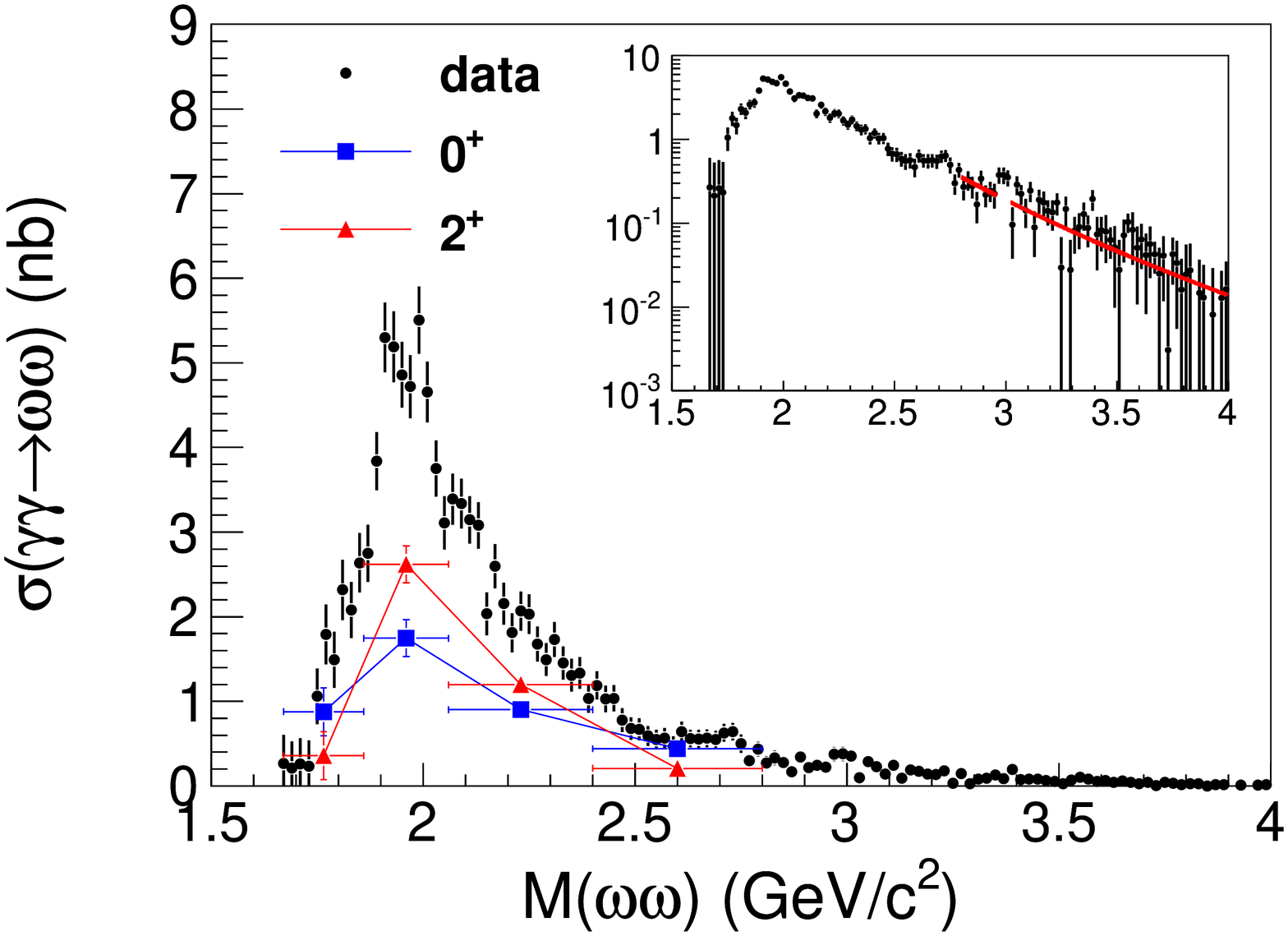}
\end{minipage}~
\vspace*{3mm}

\hspace*{1cm} Fig.20.\quad  The total cross section $\sigma(\omega\omega)$
in the c.m. angular region $|\cos\theta|< 1.0$ \cite{VV-12}.\\
W-dependence:\, $\sigma_{\omega\omega}(W)\sim W^{-n},\,\, n=(8.4\pm 1.1_{stat}),\,\,\sigma_{\omega\omega}(W=4\,GeV)\simeq 15\cdot 10^{-3}\,nb$\,.\\
\hspace*{0.8cm} Theory:\,\,$\sigma_{\omega\omega}(W=4\,GeV,\,|\cos\theta|\leq 0.8)\simeq 1\cdot 10^{-3}\,nb$\,.
\vspace*{3mm}

We obtain then from (4.1)-(4.5) at $W=4\,GeV\,,\,\, |\cos\theta|\leq 0.8\,\,:$
\be
\sigma(\omega\phi)\sim 3\,\sigma(\omega\omega)\sim 1.5\,\sigma(\phi\phi),\quad
\sigma(\rho^o\rho^o)\simeq 56\cdot 10^{-3}\,nb \,,
\ee
\be
\sigma(\omega\omega)\simeq 0.7\cdot 10^{-3}\,nb\,,\quad \sigma(\omega\phi)\sim 2\cdot 10^{-3}
\,nb,\quad
\sigma(\phi\phi)\sim 2\,\sigma(\omega\omega)\sim 1.5\cdot 10^{-3}\,nb\,. \nonumber
\ee

It is seen that there is a reasonable agreement between the predicted and measured cross sections $\sigma
(\omega\phi,\,|\cos\theta|\leq 0.8)$ and $\sigma(\phi\phi,\,|\cos\theta|\leq 0.8)$ at $W=4\,GeV$.

The much larger measured cross section \cite{VV-12}
\be
\sigma(\omega\omega,\,|\cos\theta|<1)\gg \sigma(\phi\phi,\,|\cos\theta|\leq 0.8)\sim \sigma(\omega\phi,\,|\cos\theta|\leq 0.8)
\ee
looks natural at the first sight as one expects that it is dominated by the forward region. But then it is strange that it decays so quickly, $\sigma(\omega\omega,\,|\cos\theta|<1)\sim 1/W^8$.

But the authors said \cite{LSY}\,:

"We do not observe many events in $0.9<|\cos\theta|<1.0$ angle range for $\omega\omega$. Our measured cross section in the paper is for the whole angle range in Fig.2c in the paper \cite{VV-12}.
\footnote{\,
Fig.20 here
}
If we require $|\cos\theta|<0.8$ for $\omega\omega$, the observed cross section dependence on the energy in the high energy region is similar. This is only our experimental observation. Due to very limited statistic, we cannot measure cross section for the range of $0.9<|\cos\theta|<1$ in high energy region".

Also, it is said in the article \cite{VV-12} that there are no detected events at $|\cos\theta|>0.8$ for $\omega\phi$ and $\phi\phi$.

But if, due to experimental restrictions, $\sigma(\omega\omega,\,|\cos\theta|<1)$ should be understood as
\\ $\sigma(\omega\omega,\,|\cos\theta| <0.8)$, why then is it so large\,:
\be
\sigma(\omega\omega,\,|\cos\theta|<0.8)\sim 6\Biggl (\frac{\sigma(\omega
\phi,\,|\cos\theta|<0.8)+\sigma(\phi\phi,\,|\cos\theta|<0.8)}{2} \Biggr),
\ee
while the theory predicts $\sigma(\omega\omega)<\sigma(\phi\phi)<\sigma(\omega\phi)$\,? There is no answer to this question at present.

\section{The cross sections \boldmath{$\gamma\gamma\to {\ov B} B$}}

\hspace*{3mm} The leading term QCD predictions for $\gamma\gamma\to {\ov B} B$ were calculated first in \cite{FMN,FZOZ} using the leading twist nucleon wave functions obtained from QCD sum rules in \cite{cz-84,coz} (see \cite{FMN,FZOZ} for all details). The calculated cross sections appeared  about one order smaller than data, see Fig.21. This is not so surprising as the energies $2<W=\sqrt{s}<4\,GeV$ at which these cross sections were measured look too small for the leading QCD terms to be really dominant for these processes (because only threshold energies here are $W\gtrsim 2\,GeV$). Numerically, at these energies too close to thresholds for baryon pair production, the phenomenological models have more chances to describe data. Examples of such models are the diquark model, see e.g. \cite{JKSS,Berger} and references therein, and the handbag one, see e.g. \cite{DKV-03,Kr-Schaf-05,Kroll-12,Kroll-13}. However, the qualitative difference is that the leading term QCD calculations {\it predict} cross sections without free parameters (once the leading twist baryon wave functions are determined independently from elsewhere, say, from QCD sum rules \cite{cz-84,coz}), while the phenomenological models have a number of free parameters which are {\it fitted} to data.

The basic idea of the diquark model was that two out of three quarks of the octet baryon form tightly bound diquark. Therefore, at intermediate momentum transfers, the composite nature of this hard diquark is not resolved (or resolved in part only), and the baryon behaves like a two-particle composite state (i.e. like a meson).
\footnote{\,
The diquark model introduces also the model form factors $F_{S}(Q^2)$ and $F_{V}(Q^2)$ for the scalar and vector diquarks, adjusted so that the leading term QCD behavior $F_1^{B}(Q^2)\sim 1/Q^4$ is reproduced at $Q^2\ra \infty$.
}
If so, one can naturally expect that the qualitative predictions of this model for intermediate energies will look more like those for mesons, i.e.
\be
F_1^{B,\,\rm diquark}(Q^2)\sim \frac{1}{Q^2}\,\,, \quad \frac{d\sigma (\gamma\gamma\to {\ov B}B)_{\rm diquark}}{d\cos\theta}\sim \frac{f(\theta)}{W^6}=\frac{f(\theta)}{s^3}\,,
\ee
rather than the leading QCD term behavior at sufficiently large $Q^2$
\be
F_1^{B,\,{\rm lead}}(Q^2)\sim \frac{1}{Q^4}\,\,, \quad \frac{d\sigma (\gamma\gamma\to {\ov B}B)_{\rm lead}}{d\cos\theta}\sim \frac{f(\theta)}{s^5}\,.
\ee

But it is well known that the proton form factor is sufficiently well described by the phenomenological dipole form $G_M^p(Q^2)\simeq 1/(Q^2+0.71\,GeV^2)^2$ starting with $Q^2\sim 1\,GeV^2$. Nevertheless, fitting a number of free parameters the diquark model also describes $G_M^p(Q^2)$ in the whole range $1<Q^2<35
\,GeV^2$\, \cite{JKSS}.

As for the cross section $\sigma (\gamma\gamma\to {\ov p}p)$ at $|\cos\theta|<0.6$, the data \cite{CLEO-94,CLEO-97,Venus,Opal,L3-02,Belle-p-05,Belle-p-06} for the energy behavior are shown in Fig.21. It is seen that the behavior in the lower energy region $2.2<W<2.8\, GeV$ is $\sigma\sim s^{-n},\, n\simeq 7.5$, but "n" decreases with energy and $n\simeq 6-5$ at $W>3.2\,GeV$. Therefore, there is no indication from data on the hard diquark inside the nucleon and the meson like behavior (5.1). Vice versa, the data are not in contradiction with a qualitative picture that the non-leading hard term $M^2_{o}\, b(\theta)/s^3$ in the amplitude
\be
M(\gamma\gamma\to {\ov p}p)={\it const}\Bigl (\frac{a(\theta)}{s^2}+\frac{M^2_{o}\, b(\theta)}{s^3}+\cdots \Bigr )
\ee
is larger numerically at lower energies resulting in $\sigma (\gamma\gamma\to {\ov p}p)\sim s^{-7}$, while the QCD leading term $a(\theta)/s^2$ becomes more and more important with increasing energy. In this respect, the situation is qualitatively similar to those in $\sigma (\gamma\gamma\to K_S K_S)$, see above. The reasons are different, however. In $K_S K_S$ the leading power term was small due to a zero electric charge of $K_S$, while for ${\ov B} B$ the reason rather may be the large threshold energy $\gtrsim 2\,GeV$.

As it is pointed out in \cite{Belle-p-05} (see Fig.21): "At higher energies, the data fall below the diquark predictions and exhibit a gradual approach to the three-quark model predictions \cite{FMN}. At medium energies between 2.5 and 4.0 GeV, a steeper fall of the total cross section in $W$ is observed".\\
\hspace*{3mm} The paper \cite{Belle-p-05} concludes with: "However, the diquark and handbag models were developed in order to describe the intermediate energy region at the price of introducing model form factors, etc. The disagreement of the data at $W = 2.5 - 4.0$ GeV with their predictions (see Fig.21) obviously necessitates their improvement".\\
\vspace*{1mm}

\begin{minipage}[c]{.33\textwidth}\includegraphics
[trim=0mm 0mm 0mm 0mm, width=0.95\textwidth,clip=true]{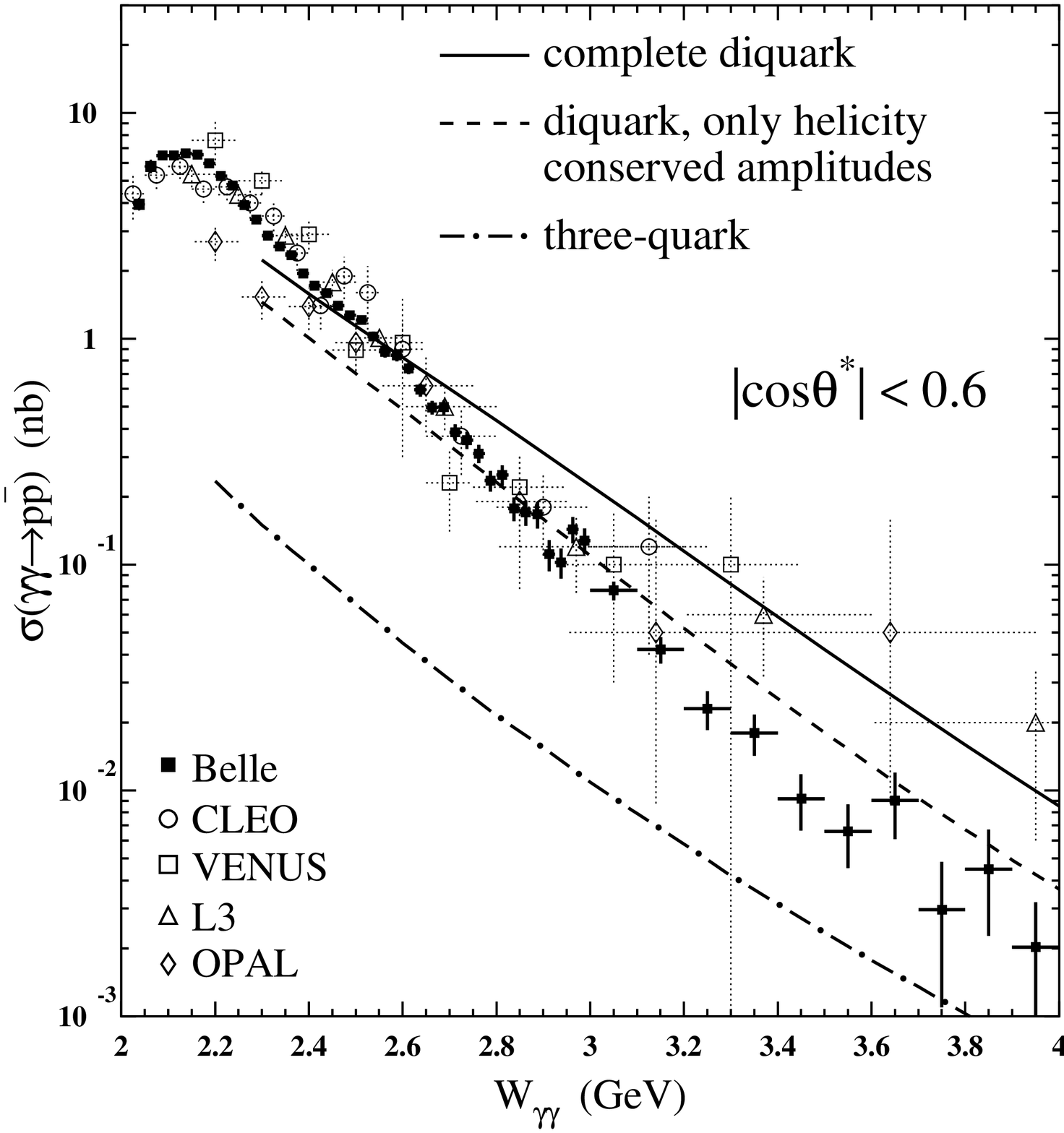}
\end{minipage}~
\begin{minipage}[c]{.33\textwidth}\hspace*{-5mm}\includegraphics
[trim=0mm 0mm 0mm 0mm, width=0.95\textwidth,clip=true]{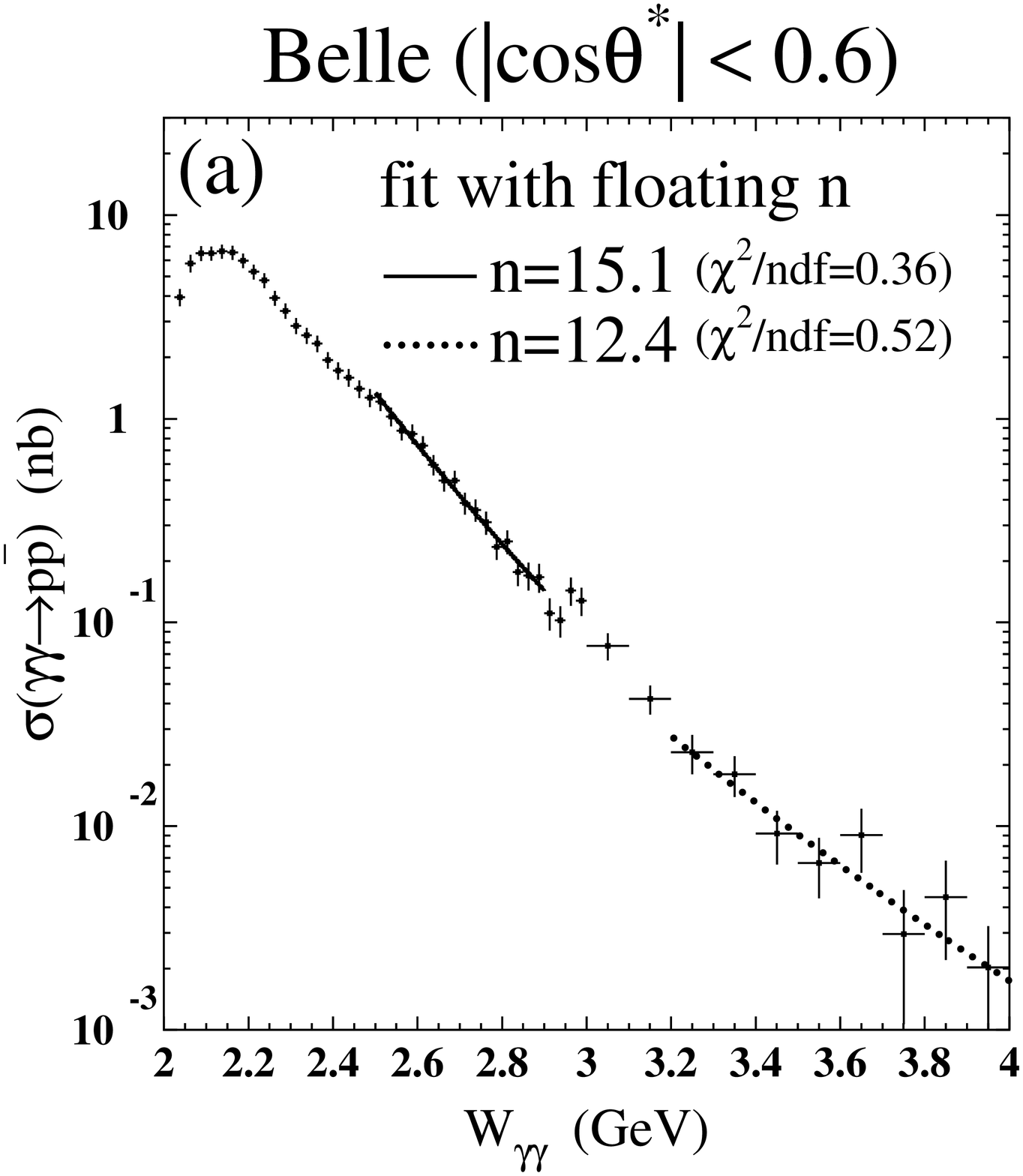}
\end{minipage}~
\begin{minipage}[c]{.33\textwidth}\hspace*{-5mm}\includegraphics
[trim=0mm 0mm 0mm 0mm, width=0.95\textwidth,clip=true]{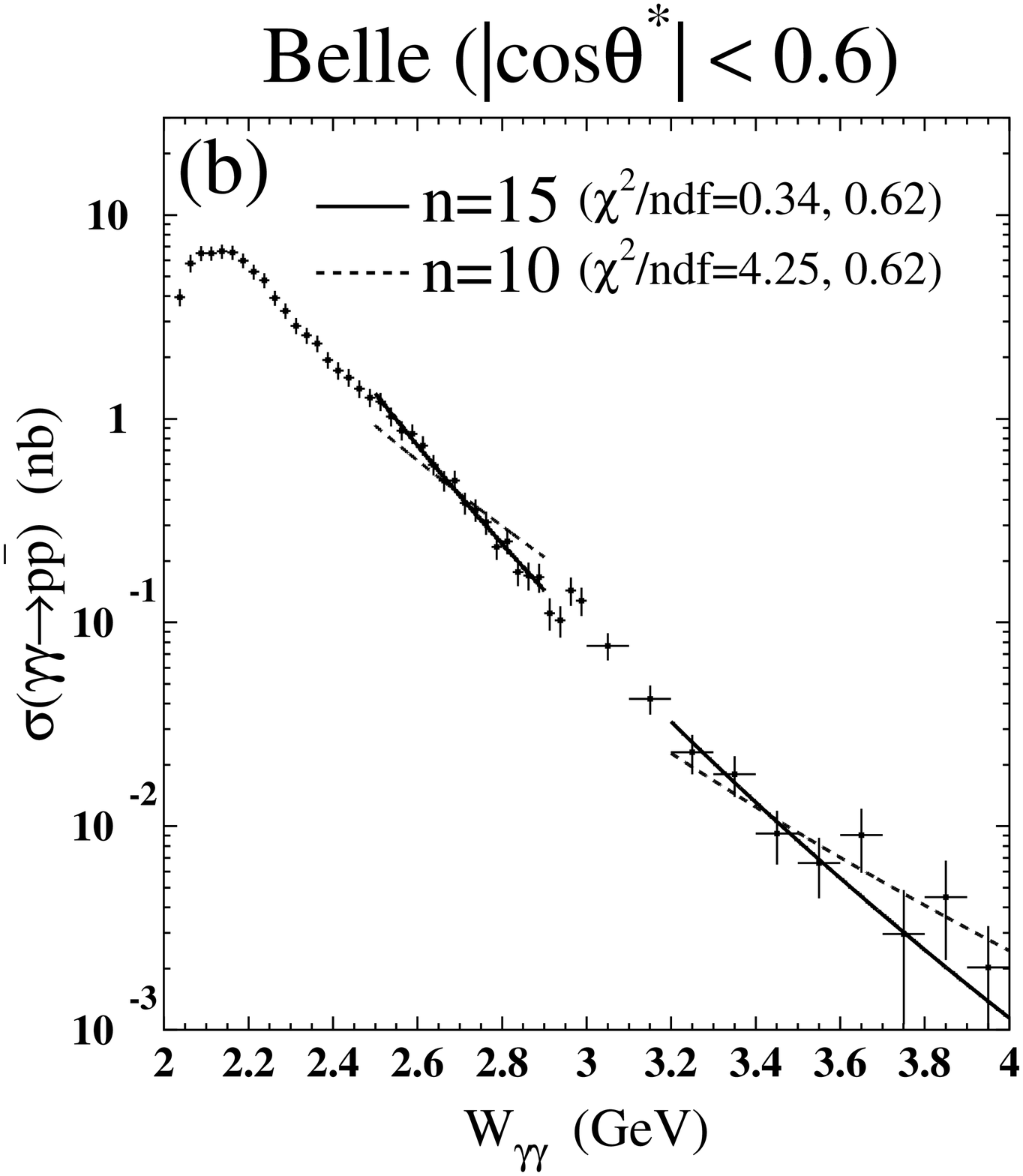}
\end{minipage}~
\vspace*{1mm}

Fig.21 (left)\, \cite{Belle-p-05}. Measured cross section $\sigma(\gamma\gamma\to{\ov p}p),\, |\cos\theta|
<0.6$. For the Belle \cite{Belle-p-05}, CLEO \cite{CLEO-94} and VENUS \cite{Venus} results, the error bars are purely statistical, while for OPAL \cite{Opal} and L3 \cite{L3-02}, both statistical and systematic uncertainties are included. Theoretical prediction curves shown are the leading term QCD calculations \cite{FMN} with the nucleon wave functions obtained in \cite{cz-84} from the QCD sum rules (three-quark) and the phenomenological diquark model (diquark) for which \cite{Berger} was chosen as a typical example.\\
\hspace*{5mm} Fig.21 (center and right). Separate fits of $\sigma\sim (W=\sqrt s)^{-n}$ to the data in the range of $W = 2.5 - 2.9$ GeV and $W=3.2 - 4.0$ GeV, with (a) n floating; (b) n = 10 and n = 15. The error bars include
statistical and systematic errors. The charmonium region between 2.9 and 3.2 GeV is excluded.\\~
\vspace*{1mm}

\begin{minipage}[c]{.22\textwidth}\includegraphics
[trim=0mm 0mm 0mm 0mm, width=1.1\textwidth,clip=true]{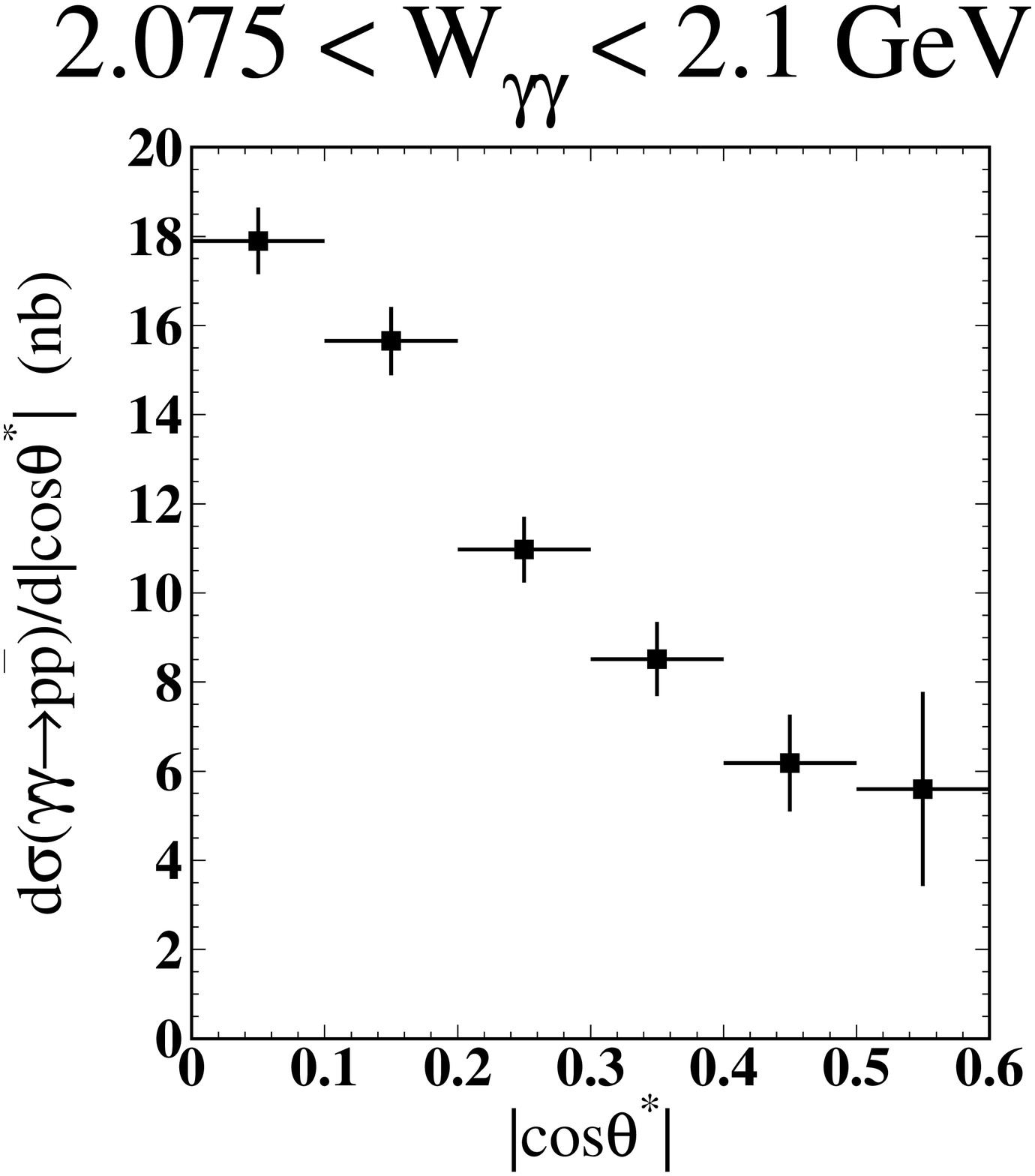}
\end{minipage}~
\begin{minipage}[c]{.22\textwidth}\includegraphics
[trim=0mm 0mm 0mm 0mm, width=1.1\textwidth,clip=true]{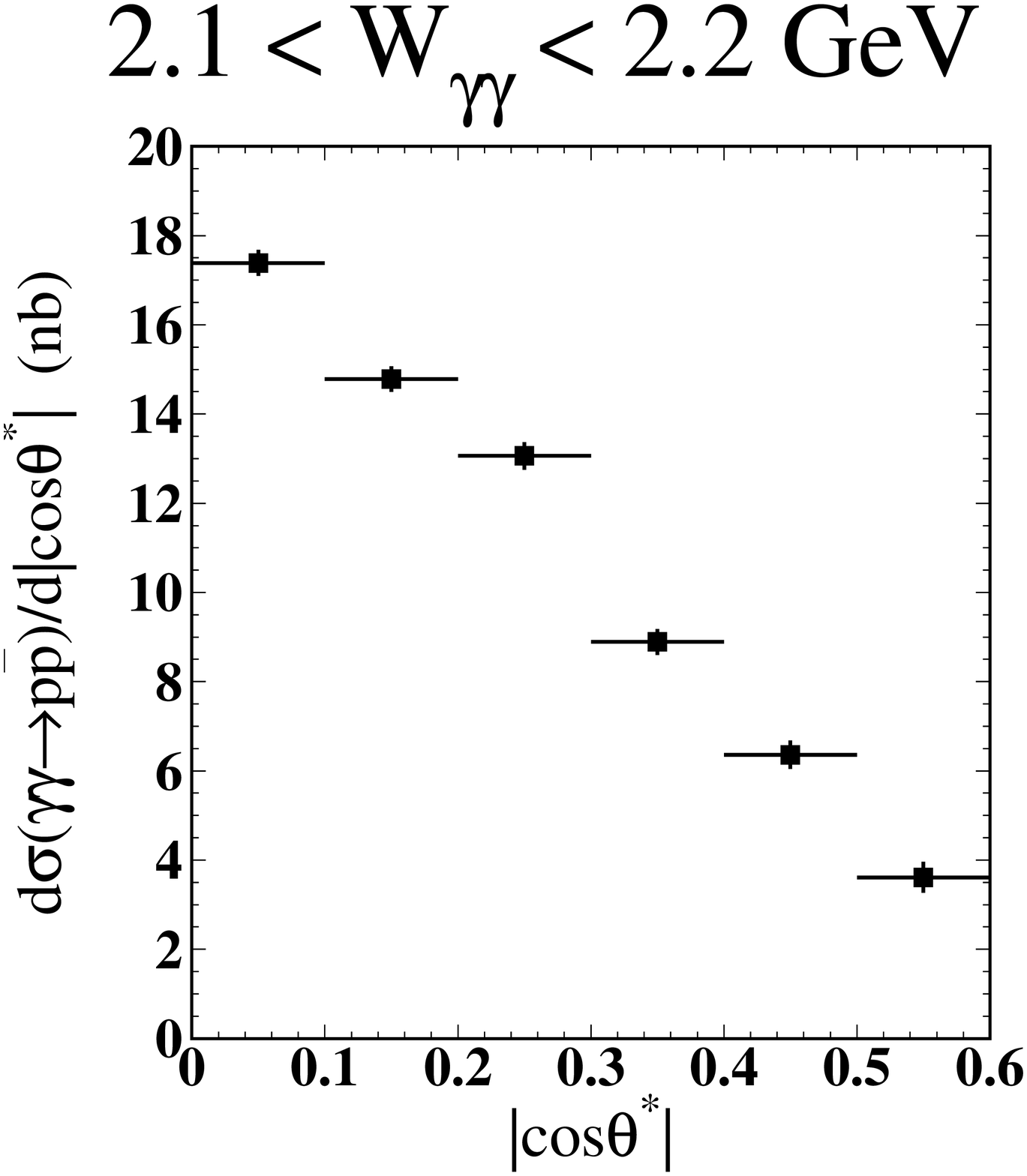}
\end{minipage}~
\begin{minipage}[c]{.22\textwidth}\includegraphics
[trim=0mm 0mm 0mm 0mm, width=1.1\textwidth,clip=true]{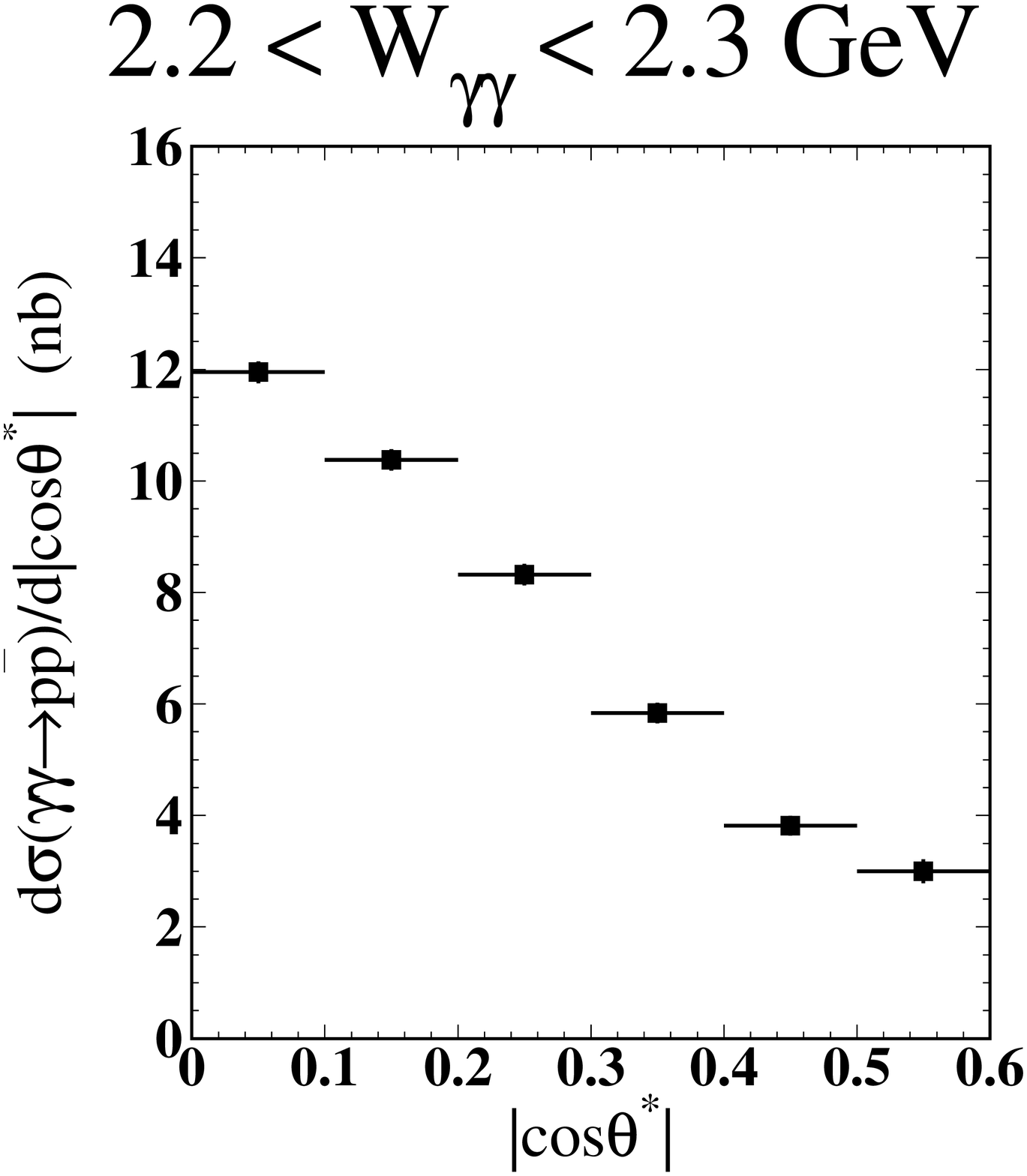}
\end{minipage}~
\begin{minipage}[c]{.22\textwidth}\includegraphics
[trim=0mm 0mm 0mm 0mm, width=1.1\textwidth,clip=true]{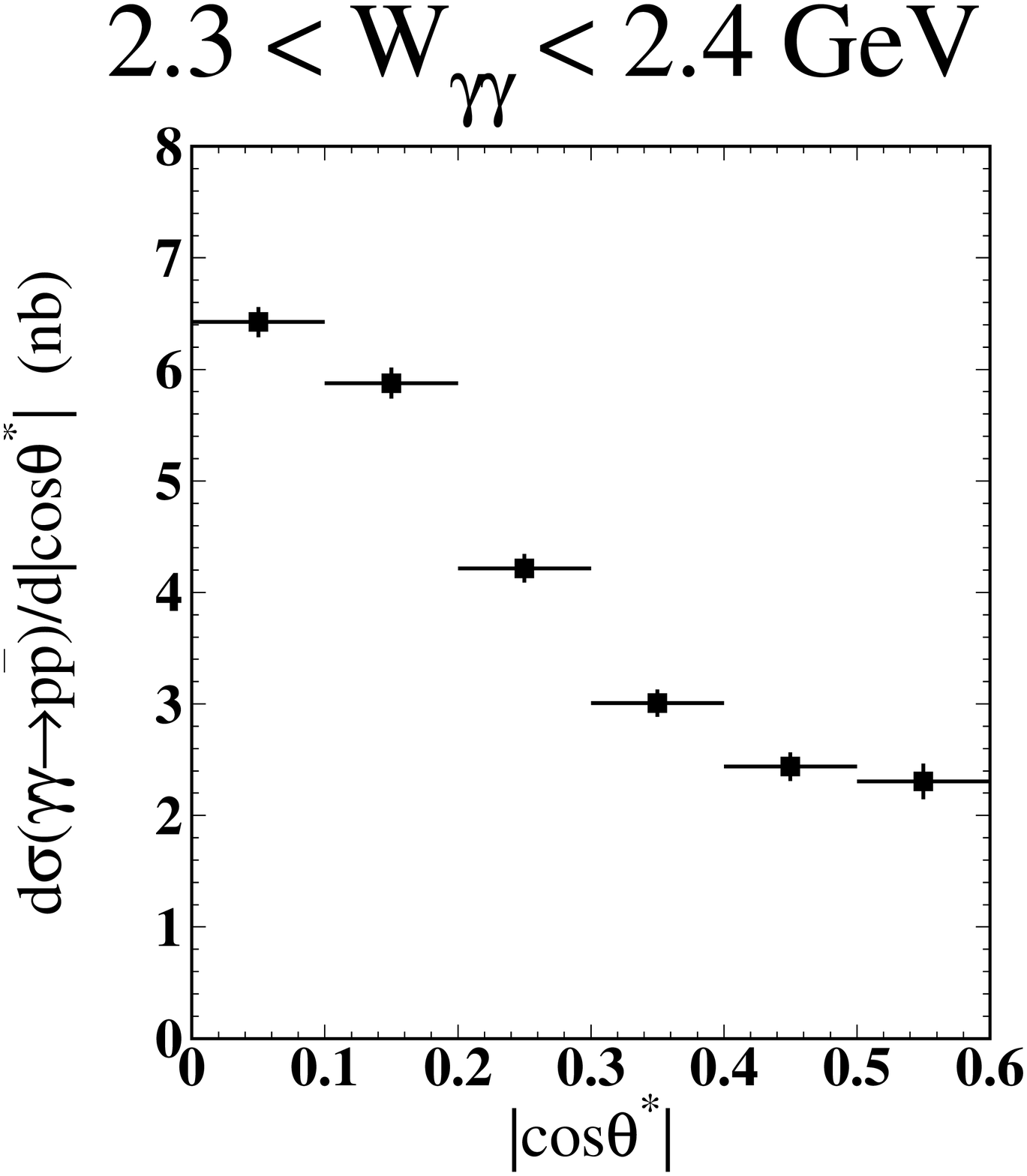}
\end{minipage}~
\vspace*{4mm}

\begin{minipage}[c]{.22\textwidth}\includegraphics
[trim=0mm 0mm 0mm 0mm, width=1.1\textwidth,clip=true]{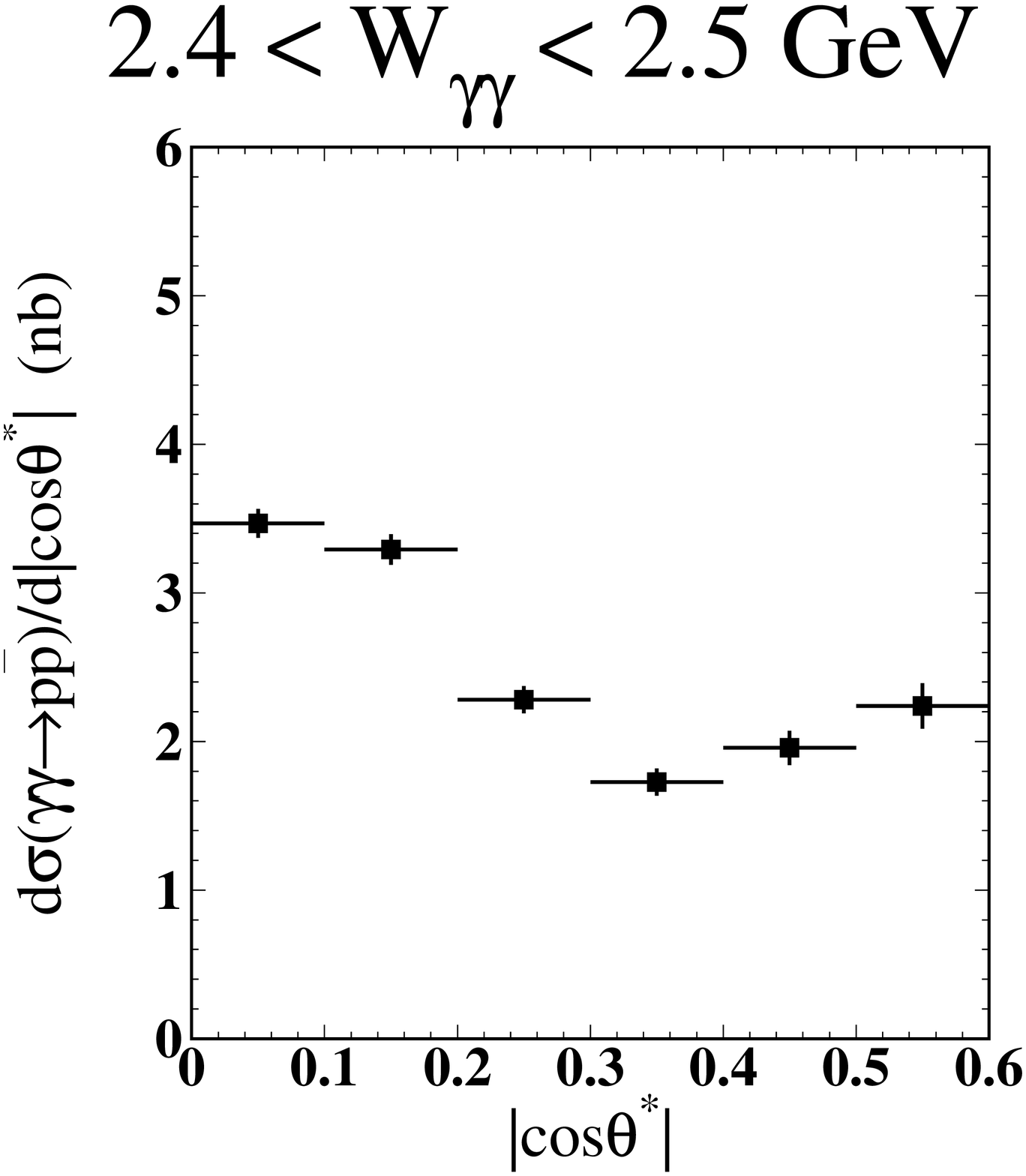}
\end{minipage}~
\begin{minipage}[c]{.22\textwidth}\includegraphics
[trim=0mm 0mm 0mm 0mm, width=1.1\textwidth,clip=true]{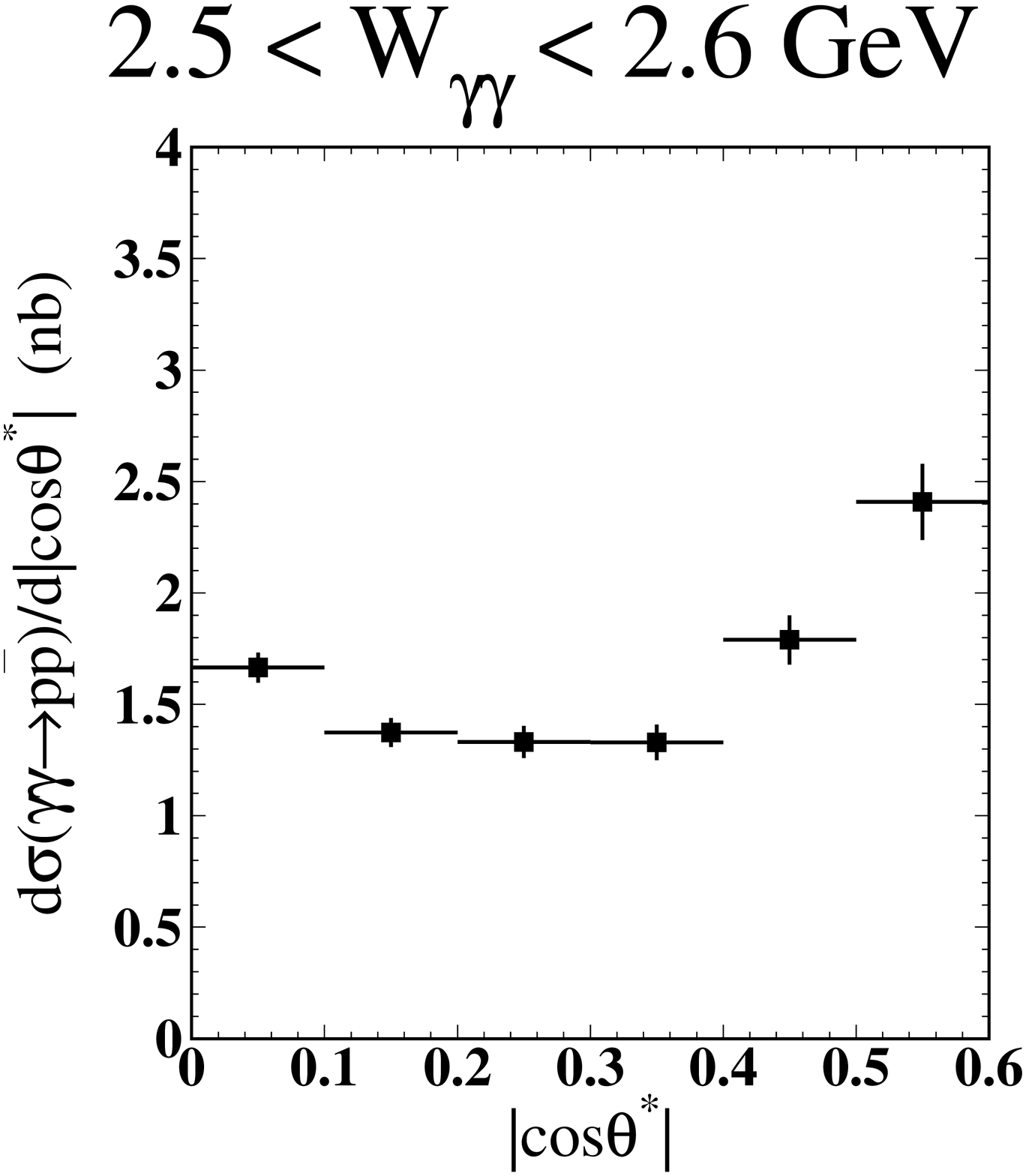}
\end{minipage}~
\vspace*{1mm}
\begin{minipage}[c]{.22\textwidth}\includegraphics
[trim=0mm 0mm 0mm 0mm, width=1.1\textwidth,clip=true]{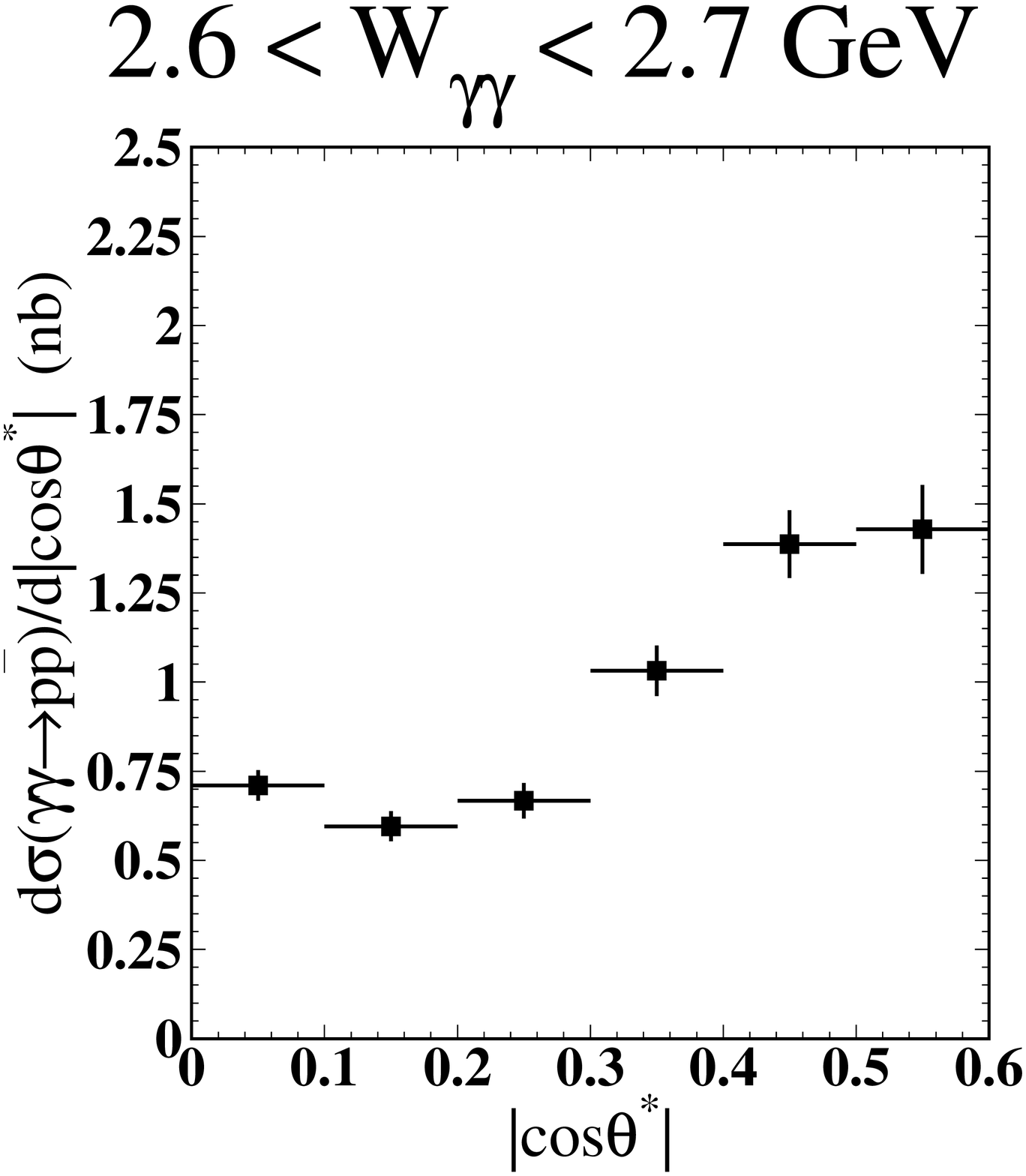}
\end{minipage}~
\begin{minipage}[c]{.22\textwidth}\includegraphics
[trim=0mm 0mm 0mm 0mm, width=1.1\textwidth,clip=true]{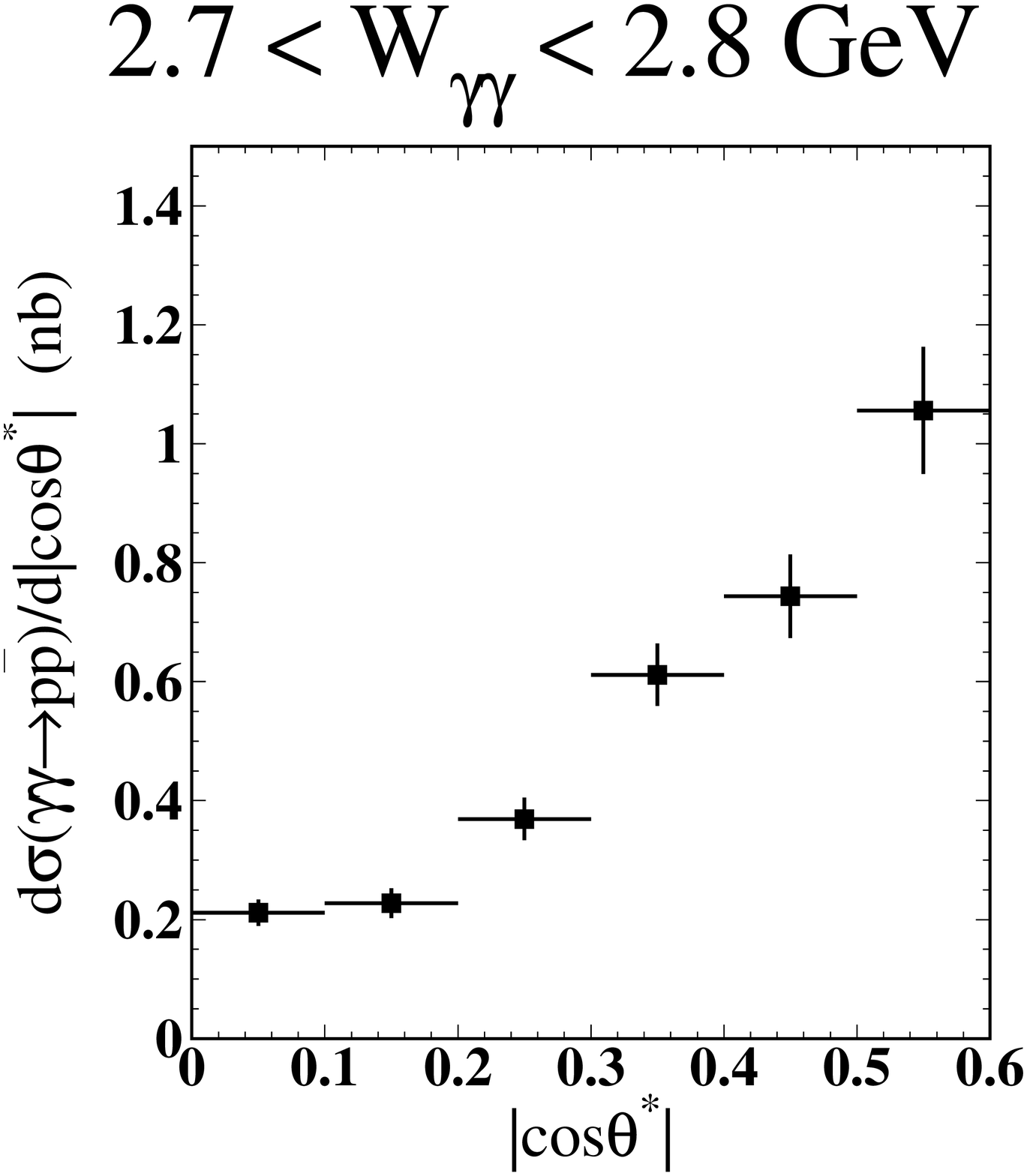}
\end{minipage}

\begin{minipage}[c]{.22\textwidth}\includegraphics
[trim=0mm 0mm 0mm 0mm, width=1.\textwidth,clip=true]{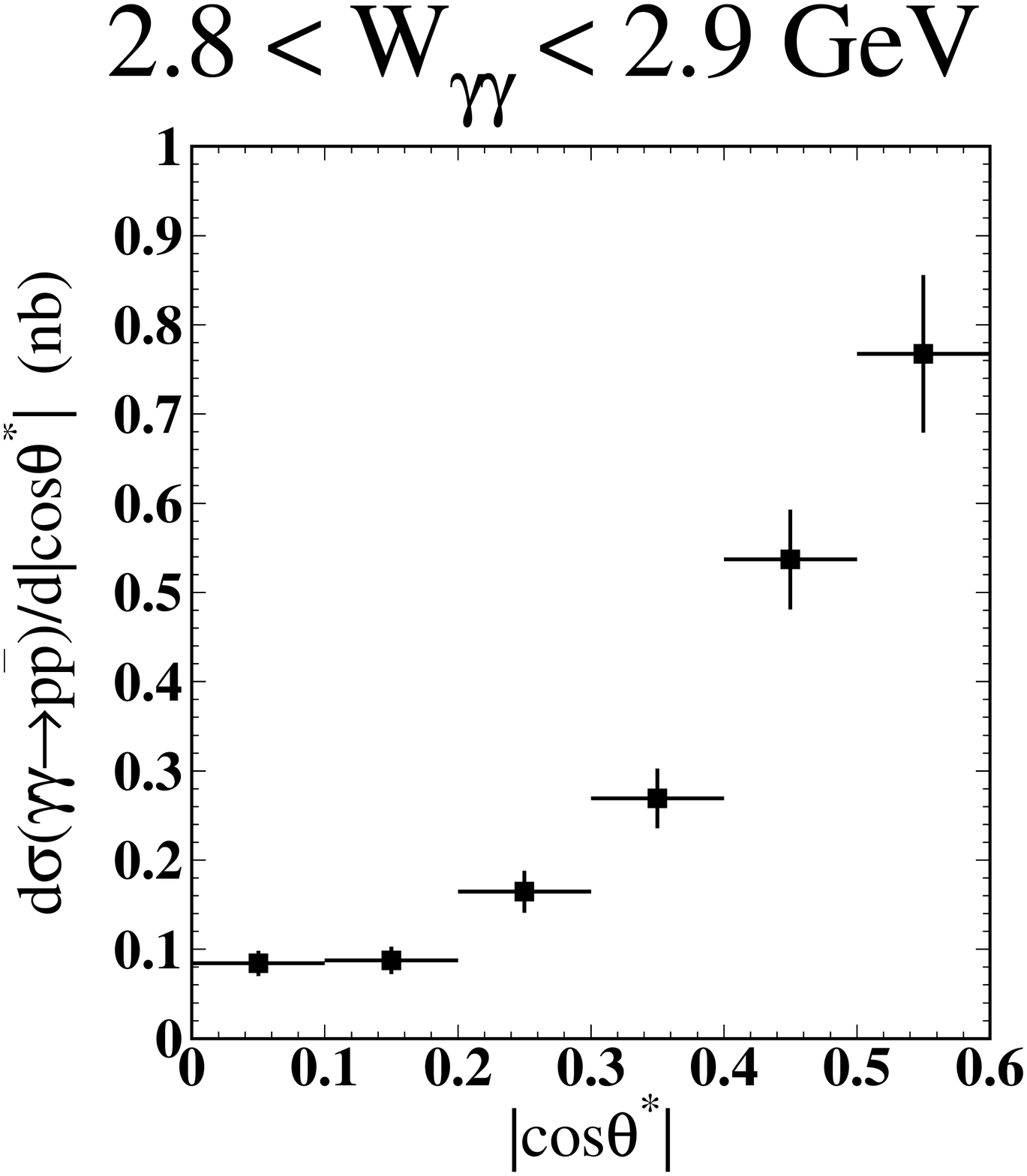}
\end{minipage}~
\vspace*{1mm}
\begin{minipage}[c]{.22\textwidth}\includegraphics
[trim=0mm 0mm 0mm 0mm, width=1.\textwidth,clip=true]{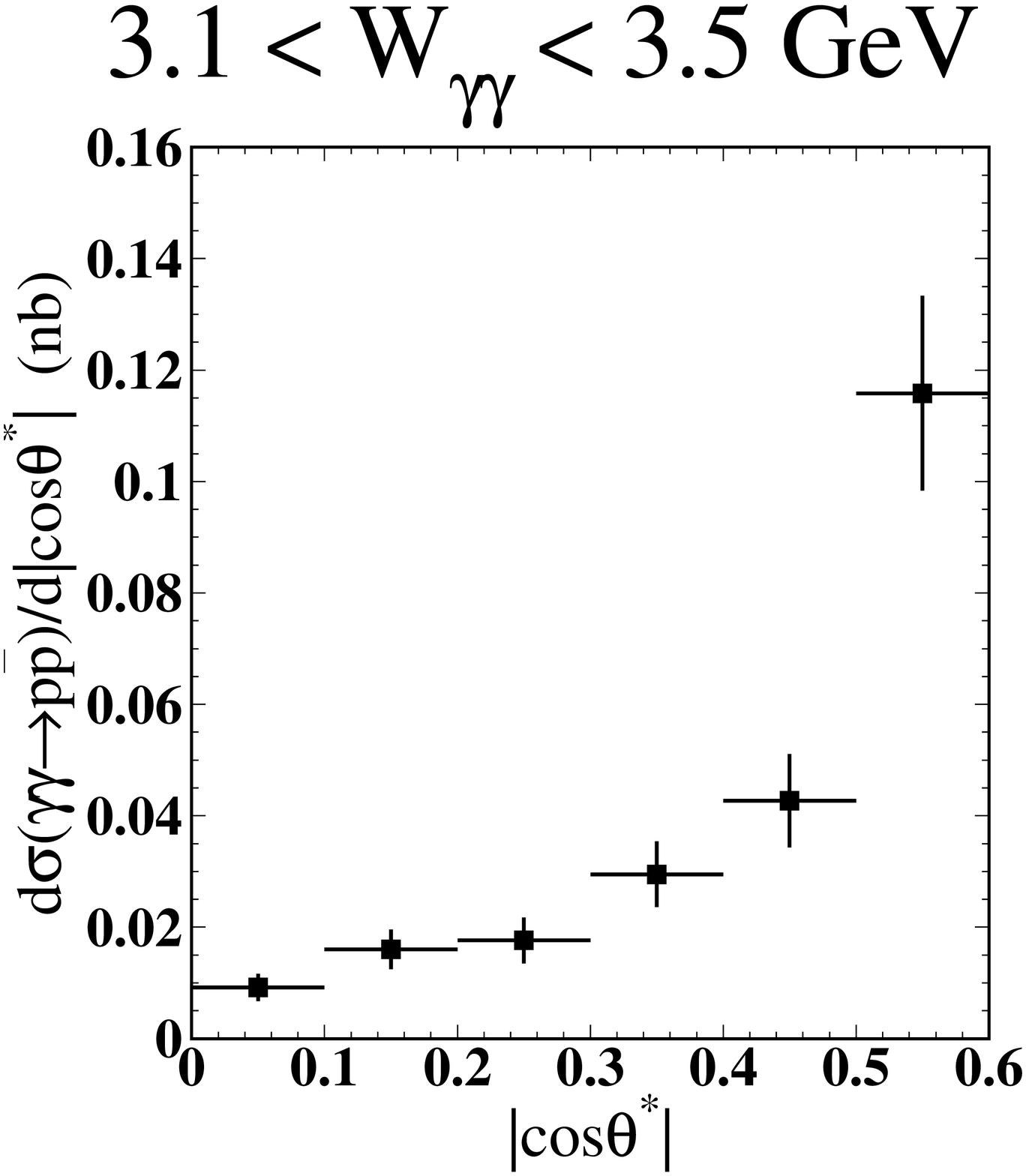}
\end{minipage}~
\begin{minipage}[c]{.22\textwidth}\includegraphics
[trim=0mm 0mm 0mm 0mm, width=1.\textwidth,clip=true]{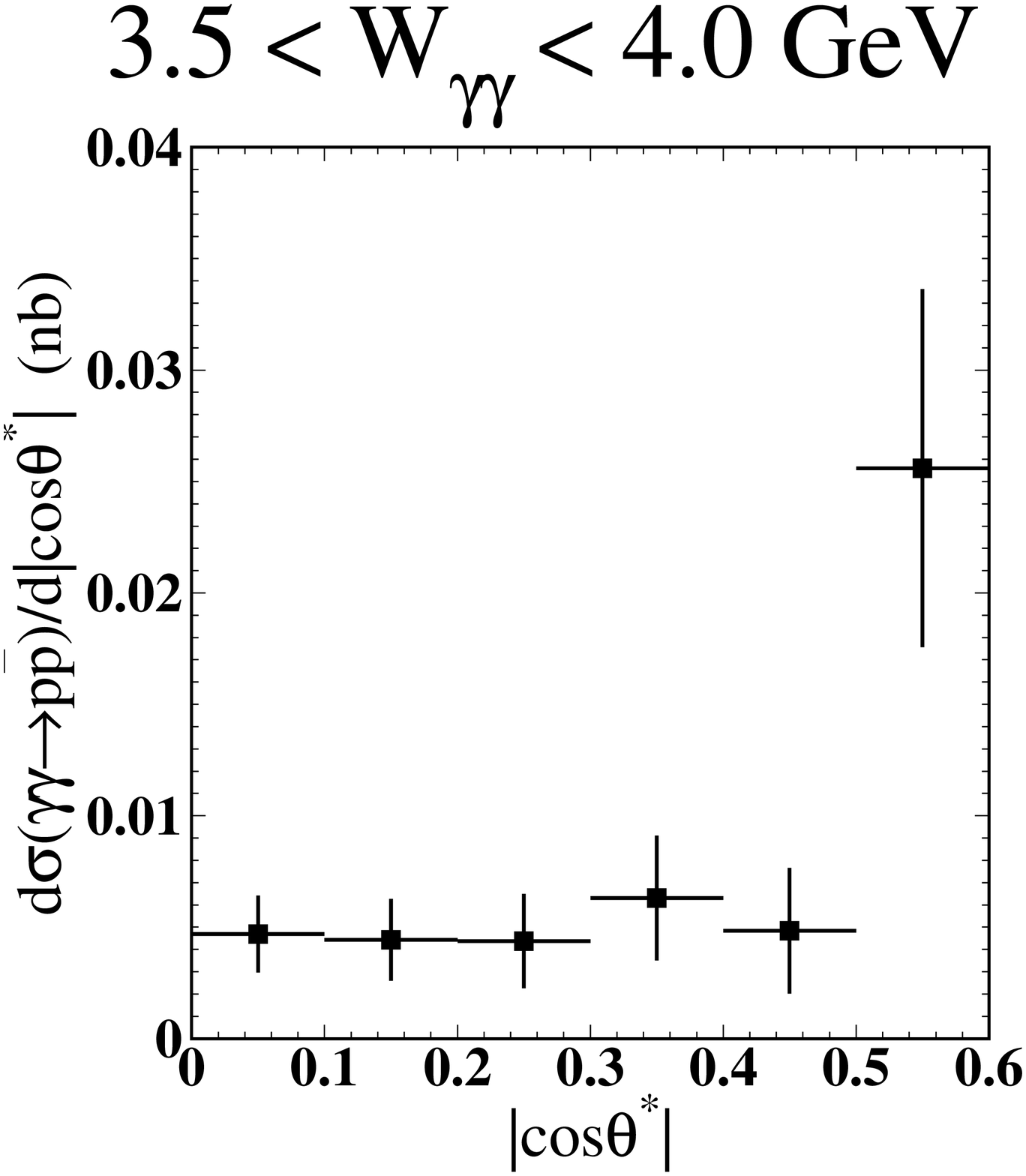}
\end{minipage}~

Fig.22\, \cite{Belle-p-05}. Measured differential cross sections in 11 ranges of $W$ as a function of $|\cos\theta|$. The $\eta_c$ region (2.9 - 3.1 GeV) is skipped. The error bars are statistical only.\\

The angular distributions of $d\sigma (\gamma\gamma\to {\ov p} p)/d |\cos\theta|$ are shown in Fig.22. It is seen that for $W< 2.4$ GeV the differential cross sections decrease as $|\cos\theta$| increases, while for
$W> 2.6$ GeV the opposite trend is observed. The transition occurs around $W= 2.5$ GeV. The natural explanation is that the lower energy region is better described by the s-channel resonance contributions while the high energy region - by the quark-gluon diagrams.

As it is emphasized in \cite{Belle-p-05}: "All existing models based on the constituent scattering picture, as expected, predict an ascending trend, which is in agreement with the data for $W> 2.5$ GeV. This is due to the factor $1/\sqrt{tu}\sim 1/\sin\theta$ contained in the hard scattering amplitudes. The same trend is obtained from naive QED estimates: $d\sigma/d |\cos\theta|\sim (1 + cos^{2}\theta)/(1- \cos^{2}\theta)$, in the massless limit. A simplified picture with diquarks would follow the naive QED expectation above, if all quark masses are neglected and only scalar diquarks are considered".

As for the handbag model, it is more formal in comparison with the diquark model and its basic idea (i.e. the endpoint Feynman mechanism) was described in detail in Section 3 on examples with mesons, $\gamma\gamma\to {\ov M} M$. For baryons, $\gamma\gamma\to {\ov B} B$, the differential cross section looks in the handbag model as \cite{DKV-03}
\vspace*{-1.7mm}
\be
\frac{d\sigma(\gamma\gamma\to {\ov p}p)}{dt}=\frac{4\pi\alpha}{s^6\sin^2\theta}\biggl \{ |s^2 R^{\rm eff}_{{\ov B}B}(s)|^{2}+\cos^2\theta\, |s^2 R^{V}_{{\ov B}B} (s)|^{2}\biggr \},\vspace*{-2mm}
\ee
\be
|R^{\rm eff}_{{\ov B}B}(s)|^{2}=|R^{A}_{{\ov B}B}(s)+R^{P}_{{\ov B}B}(s)|^2+\frac{s}{4 m_N^2}\,|R^{P}_{{\ov B}B}(s)|^2\,, \quad R^i_{{\ov B}B}=\sum_{q=u,d,s} e^2_q F^{i,q}_{{\ov B}B}\,,\,\,\, i=V, A, P \nn
\vspace*{-1.7mm}
\ee
The form factors in (5.4) were fitted in \cite{Kr-Schaf-05} (see Figs.23,24) to the Belle data \cite{Belle-p-05} in the range $3<W<4\,GeV$ as
\vspace*{-2mm}
\be
s^2|R^{\rm eff}_{{\ov B}B}(s)| = (2.90 + 0.31)\,GeV^4 \Bigl (\frac{s}{s_o}\Bigr )^{(-1.10+0.15)},\vspace*{-2mm}
\ee
\be
s^2|R^{V}_{{\ov B}B}(s)| = (8.20 + 0.77)\,GeV^4 \Bigl (\frac{s}{s_o}\Bigr )^{(-1.10+0.15)}, \quad s_o=10.4\,GeV. \nn
\ee
\vspace*{-2mm}
\begin{minipage}[c]{.5\textwidth}\includegraphics
[trim=0mm 0mm 0mm 0mm, width=0.8\textwidth,clip=true]{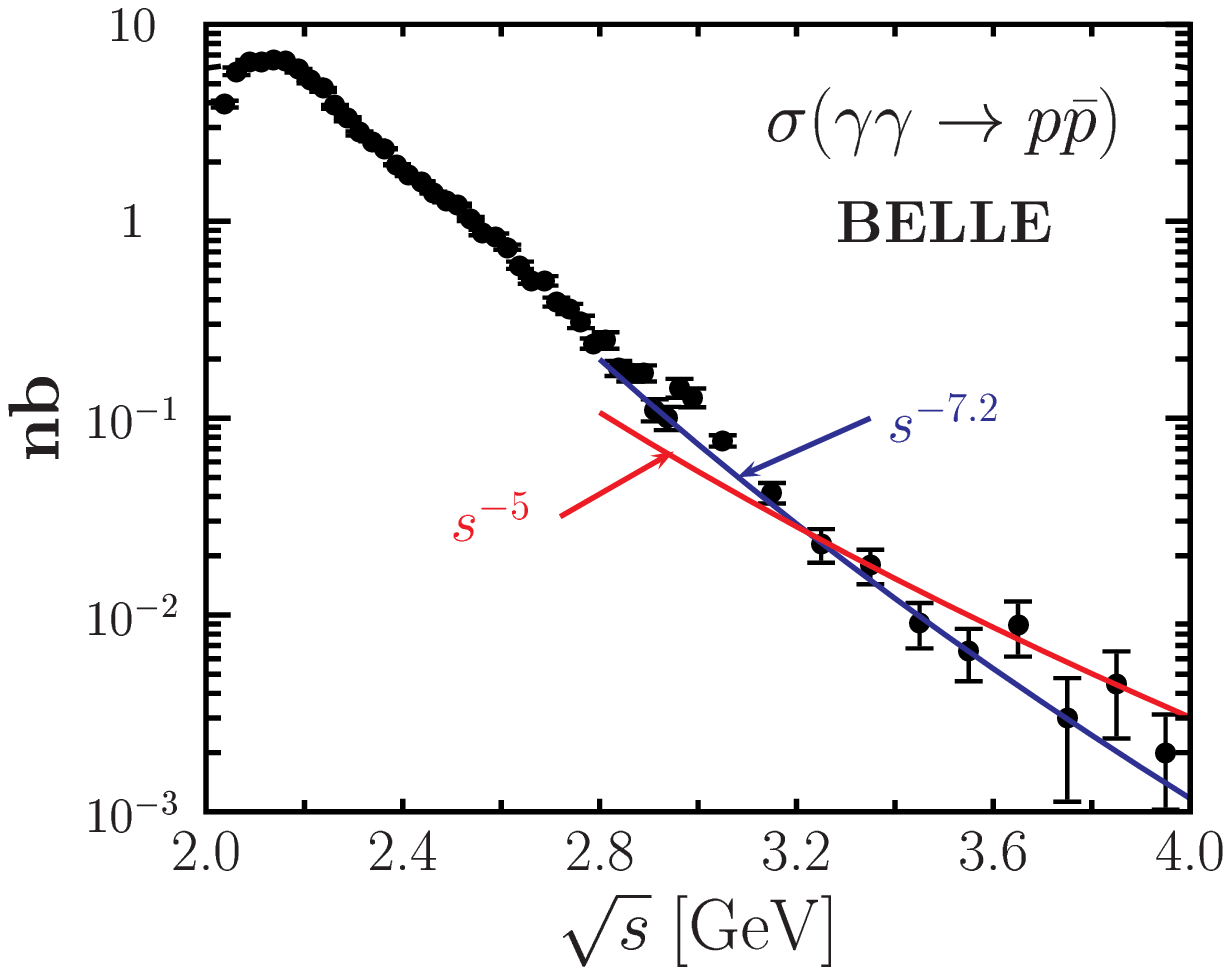}
\end{minipage}~
\begin{minipage}[c]{.5\textwidth}\includegraphics
[trim=0mm 0mm 0mm 0mm, width=0.65\textwidth,clip=true]{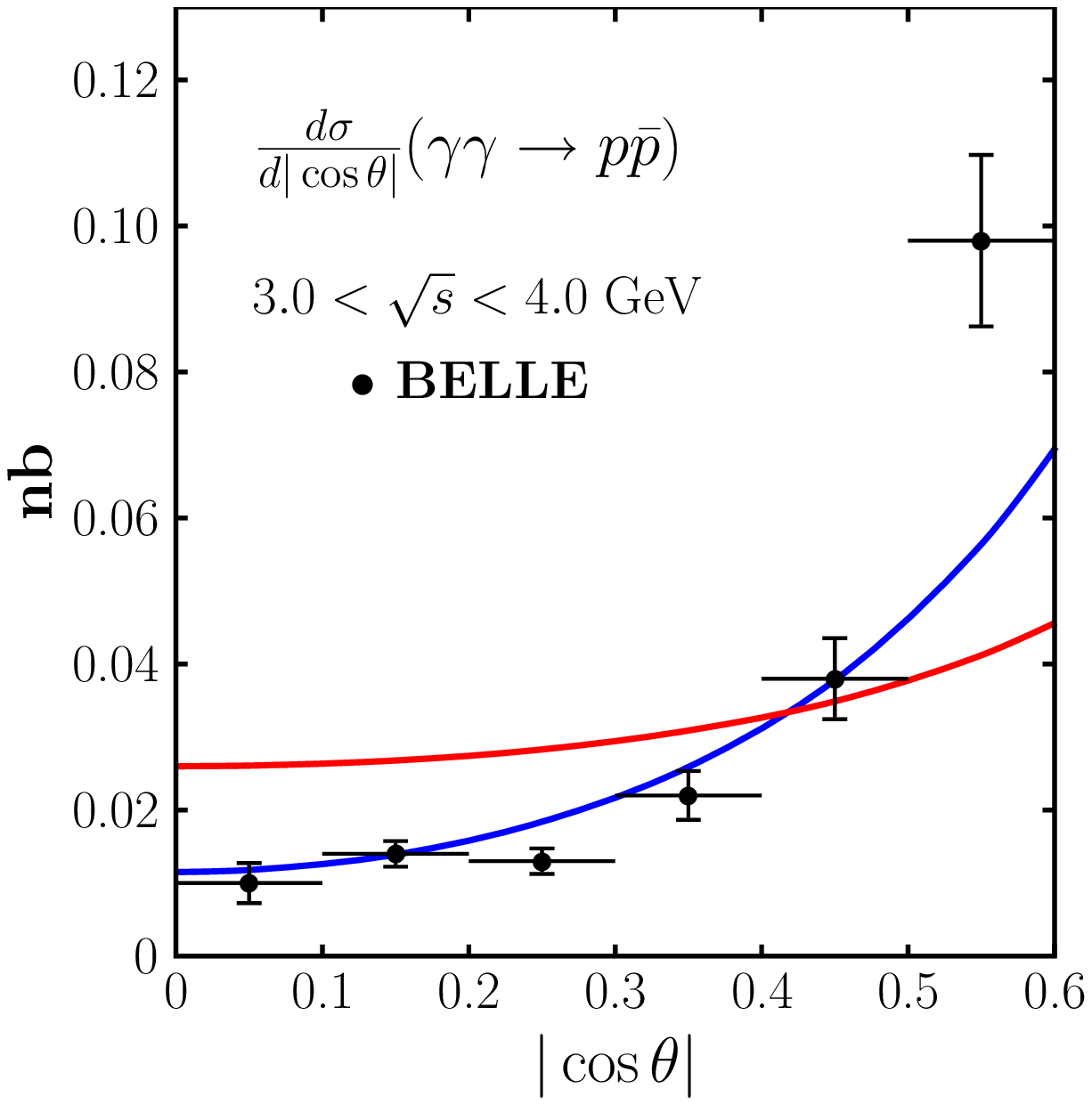}
\end{minipage}~
\vspace*{2mm}

\hspace*{3mm} Fig.23 (left)\,\,\cite{Kroll-12}. The cross section integrated over $|\cos\theta<0.6|$. The curve $\sim s^{-7.2}$ represents the fit (5.4),(5.5) from \cite{Kr-Schaf-05} (the energy dependence at $W=2.3-2.8\,GeV$ is slightly steeper, see Fig.21), the curve $\sim s^{-5}$ is the leading term QCD prediction of the energy dependence.\\
\hspace*{3mm} Fig.23 (right)\,\,\cite{Kroll-12}. The differential cross section. The curve going through points represents the fit from \cite{Kr-Schaf-05}. For comparison, another curve is for $|R^{\rm eff}_{{\ov B}B}|=|R^{V}_{{\ov B}B}|$. Data are from \cite{Belle-p-05}.\\

\begin{minipage}[c]{.5\textwidth}\includegraphics
[trim=0mm 0mm 0mm 0mm, width=0.75\textwidth,clip=true]{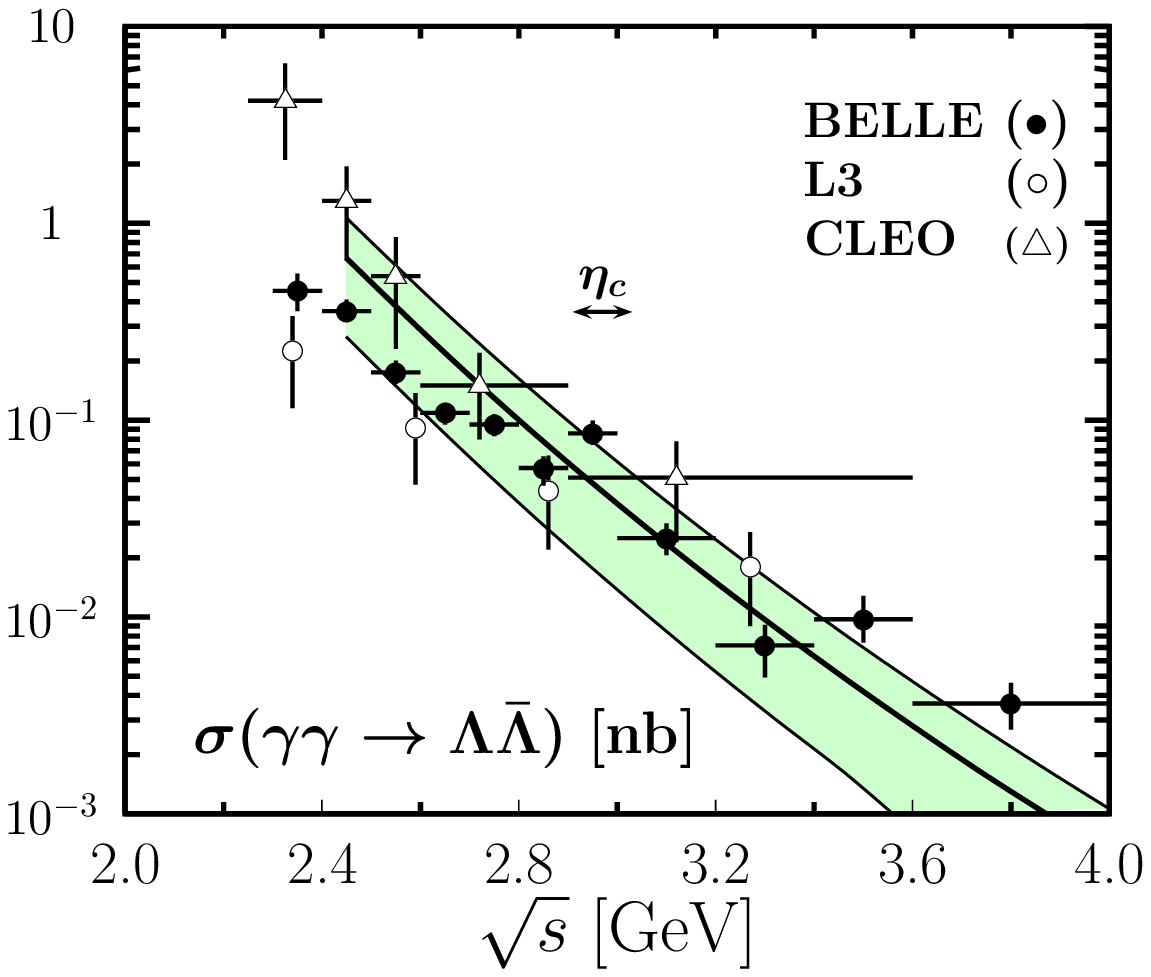}
\end{minipage}~
\begin{minipage}[c]{.5\textwidth}\hspace*{-1cm}\includegraphics
[trim=0mm 0mm 0mm 0mm, width=0.8\textwidth,clip=true]{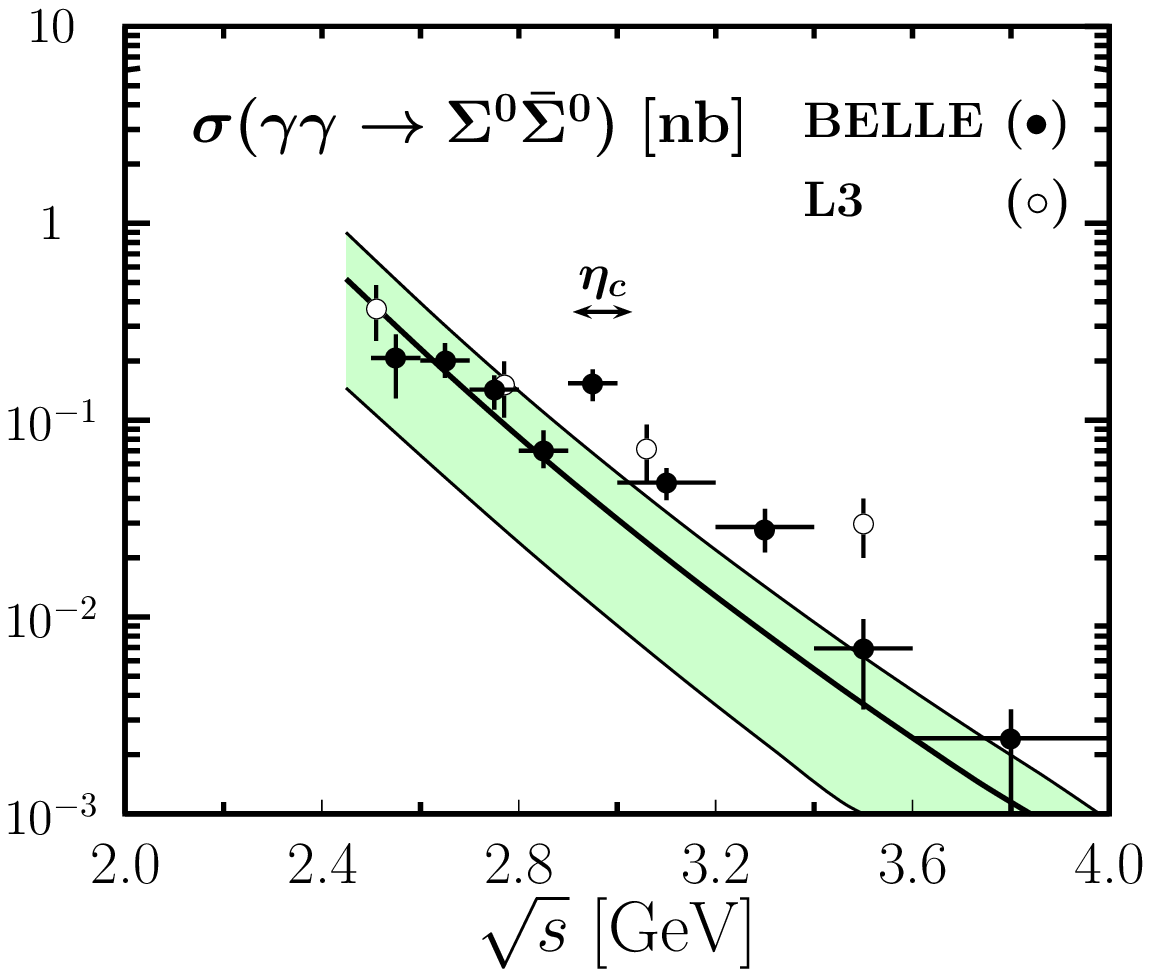}
\end{minipage}~

\hspace*{3mm} Fig.24. The integrated cross sections for $\gamma\gamma\to {\ov\Lambda}\Lambda$ (left) and
$\gamma\gamma\to \ov{\Sigma^o}\Sigma^o$ (right), $|\cos\theta|<0.6$. Data are taken from  CLEO \cite{CLEO-97}, L3 \cite{L3-02} and Belle \cite{Belle-p-06}. The solid lines represent the fits from \cite{Kroll-13}, the shaded bands are their uncertainties.\\

\hspace*{3mm} We recall now that the power suppressed behavior of the fitted form factors $|R^{\rm eff,\,V}_{{\ov B}B}(s)|\sim 1/s^{3}$ (5.5) (in comparison with the leading term QCD behavior $|F^{QCD,\,\rm lead}_{{\ov B}B}(s)|\sim 1/s^2$) is, at least qualitatively, in accordance with a general idea of the handbag model that the non-leading terms may be more important numerically at intermediate energies, while the leading QCD terms will become dominant at higher energies. However, there is a problem. The handbag model specifies additionally what are the non-leading terms it deals with. These are the contributions from endpoint regions (Feynman's mechanism) where two photons interact with the same active quark only and this active quark carries nearly the whole momenta of initial and final hadrons. All other passive quarks are wee partons which are picked out from the vacuum by soft nonperturbative interactions (see Sect.3 and Fig.10a). It was described in Sect.3 how one can determine the energy behavior of such endpoint region contributions (see (3.6),(3.8) and Figs.13 and 15). When the hadron is a baryon, there is one active quark and two wee partons. Therefore, the handbag model form factor of the baryon behaves parametrically as ($\,\phi_{B}(x_1,x_2,x_3)\sim x_1 x_2 x_3,\,\,\, x_1+x_2+x_3=1,\,\,\, x_1\ra 1,\,\,\, 0\leq x_{2,3}\leq\delta\,$)
\be
R^{\rm handbag}_{{\ov B}B}(s)\sim M^{\rm valence}_{3,\,\rm endpoint}(\gamma\gamma\to {\ov B}B)\sim
\ee
\be
\sim\int_0^{\delta}d x_2\int_0^{\delta}d x_3  \,\phi_B(x)\int_0^{\delta}d y_2 \int_0^{\delta} d y_3\,\phi_B(y)\Bigl [A_{\rm hard}(\gamma\gamma\to{\ov q}q)\sim 1 \Bigr ]\sim\nn
\ee
\be
\sim \Bigl [\,\int_0^{\delta}dx\, x\int_0^{\delta}dy\, y\,\Bigr ]^{\rm n_{\rm wee}=2}\sim \Bigl (\,\delta^{4}\,\Bigr )^{n_{\rm wee}=2}\sim\Lambda_{QCD}^{8}/s^{4},\quad \delta\sim\frac{\Lambda_{QCD}}{\sqrt s}\,,\nn
\ee
and this shows that the fitted behavior $|R^{\rm eff,\,V}_{{\ov B}B}(s)|\sim 1/s^{3}$ (5.5) originates not from the handbag model endpoint region but from the hard non-leading QCD term $M^2_o\,b(\theta)/s^3$ in the amplitude (5.3), see footnote 2.

We conclude this section with a short additional comment. It was shown in \cite{Mueller-80,Fadin-81} that there appears non-RG logarithmic contribution in the baryon form factor $F_B(Q^2)$ starting with the two loop correction to the Born term. We mention about it because one can expect naturally that similar correction will be present in $M(\gamma\gamma\to {\ov B} B)$ also. In comparison with the Born diagrams with two gluon exchanges for $F_B(Q^2)$, the numerical estimate of this two-loop correction for the model nucleon wave function from \cite{coz} and at $10<Q^2<100\,GeV^2$ looks as:\, $\Delta F_B(Q^2)/F_B^{\rm Born}(Q^2)\sim (\ov\alpha_s/2\pi)^2\ln (Q^2/\mu^2_o)= O(10^{-2})$. Besides, as was shown in \cite{Fadin-82}, higher loop contributions multiply this correction by the Sudakov form factor, so that this nonstandard contribution can be neglected with a good accuracy both in $F_B(Q^2)$ and $M(\gamma\gamma\to {\ov B} B)$. We would like to emphasize also that this nonstandard contribution arise from the region where all target valent quarks carry fractions of the hadron momenta $x_i, y_i=O(1)$, and this is why it is not additionally power suppressed (except for the Sudakov form factor). Therefore, it has no relevance to the power suppressed handbag model contribution from the endpoint region $x_{2,3}, y_{2,3}\ra 0$, see (5.6).

\section{Form factors \boldmath{$F_{\gamma P}(Q^2),\,\,P=\{\pi^o\,,\,\eta\,,\,\eta'\}$},
and \boldmath{$F_{\gamma\eta_c}(Q^2)$}}

\subsection{Asymptotic behavior of $F_{\gamma P}(Q^2)$ in QCD}

\hspace*{3mm}As for the form factor $F_{\gamma \pi}(Q^2)=F_{\gamma \pi}(Q^2=-q_1^2,\,q_2^2=0)$, the QCD prediction for its asymptotic behavior in the formal limit $Q^2\to \infty$ looks as (see e.g. \cite{LB-80})
\footnote{\,
The form factor $F_{\gamma \pi}(Q_1^2,\,Q_2^2=Q_1^2)$ with large equal virtualities of two photons (the local limit of (6.1)) was calculated first long before in \cite{WZ-72}.
}$^{,}$
\footnote{\,
Really, unlike e.g. $F_{\pi}(Q^2)$ \cite{c-2}, the leading asymptotic behavior of $F_{\gamma\pi}(Q^2)$ can be directly obtained from the standard Wilson operator expansion of $T (J_{\mu}(z)J_{\nu}(0))\to \sum_n C_n(z) O_n(0)$ in (6.1) \cite{Ch3}, as in calculations of the deep inelastic scattering. The only difference is that the forward matrix elements $\langle p|O_n|p\rangle$ are taken in the deep inelastic scattering, while these are $\langle p|O_n|0\rangle$ in the case $\gamma^*\gamma\to \pi$, but the anomalous dimensions of the leading twist operators are the same.
}
\,:
\be
\int dz\, e^{iq_1 z}\langle \pi(p)|T\{J_{\mu}(z)J_{\nu}(0) \}|0\rangle=\Bigl (i\epsilon_{\mu\nu\lambda\sigma} q_{1}^{\lambda}q_{2}^{\sigma}\Bigr )\,F_{\gamma \pi}(Q^2),
\ee
\be
\Phi_{\gamma\pi}(Q^2)\equiv Q^2 F_{\gamma \pi}(Q^2)=\frac{\sqrt 2\,f_\pi}{3}\int_0^1 dx\,\frac{\phi_{\pi}(x,\mu\sim Q)}{x}\Biggl [ T_{\rm hard}=1+O (\alpha_s ) \Biggr ]\ra\sqrt 2\,f_\pi\,\quad {\rm at}\quad Q^2\to\infty\,.
\ee

The status of theoretical calculations of $\Phi(Q^2)$ looks at present as follows.\,-\\
\,a)\,\, The one-loop logarithmic correction to the hard kernel was calculated in \cite{Aguila-81, Braaten,
Radyu-86} in the $\ov{MS}$-scheme and looks as
\be
I_{\rm rad}=1+C_F\frac{\alpha_s(\mu)}{4\pi}\Bigl [\ln^2 x+(3+2\ln x)\ln\frac{Q^2}{\mu^2}-\frac{x\ln x}{1-x}
-9 \Bigr ]\,.
\ee
b)\,\, The next to leading order loop corrections were calculated (in part) in \cite{MNP-01} (see also \cite{MNP-02}).\\
c)\,\, The power correction $\Delta\Phi_{\gamma\pi}(Q^2)\sim 1/Q^2$ originating from the asymptotic (i.e. lowest partial waves in the conformal partial wave expansion) twist-4 two- and three-particle wave functions \cite{Br-Fil-90} was calculated in \cite{Khodja-99} and looks as
\be
\Phi_{\gamma\pi}(Q^2)\ra \sqrt 2\,f_\pi\Biggl [\frac{1}{3}\int_0^1 dx\,\frac{\phi_{\pi}(x,\mu\sim Q)}{x}
-\frac{80}{27}\frac{\delta^2}{Q^2}\Biggr ]\,,
\ee
where the parameter $\delta^2$ is connected by the QCD equations of motion to the characteristic value $\langle {\vec k}^{\,2}_{\perp}\rangle$ of the quark transverse momentum inside the pion \cite{czz-83,cz-rev}
\be
\langle 0|{\ov d}\,\gamma_{\mu}\gamma_5 \Bigl ( i{\vec D}_{\perp}\Bigr )^2\, u |\pi(p)\rangle\equiv i f_{\pi} p_{\mu}\langle {\vec k}^{\,2}_{\perp}\rangle,\quad \langle 0|{\ov d}\,g_s {\tilde G}^A_{\mu\nu}\frac{\lambda^A}{2}\gamma_{\nu}\, u |\pi(p)\rangle\equiv i f_{\pi} p_{\mu}\,\delta^2\,,\quad
\langle {\vec k}^{\,2}_{\perp}\rangle=\frac{5}{9}\,\delta^2\,.
\ee
The numerical value $\delta^2\simeq 0.2\,GeV^2$ was estimated in \cite{czz-83,cz-rev,NSVVZ-84}, resulting in
$\langle {\vec k}^{\,2}_{\perp}\rangle\simeq (330\,MeV)^2$ and
\be
\Phi(Q^2)_{\gamma\pi}\ra \sqrt 2\,f_\pi\Biggl [\frac{1}{3}\int_0^1 dx\,\frac{\phi_{\pi}(x,\mu\sim Q)}{x}
-\frac{0.6\,GeV^2}{Q^2}\Biggr ]\,.
\ee
d)\,\,The next to lowest partial waves of these wave functions were studied in \cite{Br-Fil-90} using the QCD sum rules, but their total contribution is zero. However, the separate terms (connected with each other by the QCD equations of motion) which cancel each other give corrections of the same size as in (6.6). No physical reason is seen for this cancellation and it looks as accidental. This implies then that higher conformal partial waves contributions may well be also of the same size $\sim 0.6\,GeV^2$ as in (6.6). Indeed, e.g. estimates of these twist four contributions in the renormalon model give $0.9-1.2\,GeV^2$ instead $0.6\,GeV^2$ in (6.6) \cite{Agaev-05}.\\
d)\,\, There are also corrections $\Delta\Phi_{\gamma\pi}(Q^2)\sim 1/Q^2$ from the twist-4 four-particle wave functions (two quarks-two gluons and four-quarks). These are unknown at present.\\
e)\,\, Starting from the twist-6 operators their contributions to $\Phi_{\gamma\pi}(Q^2)$  become infrared sensitive and the twist and power expansions differ, i.e. all twist $\geq 6$ operators contribute $\Delta\Phi_{\gamma\pi}(Q^2)\sim 1/Q^2$ corrections to the leading twist-2 term. For this reason, it is impossible (at least at present) to write the total $\sim 1/Q^2$ (and higher order) corrections to $\Phi_{\gamma\pi}(Q^2)$ in a model independent form.

It was proposed in \cite{Khodja-99} how one can "avoid" this problem using the approach of the light-cone sum rules \cite{Balitsky,Br-Fil-89,cz-90}. What was really proposed in \cite{Khodja-99} looks as follows:\\
1) write the dispersion relation in $q^2$ (the virtuality of the soft photon) for $F_{\gamma\pi}(Q^2,q^2\ra 0)$
\be
F_{\gamma\pi}(Q^2,q^2\ra 0)=\int_0^\infty\frac{ds\,\rho_{\gamma\pi}(s,Q^2)}{s-q^2}=\int_0^{s_o^{\rm eff}}\frac{ds\,\rho_{\gamma\pi}(s,Q^2)}{s-q^2}+\int_{s_o^{\rm eff}}^\infty\frac{ds\,\rho_{\gamma\pi}(s,Q^2)}{s-q^2}\,;
\ee
2) the spectral density $\rho_{\gamma\pi}(s,Q^2)$ at fixed $s$ is calculated through the standard operator expansions, $s_o^{\rm eff}$ serves as the infrared cutoff in the last term in (6.7) and is supposed to be sufficiently large for the higher loop logarithmic corrections and higher twist power corrections to $\Phi_{\gamma\pi}(Q^2)$ (these last contain coefficients like $C_k/(s_o^{\rm eff})^k$, where the dimensional quantities $C_k$ originate from the higher twist pion wave functions) to be sufficiently small;\\
3) the soft term with $s\leq s_o^{\rm eff}$ in (6.7) is assumed to be well approximated by the $(\rho+ \omega)$-meson contributions only, $\sim F_{\gamma^{*}\rho\pi}(Q^2)$, (this is a simplest model for the soft part of the photon wave~ function)\,;\\
4) in its turn, the form factor $F_{\gamma^{*}\rho\pi}(Q^2)$ is assumed to be sufficiently well approximated by the simplest duality relation which ignores, in particular, all nonperturbative corrections to the vector meson wave function ($M^2$ is the parameter of the additional Borel transformation)
\be
F_{\gamma^{*}\rho\pi}(Q^2)\sim \int_0^{s_o^{\rm eff}}\frac{ds\,\rho_{\gamma\pi}(s,Q^2)}{m^2_{\rho}}\, e^{(m^2_{\rho}-s)/M^2}\,.
\ee

Finally, all this is a way to calculate approximately some especially chosen subset of power corrections $\Delta\Phi_{\gamma\pi}(Q^2)\simeq\sum_{\rm n\geq 1} A_n/Q^{\rm 2n}$ to the leading term $\Phi^{\rm lead}_{\gamma\pi}(Q^2)\sim {\rm const}$. Unfortunately, this introduces a model dependence with a poorly controlled accuracy. Nevertheless, following \cite{Khodja-99}, a number of papers were published in which this approach was used for numerical calculations of $\Phi_{\gamma\pi}(Q^2)$ (see section 6.3 below). In particular, some subset of the twist-6 contributions $\sim\langle q{\ov q}\rangle^2$ was estimated within this approach in the recent paper \cite{Braun-11}.\\
\hspace*{3mm}Because the process $\gamma^{*}\gamma\to \pi$ is, in a sense, a simplest one (as it contains only one form factor and one hadron), it is considered at present as the best process to study the properties of the leading twist pion wave function $\phi_{\pi}(x,\mu)$, and a large number of papers were published in which various models for $\phi_{\pi}(x,\mu)$ were proposed. Therefore, we consider in the next subsection a status of some of these models.

\subsection{Once more on the pion wave function}

\hspace*{3mm} We start with a discussion of the value of $\langle\xi^2\rangle_{\pi}$ (the second moment of the pion wave function, $iD_{\mu}$ is a covariant derivative)
\be
\langle 0|{\ov d}(0)\gamma_{\mu}\gamma_5\,\overleftrightarrow {iD_{\nu}}\,\overleftrightarrow {iD_{\lambda}}\,u(0)|\pi(p)\rangle= i f_{\pi}\, p_{\mu} p_{\nu}p_{\lambda}\langle\xi^2\rangle_{\pi}+\dots,\quad \langle\xi^2\rangle_{\pi}=\int_0^1 dx\,\phi_{\pi}(x)(2x-1)^2\,.
\ee

Predictions for $\langle\xi^2\rangle_{\pi}$ from a number of papers are presented in Table 3. It is seen from Table 3 that the values of the matrix element $\langle\xi^2\rangle_{\pi}$ obtained in different papers and with different methods vary significantly. Therefore, we comment below in this subsection on possible reasons for some of these differences.

\subsubsection{Standard QCD sum rules}

\hspace*{3mm} We return once more to the standard QCD sum rules for the zeroth and second moments. The original QCD sum rule for the zeroth moment looks as \cite{SVZ}\,:
\be
L_o(M^2)=R_o(M^2)\equiv\Biggl [R^{\rm loop}_{o}(M^2)+R^{\rm power}_{o}(M^2)\Biggr ],\,\,\,R^{\rm power}_{o}(M^2)=\frac{1}{12 M^2}\langle\frac{\alpha_s}{\pi}G^2\rangle+11\frac{16\pi}{81 M^4}
\langle{\sqrt\alpha_s}u{\ov u}\rangle^2\,,\nn
\ee
\be
L_o(M^2)=\Bigl ( f_{\pi}^2+f_{a}^2 e^{-m^2_{a}/M^2}\Bigr ),\quad R^{\rm loop}_{o}(M^2)=\frac{1}
{4\pi^2}\Bigl (1+\frac{{\ov\alpha}_s}{\pi}\Bigr )M^2\Bigl (1-e^{-s_o/M^2}\Bigr )\equiv R_{o}^{o}\Bigl (1-e^{-s_o/M^2}\Bigr ).
\ee

In dealing with (6.10) we take the values of $f_{\pi}=(130.4\pm 0.2)\,MeV$ and the $a_1$ meson mass, $m_a=1.23\,GeV$, from PDG, while the coupling $f_{a}$ and the effective duality interval $s_o$ are fitted. The used value of the gluon condensate is the standard one: $\langle\alpha_s G^2/\pi\rangle=1.2\cdot 10^{-2}\,GeV^4$. The value of the quark condensate $\langle{\sqrt\alpha_s}u{\ov u}\rangle$ in (6.10) can be obtained from
\be
(m_u+m_d)\, 2\, \langle u{\ov u}\rangle=f^2_{\pi}m^2_{\pi^+}
\ee
and from the latest $N_F=(2+1)$ lattice calculations \cite{Data-13}
\be
\frac{1}{2}(m_u+m_d)(\mu=2\,GeV)=(3.42\pm 0.06)\,MeV\quad\ra\quad \langle u{\ov u}\rangle(\mu=2\,GeV)\simeq (0.289\,MeV)^3\,.
\ee
Besides, $\alpha_s(\mu=2\,GeV)\simeq 0.31$ and, with a good accuracy, $\langle{\sqrt\alpha_s}u{\ov u}\rangle$ is a RG-invariant quantity. Therefore
\be
\langle{\sqrt\alpha_s}u{\ov u}\rangle\simeq 1.34\cdot 10^{-2}\,GeV^3\,,
\ee
this value coincides with $\langle{\sqrt\alpha_s}u{\ov u}\rangle\simeq 1.35\cdot 10^{-2}\,GeV^3$ used originally in \cite{SVZ} (and in \cite{cz-82}).\\

\begin{minipage}[c]{0.75\textwidth}
\renewcommand{\arraystretch}{1.2}
\begin{center}\hspace*{1mm}
\begin{tabular}{@{}l|l|l|l@{}} \hline
\quad Method  &$\langle\xi^2\rangle(\mu=1\,GeV)$  & $a_2(\mu=1\,GeV)$  & Refs
\\ \hline
 QCDSR, CZ model & $\simeq 0.40$\, $(\mu^2\simeq 1.5)$ & $\simeq 0.58\,(\mu^2\simeq 1.5),\, (a_4\simeq 0)
$ &  \cite{cz-82,cz-rev}
\\ \hline
QCDSR, BF-1 model & $\simeq 0.35$ & $ 0.44\,\,\quad (a_4= 0.25)$   &\cite{Br-Fil-89}\\
QCDSR, BF-2 model & $\simeq 0.43$ & $2/3\,\,\quad (a_4= 0.43)$   &\cite{Br-Fil-89}
\\\hline
QCDSR & $0.29^{+0.07}_{-0.03}$ & $0.26^{+0.21}_{-0.09}$   &\cite{KMM-04}\\
QCDSR & $0.30\pm 0.03$ & $0.28\pm 0.08$ &   \cite{BBL-06} \\
\hline
QCDSR, NLC  &  $\simeq 0.32$ & $\simeq 0.35$ $\quad (a_4\simeq 0.23)$ & \cite{M-Rad-90}\\
QCDSR, NLC  & $\simeq 0.13$ & $\simeq -\,0.20$ &   \cite{Rad-94}\\
QCDSR, NLC  & $0.27\pm 0.02$ & $0.19\pm 0.06$ $\quad (a_4\simeq -0.14)$   & \cite{BMS-01}
\\
\hline
$F_{\pi\gamma\gamma^*}$, LCSR  & $0.30\pm 0.03$ & $0.30 \pm 0.08 $ & \cite{Yakov-00}\\
&  & $a_4(\mu\simeq 2.4\,GeV)\simeq -0.14$   &  \\
\hline
$F_{\pi\gamma\gamma^*}$, LCSR    & $0.31$  &  $0.32$  &   \cite{BMS-03} \\
$F_{\pi\gamma\gamma^*}$, LCSR, R & $0.29$  & $0.27$   &   \cite{Agaev-05}\\
$F_{\pi\gamma\gamma^*}$, LCSR, R & $0.35$  & $0.44$   &   \cite{BMS-06}\\
$F_{\pi\gamma\gamma^*},\,\,\phi_{\pi}(x,\mu\simeq 1\,GeV)=1$ & $1/3$ & $0.39\quad (a_4=0.24,\, a_6=0.18)$  &   \cite{Rad-09,Polyakov}\\
\hline
$F^{\rm em}_{\pi}$, LCSR & $0.23\pm 0.03$ & $0.1\pm 0.1$ &   \cite{BKM-00}  \\
$F^{\rm em}_{\pi}$, LCSR & $0.28\pm 0.05\pm 0.03$ & $0.24\pm 0.14\pm 0.08$ &   \cite{BKh-02}\\
$F^{\rm em}_{\pi}$, LCSR, R & $0.27\pm 0.01$ & $0.20\pm 0.03$ &   \cite{Agaev-05-2}
\\ \hline
$F_{B\to\pi\ell\nu}$, LCSR  & $0.27\pm 0.06$ & $0.20\pm 0.19$ &   \cite{BZ-05}
\\
$F_{B\to\pi\ell\nu}$, LCSR  & $0.25\pm 0.06$ & $0.15$ &   \cite{DKMMO}
\\ \hline
AdS/QCD,\,\,$\phi_{\pi}(x)\sim\sqrt{x(1-x)}$  & $0.25$  & $0.146$ \quad $(a_4=0.057)$ & \cite{BT-08}
\\ \hline
SDE,\,\,$\phi_{\pi}(x)\sim [x(1-x)]^{0.3}$  & $0.28$  & $0.23$  & \cite{Roberts-13-1}
\\ \hline
LQCD, {\footnotesize $N_f=2$}, CW & $0.300\pm 0.057$ & $0.289\pm 0.166$ &
\cite{Braun-latt}
\\ LQCD, {\footnotesize $N_f=2\!+\!1$}, DWF & $0.314\pm 0.044$ & $0.334 \pm 0.129$ &
\cite{Arthur-latt}
\\ \hline
\end{tabular}
\end{center}
\renewcommand{\arraystretch}{1.0}
\end{minipage}
\vspace*{6mm}

Table 3. The values of $\langle\xi^2=(x_d-x_u)^2\rangle_{\pi}$ and of the second Gegenbauer moment, \\ $a_2=35(\langle\xi^2\rangle-0.2)/12$, for the leading twist pion wave function $\phi_{\pi}(x,\mu)$. The abbreviations stand for: QCDSR - QCD sum rules; NLC - non-local condensates; LCSR - light-cone sum rules;
R - renormalon model for twist-4 corrections; SDE - Schwinger-Dyson equations; LQCD - lattice calculation; $N_f=2(+1)$ - calculation using  $N_f=2(+1)$ dynamical quarks; CW - nonperturbatively ${\mathcal O}(a)$ improved Clover--Wilson fermion action; DWF - domain wall fermions.\\

The fit of (6.10) in the range $0.75<M<1.45\,GeV$ is shown in Fig.25 (left)  and gives
\be
f_a\simeq 200\,MeV,\quad s_o\simeq 1.85\,GeV^2\,.
\ee

The QCD sum rule for the second moment look as \cite{cz-82} (the one-loop correction calculated in \cite{Gorsky-85} is accounted for)\,:
\be
L_2(M^2)=R_2(M^2)\equiv\Biggl [R^{\rm loop}_{2}(M^2)+R^{\rm power}_{2}(M^2)\Biggr ]\,,
\ee
\be
L_2(M^2)=\Bigl (f_{\pi}^2\langle\xi^2\rangle_{\pi}+f_{a}^2\langle\xi^2\rangle_{a}\, e^{-m^2_{a}/M^2}\Bigr )\,,\nn
\ee
\be
R^{\rm loop}_{2}(M^2)=\frac{1}{20\pi^2}\Bigl (1+\frac{5}{3}\frac{{\ov\alpha}_s}{\pi}\Bigr )M^2\Bigl (1-e^{-s_2/M^2}\Bigr )\equiv R^{o}_{2}\Bigl (1-e^{-s_2/M^2}\Bigr ),\nn
\ee
\be
R^{\rm power}_{2}(M^2)=\frac{1}{12 M^2}\langle\frac{\alpha_s}{\pi}G^2\rangle+19\frac{16\pi}{81 M^4}
\langle{\sqrt\alpha_s}u{\ov u}\rangle^2\,.\nn
\ee
\vspace*{2mm}

\begin{minipage}[c]{.5\textwidth}\vspace*{3mm}\hspace*{1cm}
\includegraphics[width=0.8\textwidth,clip=true]{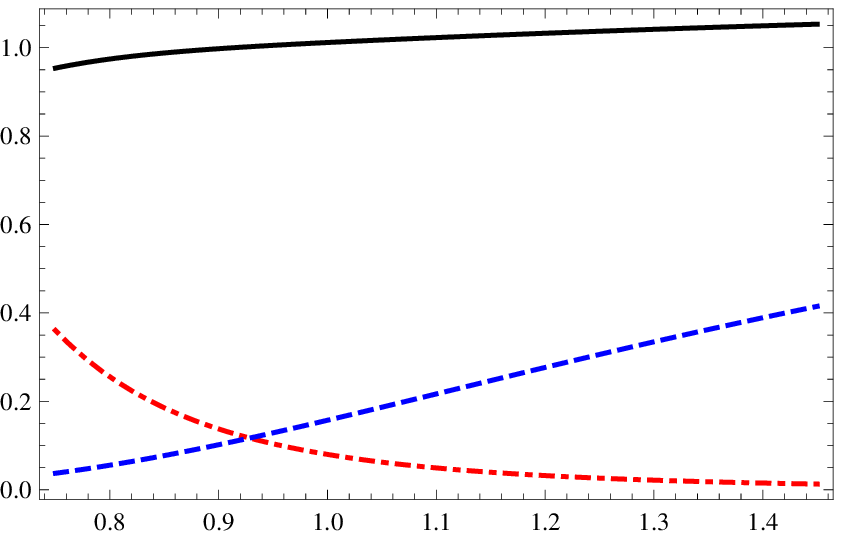}
\put (5,3) {$M$}
\end{minipage}~
\begin{minipage}[c]{.5\textwidth}\hspace*{2mm}
\includegraphics[width=0.8\textwidth,clip=true]{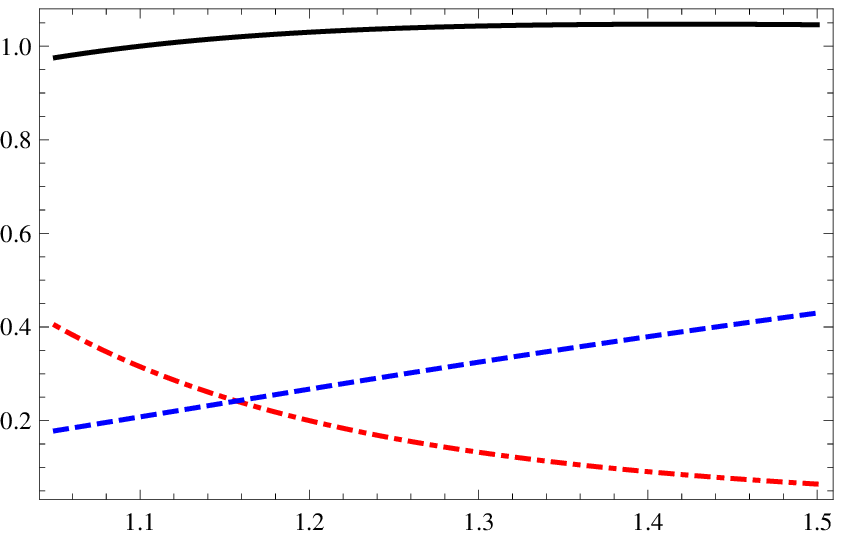}
\put (5,3) {$M$}
\end{minipage}
\vspace*{2mm}

Fig.25 (left):\, the solid line is the ratio $L_o(M^2)/R_o(M^2)$ of the left and right sides of (6.10) as a function of $M$;\, the dot-dashed line is the ratio $R^{\rm power}_{o}(M^2)/R^o_{o}$ which shows the relative value of non-perturbative power corrections;\, the dashed line is $e^{-s_o/M^2}$ which shows the relative role of the continuum\,.\\
\vspace*{2mm}
\hspace*{4mm}Fig.25 (right):\, the solid line is the ratio $L_2(M^2)/R_2(M^2)$ of the left and right sides of (6.15) as a function of $M$;\, the dot-dashed line is the ratio $R^{\rm power}_{2}(M^2)/R^{o}_{2}$;\, the dashed line is $e^{-s_2/M^2}$.\\
\vspace*{2mm}

The fit of (6.15) in the range $1.05<M<1.45\,GeV$ is shown in Fig.25 (right) and gives
\be
\langle\xi^2\rangle_{\pi}\simeq 0.40\,\,\,\ra\,\,\, a_2^{\pi}=\frac{35}{12}\langle
\xi^2-0.2\rangle_{\pi}\simeq 0.58\quad {\rm at}\quad \mu\simeq{\ov M}\simeq 1.25\,GeV\,,
\ee
\be
\langle\xi^2\rangle_{a}\simeq 0.10\,\,\,\ra\,\,\, a_2^{a}\simeq -\,0.30\,\,\,{\rm at}\quad \mu\simeq{\ov M}\simeq 1.25\,GeV,\,\,\, s_2\simeq 1.9\,GeV^2\,.\nn
\ee

Therefore, we confirm the results for $\langle\xi^2\rangle_{\pi}$ obtained in \cite{cz-82}. As for the accuracy of the result (6.16) for $\langle\xi^2\rangle_{\pi}$, it is known that it is difficult to estimate very reliably the accuracy of results obtained from the QCD sum rules (and different types of sum rules have significantly different accuracies). The stability of the sum rule (6.15) in $M^2$ is very good, see Fig.25 (right), and the accuracy of the quark condensate value (6.12),(6.13), which is most important numerically in (6.15), is sufficiently good at present. On the whole, we estimated the accuracy of (6.16) as $(0.40\pm 0.03)$.

Recall also a simple meaning of (6.16), see e.g. \cite{cz-rev}. The naive perturbative {\it local} duality (i.e. the limit $M\ra\infty$ in sum rules for each resonance {\it separately}) predicts that all meson wave functions are close to the asymptotic one, i.e. $\langle\xi^2\rangle_{\pi}^{asy}\simeq \langle\xi^2
\rangle_{a}^{asy}\simeq 0.20$. The nonperturbative effects change significantly, in general, the wave function forms of separate mesons.  In the case considered, the pion wave function is wider than the asymptotic one, while the $a_1$-meson wave function is narrower. But, on the average, "the mean wave function" is not much different from the asymptotic one. Indeed, using the numbers from (6.14) and (6.16) one obtains
\be
\frac{f^2_{\pi}\langle\xi^2\rangle_{\pi}+f_a^2\langle\xi^2\rangle_{a}}{f^2_{\pi}+f^2_{a}}\simeq 0.19\,.
\ee
\vspace*{2mm}

Let us trace now, for example, the origin of the result $a_2^{\pi}(\mu\simeq 1\,GeV)=(0.28\pm 0.08)$ in \cite{BBL-06} (the same concerns also \cite{KMM-04} and also the value $a_2^{K}(\mu\simeq 1\,GeV)$ for the K-meson in \cite{KMM-04,BBL-06}), see Table 3. There are two main reasons why the result for $a_2^{\pi}$ in (6.16) is twice larger.\\
1)\, The first reason is pure numerical. The value of the quark condensate used in \cite{BBL-06} is considerably smaller. It was used in \cite{BBL-06} (see Table A in the Appendix B therein)
\be
\langle u{\ov u}\rangle(\mu=1\,GeV)\simeq (0.240\,MeV)^3\,,\quad \alpha_s(\mu=1\,GeV)\simeq 0.53\,,
\ee
and therefore
\be
\langle\sqrt{\alpha_s} u{\ov u}\rangle^2\simeq 1.0\cdot 10^{-4}\,GeV^6\,\, {\rm used \,\, in}\,\, \cite{BBL-06} \quad \leftrightarrow\quad\langle\sqrt{\alpha_s} u{\ov u}\rangle^2\simeq 1.8\cdot 10^{-4}\,GeV^6\,\,{\rm used \,\, in}\,\, \cite{cz-82}\,.
\ee
The value of this condensate is very important in sum rules (6.10),(6.15).\\
2) The second reason is of a qualitative nature. The $a_{1}(1230)$-meson contribution was considered in \cite{BBL-06} as a part of the effective continuum in sum rules (6.10) and (6.15), so that they were taken in the form.~ -\\

Instead of (6.10):
\be
\Biggl [ L^{\rm Braun}_o(M^2)= f_{\pi}^2\Biggr ]=R^{\rm Braun}_o(M^2)\equiv \Biggl [R^{\rm Braun,\,\rm loop}_{o}(M^2)+R^{\rm power}_{o}(M^2)\Biggr ]\,,
\ee
\be
R^{\rm Braun,\,\rm loop}_{o}(M^2)=\frac{1}{4\pi^2}\Bigl (1+\frac{{\ov\alpha}_s}{\pi}\Bigr )M^2\Bigl (1-e^{-{\ov s}_{o}/M^2}\Bigl )\equiv R_{o}^{\rm Braun,\, o}\Bigl (1-e^{-{\ov s}_{o}/M^2}\Bigr ),\nn
\ee
\be
R^{\rm power}_{o}(M^2)=\frac{1}{12 M^2}\langle\frac{\alpha_s}{\pi}G^2\rangle+11\frac{16\pi}{81 M^4}
\langle{\sqrt\alpha_s}u{\ov u}\rangle^2\,.\nn
\ee

Instead of (6.15):
\be
\Biggl [L^{\rm Braun}_{2}(M^2)= f_{\pi}^2\Bigl (\langle\xi^2\rangle_{\pi}=0.2+\frac{12}{35}a_2^{\pi}\Bigr )
\Biggr ] =R^{\rm Braun}_2(M^2)\equiv\Biggl [R^{\rm Braun,\,\rm loop}_{2}(M^2)+R^{\rm power}_{2}(M^2)\Biggr],
\ee
\be
R^{\rm Braun,\,\rm loop}_{2}(M^2)=\frac{1}{20\pi^2}\Bigl (1+\frac{5}{3}\frac{{\ov\alpha}_s}{\pi}\Bigr )M^2\Bigl (1-e^{-{\ov s}_{2}/M^2}\Bigr )\equiv R_{2}^{\rm Braun,\,\rm o}\Bigl (1-e^{-{\ov s}_{2}/M^2}\Bigr ),\nn
\ee
\be
R^{\rm power}_{2}(M^2)=\frac{1}{12 M^2}\langle\frac{\alpha_s}{\pi}G^2\rangle+19\frac{16\pi}{81 M^4}
\langle{\sqrt\alpha_s}u{\ov u}\rangle^2\,.\nn
\ee

It is well known that the sum rule (6.20) requires ${\ov s}_{o}\simeq 0.72\,GeV^2$ for the pion duality interval and, moreover, the appropriate stability region is very narrow here: $0.8<M<0.9\,GeV$.\\

The fit of (6.21) with the quark condensate (6.13) is shown in Fig.26 (left) (the meaning of curves here is the same as in Fig.25) and gives
\be
\langle\xi^2\rangle_{\pi}(\mu\simeq{\ov M}\simeq 1.2\,GeV)\simeq 0.41\,\,\,\ra\,\,\, a_2^{\pi}(\mu\simeq 1.2\,GeV) \simeq 0.61\,,\quad {\ov s}_{2}\simeq 1.35\,GeV^2\,,
\ee
and this value of $\langle\xi^2\rangle_{\pi}$ agrees with (6.16). But the value of the effective continuum threshold is much larger in (6.21) than in (6.20),\, ${\ov s}_{2}\simeq 2\, {\ov s}_{o}$ and, what is even more important, the appropriate stability region of (6.21) is $1.05<M<1.35\,GeV$ and it does not overlap with those of (6.20). Unfortunately, these facts were ignored in \cite{BBL-06} where the sum rules (6.20) and (6.21) were formally combined with the common effective threshold $s^{\rm Br}_{2}$\,:
\be
a_2^{\pi,Braun}=\Biggl [{\tilde R}^{\,\rm loop}_2=\frac{7}{72\pi^2 f^2_{\pi}}\frac{{\ov\alpha}_s}{\pi}M^2\Bigl (1-e^{-s^{\rm Br}_{2}/M^2}\Bigr )\Biggr ]+\Biggl [{\tilde R}^{\,\rm power}_2=\frac{7}{36 f^2_{\pi} M^2}\langle\frac{\alpha_s}{\pi}G^2\rangle+\frac{784\pi}{81 f^2_{\pi} M^4}\langle{\sqrt\alpha_s}u{\ov u}\rangle^2\Biggr ]\,.
\ee
\vspace*{2mm}

\begin{minipage}[c]{.5\textwidth}\hspace*{1cm}
\includegraphics[width=0.75\textwidth,clip=true]{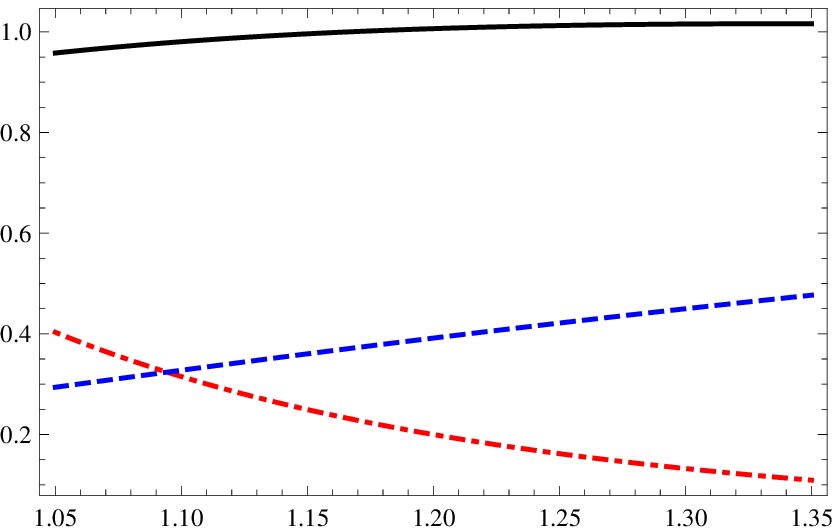}
\put (5,3) {$M$}
\end{minipage}~
\begin{minipage}[c]{.5\textwidth}\hspace*{2mm}
\includegraphics[width=0.75\textwidth,clip=true]{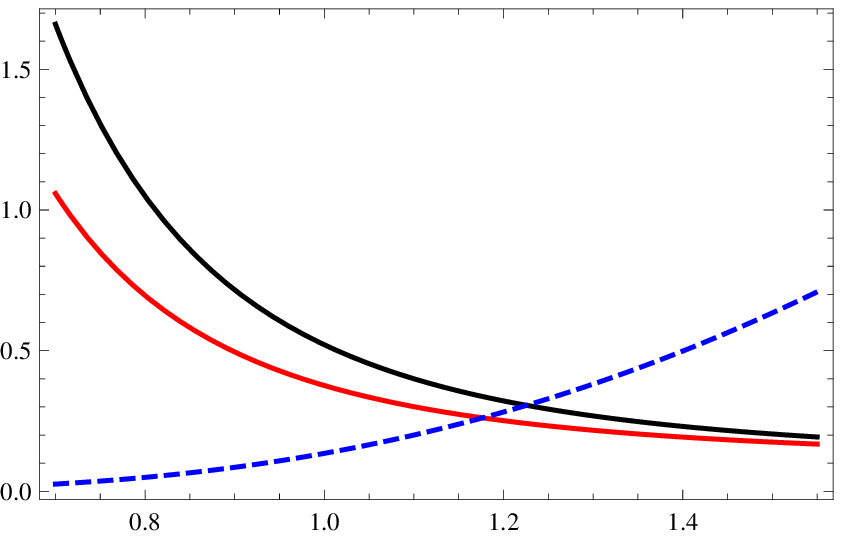}
\put (-215,75) {\rotatebox{90} {$a_2^{\pi,\,Braun}$}}\put (5,3) {$M$}
\end{minipage}
\vspace*{2mm}

Fig.26 (left): the fit of the sum rule (6.21), the meaning of curves is the same here as in Fig.25.\\
\hspace*{6mm}Fig.26 (right): the fit of the sum rule (6.23), the upper and lower solid curves correspond to the quark condensates (6.13) and (6.18). The dashed curve here is the ratio
${\tilde R}^{\,\rm loop}_2/{\tilde R}^{\,\rm power}_2$ in (6.23).\\
\vspace*{1mm}

The value of $a_2^{\pi,Braun}$ from (6.23) as a function of $M$ is shown in Fig.26 (right). It is seen that there is no stability.
\footnote{\,
Because the coefficient of the leading power term $\sim \alpha_s M^2$ in (6.23) is very small, this sum rule is not very sensitive to the value of $s^{\rm Br}_{2}\simeq 1.0\, GeV^2$.
}

Let us choose e.g. the region $0.8<M<1.35\,GeV$. The upper solid curve in Fig.26 (right) is for the quark condensate $\langle{\sqrt\alpha_s}u{\ov u}\rangle^2=1.8\cdot 10^{-4}\,GeV^6$ from (6.13) and we obtain: $a_2^{\pi}\simeq (0.65\pm 0.40)$. The lower solid curve in Fig.26 (right) is for the quark condensate $\langle{\sqrt\alpha_s}u{\ov u}\rangle^2=1.0\cdot 10^{-4}\,GeV^6$
from (6.19) used in \cite{BBL-06} and we obtain: $a_2^{\pi}\simeq (0.45\pm 0.25)$.

Let us choose now the region $1.0<M<1.35\,GeV$. We obtain: $a_2^{\pi}\simeq (0.38\pm 0.14)$ from the upper curve and $a_2^{\pi}\simeq (0.29\pm 0.09)$ from the lower one, this last in agreement with \cite{BBL-06}.\\

Now, the reasons for a difference between the larger value $a_2^{\pi}\simeq 0.60$ obtained from (6.15) or
(6.21) and the smaller value $a_2^{\pi}\simeq 0.30$ obtained in \cite{BBL-06} are clear.\,-\\
a)\,\, In principle, one can combine  as $[\,(6.15)-0.2\,(6.10)\,]$ the sum rules (6.10) and (6.15) which both account for the $a_1(1230)$-meson explicitly (the effective continuum thresholds in (6.10) and (6.15) are practically the same), and obtains then the analog of (6.23) for $a_2^{\pi}$. Clearly, inside the appropriate region of stability for this combination, e.g. $1.05<M<1.35\,GeV$, one obtains the same result $a_2^{\pi}\simeq 0.60$, see Fig.26 (left). The main difference with (6.23) will be {\it the additional negative contribution} of $a_1(1230)$-meson in the left hand side of (6.23) because $a_2^{a_1}\simeq -0.30$, see (6.16). This negative $a_1(1230)$-meson contribution improves greatly the stability of this sum rule and results in a larger value of $a_2^{\pi}\simeq 0.60$.
\\
b)\,\, Instead, one can combine as $[\,(6.21)-0.2\,(6.20)\,]$ the sum rules (6.20) and (6.21). But in this case it is impossible to obtain a meaningful result for $a_2^{\pi}$ because the regions of appropriate stabilities are not overlapping in (6.20) and (6.21), see Fig.26 (right), and varying the value of the effective threshold $s_2^{\rm Br}$ in (6.23) it is impossible to achieve a reasonable stability.\\
\vspace*{2mm}

\subsubsection{"Improved" QCD sum rules with nonlocal condensates}

\hspace*{3mm} A large number of papers (see e.g. the last paper \cite{MPS-14} and references therein) have been published with calculations of $F_{\gamma \pi}(Q^2)$ using the light cone sum rule proposed in \cite{Khodja-99}. The $\chi^2$-values were carefully calculated in these papers in comparing with data the results for the $\phi_{\pi}^{\rm BMS}(x)$ (BMS=Bakulev-Mikhailov-Stefanis) model pion wave function obtained from the "improved" QCD sum rules with non-local condensates, and for other model pion wave functions. This approach with non-local condensates was proposed and developed originally in \cite{M-Rad-86,M-Rad-90,M-Rad-92}.

The original (and standard) approach \cite{SVZ} for obtaining QCD sum rules calculates the (Euclidean space) correlator of two local currents at small distances expanding it into a power series of local vacuum condensates of increasing dimension
\footnote{\,\,
\,In practice, the Fourier transform of (6.24) is usually calculated, supplied in addition by
a special "Borelization" procedure which suppresses contributions of poorly known higher
dimension terms. This is not of principle importance, but is a matter of technical convenience
and improving the expected accuracy. On account of loop corrections the coefficients $C_n$ depend logarithmically on the scale $z^2$ (or $q^2$).}
\be
\langle 0|T\, J_1(z) J_2(0)|0\rangle =
\Pi(z)\sum_{n=0}^{\infty} (z^2)^n\,C_n(z^2,\mu_o^2)\, \langle 0|O_n(0)|0 \rangle_{\mu_o}\, ,
\ee
where the overall factor $\Pi(z)$ describes the singularity of the leading $n=0$ term. In practical applications this series of power corrections is terminated after first several terms, so that only a small number of phenomenological parameters $\langle 0|O_n|0\rangle,\, n<n_o$ determine the behavior of many different correlators. This standard approach was used in \cite{cz-82} to calculate a few lowest moments of the pion wave function, $\langle \xi^{2n}\rangle_{\pi}=\int_0^1 dx\,\phi_{\pi}(x)(2x-1)^{2n}.$ It appeared that the values of these moments are larger significantly than those for $\phi^{asy}(x)=6x(1-x)$. The most important power corrections in these sum rules originated from the local quark condensate $\sim \langle 0|{\ov q}(0)q(0)|0\rangle ^2$.

The "improved" approach \cite{M-Rad-86,M-Rad-90,M-Rad-92} proposed not to expand a few lowest dimension non-local condensates, for instance (the gauge links are always implied):
$\Phi(z)=\langle 0|{\ov q}(0)\Gamma q(z)|0\rangle,\,\langle 0|G_{\mu\nu}(0)G_{\mu
\nu}(z)|0\rangle$, into a power series in $z$, but to keep them as  whole non-local
objects, while neglecting contributions of all other higher dimension non-local condensates.
This is equivalent to keeping in QCD sum rules a definite (infinite) subset of higher order power
corrections while neglecting at the same time all other power corrections which are supposed to be small. {\it This is the basic assumption underlying this "improvement"}. In other words, it was supposed
that the numerically largest contributions to the coefficients $C_n$ in (6.24) originate
from expansion of a few lowest dimension non-local condensates, while contributions to $C_n$
from higher dimension non-local condensates are small and can be neglected. Clearly, without
this basic assumption the "improvement" has no much meaning as it is impossible to account
for all multi-local condensates. But really, no one justification of this basic assumption
has been presented in \cite{M-Rad-86,M-Rad-90,M-Rad-92} and in numerous papers of BMS.

Moreover, within this approach one has to specify beforehand not a few numbers like $\langle
0|G^2(0)|0\rangle,\\ \langle 0|{\ov q}(0)q(0)|0\rangle$, but a number of {\it functions}
describing those non-local condensates which are kept unexpanded. Really, nothing definite is
known about these functions, except (at best) their values and some of their first derivatives at
the origin. So, in \cite{M-Rad-86,M-Rad-90,M-Rad-92} and in \cite{BMS-01} definite model forms of these
functions were used, which are arbitrary to a large extent. The uncertainties introduced to
the answer by chosen models are poorly controllable. In principle, with such kind of
"improvements" the whole approach nearly loses its meaning, because to find a few {\it pure
numbers} $\langle \xi^{2n}\rangle_{\pi}$ one has to specify beforehand a number of {\it
poorly known functions}.

As for $\langle \xi^{2n}\rangle_{\pi} $, the main "improvement" of the standard sum rules was
a replacement $\langle 0|{\ov q}(0)q(0)|0\rangle ^2\ra \langle 0|{\ov q}(0)q(x){\ov q}(y)q(z)
|0\rangle$, "factorized {\it via} the vacuum dominance hypothesis to the product of two
simplest $\langle 0|{\ov q}(0)q(z)|0\rangle $ condensates" \cite{M-Rad-86,BMS-01}.
Clearly, such a "functional factorization" looks very doubtful in comparison with the standard
"one number factorization" of $\langle 0|{\ov q}(0)q(0){\ov q}(0)q(0)|0\rangle$.

As for the above described basic assumption of "the improved approach" to the QCD sum rules,
it was emphasized in \cite{ch-06} that it can be checked explicitly using such a correlator for
which the exact answer is known in QCD. With this purpose, the correlator of the axial vector $A_{\mu}(0)={\ov d}(0)\gamma_{\mu}\gamma_5 u(0)$ and pseudoscalar $P(z)={\ov u}(z)i\gamma_5 d(z)$
currents was considered
\be
I_{\mu}(z)=\langle 0|T\, A_{\mu}(0)\,P(z)|0\rangle \equiv \frac{z_\mu}{z^4}\,I(z^2)\,,
\ee
\be
I(z^2)=\sum_{n=0}^\infty (z^2)^n\,C_n\,Z_n \langle 0| O_n(0)|0\rangle_{\mu_o},\;
C_n=\sum_{i=a,b,c...}C_n^i=\sum_i C_n^{(\rm Born,\,i)}\,\Bigl (1+f_n^i\Bigr ),\nonumber
\ee
where $Z_n=Z_n(\mu^2\sim 1/z^2,\mu_o^2)$ are the logarithmic renormalization factors of operators
$O_n$, while $f_n^i=O(\alpha_s)$ is due to non-leading terms in loop corrections to the hard
kernels (in what follows $f_n^i$ will be omitted for simplicity as they play no significant role).

The exact answer for this correlator is known in the chiral limit $m_{u,d}=0$, so that when
calculating it in various approximations one can compare which one is really better, and this
will be a clear check. So, forget for a time that we know the exact answer and
calculate this correlator using the "improved" and standard approaches.

The exact analog of the above described basic assumption of \cite{M-Rad-86,M-Rad-90,M-Rad-92} predicts here that, at each given $n$, the largest coefficients in (6.25) originate from the expansion of the lowest dimension non-local quark condensate $\Phi(z^2)=\langle 0|{\ov q}(0)q(z)|0\rangle$ shown in Fig.27 (left). Decomposing it in powers of $z^2$, this results in a tower of power corrections with "the largest
coefficients" $ {C}^{(\rm Born,\,a)}_n$:
\be
I^{(\,\rm fig.27(left)\,)}(z^2)= \sum_{n=0}^\infty (z^2)^n\,{C}^{(\rm Born,\,a)}_n
Z_n \langle 0| O_n(0)|0\rangle_{\mu_o} \,,
\ee
where $O_n$ are the corresponding local operators, $O_o\sim {\ov q}q,\,\, O_1\sim {\ov q}
\sigma G q$, etc.

The contribution of Fig.27 (right) to $I(z^2)$ in (6.25) is originally described by the
three-local higher dimension condensate $\langle 0|{\ov q}(0)g\sigma_{\mu\nu}G(x)_{\mu\nu}q(z)|0\rangle$.
Its expansion produces finally a similar series in powers of $z^2$ which starts from $\langle O_1\rangle$
and have coefficients $C^{(\rm Born,\,b)}_n$. Besides, an infinite number of other diagrams
(not shown explicitly in Fig.27) with additional emitted gluons produce similar series starting
from higher dimension condensates $\langle O_n \rangle$ with $n\geq 2$ and with the coefficients
$C^{(\rm Born,\,c)}_n$, etc.
In the framework of the above basic assumption, there should be a clear numerical hierarchy:
$C^{(\rm Born,\,a)}_n \gg C^{(\rm Born,\,b)}_n \gg C^{(\rm Born,\,i)}_n \gg C^{(\rm Born,\,
i+1)}_n\,...\,$ (all $C_n^{(\rm Born,\,i)}$ are parametrically $O(1)$), so that one can retain
only the largest terms $C^{(\rm Born,\,a)}_n$ and safely neglect all others.\\


\begin{minipage}[c]{.9\textwidth}\hspace*{2cm}
\includegraphics[width=0.75\textwidth]{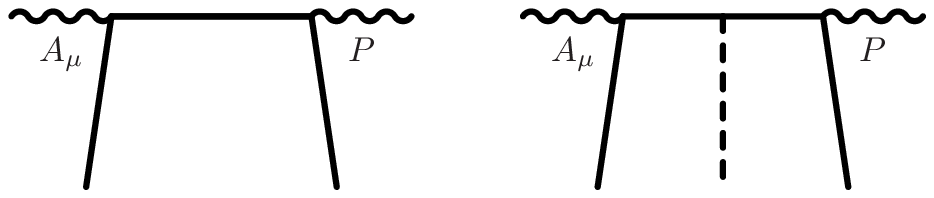}
\end{minipage}
\vspace*{2mm}

Figs.27 (left): the series of power terms originating from the bilocal quark condensate $\langle 0|{\ov q}(0)q(z)|0\rangle$.\\
\hspace*{-2mm} Figs.27 (right): the series of power terms originating from the nonlocal condensate $\langle 0|{\ov q}(0) g\sigma G(x)\, q(z)|0\rangle$.\\
\vspace*{0.5mm}

Recall now that the exact answer for this correlator (6.25) is very simple in the chiral limit (the spectral
density is saturated by the one pion contribution only), and is exhausted by the first term
with $n=0$ in (6.26). In other words, there are no corrections in powers of $(z^2)^{\rm n\,>\,0}$ in this correlator at all. The reason is that other contributions with coefficients $C^{(i\neq a)}_n$ in (6.25) neglected in the "improved" approach cancel exactly all (except for the first one) "the most important"
coefficients $C^{(a)}_n$ in (6.26). For instance, the first power correction $\sim\langle O_1 \rangle$
from Fig.27 (right) diagram cancels the second term $n=1$ in (6.26) from Fig.27 (left) diagram. And power corrections from other higher dimension multi-local condensates from next diagrams not shown explicitly in Fig.27, together with next power corrections from Fig.27 (right) diagram, cancel exactly all next $n\geq 2$ "the most important" terms $C_n^{(a)}$ in (6.26) from Fig.27 (left) diagram.

Therefore, the basic assumption of the "improved" approach clearly fails: those power corrections which are basically claimed to be "less important" in comparison with "the most important corrections", appeared to be not small and even cancel completely here all "largest corrections".

On the other hand, calculating the correlator in (6.25) in the standard approach (which is a
direct QCD calculation accounting for {\it all terms of given dimension}), one finds that the
sum of all corrections of given dimension is zero, as it should be.

The conclusion made in \cite{ch-06} was that the above described "improved approach" to QCD sum rules can easily give, in general, the misleading results.
\footnote{\,
See \cite{ch-06} for much more details.
}

As for concrete results obtained within this "improved" approach, they changed significantly over time (for a not clear reason). It was obtained in \cite{M-Rad-90}: $a_2(\mu\simeq 1\,GeV)\simeq 0.35,\,\, a_4(\mu
\simeq 1\,GeV)\simeq 0.23$, while the same author obtained $a_2(\mu\simeq 1\,GeV)\simeq (-0.20)$ in the later paper \cite{Rad-94}, while the result of \cite{BMS-01} looks as $a_2(\mu=1\,GeV)= (0.19\pm 0.06),\,\, a_4(\mu= 1\,GeV)\simeq (-0.14)$ (all these results within the same approach).

\subsubsection{Sum rules for form factors}

\hspace*{3mm} A useful example with an analog of QCD sum rules for the hadron form factors has been presented in \cite{Melikhov-09}. Considered was the form factor $F_o(q)$ of the charged particle in the ground state of the quantum mechanical oscillator . The exact solution is known in this case and one can check some qualitative properties of analogous QCD sum rules. Author's purpose was to check to what extent $z_{\rm eff}(T,q)$ (the analog of the effective continuum threshold $s_o$ in QCD sum rules) can be taken as a constant $z_c$ (this is a common practice in using various QCD sum rules for numerical calculations). Some results are presented in Fig.28 (see \cite{Melikhov-09} for all details). Here\,:\, $q$ is the momentum transfer, $\omega$ is the oscillator frequency, $m$ is the particle mass, $T$ plays the role of the inverse Borel parameter $1/M$.
\vspace*{3mm}

\begin{minipage}[c]{.5\textwidth}\hspace*{-1cm}
\includegraphics[width=0.9\textwidth,,clip=true]{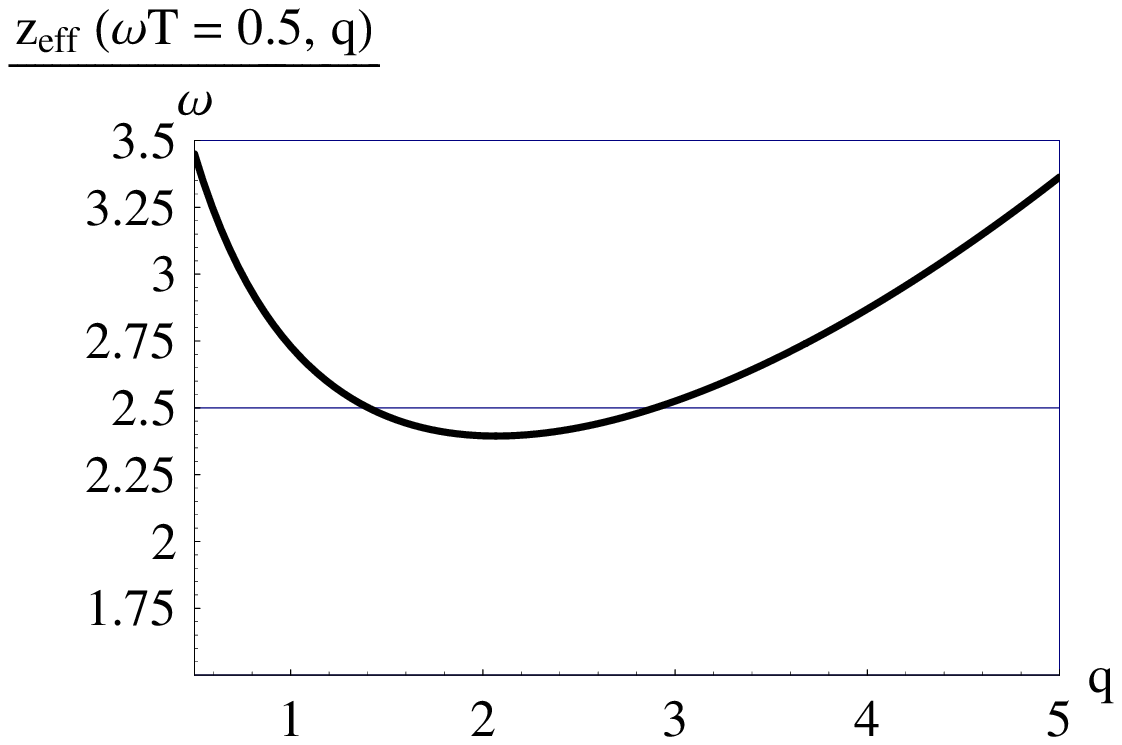}
\end{minipage}~
\begin{minipage}[c]{.5\textwidth}\hspace*{-1cm}
\includegraphics[width=0.9\textwidth,,clip=true]{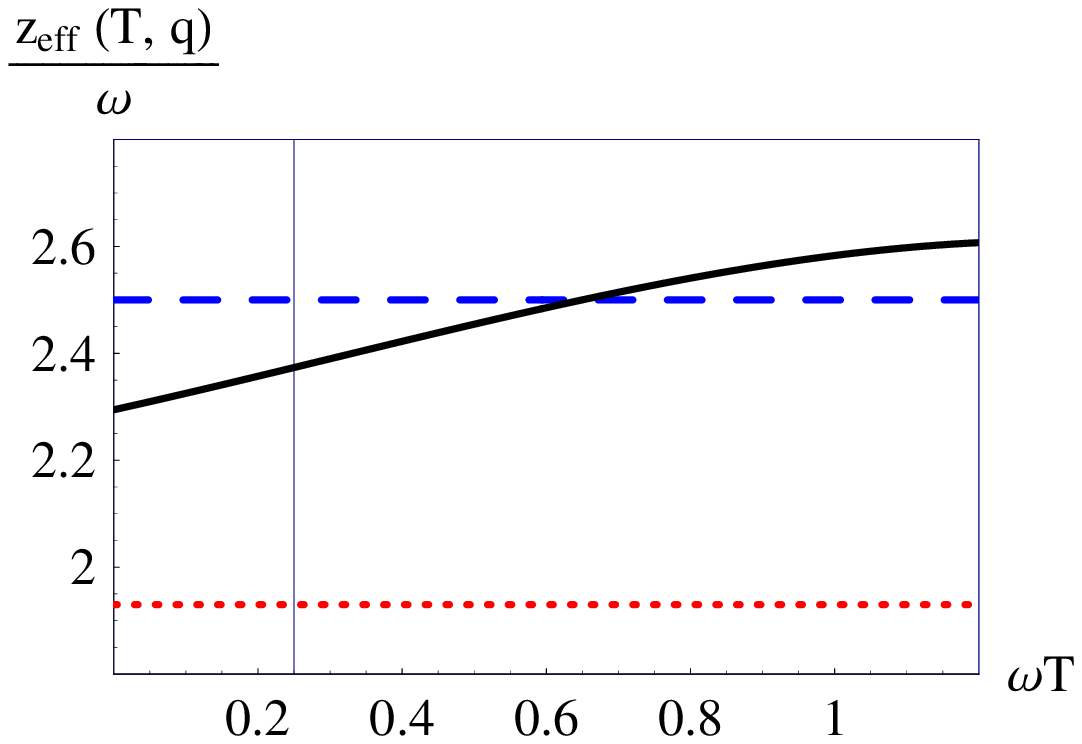}
\end{minipage}~
\vspace*{1mm}

Fig.28 (left).\, The exact effective continuum threshold $z_{\rm eff}(T,q)$ (as obtained by solving the equation for the exact bound-state parameters) vs $q$ for $\omega T=0.5$. A horizontal line at $z_c/\omega=2.5$ is given as a benchmark: this constant was shown to provide a good approximation to the exact continuum threshold in the sum rule for the two-point correlator  $\Pi(T)$.\\
\hspace*{3mm} Fig.28 (right).\, The intermediate momentum transfer $q=2\omega$ ($q^2/4m\omega=1$).
Dashed line: constant effective continuum threshold $z_c/\omega=2.5$. Dotted line: $T$-independent effective continuum threshold. Solid line: tuned $T$-dependent exact effective continuum threshold $z_{\rm eff}(T,q)$.\\

It is seen from Fig.28 that, while the dependence of $z_{\rm eff}(T,q)$ on $T$ (at fixed $q$) does not look very significant (at least here) and can be diminished to some extent by tuning its constant value and the range of $T$ (see the solid and dashed lines in Fig.28b), the dependence of $z_{\rm eff}(T,q)$ on $q$ may be highly nontrivial. Unfortunately, the common practice is to use the constant value of the effective continuum threshold in all kinds of sum rules.

This dependence of the effective continuum threshold $s_o(Q)$ on the momentum transfer may be especially dangerous e.g. in the light cone QCD sum rules (LCSR) for hadron form factors, especially at large $Q$. It is dangerous because the constant value of $s_o$ leads usually to the right parametric power behavior of the form factor at large $Q^2$, $F(Q^2)\sim (s_o/Q^2)^n$  (known beforehand from QCD \cite{c-1}), but if the appropriate numerical values of the $Q^2$-dependent effective threshold $s_o(Q)$ at intermediate and large values of $Q$ differ, say, two times, this will change the numerical value of the form factor also two times (or more, depending on the value of $n$).

\subsubsection{Lattice calculations of $\langle\xi^2\rangle_{\pi}$}

\hspace*{3mm} These have been performed in \cite{Braun-latt,Arthur-latt}. The present accuracy of these calculations is not high, but, nevertheless, they both give a clear indication that the pion wave function is significantly wider than the asymptotic one, see Table.3. It seems one of the main problems here at present is that masses of light $u$ and $d$-quarks (and so the pion mass) are not sufficiently small, see Fig.29.\\

\begin{minipage}[c]{.45\textwidth}\hspace*{-1.8cm}
\includegraphics [trim=0mm 0mm 0mm 0mm, width=1.2\textwidth,clip=true]{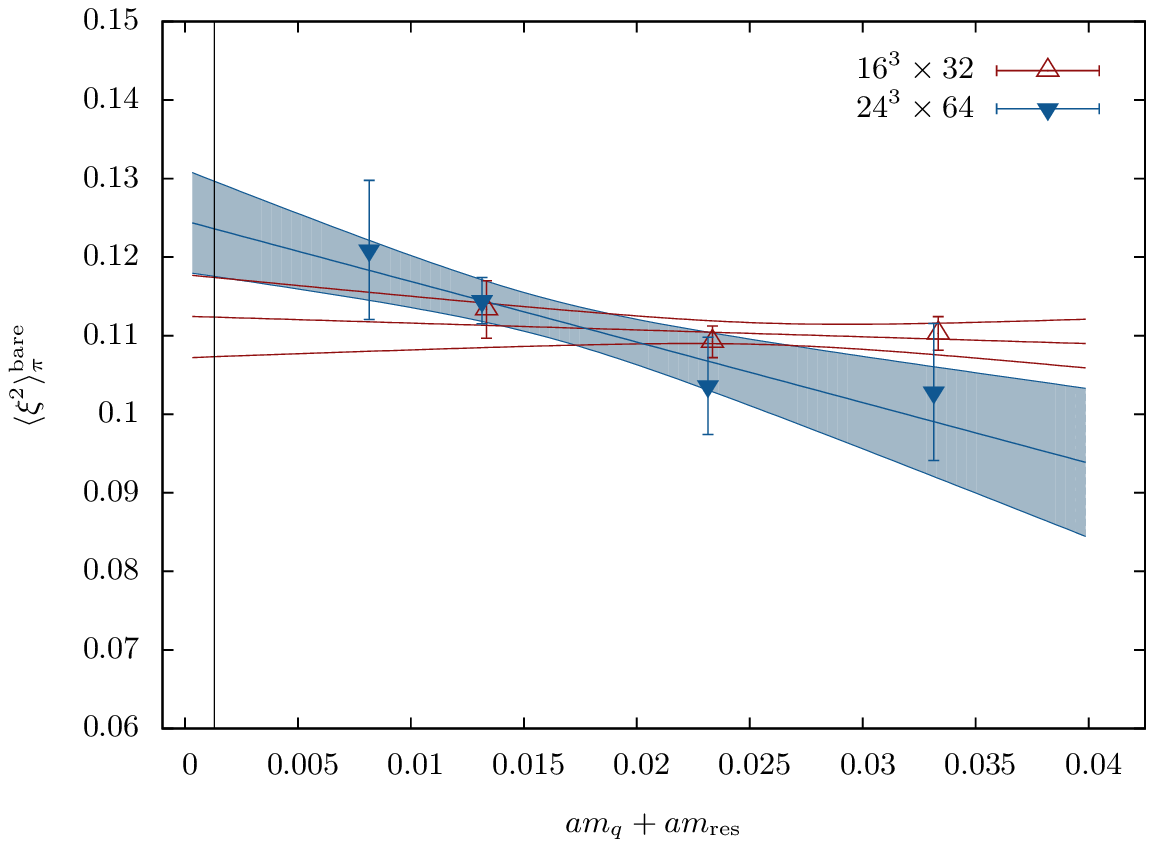}
\end{minipage}
\begin{minipage}[c]{.5\textwidth}\vspace*{-1mm} \hspace*{2cm} Fig.29 \cite{Arthur-latt}.
\,\,\,\,   Dependence of $\langle\xi^2=(x_d-x_u)^2\rangle^{\rm bare}_{\pi}$ at $\mu=2\,GeV$ (vertical axis) on the quark mass corresponding  to $330\,MeV <m_{\pi}<670\,MeV$ (horizontal axis), for two different  lattice volumes. It is seen that there is no dependence of $\langle\xi^2\rangle$ on the light quark mass value on the smaller lattice (and this is clearly a flow of not sufficiently large lattice volume), while there appears a significant slope on a larger lattice.  What is still lacking at present is a check that the slope does  not increase further for even larger lattices and smaller quark masses.
\footnote{\,Recall: the heavier is quark, the narrower is wave function, i.e. it has the smaller value of $\langle\xi^2\rangle$.
}
\\
\end{minipage}

\subsubsection{On the endpoint behavior of $\phi_{\pi}(x,\mu)$ at $x\ll 1$}

\hspace*{3mm} Some general arguments in favor that the leading twist meson wave functions $\phi(x,\mu\sim 1\,GeV)$ behave at $x\ll 1$ in the same way as the asymptotic wave function, $\phi^{\rm asy}(x)=6x(1-x)$, have been given long ago in \cite{czz-83} (see Appendix therein). At present, the experimental data appeared on the form factors $F_{VP}(q^2)$ (V=vector, P=pseudoscalar) at large $s=q^2$, and these can be used to check the parametric behavior of $\phi_{\pi}(x\ll 1,\mu\sim 1\,GeV)$.

The form factor $F_{VP}(Q^2)$ is defined by
\be
\langle V_{\lambda}(p_2)|J_{\mu}(0)|P(p_1)\rangle=\epsilon_{\mu\nu\rho\sigma}\, e_{\lambda}^{\nu}\, p_2^{\rho}\, p_1^{\sigma}\,F_{VP}(Q^2)\,,
\ee
where $e_{\lambda}^{\nu}$ is the polarization vector of $V_{\lambda}$. For kinematical reasons, the vector meson is produced in this process with the unit helicity only, $|\lambda|=1$.

Therefore, at large $Q^2$, it seems that the form factor is, at least, highly suppressed. The reason is that: a)\, both quarks have their momenta nearly parallel to the momenta of their parent hadrons;\\ b)\, spin projections of two quarks onto the hadron momentum are opposite in $|P\rangle$, while they are the same in $|V_{|\lambda|=1}\rangle$. This contradicts the helicity conservation for massless quarks in perturbative QCD.
\footnote{\,
The dynamical nonperturbative constituent mass of the quark behaves as $m_{\rm const}(Q^2)\sim \langle \alpha_{s} q{\bar q}\rangle/Q^2$ at large $Q^2$ and does not help much.
}
However, because the quark transverse momentum ${\vec k}_{\perp}$ is really nonzero in the meson, the direction of the meson momentum $\vec p$ and the quark momentum $\vec k$ are not exactly the same even at large $|{\vec p}\,|\sim Q\gg 1\,GeV$, so that there is a small but nonzero fraction $\sim \langle{|\vec k}_{\perp}|\rangle/Q\ll 1$ of the opposite quark spin projection onto direction of the meson momentum.

As a result, instead of the naive dimensional counting $F_{VP}(Q^2)\sim 1/Q^{3}$, it was predicted in \cite{c-1} that this form factor behaves really as $F_{VP}(Q^2)\sim 1/Q^{4}$, resulting in the behavior $\sigma(e^+e^- \to VP)\sim 1/s^4$.

The leading power contributions to $F_{\omega\pi}(Q^2)$ originated from the one gluon exchange diagrams were calculated in \cite{cz-rev} (see section 9.2). They are particularly sensitive to the endpoint behavior of the leading twist meson wave functions of the type $\phi^{(\alpha)}_{\pi,\,\omega}(x\ll 1,\mu\sim 1\,GeV)\sim x^{\alpha},\,\, 0\leq\alpha<1$, as there appear integrals like (and similar integrals with the leading twist $\omega_{|\lambda|=1}$ - meson wave function)
\be
Q^4 F^{(\alpha)}_{VP}(Q^2)\sim{\ov\alpha_s}\int_0^1 dx \int_0^1 dy \frac{\phi^{(\alpha)}_{\pi}(x)\,\phi^{\perp}_{V}(y)}{(x y+\tilde\delta)(x+\tilde\delta)}\,\,,\quad Q^2\tilde\delta\sim \langle {\vec k}_{\perp}^{\,2}\rangle\sim 0.1\,GeV^2\,\,,
\ee
where $Q^2\tilde\delta$ is the infrared cut off and $\phi^{\perp}_{V}(y)$ is the twist-3 two-particle wave function of the vector meson, its asymptotic form is \cite{cz-rev,czz-85}:\, $\phi^{\perp}_{V}(y,\mu\to \infty)\to 3[\,1+(2y-1)^2\,]/8$. For definiteness, we used in numerical calculations below the model form for $\phi^{\perp}_{V}(y,\mu=1\,GeV)$ obtained in \cite{Ball-Br-98} from QCD sum rules. Besides (6.28), there are also other contributions to $F^{(\alpha)}_{VP}(Q^2)$ with the similar $Q^2$-behavior, see \cite{cz-rev}. They are ignored for simplicity because our purpose here is not to calculate the absolute value of $F^{(\alpha)}_{VP}(Q^2)$ but to illustrate the strong dependence of its behavior with $Q^2$ on the parametric behavior of the leading twist meson wave functions $\phi^{(\alpha)}_{\pi,\,\omega}(x\ll 1,\mu\sim 1\,GeV)\sim x^{\alpha},\, 0\leq\alpha<1$, while for such type wave functions the omitted terms mainly change the overall normalization only.

If $\phi_{\pi}(x\ll 1,\mu=1\,GeV)\sim x$, the behavior of (6.28) will be $Q^4 F_{V\pi}(Q^2)\sim {\it const}$ (up to growing logarithmic factors), while for $\phi^{(\alpha)}_{\pi}(x\ll 1,\mu=1\,GeV)\sim x^{\alpha},\, 0\leq \alpha<1$ the integral will be dominated by the region $\quad x\sim \tilde\delta\sim 1/Q^2$ and the power behavior will be $Q^4 F^{(\alpha)}_{V\pi}(Q^2)\sim (Q^2)^{1-\alpha}$.

Therefore, for the wave functions of the type $\phi^{(\alpha)}_{\pi}(x,\mu=1\,GeV)\sim (x{\bar x})^{\alpha\,<\,1}$\,, the leading "double logarithmic" terms $\sim [\alpha_s\ln(Q/\mu)\ln(1/x)]^n$ and $\sim [\alpha_s\ln^2(1/x)]^n$ in loop corrections to (6.28) become of comparable importance. As was shown in the last paper in \cite{c-2} (see also section 3.5 in \cite{cz-rev}), these double logarithmic terms originated from loop corrections integrated over the range $\mu^2=1\,GeV^2<k^2<Q^2$ are universal and sum up into the factor
\be
I_{\rm rad}(Q^2,\sigma^2,\mu^2)=\Bigl (1-\frac{\alpha_s C_F}{4\pi}\Omega+\frac{1}{2!}(\frac{\alpha_s C_F}{4\pi}\Omega)^2+\dots\Bigr )\ra\Bigl (\frac{\mu^4}{Q^2\sigma^2}\Bigr )^{\tau(Q^2,\,\sigma^2)}\,,
\ee
\be
\Omega=\Bigl (2\ln\frac{Q^2}{\mu^2}-\ln\frac{Q^2}{\sigma^2}\Bigr )\ln \frac{Q^2}{\sigma^2}\,,\quad \tau(Q^2,\sigma^2)=\frac{C_F}{\bo}\ln\frac{\alpha_s(\sigma^2)}{\alpha_s(Q^2)}\,,\quad \sigma^2=Q^2(x y+\tilde\delta),
\ee
where $\sigma^2\ll Q^2$ is the virtuality of the exchanged gluon in the Born diagram. It is clear beforehand that in the endpoint region $x\sim 1\,GeV^2/Q^2,\, \sigma^2\lesssim 1\,GeV^2$, the "hard" kernel in (6.28) becomes really soft, with the scale of internal virtualities $\lesssim 1\,GeV^2$, and all perturbative and nonperturbative interactions inside this soft part of the amplitude will influence only the overall normalization of $F^{(\alpha)}_{VP}(Q^2)$ but not its dependence on $Q^2$. At the same time, the hard loop corrections should give the Sudakov form factor to the active quark carrying nearly the whole meson momentum, ${\ov x}=(1-x)\ra 1$. Indeed, see (6.29),(6.30): $I_{\rm rad}(Q^2,\sigma^2\sim\mu^2=1\,GeV^2)\sim S(Q^2,\mu^2=1\,GeV^2)\sim (1\,GeV^2/Q^2)^{\tau(Q^2,\, 1\,GeV^2)}$.

On the whole, on account of these leading loop corrections, (6.28) is replaced by
\be
\hspace*{-3mm}Q^4 F^{(\alpha)}_{VP}(Q^2)\sim\int_0^1 dx\int_0^1 dy\,\alpha_s(\sigma^2)I_{\rm rad}(Q^2,\sigma^2, \mu^2=1\,GeV^2)\frac{\phi^{(\alpha)}_{\pi}(x,\mu=1\,GeV)\,\phi^{\perp}
_{V}(y,\mu=1\,GeV)}{(x y+\tilde\delta)(x+\tilde\delta)}.
\ee

Formally, the expressions (6.29)-(6.30) sum all double logarithmic terms $(\alpha_s\ln A\ln B)^n$, where $A$ and $B$ are either $Q^2/\mu^2,\,\mu=1\,GeV$, or "x". The terms with powers of $\ln x$  are nonleading ones at $x=O(1)$, but become more and more important as "x" diminishes. And $\ln x$ becomes as important as $\ln Q^2/\mu^2$ at $x\sim \mu^2/Q^2$\,. Clearly, resummation of such terms is of importance for wave functions of the type $\phi^{(\alpha)}_{\pi}(x\ll 1,\mu=1\,GeV)\sim x^{\alpha\,<\,1}$ in integrals like (6.31).

Finally, we used in numerical calculations of (6.31) the simple modified perturbative form of the strong coupling
\be
\alpha_s(\mu^2)=\frac{4\pi}{\bo \ln\Bigl (
\mu^2+0.6\,GeV^2/\Lambda^2\Bigr )}\,,\quad \Lambda\simeq 200\,MeV\,.
\ee

The behavior of the ratio $F^{(\alpha)}_{VP}(Q^2)/F^{(\alpha)}_{VP}(Q_o^2)$ (6.31) in the range $Q^2_o=13.8\,GeV^2<Q^2<112\,GeV^2$ looks then as follows:
\footnote{\,
One has to realize that the power of $Q^2$ in (6.33)-(6.36) is the effective number in this range of energies. The real analytic form of the $Q^2$ dependence of $F^{(\alpha)}_{VP}(Q^2)$ in (6.31) is much more complicated. It is worth noting also that while the absolute value of $F^{(\alpha)}_{VP}(Q^2)$ depends strongly on the value of $Q^2\tilde\delta$ in (6.31) for $\phi^{(\alpha)}_{\pi}(x\ll 1,\,\,\mu=1\,GeV)\sim x^{\alpha\,<\,1}$, the ratios $F^{(\alpha)}_{VP}(Q^2)/F^{(\alpha)}_{VP}(Q_o^2)$ in (6.33)-(6.36) are only weakly sensitive e.g. to changes $0.1\,GeV^2\leq Q^2\tilde\delta\leq 0.6\,GeV^2$.
}
\\
a)\, for the flat pion wave function $\phi^{(\alpha=0)}_{\pi}(x,\mu=1\,GeV)=1$ proposed in \cite{Rad-09,Polyakov}
\be
F^{(\alpha=0)}_{VP}(Q^2)\sim \frac{1}{Q^{\,2.3}},\quad \sigma(e^+ e^-\to VP)\sim |F^{(\alpha=0)}_{VP}(s)|^2\sim\frac{1}{s^{\,2.3}}\quad {\rm for}\quad \alpha=0,
\ee
b)\, for the pion wave function $\phi^{\rm Braun-Fil}_{\pi}(x,\mu=1\,GeV)\simeq [\,0.92+2.34 x^2(1-x)^2\,]$, see (6.40) below
\be
F^{\rm Braun-Fil}_{VP}(Q^2)\sim \frac{1}{Q^{\,2.3}},\quad \sigma(e^+ e^-\to VP)\sim |F^{\rm Braun-Fil}_{VP}(s)|^2\sim\frac{1}{s^{\,2.3}}\,\,,
\ee
c)\, for the pion wave function $\phi^{(\alpha=0.3)}_{\pi}(x,\mu=1\,GeV)=N_{0.3}[x(1-x)]^{0.3},\, N_{\alpha}=\frac{\Gamma(2\alpha+2)}{\Gamma^2(\alpha+1)}$ proposed in \cite{Roberts-13-1}
\be
F^{(\alpha=0.3)}_{VP}(Q^2)\sim \frac{1}{Q^{\,2.8}},\quad \sigma(e^+ e^-\to VP)\sim |F^{(\alpha=0.3)}_{VP}(s)|^2\sim\frac{1}{s^{\,2.8}} \quad {\rm for}\quad \alpha=0.3\,,
\ee
d)\, for the "holographic" AdS/QCD model $\phi^{(\alpha=0.5)}_{\pi}(x,\mu=1\,GeV)=8\,\sqrt{x(1-x)}/\pi$ proposed in \cite{BT-08}
\be
F^{(\alpha=0.5)}_{VP}(Q^2)\sim \frac{1}{Q^{\,3.1}},\quad \sigma(e^+ e^-\to VP)\sim |F^{(\alpha=0.5)}_{VP}(s)|^2\sim\frac{1}{s^{\,3.1}}\quad {\rm for}\quad \alpha=0.5\, .
\ee

The experimental data are presented in Figs.30,\,31. Most precise measurements have been performed by the Belle Collaboration \cite{Shen} with the results (for the range $13.8\,GeV^2<s<112\,GeV^2$):
\be
\sigma(e^+ e^-\to K^{*}(892)^o {\bar K}^o)\sim 1/s^n\,,\quad\quad n=(3.83\pm 0.07)\,,
\ee
\be
\sigma(e^+ e^-\to \omega \pi^o)\sim 1/s^n\,,\quad\quad n=(3.75\pm 0.12)\,.\nn
\ee

It is seen that the data disagree with (6.33)-(6.36).
\footnote{\,
We do not substitute the wave functions like $\phi_{\pi}^{asy}(x)$ or $\phi_{\pi}^{CZ}(x)$ into (6.31)
because the integral (6.31) is only logarithmically sensitive to the endpoint region for such type wave functions with $\phi_{\pi}(x\ll 1)\sim x$, and (6.31) is not a sufficiently good approximation for $F_{VP}(Q^2)$ in this case (see section 9.2 in \cite{cz-rev}).
}

\begin{minipage}[c]{.5\textwidth}{\hspace{-10mm}}
\includegraphics[width=1.1\textwidth]{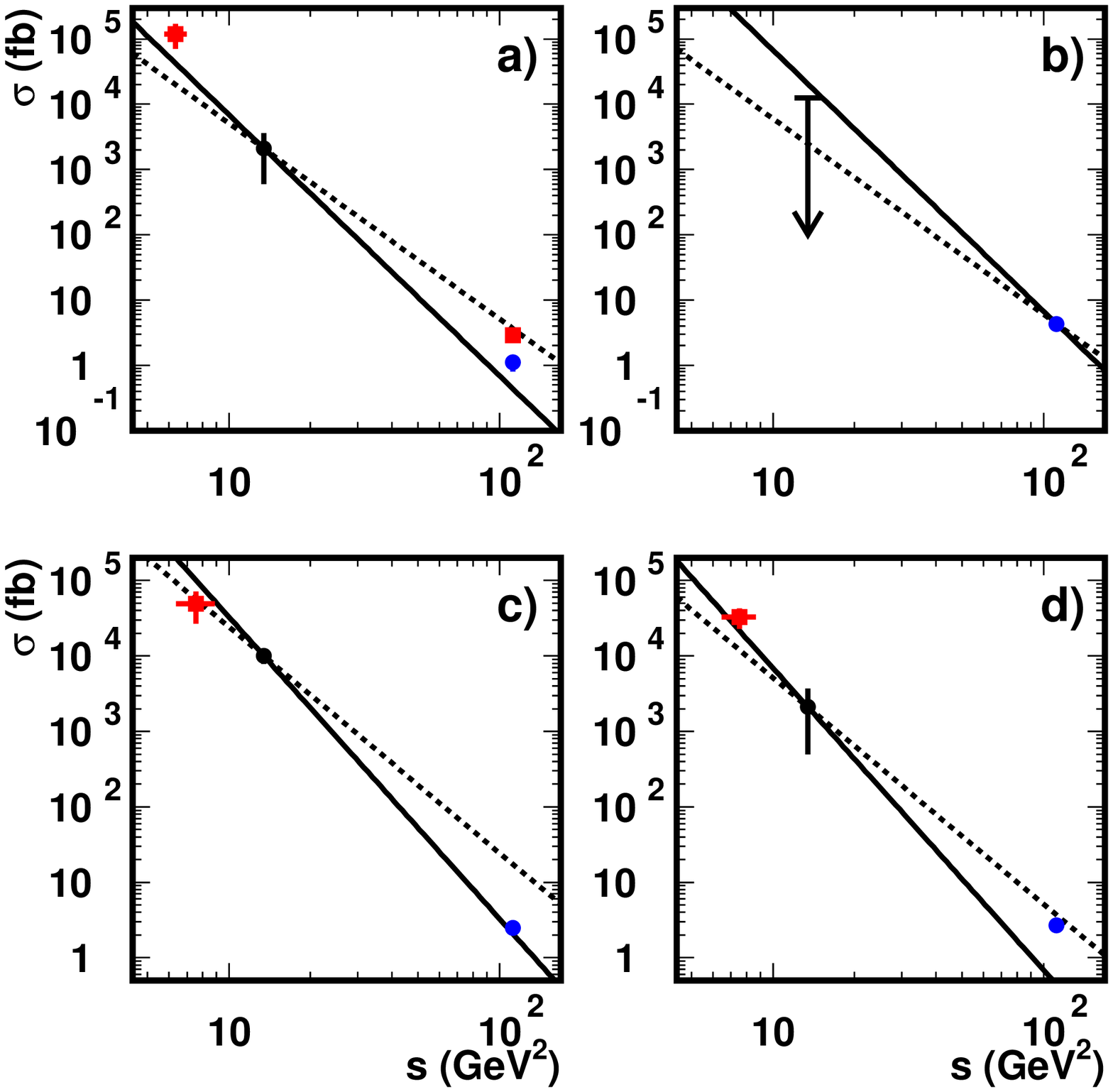}
\end{minipage}~
\begin{minipage}[c]{.6\textwidth}\vspace{-2cm}
{\hspace*{3cm} Fig.30.\\ Solid lines correspond to $1/s^4$ dependence}\\
{\hspace*{0.5cm} and  dashed ones represent $1/s^3$}.
\vspace*{1mm}

{\hspace*{3mm} {\bf a)} $\sigma(e^+e^- \to {\phi \eta})$\quad {\bf b)} $\sigma(e^+e^- \to {\phi \eta'})$}\\
{\hspace*{5mm}{\bf c)} $\sigma(e^+e^- \to {\rho \eta})$\quad {\bf d)} $\sigma(e^+e^- \to {\rho \eta'})$ }\\

{\hspace*{3mm}The measured cross sections: \\
at $\sqrt{s}\simeq 2.5,\,2.75\, GeV$ by BaBar \cite{BB-07},\\
at $\sqrt{s} = 3.67 $ GeV by CLEO \cite{CL},\\
at $\sqrt{s} = 10.58$ GeV by BaBar \cite{BB-2-06}\\ and Belle \cite{Belous}.
{\hspace*{3mm} BaBar measurements \\ are represented by squares.
}}
\end{minipage}~

\begin{minipage}[c]{.5\textwidth}\hspace*{1cm}
\includegraphics[width=0.5\textwidth,angle=-90,clip=true]{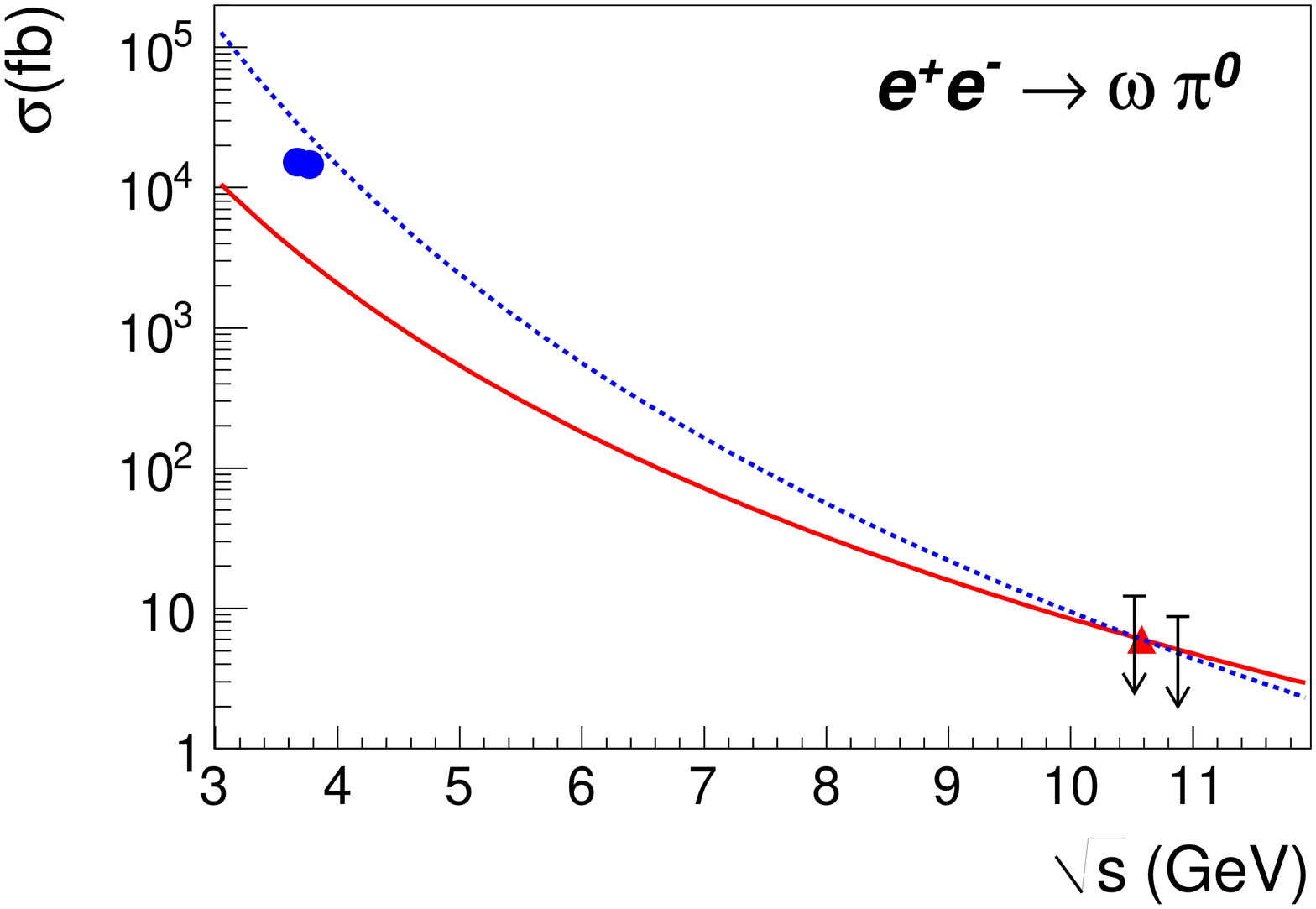}
\end{minipage}~
\begin{minipage}[c]{.5\textwidth}\hspace*{2mm}
\includegraphics[width=0.5\textwidth,angle=-90,clip=true]{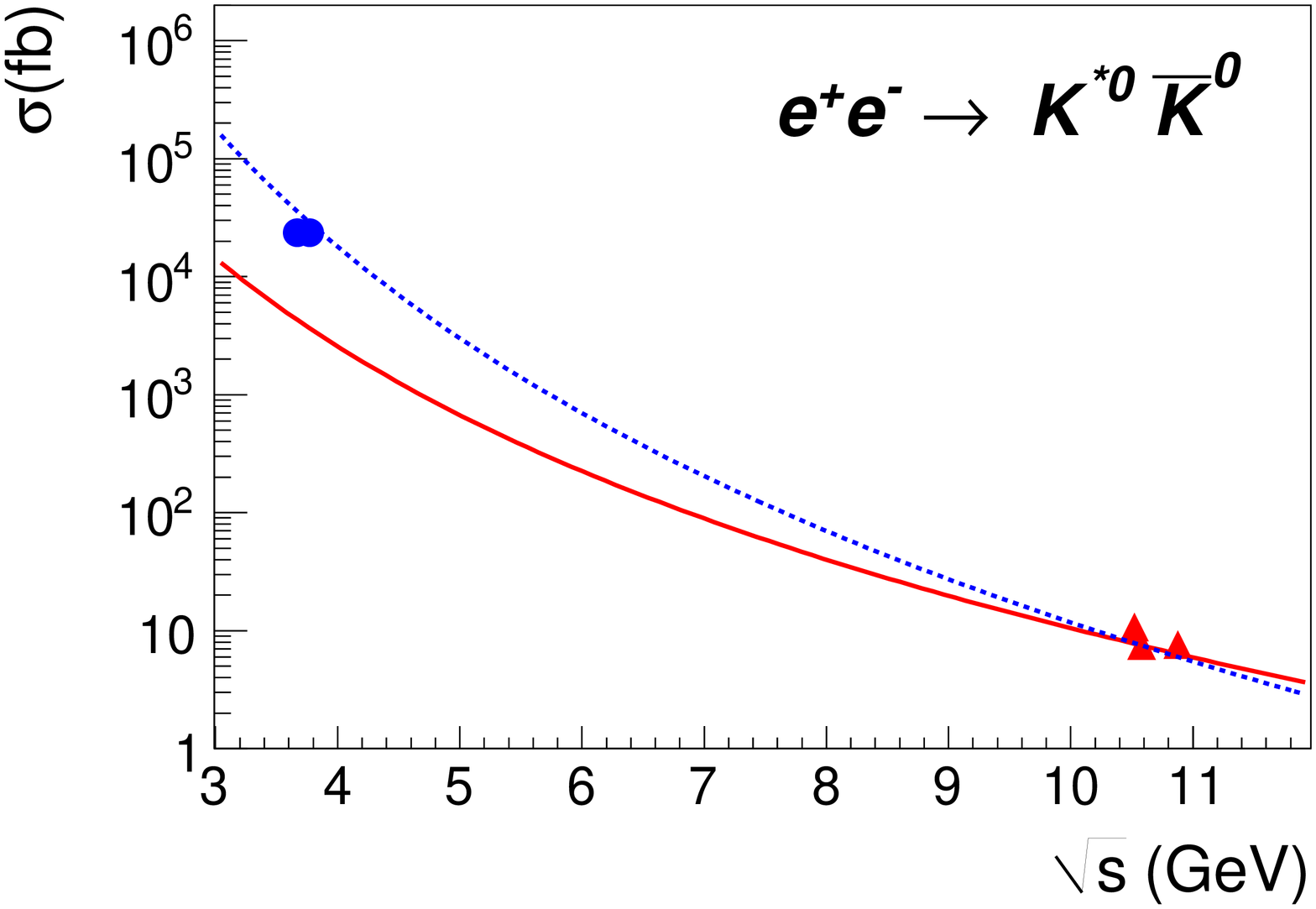}
\end{minipage}

\begin{minipage}[c]{.5\textwidth}\hspace*{1cm}
\includegraphics[width=0.5\textwidth,angle=-90,clip=true]{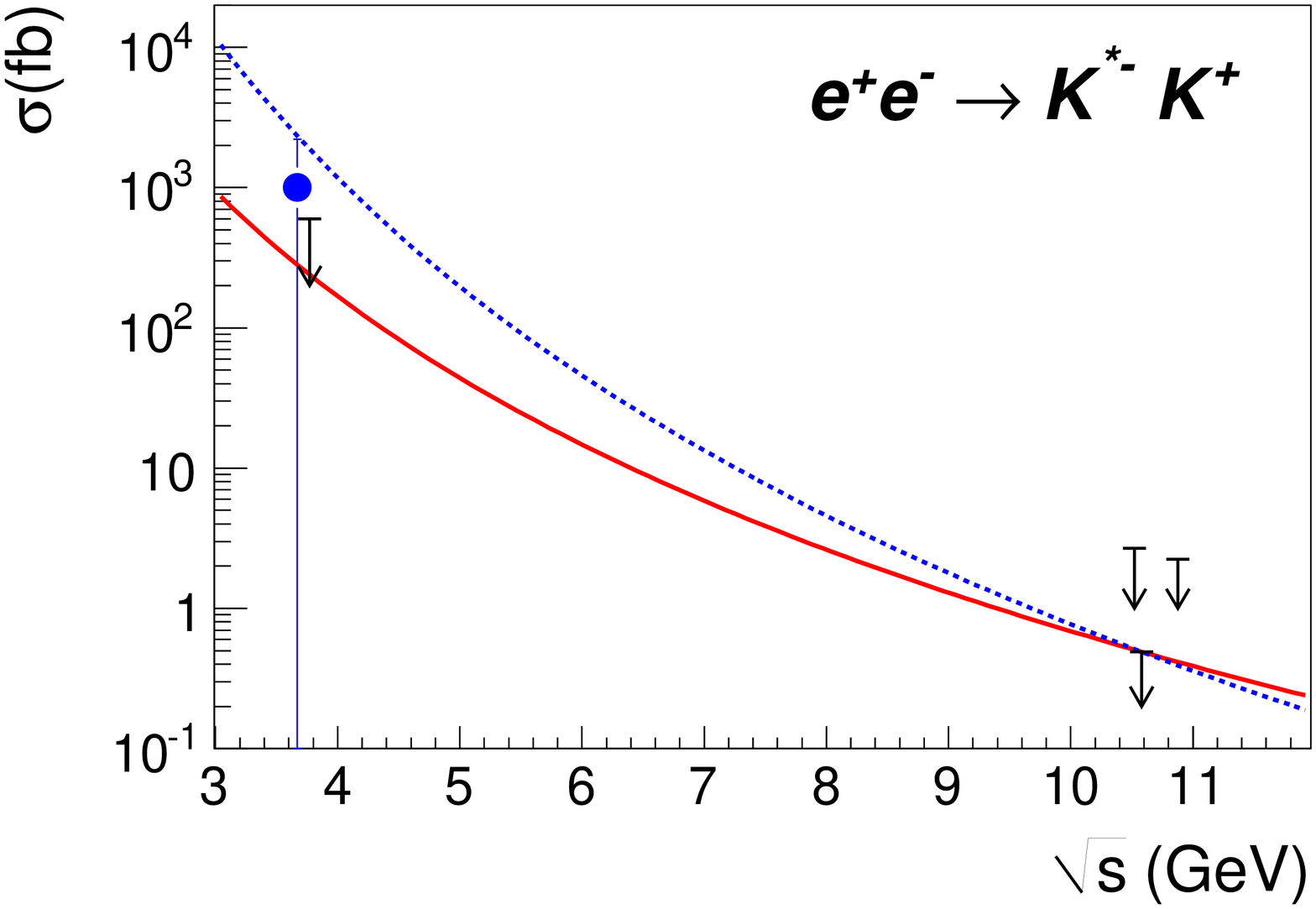}
\end{minipage}~
\begin{minipage}[c]{.5\textwidth}\hspace*{2mm}
\includegraphics[width=0.5\textwidth,angle=-90,clip=true]{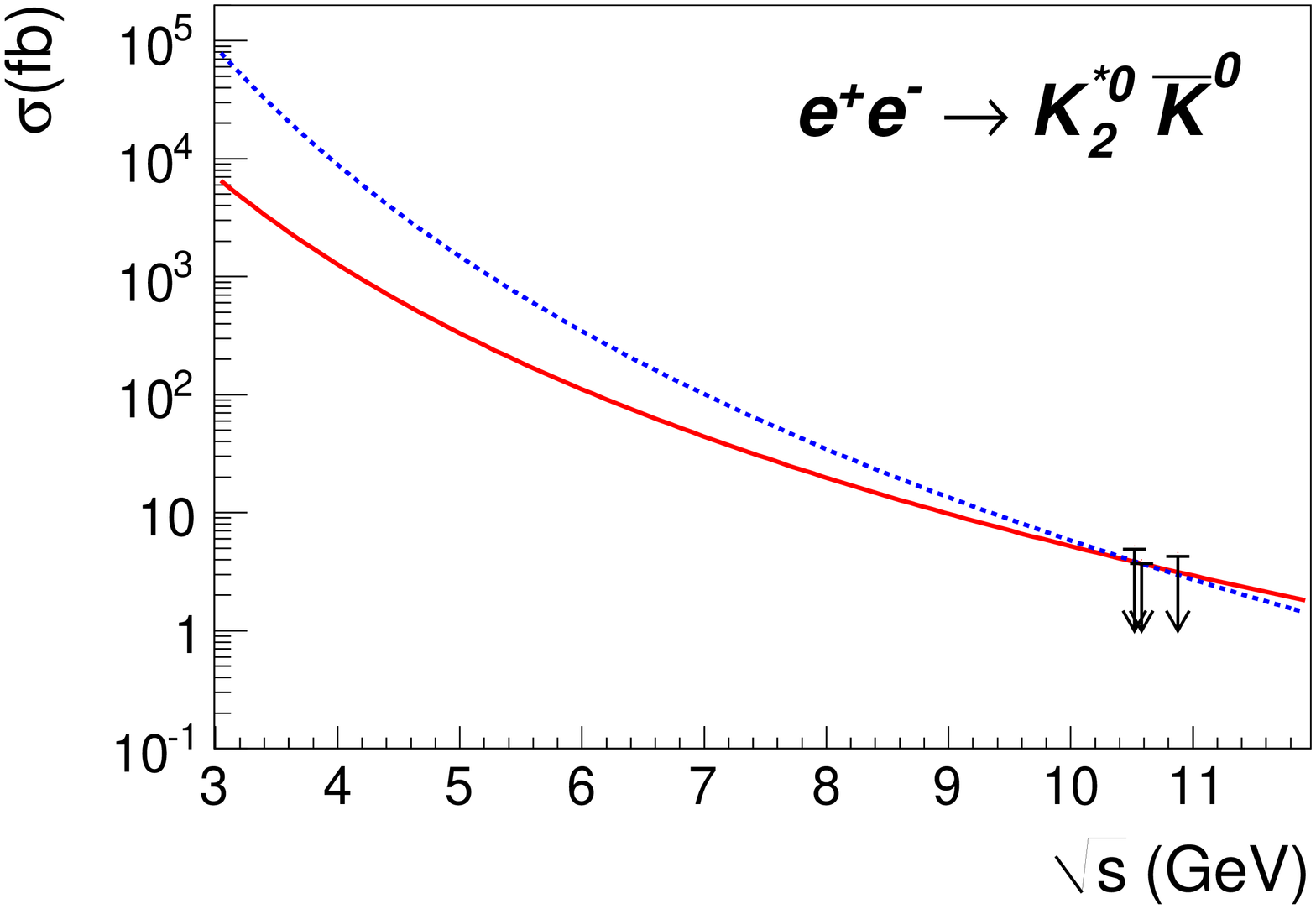}
\end{minipage}

\vspace*{0.3mm}
\quad  Figs.31 a-d.\, The cross sections for $\sigma(e^+e^-\to\omega\pi^0,\, K^*(892)\bar K,\, K_2(1430)K\,)$. The data at $\sqrt{s}=10.52\, GeV, 10.58\, GeV\, {\rm and}\, 10.876\, GeV$ are from Belle \cite{Shen}. The data at $\sqrt{s}=3.67\, GeV\, {\rm and}\, 3.77\, GeV$, where shown, are from the CLEO measurement \cite{CL}. Here, the uncertainties are the sum of the statistical and systematic uncertainties in quadrature. Upper limits are shown by the arrows. The solid line corresponds to a $1/s^3$ dependence and the dashed line to a $1/s^4$ dependence; the curves pass through the measured cross section at $\sqrt{s}= 10.58\, GeV$.\\

We show in addition in Fig.32 (left) the behavior of the ratio $F^{(\alpha)}_{VP}(Q^2)/F^{(\alpha)}_{VP}(Q_o^2=13.8\,GeV^2)$ from (6.31) for four different pion wave functions in the range $13.8<Q^2<112\,GeV^2$.\\

\begin{minipage}[c]{.55\textwidth}
\includegraphics[width=0.65\textwidth]{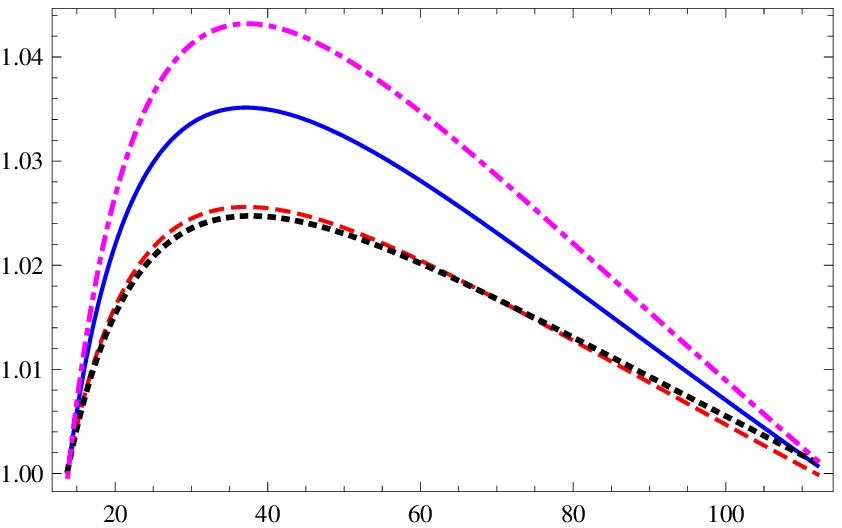}
\put (5,3) {$Q^2$}
\end{minipage}~
\begin{minipage}[c]{.5\textwidth}\hspace*{-2cm}
\includegraphics[width=0.7\textwidth,,clip=true]{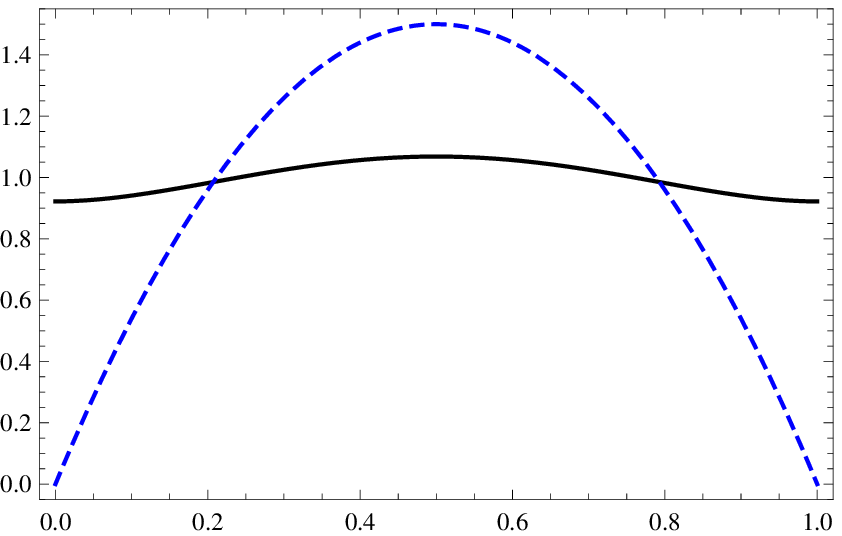}
\put (-200,25) {\rotatebox{90} {$\phi^{\rm Braun-Fil}_{\pi}(x)$}}\put (5,3) {$x$}
\end{minipage}
\vspace*{3mm}

Fig.32 (left). The behavior of (6.31) in the range $Q^2_o=13.8\,GeV^2<Q^2<112\,GeV^2$\,: \\\,\,
a) $[Q/Q_o]^{2.3} F^{(\alpha=0)}_{VP}(Q^2)/F^{(\alpha=0)}_{VP}(Q_o^2)$  with  $\phi^{(\alpha=0)}_{\pi}(x,\mu=1\,GeV)=1$ \cite{Rad-09,Polyakov} (dashed);\\
b) $[Q/Q_o]^{2.3} F^{(\alpha=0)}_{VP}(Q^2)/F^{(\alpha=0)}_{VP}(Q_o^2)$  with $\phi^{\rm Braun-Fil}_{\pi}(x,\mu=1\,GeV)\simeq$ \\ $\hspace*{1cm} \simeq [\,0.92+2.34\, x^2(1-x)^2\,]$ (dotted), see (6.40) below ;\\
c) $[Q/Q_o]^{2.8} F^{(\alpha=0.3)}_{VP}(Q^2)/F^{(\alpha=0.3)}_{VP}(Q_o^2)$ with $\phi^{(\alpha=0.3)}_{\pi}(x,\mu=1\,GeV)=N_{0.3}(x\bar x)^{0.3}$ \cite{Roberts-13-1} (solid).\\
d) $[Q/Q_o]^{3.1} F^{(\alpha=0.5)}_{VP}(Q^2)/F^{(\alpha=0.5)}_{VP}(Q_o^2)$ with $\phi^{(\alpha=0.5)}_{\pi}(x,\mu=1\,GeV)=8(x\bar x)^{0.5}/\pi$ \cite{BT-08} (dot-dashed).\\
\hspace*{5mm} Fig.32 (right). The wave function (6.40):\,\,$\phi^{\rm Braun-Fil}_{\pi}(x,\mu\simeq 1\,GeV)\simeq [0.92+2.34\,x^2(1-x)^2]$  (solid). The asymptotic wave function $\phi_{\pi}^{\rm asy}(x)=6x(1-x)$ is shown for comparison (dashed).\\

In the second part of this section we comment in short on the paper \cite{Br-Fil-89} where the LCSR (light cone sum rule) was obtained for the leading twist pion wave function $\phi_{\pi}(x,\mu\simeq 1\,GeV)$. It has the form (the term $O(1/t)$ was calculated really)
\be
C_o e^{-m^2_{\rho}/t}=t\Bigl (1- e^{-s_{\rm eff}/t}\Bigr)\phi_{\pi}(x)-\frac{8}{9}\delta^2\phi_4
(x)+O(1/t),\quad C_o=\frac{f^2_{\rho}m^2_{\rho}}{\sqrt{2}\,f_{\pi}}g_{\omega\rho\pi}\simeq 2.1\,GeV^2\,,
\ee
\be
\delta^2\simeq 0.2\,GeV^2\,,\quad t=\frac{M_1^2 M_2^2}{M^2}=x(1-x) M^2,\quad x=\frac{M_1^2}{M^2},\quad M^2=M_1^2+M_2^2\,,\quad \phi_4(x)=30\,x^2(1-x)^2,\nn
\ee
where $M_{1}^2$ and $M_2^2$ are the two Borel parameters from two independent Borel transforms and $s_{\rm eff}$ is the effective continuum threshold. Taking $M_1=M_2$ and the commonly used value $s_{\rm eff}=1.5\,GeV^2$ for the $\rho$-meson duality interval, it was obtained in \cite{Br-Fil-89}\,: $\phi_{\pi}(x=1/2,\mu\simeq 1\,GeV)\simeq 1.2\pm 0.3$\,.

We would like to emphasize that much more can be obtained from (6.38). Let us take the limit $\,\,\,x={\it const},\,\,\,M^2\to \infty$ (the limit of local duality). As was discussed above in Section 6.2.3, $s_{\rm eff}$ can really depend weakly on $M^2$, so that we introduce the number $s^{(\infty)}_{\rm eff}=s_{\rm eff}(M^2\to \infty)$. Then
\be
s^{(\infty)}_{\rm eff}\phi_{\pi}(x)=C_o+\frac{8}{9}\delta^2\,\phi_4(x)\,.
\ee
Usually the number $s^{(\infty)}_{\rm eff}$ is not known beforehand and similar local duality sum rules are not very useful. But in this case we can use the normalization conditions of both wave functions
\be
\hspace*{-4mm}\int_0^1 dx\,\phi_{\pi}(x)=\int_0^1 dx\,\phi_4(x)=1\,\,\to\,\, s^{(\infty)}_{\rm eff}\simeq 2.28\,GeV^2\,\,\to\,\, \phi^{\rm Braun-Fil}_{\pi}(x)\simeq [\,0.92+2.34\,x^2(1-x)^2\,]\,.
\ee

This pion wave function $\phi^{\rm Braun-Fil}_{\pi}(x,\mu\simeq 1\,GeV)$ is shown in Fig.32 (right). Its most characteristic features: it is convex, $\phi^{\rm Braun-Fil}_{\pi}(x\to 0,\mu\simeq 1\,GeV)\to 0.92$ and $\phi^{\rm Braun-Fil}_{\pi}(x=1/2,\mu\simeq 1\,GeV)=1.07$\,.

We do not discuss here to what extent LCSR like (6.38) dealing with the double spectral density $\rho(s_1,s_2)$ and doing two independent Borel transforms are reliable. This is not a simple question. But as was shown above in this section, the pion wave functions with the behavior $\phi^{(\alpha=0)}_{\pi}(x\to 0,\,\mu\sim 1\,GeV)\to {\it const}$ lead typically to $F_{VP}(Q^2)\sim 1/Q^{2.3}$, in contradiction with data.\\

The dependence of the form factor $F_{VP}(Q^2)$ on $Q^2$ is predicted also from the more standard LCSR, see e.g. \cite{Khodja-99}, using the simplified form of the duality which , in particular, ignores nonperturbative corrections to the vector meson wave function (see the point "e" in Section 6.1). It uses the Borel transform of the single variable spectral density $\rho_{\gamma\pi}(s,Q^2)$ (but dependent on the external parameter $Q^2$, see Section 6.2.3, $M^2$ is the parameter of the Borel transform),
\be
F^{(LCSR)}_{VP}(Q^2)\sim\int_0^{x_o} dx\,\frac{\phi_{\pi}(x)}{1-x}\exp\Bigl \{\frac{m^2_{
\rho}}{M^2}-\frac{x\,Q^2}{(1-x)M^2}\Bigr \},\quad x_o=\frac{s^{(\rho)}_o}{Q^2+s^{(\rho)}_o},\quad s^{(\rho)}_o\simeq 1.5\,GeV^2\,.
\ee

This form, as it is, ignores both loop- and power corrections. It is important to remember however that the expression (6.41) corresponds at large $Q^2$ to the Feynman endpoint mechanism with the small momentum fraction, $x_o\sim 1/Q^2\to 0$, carried by quark and with small virtuality of the quark propagator, $\sigma^2\simeq x Q^2\leq s_o$, in the handbag diagram (resulting in $0\leq x\leq x_o\simeq s_o/Q^2$). In other words, the virtuality of the "hard" kernel is really small in this endpoint region $\sigma^2<s_o\sim 1\,GeV^2$, so that the pion wave function entering (6.41) is $\phi_{\pi}(x,{\mu\simeq 1\,GeV})$, it does not evolve really here to $\phi_{\pi}(x,\mu\sim Q)$ at large $Q^2$. On the other hand, the most important effect of loop corrections in this endpoint region is the appearance of the Sudakov form factor $S(Q^2,\mu^2\simeq 1\,GeV^2)$ of the active quark carrying nearly the whole meson momentum, ${\ov x}=(1-x)\ra 1$. Therefore, on account of all these, (6.41) will be replaced by
\be
F^{(LCSR)}_{VP}(Q^2)\sim S(Q^2,\mu^2=1\,GeV^2)\int_o^{x_o} dx\,\frac{\phi_{\pi}(x,\mu=1\,GeV)}
{1-x}\exp\Bigl \{\frac{m^2_{\rho}}{M^2}-\frac{xQ^2}{(1-x)M^2}\Bigr \}\,,
\ee
\be
S(Q^2,\mu^2=1\,GeV^2)=\Bigl (\frac{1\,GeV^2}{Q^2}\Bigr )^{\tau_1},\quad \tau_1=\frac{C_F}{\bo}\ln\Bigl (\frac{\alpha_s(1\,GeV^2)}{\alpha_s(Q^2)}\Bigr )\,.\nn
\ee

Without the Sudakov form factor and with $\phi_{\pi}(x\ll 1,\mu\simeq 1\,GeV)\sim x$, (6.42) results in $F^{(LCSR)}_{VP}(Q^2)\sim s_o^2/Q^4$ as expected \cite{c-1} (put attention to the strong dependence $\sim s^2_o$ of the numerical value on the effective continuum threshold $s_o$, see Section 6.2.3), but the behavior will be $F^{(LCSR)}_{VP}(Q^2)\sim s_o/Q^2$ for $\phi^{(\alpha=0)}_{\pi}(x\ll 1,\mu\simeq 1\,GeV)\sim {\it const}$.

To trace the role of the Sudakov form factor we calculated the ratio $F^{(LCSR)}_{VP}(Q^2)
/F^{(LCSR)}_{VP}(Q_o^2)$ in the range $Q_o^2=13.8\,GeV^2<Q^2<112\,GeV^2$ for different wave functions
($M^2=1\,GeV^2$ was taken in (6.42) as a characteristic value and $\alpha_s(\mu^2)$ from (6.32)). The results look as (see footnote 13):\\

\hspace*{-9mm} a)\,\,for the flat model $\phi^{(\alpha=0)}_{\pi}(x,\mu= 1\,GeV)=\phi_{\pi}^{\rm Radyush-Polyak}=1$ \cite{Rad-09,Polyakov}
\be
F^{(LCSR,\, \rm Radyush-Polyak)}_{VP}(Q^2)\sim \frac{1}{Q^{\,2.3}}\,,\quad \sigma(e^+ e^-\to VP)\sim |F^{(LCSR,\,\rm Radyush-Polyak)}_{VP}(s)|^2\sim\frac{1}{s^{\,2.3}}\,;
\ee
b)\,\,for $\phi_{\pi}(x,\mu= 1\,GeV)=\phi_{\pi}^{\rm Braun-Fil}\simeq [\,0.92+2.34 x^2(1-x)^2\,]$, see (6.40)
\be
F^{(LCSR,\, \rm Braun-Fil)}_{VP}(Q^2)\sim \frac{1}{Q^{\,2.3}}\,,\quad \sigma(e^+ e^-\to VP)\sim |F^{(LCSR,\, \rm Braun-Fil)}_{VP}(s)|^2\sim\frac{1}{s^{\,2.3}}\,;
\ee
c)\,\, for $\phi^{(\alpha=0.3)}_{\pi}(x,\mu=1\,GeV)=N_{0.3}\,[x(1-x)]^{0.3},\quad N_{\alpha}={\Gamma(2\alpha+2)}/{\Gamma^2(\alpha+1)}$ \cite{Roberts-13-1}
\be
F^{(LCSR,\, \alpha=0.3)}_{VP}(Q^2)\sim \frac{1}{Q^{\,2.9}}\,,\quad \sigma(e^+ e^-\to VP)\sim |F^{(LCSR,\, \alpha=0.3)}_{VP}(s)|^2\sim\frac{1}{s^{\,2.9}}\,;
\ee
d)\,\, for $\phi^{(\alpha=0.5)}_{\pi}(x,\mu= 1\,GeV)=\phi_{\pi}^{\rm Brodsky-Ter}=8[x(1-x)]^{0.5}/\pi$ \cite{BT-08}
\be
F^{(LCSR,\, \rm Brodsky-Ter)}_{VP}(Q^2)\sim \frac{1}{Q^{\,3.2}}\,,\quad \sigma(e^+ e^-\to VP)\sim |F^{(LCSR,\, \rm Brodsky-Ter)}_{VP}(s)|^2\sim\frac{1}{s^{\,3.2}}\,;
\ee
e) \,\,for $\phi_{\pi}(x,\mu= 1\,GeV)=\phi_{\pi}^{asy}=6x(1-x)$
\be
F^{(LCSR,\, asy)}_{VP}(Q^2)\sim \frac{1}{Q^{\,4.2}}\,,\quad \sigma(e^+ e^-\to VP)\sim |F^{(LCSR,\, asy)}_{VP}(s)|^2\sim\frac{1}{s^{\,4.2}}\,;
\ee
f)\,\,for $\phi_{\pi}(x,\mu= 1\,GeV)=\phi_{\pi}^{CZ}=30x(1-x)(2x-1)^2$
\be
F^{(LCSR,\, CZ)}_{VP}(Q^2)\sim \frac{1}{Q^{\,4.0}}\,,\quad \sigma(e^+ e^-\to VP)\sim |F^{(LCSR,\, CZ)}_{VP}(s)|^2\sim\frac{1}{s^{\,4.0}}\,.
\ee
\vspace*{3mm}

In any case, it is seen from (6.33)-(6.36) and (6.43)-(6.48) that the data (6.37) from Belle \cite{Shen} and CLEO \cite{CL} for the cross sections $\sigma(e^+ e^-\to K^*(892)^o{\bar K}^o)$ and $\sigma(e^+e^-
\to\omega\pi^0)$ are in contradiction with the energy dependence of $F_{VP}(Q^2)$ for the leading twist meson wave functions with, say, the behavior  $\phi^{(\alpha)}_{M}(x\ll 1,\mu\simeq 1\,GeV)\sim x^{\,\alpha\,\leq\,0.5}$ and prefer the behavior $\phi_{M}(x\ll 1,\mu\simeq 1\,GeV)\sim x$.

\subsection{Some calculations of $F_{\gamma\pi}(Q^2)$}

\hspace*{3mm} Predictions for $F_{\gamma \pi}(Q^2)$  were given in a large number of theoretical papers, using many different models for the leading twist pion wave function $\phi_{\pi}(x,\mu)$, see Table 3, and many different approaches to calculate $F_{\gamma \pi}(Q^2)$. The previous data for $F_{\gamma \pi}(Q^2)$ from CLEO \cite{Savin,CL-98} covered the space-like region $0<Q^2<8\,GeV^2$ only. The recent data from BaBar \cite{BB-09, Druzh-10} and then from Belle \cite{Uehara-12} extended this one to $Q^2\lesssim 40\,GeV^2$.

We present below, as most elaborated examples, mainly the results of two papers \cite{Braun-11,Braun-12} where the form factor $F_{\gamma\pi}(Q^2)$ was calculated within the LCSR (light cone sum rules) approach \cite{Khodja-99} for a number of various pion wave functions and with account of some logarithmic and power corrections.

The coefficients $a_n$ of the Gegenbauer polynomials of four model pion wave functions considered in \cite{Braun-11,Braun-12} are given in Table 4. First three were fitted in \cite{Braun-11} to the BaBar data \cite{BB-09} and the last one was fitted in \cite{Braun-12} to the Belle data \cite{Uehara-12}. The forms of some of these wave functions are presented in Fig.33 (left).
\footnote{\,
The model forms presented in Table 4 were additionally constrained to satisfy $\phi_{\pi}(x=1/2,\mu\simeq 1\,GeV)= (1.2\pm 0.3)$ obtained in \cite{Br-Fil-89}, see the text under (6.38).
}

Some results from \cite{Braun-11} and \cite{Braun-12} are presented in Fig.33 (right) and Fig.34.

\vspace*{1mm}
\begin{center}
\begin{tabular}{|c|c|c|c|c|c|c|c|c|} \hline
  Model & scale & $a_2$ & $a_4$ & $a_6$ & $a_8$ & $a_{10}$ & $a_{12}$ & Ref
\\ \hline
{  I } & $\mu=1$~GeV
& 0.130 & 0.244 & 0.179 & 0.141 & 0.116 & 0.099 & \cite{Braun-11}
\\ \hline
{ II } & $\mu=1$~GeV
& 0.140 & 0.230 & 0.180 & 0.05 & 0.0 & 0.0 & \cite{Braun-11}
\\ \hline
{ III } & $\mu=1$~GeV
 & 0.160 & 0.220 & 0.080 & 0.0 & 0.0 & 0.0 & \cite{Braun-11}
\\ \hline
{IV} & $\mu=1$~GeV & 0.10 & 0.10  & 0.10 & 0.034 & 0.0 & 0.0 & \cite{Braun-12}
\\ \hline
\end{tabular}
\vspace*{5mm}

Table 4. The Gegenbauer coefficients of three sample models I-III of the leading twist pion wave function, $\phi_{\pi}(x,\mu)$, that are fitted in \cite{Braun-11} to the BaBar measurements \cite{BB-09} of the transition form factor $F_{\gamma\pi}(Q^2)$. The model IV is fitted in \cite{Braun-12} to the Belle data \cite{Uehara-12} for $F_{\gamma\pi}(Q^2)$.
\end{center}
\vspace*{5mm}

For calculations with the pion wave functions of the type $\phi^{(\alpha)}_{\pi}(x,\mu=1\,GeV)\sim [x(1-x)]^{\alpha}$ with $\alpha=0$ \cite{Rad-09,Polyakov} or $\alpha=0.5$ \cite{BT-08} in Fig.34 (left) the infinite series of Gegenbauer polynomials was approximated in \cite{Braun-11} by several first terms. We would like to point out that it is not so difficult to calculate the endpoint behavior of such type wave functions $\phi^{(\alpha<1)}_{\pi}(x,\mu\gg 1\,GeV)$.
\be
\phi^{(\alpha)}_{\pi}(x,\mu=1\,GeV)=N_{\alpha} (x{\bar x})^{\alpha}\,,\quad \int_0^1 dx\,\phi_{\pi}^{(\alpha)}(x,\mu)=1\,,\quad N_{\alpha}=\frac{\Gamma(2\alpha+2)}{\Gamma^2(\alpha+1)}\,,\quad 0\leq\alpha<1\,.
\ee

The RG-evolution of $\phi^{(\alpha)}_{\pi}(x,\mu)$ looks as (${\ov x}=1-x$)
\vspace*{-2mm}
\be
\phi_{\pi}^{(\alpha)}(x,\mu)=6x{\bar x}\sum_{n=2 k}a^{(\alpha)}_{n}(\mu)\,C^{3/2}_{n}(x-{\ov x})\,,\quad a^{(\alpha)}_{n}(\mu)=a^{(\alpha)}_{n}(\mu=1\,GeV)\Biggl (\frac{\alpha_s(\mu)}{\alpha_s(\mu=1\,GeV)}\Biggr )^{\gamma_n/\bo}\,,
\ee
where $\gamma_n$ is given in (1.1). As far as $\phi_{\pi}^{(\alpha<1)}(x,\mu)/x$ is singular at $x\ra 0$, its endpoint behavior is determined by the $n\gg 1$ tail of the series in (6.50). At large $n\gg 1\,:\, \gamma_{n}\simeq 4\, C_F\ln n$, while $a^{(\alpha)}_{\rm n}(\mu=1\,GeV)\sim (1/n)^{1+2\alpha}$. Therefore, we can write with a reasonable accuracy
\be
\phi^{(\alpha)}_{\pi}(x,\mu>1\,GeV)\simeq N_{\mu} (x{\bar x})^{\alpha+2 \tau_{\mu}}\,,\quad \tau_{\mu}=\frac{C_F}{\bo}\ln\Bigl (\frac{\alpha_s(\mu=1\,GeV)}{\alpha_s(\mu)}\Bigr ),\,\, N_{\mu}=\frac{\Gamma(2\alpha+2+4\tau_{\mu})}{\Gamma^2(\alpha+1+2\tau_{\mu})}
\ee
as far as $(\alpha+2\tau_{\mu})<1$, while when $(\alpha+2\tau_{\mu})$ becomes $\geq 1$ at sufficiently large $\mu$, the series in (6.50) becomes convergent at $x\ra 0$ and the wave function behaves then as $\phi^{(\alpha)}_{\pi}(x\ll 1,\mu)\sim x$.\\

\begin{minipage}[c]{.5\textwidth}\vspace*{-5mm}
\includegraphics [trim=0mm 0mm 0mm 0mm, width=0.8\textwidth,clip=true]{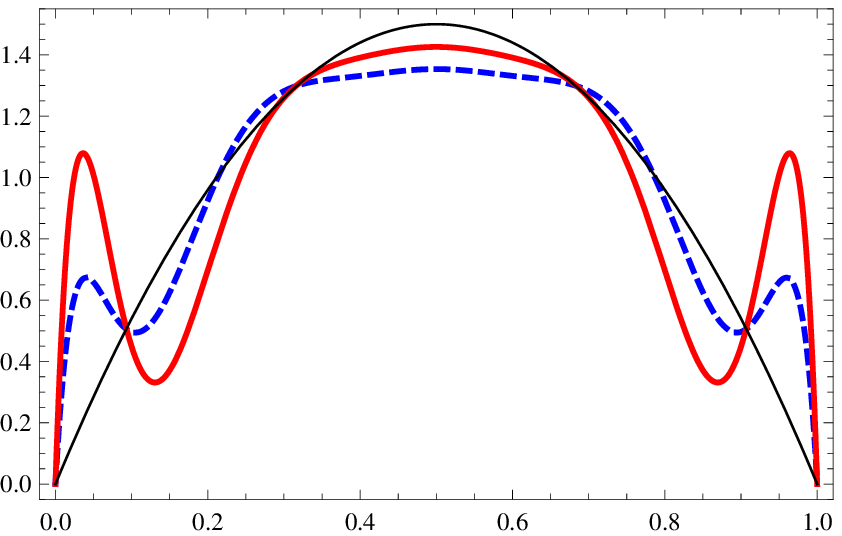}
\put (-225,50){\rotatebox{90} {$\phi_{\pi}(x)$}} \put (5,3) {$x$}
\end{minipage}
\begin{minipage}[c]{.4\textwidth}
\centerline{
\begin{picture}(210,140)(0,0)
\put(-5,0){\epsfxsize7.8cm\epsffile{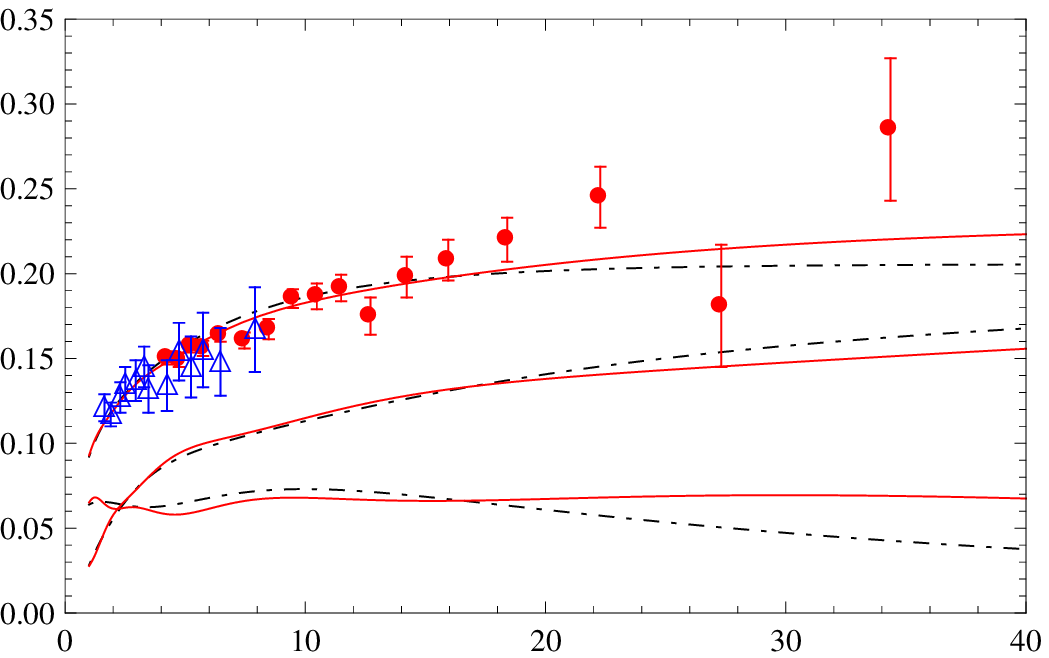}}
\put(20,120){$Q^2F_{\pi^0\gamma^*\gamma}(Q^2)$}
\renewcommand{\arraystretch}{0.7}
\put(180,27){soft}
\put(175,56){hard}
\put(165,95){soft+hard}
\put (220,3) {$Q^2$}
\end{picture}\vspace*{3mm}
\renewcommand{\arraystretch}{1.0}
}
\end{minipage}

\hspace*{-2mm} Fig.33 (left). Thick solid line: the model II from Table 4 for $\phi_{\pi}(x,\mu=1\,GeV)$;\,\, dashed line: the model IV from Table 4;\,\,thin solid line: the asymptotic wave function $\phi^{\it asy}(x)=6x(1-x)$;\\
\hspace*{4mm} Fig.33 (right) \cite{Braun-11}. Contributions to the form factor $Q^2 F_{\gamma\pi}(Q^2)$ from large (``hard'') and small (``soft'') invariant masses in the LCSR: for model I (solid curves) and model III (dash-dotted curves) from Table 4 (see \cite{Braun-11} for details). The experimental data are from \cite{CL-98} (open triangles) and \cite{BB-09} (full circles).
\vspace*{2mm}

\begin{minipage}[c]{.5\textwidth}
\includegraphics[width=0.8\textwidth,,clip=true]{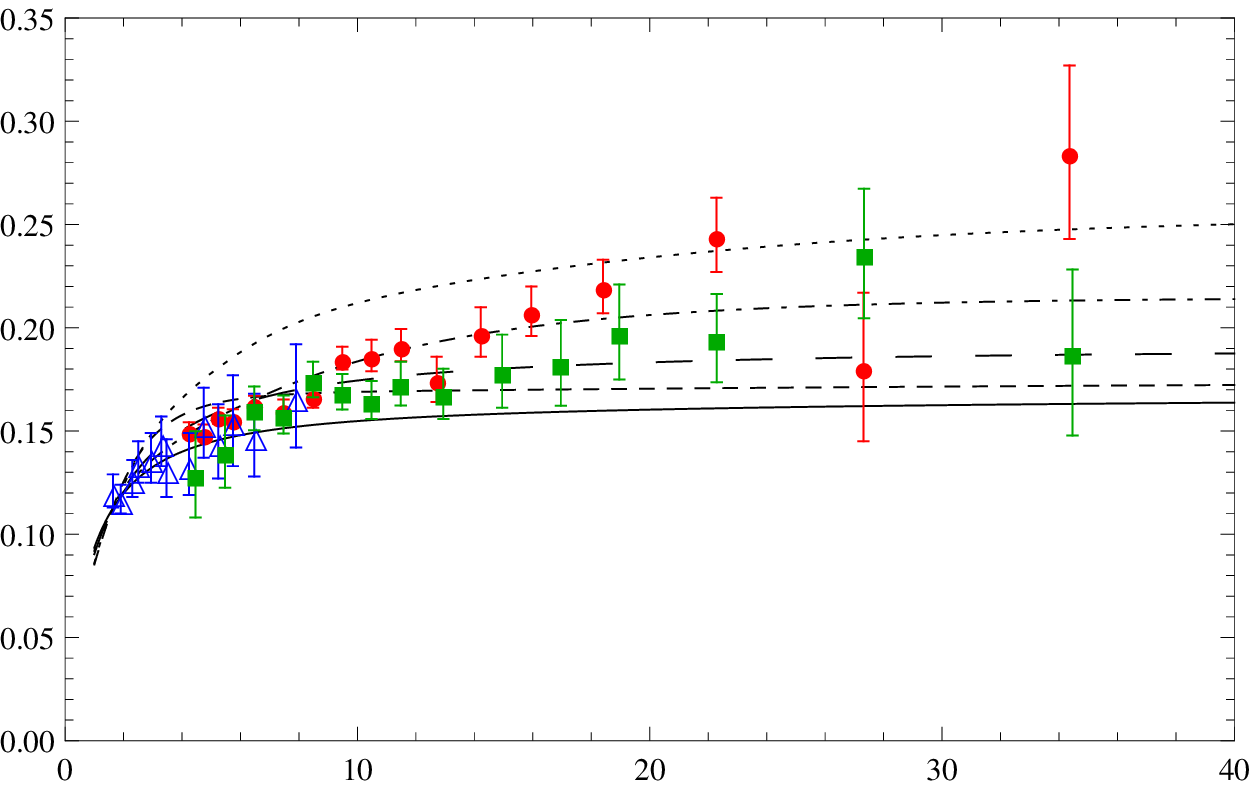}
\put (-235,50){\rotatebox{90} {$Q^2\,F_{\gamma\pi}(Q^2)$}} \put (10,3) {$Q^2$}
\end{minipage}~
\begin{minipage}[c]{.5\textwidth}
\includegraphics[width=0.8\textwidth,,clip=true]{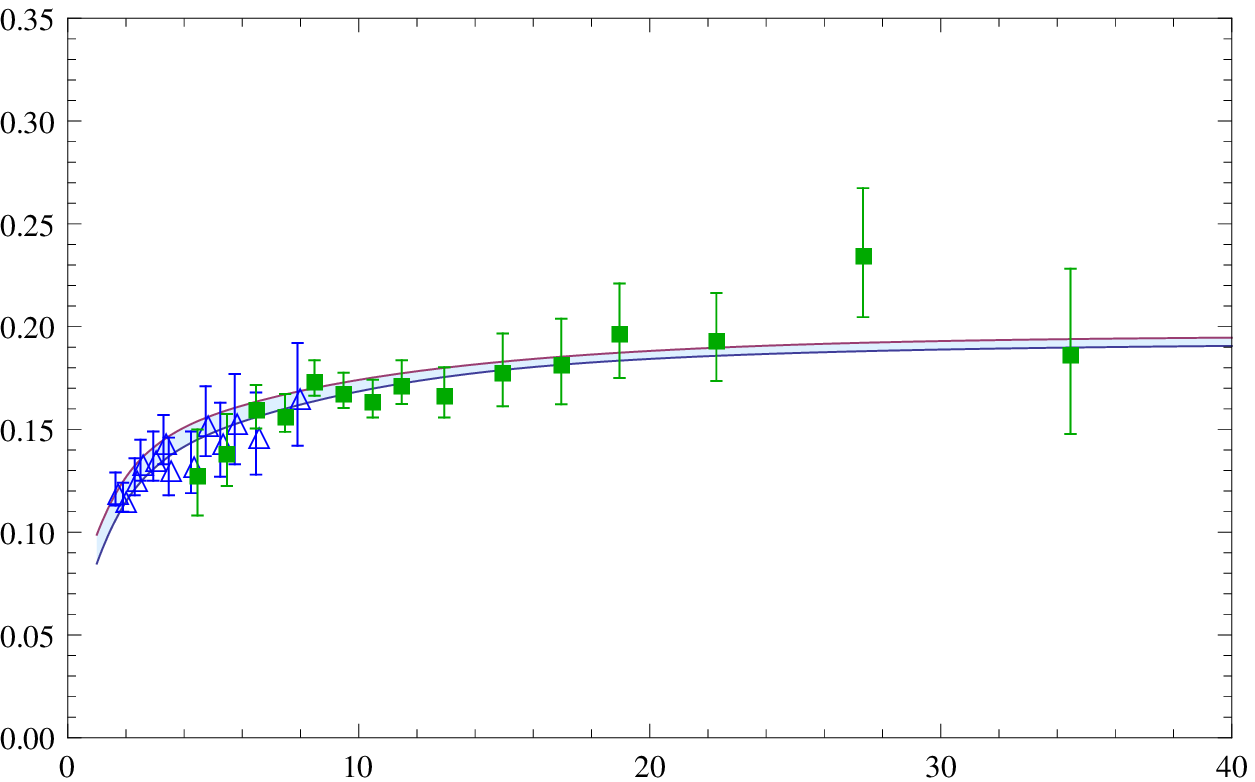}
\put (-235,50){\rotatebox{90} {$Q^2\,F_{\gamma\pi}(Q^2)$}} \put (9,3) {$Q^2$}
\end{minipage}
\vspace*{3mm}

\hspace*{2mm} Fig.34 (left) \cite{Braun-12}.\,\, Results from the LCSR-approach for the pion transition form factor $Q^2 F_{\gamma\pi}(Q^2)$ for the "asymptotic"\, (solid line), "BMS" \cite{BMPS-12} (short dashes), "holographic" $\phi_{\pi}(x)=8\sqrt{x(1-x)}/\pi$ \,\cite{BT-08} (long dashed), "model II" from Table 4 \cite{Braun-11} (dash-dotted), and "flat" $\phi_{\pi}(x)=1$ \cite{Rad-09,Polyakov} (dots) pion wave functions. The data are from CLEO \cite{CL-98} (open triangles), BaBar \cite{BB-09} (circles), and Belle \cite{Uehara-12} (squares).\\
\hspace*{8mm} Fig.34 (right) \cite{Braun-12}. \,\,\, The same for the model IV pion wave function (see Table 4). The estimated theoretical uncertainty is shown by the shaded area. The data are from CLEO \cite{CL-98} (open triangles) and Belle \cite{Uehara-12} (squares).\\

\begin{minipage}[c]{.5\textwidth}\vspace*{2mm}\hspace*{-1.1cm}
\includegraphics[width=0.85\textwidth,,clip=true]{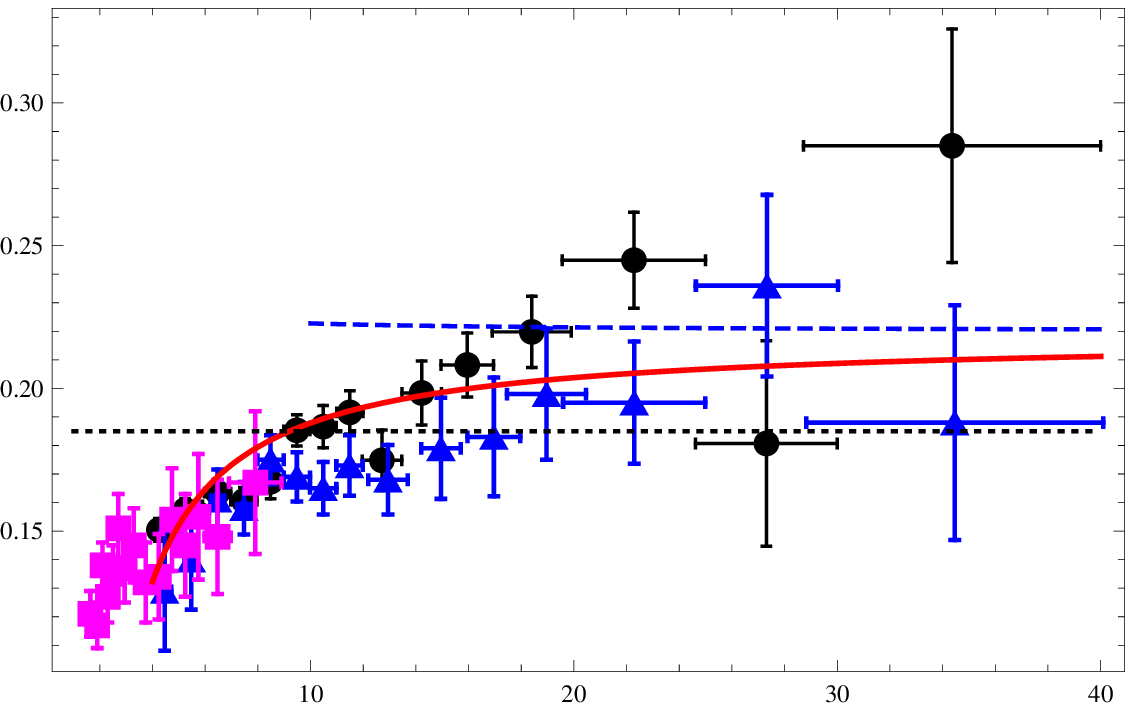}
\put (-190,120) {$Q^2 F_{\gamma\pi}(Q^2)$}\put (10,3) {$Q^2$}
\end{minipage}~
\begin{minipage}[c]{.6\textwidth}\vspace*{-5mm}
\hspace*{-20mm} {Fig.35. The pion transition form factor $Q^2 F_{\gamma\pi}(Q^2)$ data:}\\
\hspace*{-20mm} {BaBar \cite{BB-09} - disk, Belle \cite{Uehara-12} - triangle, CLEO  \cite{CL-98} -}\\
\hspace*{-20mm} {square. The point straight line: the limit $Q^2\ra\infty$\,;}\\
\hspace*{-20mm} {the dashed line: the perturbative contribution}\\
\hspace*{-20mm} {$\simeq (\sqrt{2}\,f_{\pi})1.19$ for the CZ - pion wave function as it is}\\
\hspace*{-20mm} {calculated in \cite{MNP-01}; the solid curve: the same with some}\\
\hspace*{-20mm} {sample power corrections \cite{ch-09}:}\\
\hspace*{-10mm} {$\,\sqrt{2}\,f_{\pi}[\,\,1.19-(1.5\,GeV^2/Q^2)-(1.2\,GeV^2/Q^2)^2\,\,]$.}
\end{minipage}
\vspace*{2mm}

In Fig.35 the results for the CZ- pion wave function, $\phi^{CZ}(x,\mu=1\,GeV)$ \cite{cz-82}, are presented, both pure perturbative \cite{MNP-01} and, as an example, with addition of some power corrections \cite{ch-09}. This is to illustrate that (unknown really at present) power corrections of admissible size may be noticeable up to sufficiently large values of $Q^2$. Therefore, more precise experimental data at, say, $Q^2\,>\,20\,GeV^2$ are needed to discriminate clearly between various models of the pion wave function. Hopefully, the data from the next run of Belle will help greatly.

\subsection{Form factors $F_{\gamma\eta}(Q^2)$ and $F_{\gamma\eta^\prime}(Q^2)$ }

\hspace*{3mm} The form factors of the $\eta$ and $\eta'$ mesons, $F_{\gamma \eta}$ and $F_{\gamma \eta'}$,
look similarly to $F_{\gamma \pi}$. For instance, a simplified description of $|\eta\rangle,\,|\eta'\rangle$ states in the quark flavor basis (neglecting here a possible admixture of the two-gluon basic state $|gg\rangle$) looks as follows \cite{FKS-1,FKS-2}:
\be
|\pi^o \rangle \to |({\ov u}u-{\ov d}d)/{\sqrt 2} \rangle,\quad
|n\rangle\to |({\ov u}u+{\ov d}d)/{\sqrt 2} \rangle,\quad  |s \rangle\to |{\ov s}s \rangle,
\ee
\be
|\eta\rangle\simeq\cos\phi \,|n\rangle-\sin\phi \, |s\rangle\,, \quad |\eta'\rangle\simeq \sin\phi \,|n\rangle+\cos\phi \,|s\rangle\,,\nn
\ee
\be
f_{\pi}\simeq 130.4\, MeV,\quad f_{\rm n}\simeq f_{\pi},\quad f_{s}\simeq 1.3\,f_{\pi},\quad \phi\simeq 38^{o}\,.\nn
\ee

Then the (simplified) description of form factors looks as
\be
F_{\gamma\pi}(Q^2)=\frac{\sqrt 2 (e_u^2-e_d^2)\, f_{\pi}}{Q^2}\int_0^1 dx \,\frac{\phi_{\pi}(x,\mu\sim Q)}{x}I_{\rm rad}\,; \,\, F_{\gamma \rm n}(Q^2)\simeq\frac{\sqrt 2 (e_u^2+e_d^2)\, f_{\pi}}{Q^2}\int_0^1 dx \,\frac{\phi_{\pi}(x,\mu\sim Q)}{x}I_{\rm rad},\nonumber
\ee
\be
F_{\gamma s}(Q^2)\simeq\frac{2 e_s^2 \, f_{s}}{Q^2}\int_0^1 dx \,\frac{\phi_{s}(x,\mu\sim Q)}{x}I_{\rm rad}\,; \quad \, Q^2\gg 1\,GeV^2\,,
\ee
\be
F_{\gamma\eta}(Q^2)\simeq\Biggl ( \cos \phi\, F_{\gamma \rm n}(Q^2)-\sin \phi \,F_{\gamma s}(Q^2)\Biggr ),\,\,
F_{\gamma\eta'}(Q^2)\simeq \Biggl (\sin \phi \,F_{\gamma \rm n}(Q^2)+\cos \phi \, F_{\gamma s}(Q^2)\Biggr ),\nonumber
\ee
where the factor $I_{\rm rad}$ accounts for the one-loop corrections \cite{Aguila-81, Braaten,
Radyu-86,MNP-01}.

With these simplifications, the wave function of $|n\rangle$ will be the same as $|\pi\rangle$, but, naturally, the wave function of $|s\rangle$ consisting of two heavier strange quarks will be significantly narrower (recall: the heavier is quark, the narrower is wave function). In Table 5 some results are presented \cite{ch-09} which follow then from (6.52)-(6.53).

The form factors $Q^2 F_{\gamma\eta}$ and $Q^2 F_{\gamma\eta^\prime}$ have been measured at lower energies $|Q^2|<40\,GeV^2$ in \cite{CELLO,Savin,CL-98,CL-09,L3-98,BB-1101}. The results from CLEO \cite{CL-98} and BaBar \cite{BB-1101} in the space-like region and from CLEO \cite{CL-09} in the time-like region are shown in Fig.36, together with fits from the recent paper \cite{Kr-Pass} (which attempted to account in addition for the quark-gluon mixing in the RG-evolution, see e.g. Section 3.7 in \cite{cz-rev} and references therein about this mixing ).
\footnote{\,
see also the very recent paper \cite{Braun-14}.
}
\\

\begin{center} Table 5. Form factors at $q^2=112\,GeV^2$\end{center}
\vspace*{-3mm}\hspace*{0.5cm}{
\renewcommand{\arraystretch}{2.0}
\begin{footnotesize}
\begin{tabular}{l|c|c|c|c} \hline \hline

Wave functions & $|q^2 F_{\gamma\pi}(q^2)|$ & $|q^2 F_{\gamma\eta}(q^2)|$ & $|q^2 F_{\gamma\eta^\prime}(q^2)|$  &
Ref.\\ \hline \hline
$\phi_{n}(x)\simeq \phi_{s}(x)\simeq \phi^{asy}(x)=6 x(1-x)$ &  $0.14$  & $0.13$ & $0.21$ & \cite{ch-09}
\\ \hline
$\phi_{n}(x)\simeq \phi_{s}(x)\simeq \phi_{\pi}^{cz}(x)$  & $0.22$ & $0.21$ & $0.33$ & \cite{ch-09}
\\ \hline
$\phi_{n}(x)\simeq\phi_{\pi}^{cz}(x);\quad \phi_{s}(x)\simeq\phi^{asy}(x)$ & \boldmath $ {0.22}$ & \boldmath ${0.24}$ & \boldmath ${ 0.29 }$ & \cite{ch-09}
\\ \hline \hline
experiment at $q^2=s=112\,GeV^2$ & --- & ${0.23\pm 0.03}$ & ${0.25\pm 0.02}$ & \cite{BB-1}\\
\hline\hline
\end{tabular}
\end{footnotesize}
\vspace*{3mm}

\begin{minipage}[c]{.5\textwidth}\hspace*{0.1cm}
\includegraphics[width=0.7\textwidth,,clip=true]{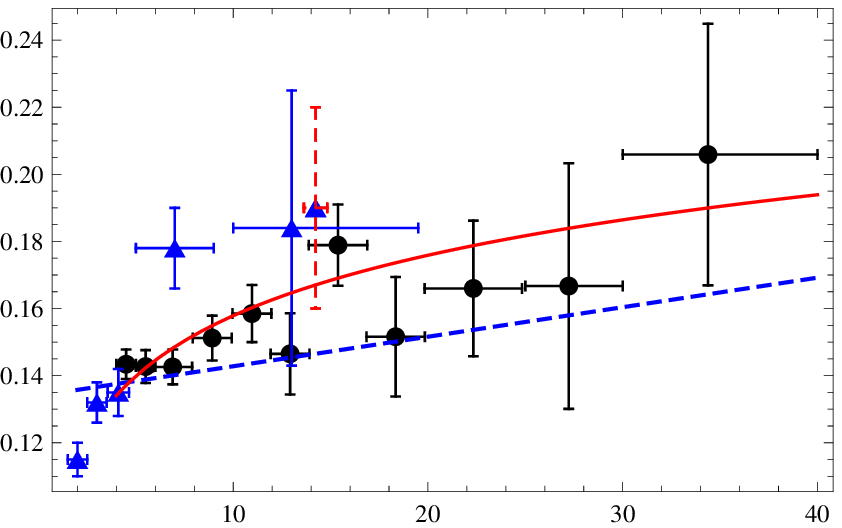}
\put (-230,50){\rotatebox{90} {$Q^2\,F_{\gamma\eta}(Q^2)$}} \put (10,3) {$Q^2$}
\end{minipage}~
\begin{minipage}[c]{.5\textwidth}\hspace*{-0.7cm}
\includegraphics[width=0.7\textwidth,,clip=true]{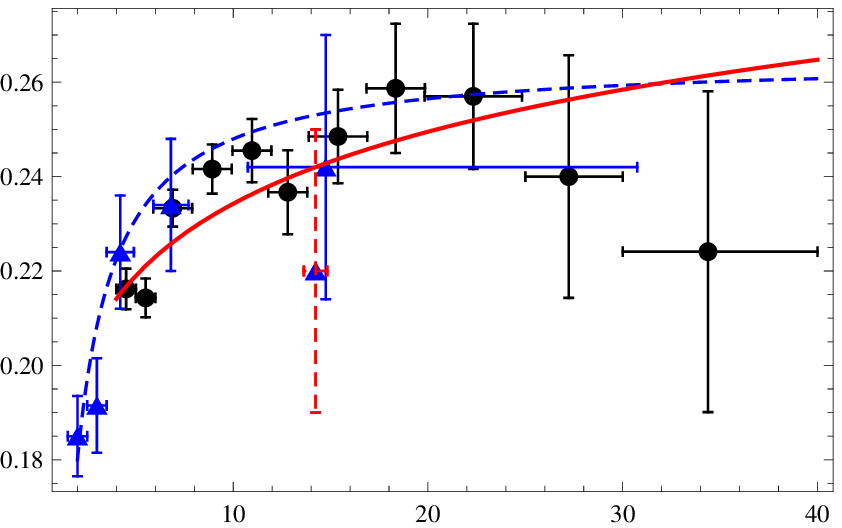}
\put (-230,50){\rotatebox{90} {$Q^2\,F_{\gamma\eta^\prime}(Q^2)$}} \put (10,3) {$Q^2$}
\end{minipage}~

Fig.36. The data on the $\eta$ (left) and $\eta^\prime$ (right) transition form factors $|Q^2 F_{\gamma P}(Q^2)|$ : \\ BaBar \cite{BB-1101} - disk, CLEO \cite{CL-98} - triangle. Two points at $q^2=s=-Q^2=14.2
\,GeV^2$ with dashed error bars
are from CLEO \cite{CL-09}. Solid curves - BaBar fits, dashed curves - fits from \cite{Kr-Pass}.

\subsection {Form factor $F_{\gamma\eta_c}(Q^2)$ }

\hspace*{4mm} The form factor $F_{\eta_c\gamma}(Q^2)$ is analogous to $F_{\pi\gamma}(Q^2)$, the qualitative difference is that the charm quark is sufficiently heavy and so the leading twist wave function of $\eta_c$\,, $\phi_{\eta_c}(x)$, is much narrower than $\phi_{\pi}(x)$ and, besides, the masses of the quark $M_c$ and $M_{\eta_c}\simeq 2 M_c$ cannot be neglected at available values of $Q^2$.

The best measurements of the form factor $F_{\eta_c\gamma}(Q^2)$ at $2<Q^2<50\,GeV^2$ has been done by the BaBar Collaboration \cite{Lees}, see Fig.37. The fit to the data points looks as \cite{Lees}
\be
\frac{F_{\eta_c\gamma}(Q^2)}{F_{\eta_c\gamma}(0)}=\frac{1}{1+Q^2/\Lambda^2}\,\,,\quad \Lambda^2
=(8.5\pm 0.6\pm 0.7)\,GeV^2\,.
\vspace*{-3mm}
\ee

\begin{minipage}[c]{.5\textwidth}
\includegraphics[width=0.7\textwidth,,clip=true]{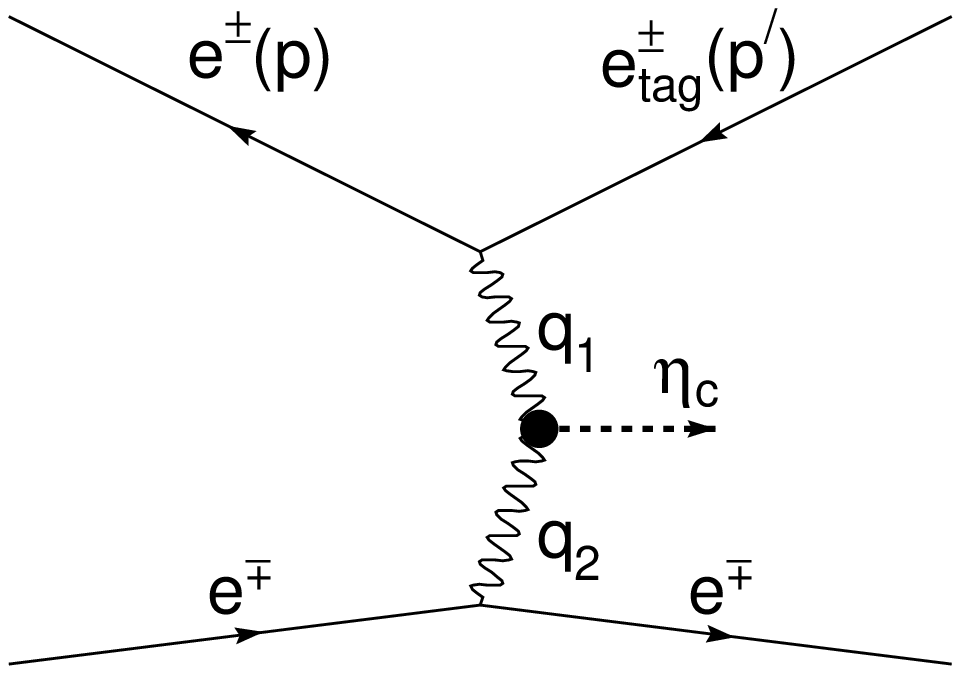}
\end{minipage}~
\begin{minipage}[c]{.5\textwidth}
\includegraphics[width=0.7\textwidth,,clip=true]{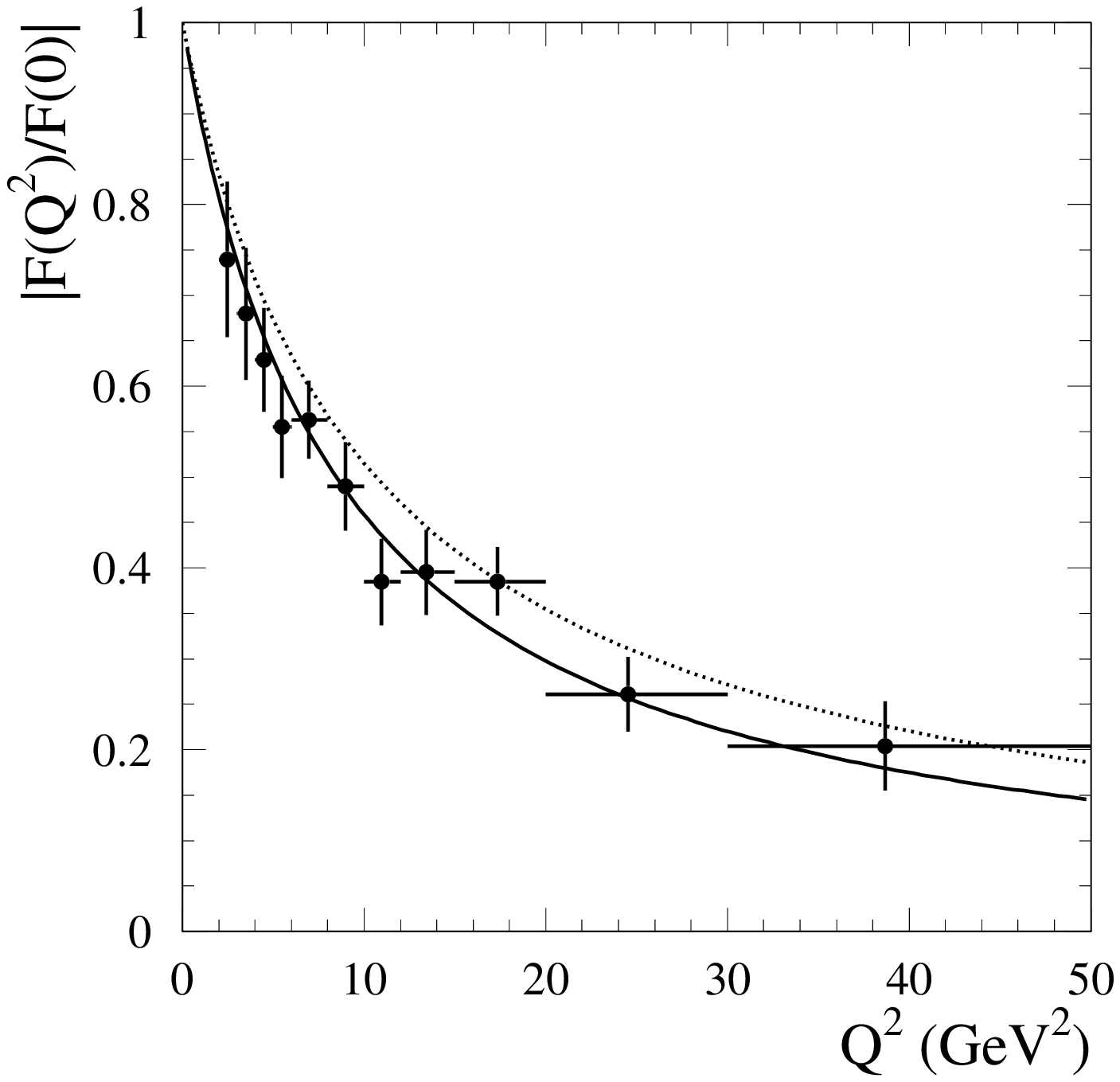}
\vspace*{-4mm}
\end{minipage}~

\hspace*{1mm} Fig.37 from \cite{Lees} (left). The process $e^+e^-\to e^+e^{-}+\eta_c\,,\,\,\eta_c\to K_S K^{\pm}\pi^{\mp},\, q_1^2=-Q^2<0,\,\, q_2^2\simeq 0$.\\
\hspace*{7mm} Fig.37 from \cite{Lees} (right). \,\, The $\gamma\gamma^{*}\to \eta_c$ transition form factor $F_{\gamma\eta_c}(Q^2)$ normalized to $F_{\gamma\eta_c}(0)$ (points with error bars). The solid curve shows the fit (6.54). The dotted curve shows the leading order pQCD prediction from \cite{FK-97}. \\

The leading twist contribution to $F_{\eta_c\gamma}(Q^2)$ looks as:
\footnote{\,
The one loop correction to (6.55) in the strict non-relativistic limit, $\phi_{\eta_c}(x,v^2\to 0)\to\delta(x-0.5)$, see Fig.4, was calculated in \cite{Sh-V-81}. The lattice calculations of $F_{\eta_c\gamma}(Q^2)$ in the range \,\, $(-4<Q^2<4)\,GeV^2$ see in \cite{latt-eta-c}.
}
\be
F_{\eta_c\gamma}(Q^2)=\frac{8 f_{\eta_c}}{9}\int_0^1 \frac{dx \,\phi_{\eta_c}(x)}
{x Q^2+(x{\ov x}+0.25)M^2_{\eta_c}}\,,\quad \int_0^1 dx\,\phi_{\eta_c}(x)=1\,,\quad 0.25\simeq M^2_c/M^2_{\eta_c}\,,
\ee
\be
\langle 0|{\ov c}(0)\gamma_{\mu}\gamma_5 c(z)|\eta_c(p)\rangle_{\rm lead. twist}=i f_{\eta_c}p_{\mu}\int_0^1 dx e^{-ix (zp)}
\phi_{\eta_c}(x),\quad \phi_{\eta_c}(x)=\phi_{\eta_c}(1-x)\,.\nn
\ee

For comparison, we substituted into (6.55) the heavy quarkonium leading twist wave function $\phi(x,v^2)$ proposed in \cite{bc-05} in connection with a theoretical calculation of the process \, "$e^+ e^- \to J/\psi+\eta_c$" measured by the Belle Collaboration \cite{Belle-0205}. It looks as (see \cite{bc-05} for more details, $c(v^2)$ is the normalization constant)
\be
\phi(x,v^2)=c(v^2)\, 6x{\ov x}\Biggl (\frac{x{\ov x}}{1-4x{\ov x}(1-v^2)}\Biggr )^{1-v^2}\,,\quad {\ov x}=(1-x)\,, \quad \int_0^1 dx\,\phi(x,v^2)=1\,.
\ee

For the ground state charmonium with $v^2\simeq 0.3$ this looks as, see Fig.4,
\be
\phi_{\eta_c}(x,v^2=0.3)=9.6\, x{\ov x}\Biggl (\frac{x{\ov x}}{1-2.8\, x{\ov x}}\Biggr )^{0.7}\,.
\ee

The results of substituting the wave function (6.57) into (6.55) are shown in Fig.38. Somewhat surprisingly, the agreement with data is good (maybe too).
\vspace*{3mm}

\begin{minipage}[c]{.5\textwidth}
\includegraphics[width=0.8\textwidth,,clip=true]{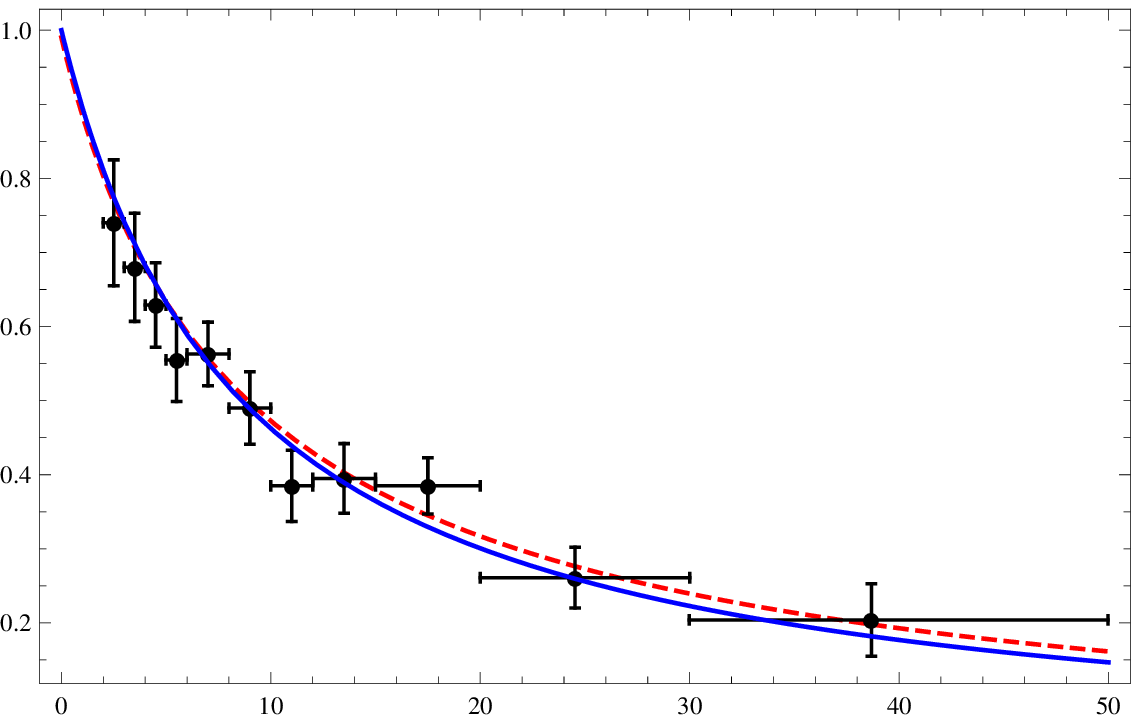}
\put (-230,35){\rotatebox{90} {$F_{\gamma\eta_c}(Q^2)/F_{\gamma\eta_c}(0)$}}\put (10,3) {$Q^2$}
\end{minipage}
\begin{minipage}[c]{.4\textwidth}
Fig.38.  Dashed line: the ratio $F_{\gamma\eta_c}(Q^2)/F_{\gamma\eta_c}(0)$ from (6.55) with the wave function (6.57). Solid line: the fit (6.54) from BaBar. The data points from BaBar \cite{Lees}.
\end{minipage}

\section{Conclusions}

\hspace*{3mm} Extensive studies of the processes $\gamma\gamma\to M_1 M_2$ in the Belle experiments
discussed in this review allowed  tests of various predictions of perturbative QCD and phenomenological models. However, from the theoretical viewpoint, in many cases it is clear that measurements at higher energy and/or higher statistics would be very helpful. The reason is that one of the main theory problems at present is lack of the real control over power corrections in QCD calculations. Clearly, at higher energies and momentum transfers the role of power corrections will be diminished.

In some respects, the processes $e^+ e^-\to e^+ e^- +\pi^o$ and $e^+ e^-\to \gamma\pi^o$ which allow to measure the $\gamma^{*}\gamma\pi^o$ transition form factor $F_{\gamma\pi}$ in the space-like and time-like regions, are the simplest ones for the theory. They are considered at present as the best way to obtain reliable information about the leading twist pion wave function (=distribution amplitude), $\phi_{\pi}(x,\mu)$, which determines the distribution over momentum fractions of two quarks in the pion. This will then allow to choose between many different models for $\phi_{\pi}(x,\mu)$ available in the literature. Therefore, badly needed is a new high-precision study of the transition form factor $F_{\gamma\pi}(Q^2)$ at, say, $Q^2>20\,GeV^2$ for which the results from BaBar \cite{BB-09,Druzh-10} and Belle \cite{Uehara-12} do not show good agreement (within large error bars).

Besides, very useful are the data in the time-like region $q^2\simeq 13.8\,GeV^2$ from CLEO \cite{CL} and at $q^2\simeq 112\,GeV^2$ from Belle \cite{Shen} for the electromagnetic form factors $e^+e^-\to\gamma^*\to K^{*}(892)^o {\bar K}^o$ and $e^+e^-\to\gamma^*\to\omega\pi^o$ which allowed to measure the $q^2$-dependence of the form factor $F_{VP}(q^2)$. As described in some details in section 6.2.5, the dependence of $F_{VP}(q^2)$ on $q^2$ is very sensitive to the endpoint behavior of the leading twist meson wave function
$\phi^{(\alpha)}(x\ll 1,\mu\simeq 1\,GeV)\sim x^{0\,\leq\alpha\,<1}$. Therefore, these data \cite{CL,Shen} allow to put strong restrictions on the behavior of $\phi^{(\alpha)}(x\ll 1,\mu\simeq 1\,GeV)$, see section 6.2.5.

At the moment there is a number of measurements of transition form factors $F_{\gamma P}(Q^2)$ for other pseudoscalars, $\eta,\, \eta^{\prime},\, \eta_c$. In the space-like region:\,\,a) at lower energies for
$F_{\gamma\eta}$ and $F_{\gamma\eta^\prime}$ \cite{CELLO,Savin,CL-98,L3-98};\,\, b) the single measurement of $F_{\gamma\eta}$ and $F_{\gamma\eta^\prime}$ by BaBar at higher energies $Q^2<40\,GeV^2$ \cite{BB-1101};\,\, c) the recent measurement of $F_{\gamma\eta_c}$ by BaBar up to $Q^2<40\,GeV^2$ \cite{Lees}. 
There are also measurements of $F_{\gamma\eta}$ and $F_{\gamma\eta^\prime}$ in the time like region by CLEO
\cite{CL-09} at $q^2=14.2\,GeV^2$ and by BaBar at $q^2=112\,GeV^2$ \cite{BB-1}, see Fig.36 and Table 5. Clearly, independent similar measurements by Belle using its larger data sample will be a very helpful complement before experiments with Belle-II start. It will be also interesting to measure such form factors for mesons with other quantum numbers, e.g. for C-even axial vectors, tensors, etc., this will give constraints on possible forms of their wave functions.

A real breakthrough can be expected from experiments with the Belle-II detector \cite{Abe-10} at SuperKEKB, the upgraded KEKB $e^+e^-$ collider, which will provide such a possibility due to an expected data sample 50 times that at Belle. However, even with the existing data samples already collected at Belle one could benefit from analyzing the $\pi^+\pi^-,\, K^+K^-$ and $p{\ov p}$ final states with full statistics. Potentially interesting is a high-statistics study of the final states like $\gamma\gamma\to\rho^o\rho^o,
\,\rho^+\rho^-$,\, $\rho^o\omega,\, \rho^o\phi,\, K^{*o}{\ov K^{*o}},\, K^{*+}K^{*-}$, for some of which, at best, only old measurements at ARGUS in the resonance region $W < 2.2\, GeV$ exist
\cite{Arg-1,Arg-2,Arg-3,Arg-4}.

From a theoretical viewpoint,  most clear and so most useful will be the measurement with a sufficient accuracy (if possible) of $F_{\gamma\pi}(q^2)$ in the time-like region at $q^2\simeq 112\,GeV^2$  by Belle-II, because the poorly controllable power corrections are clearly small at so large energies, while theoretical calculations of $F_{\gamma\eta}(q^2)$ and $F_{\gamma\eta^\prime}(q^2)$ are complicated by the quark-gluon mixing. From the theory side, it well may be that loop corrections and the RG-evolution of $F_{\gamma\pi}(Q^2)$ will be calculated completely at the NNLO in the near future.\\

On the whole, good perspectives are seen in the near future for further developments of QCD in the region of hard exclusive processes, both from the experimental and theoretical sides.

\end{document}